%% file: Lavoie_Thesis.tex
\documentclass[]{ucalgthes1}
\usepackage[letterpaper,top=1in,bottom=1in,left=1in,right=1in]{geometry}

\usepackage{graphicx}
\usepackage{amsfonts}
\usepackage{amsmath}
\usepackage{amssymb}
\usepackage{wasysym}
\usepackage{dsfont}
\usepackage{indentfirst}
\usepackage[hypcap]{subfig}
\usepackage{bm}
\usepackage[numbers,sort&compress]{natbib}
\usepackage{doi}
\usepackage{hyperref}
\usepackage[nottoc]{tocbibind}
\usepackage{dcolumn}
\usepackage{tikz}
\usetikzlibrary{shapes,arrows}

\newcolumntype{d}[1]{D{.}{.}{#1}}

\tikzset{
decision/.style = {diamond, draw, fill=none, 
  text width=4.5em, text badly centered, node distance=3cm, inner sep=0pt},
block/.style = {rectangle, draw, fill=none, 
  text width=5em, text centered, rounded corners, minimum height=2cm},
line/.style = {draw, -latex'},
cloud/.style = {draw, ellipse,fill=none, node distance=3cm,
  minimum height=1cm},
  inout/.style={draw, trapezium,trapezium left angle=70,trapezium right angle=-70, fill=none, node distance=3cm},
subroutine/.style = {draw,rectangle split, rectangle split horizontal,
  rectangle split parts=3,minimum height=1cm,
  rectangle split part fill={red!50, green!50, blue!20, yellow!50}},
connector/.style = {draw,circle,node distance=3cm,fill=yellow!20},
data/.style = {draw, trapezium,node distance=3cm,fill=none}
}

\newcommand{\mathsym}[1]{{}}
\newcommand{\unicode}[1]{{}}

\newcommand{\e}{\,\hbox{\rm e}}                                      

\renewcommand{\Re}[1]{\hbox{\rm Re}\left(#1\right)}         
\renewcommand{\Im}[1]{\hbox{\rm Im}\left(#1\right)}         

\title{Slow Light in Metamaterial Waveguides}
\author{Benjamin R. Lavoie}
\thesisyear{2013}
\thesis{thesis}    

\monthname{December}
\dept{PHYSICS AND ASTRONOMY}
\degree{DOCTOR OF PHILOSOPHY}
\begin{document}
\makethesistitle
\pagenumbering{roman}     
\setcounter{page}{1}

%
%
%
\newpage
\phantomsection
\altchapter{\bf{Abstract}}

Metamaterials, which are materials engineered to possess novel optical properties, have been increasingly studied. The ability to fabricate metamaterials has sparked an interest in determining possible applications. We investigate using a metamaterial for boundary engineering in waveguides.

A metamaterial-clad cylindrical waveguide is used to provide confinement for an optical signal, thereby increasing the local electromagnetic energy density. We show that metamaterial-clad dielectric waveguides have unique optical properties, including new modes, which we call hybrid modes. These modes have properties of both ordinary guided modes and surface plasmon-polariton modes.

We show that for certain metamaterial parameters, the surface plasmon-polariton modes of a metamaterial-clad guide have less propagation loss than those of a metal-clad waveguide with the same permittivity. This low-loss mode is exploited for all-optical control of weak fields. Embedding three-level $\Lambda$ atoms in the dielectric core of a metamaterial-clad waveguide allows the use of electromagnetically induced transparency to control an optical signal propagating through the guide. Adjusting the pump field alters the group velocity of the signal, thereby controllably delaying pulses. The signal can even be stopped with applications to optical memory. 

Using the low-loss surface mode of a metamaterial-clad guide reduces losses by 20\% over a metal cladding without sacrificing the group velocity reduction or confinement. In addition, we show that losses can be reduced by as much as 40\% with sufficient reduction of the magnetic damping constant of the metamaterial. 

As this work aims for applications, practical considerations for fabricating and testing metamaterial-clad waveguides are discussed. An overview of the benefits and drawbacks for two different dielectric core materials is given. Also, a short discussion of other modes that could be used, along with some issues that may arise with their use, is given.

\newpage
\phantomsection
\altchapter{\bf{Acknowledgements}}
This project was partially funded by the Natural Sciences and Engineering Research Council of Canada (NSERC) through the Postgraduate Scholarship program. Funding was also provided by the Informatics Circle of Research Excellence (iCORE), and its successor Alberta Innovates Technology Futures (AITF). Additional funding was provided by the Department of Physics and Astronomy at the University of Calgary. I would like to thank all of these wonderful institutions for their support.

I would like to thank my supervisor Dr. Barry C. Sanders for recommending this project to me and for his guidance throughout. It has been a pleasure to work on this project. Thank you for all of the helpful advice and many insightful conversations. I would also like to thank Dr. Patrick Leung for providing advice and lending his expertise to this project. 

Finally, I would like to thank Elena, my wife, for her continued support and encouragement, not only regarding my studies but in every aspect of our lives.

\begin{singlespace}
\newpage
\phantomsection
\tableofcontents
\pagestyle{plain}
\newpage
\phantomsection
\listoftables
\pagestyle{plain}
\newpage
\phantomsection
\listoffigures
\pagestyle{plain}
\newpage
\phantomsection
\addcontentsline{toc}{chapter}{Citations to previously published work}
\include{citations}

\pagestyle{plain}
\clearpage
\clearpage          
\end{singlespace}
\newpage
\phantomsection
\chapter*{\bf{List of Symbols, Abbreviations and Nomenclature}\hfill} \addcontentsline{toc}{chapter}{List of Symbols}
\listofsymbols
\pagestyle{plain}
\clearpage

\pagenumbering{arabic}
\include{Introduction}

\include{Background}
\include{metamaterials}

\include{LLmodesMMwgs}

\include{Slowlight}

\include{Practicalconsid}

\include{Summary}
\appendix
\include{appendix1}

\bibliographystyle{unsrtnat}
\bibliography{Lavoie-bib}

\end{document}

%% file: citations.tex
\chapter*{Citations to Previously Published Work}

Two of the chapters contained herein are largely reproduced from my work that has been published previously. To maintain continuity within this thesis, some notation has been changed from the original and some figures and equations have been moved to earlier sections. Equations that have been moved are accompanied by the text containing relevant definitions. Figures that have been moved are cited accordingly. 

Chapter~\ref{ch:mmwgs} is published as:
\begin{itemize}
\item[]Benjamin R.\ Lavoie, Patrick M.\ Leung, and Barry C.\ Sanders. ``Low-loss surface modes and lossy hybrid modes in metamaterial waveguides'' \emph{Photon.\ Nanostruct:\ Fundam.\ Appl.,\ }10:602--614, 2012. ``Copyright (2012) Elsevier Limited.''
\end{itemize}

The major changes to Chapter~\ref{ch:mmwgs} from the original publication are as follows:
\begin{itemize}\label{Ch3 citation start}

\item Eq.~(1) in original moved and modified: Now Eqs.~(\ref{generalwaveeqCh2}) and (\ref{generalwaveeqCh2-B})

\item Eq.~(2) in original moved and modified: Now Eq.~(\ref{fieldpropform})

\item Eq.~(3) in original removed

\item Eq.~(4) in original moved and modified: Now Eq.~(\ref{epsilonMM})

\item Eq.~(5) in original moved and modified: Now Eq.~(\ref{muMM})

\item Fig.~1 in original moved: Now Fig.~\ref{fig:guidediagramCh2}

\item Eq.~(7) in original moved: Now Eq.~(\ref{slabdisprelCh2})

\item Eq.~(9) in original moved: Now Eq.~(\ref{cyldisprel})

\item The third and fourth paragraphs of Sec.~\ref{sec:LLWGmodes} were added

\item Fig.~\ref{fig:TMjdef} was added

\item The fourth paragraph of Sec.~III in original was split into two paragraphs:\ Now sixth and seventh paragraphs of Sec.~\ref{sec:LLWGmodes}

\item The last two sentences of the eighth paragraph of Sec.~\ref{sec:LLWGmodes} were added

\item The third sentence of the seventh paragraph of Sec.~\ref{sec:LLWGmodes} was added

\item Fig.~\ref{fig:surfTMdef} was added

\item The last sentence of the eighth paragraph of Sec.~\ref{sec:LLWGmodes} was added

\item Fig.~\ref{fig:TMhyb} was added

\item The ninth paragraph of Sec.~\ref{sec:LLWGmodes} was added

\item The 12\textsuperscript{th} and 13\textsuperscript{th} paragraphs of Sec.~\ref{sec:LLWGmodes} were added

\item Fig.~\ref{fig:TMjcyldef} was added

\item The last two paragraphs of Sec.~\ref{sec:LLWGmodes} were added

\item Fig.~\ref{fig:TMcylsurf-hyb} was added

\item The second sentence of the 16\textsuperscript{th} paragraph of Sec.~\ref{sec:LLWGcharacter} was added

\item The fourth-last to second-last paragraphs of Sec.~\ref{sec:LLWGcharacter} were added

\item The seventh paragraph of Sec.~V in original was split into two paragraphs:\ Now seventh and eighth paragraphs of Sec:~\ref{sec:LLWGdiscussion}

\item Last five paragraphs of Sec.~\ref{sec:LLWGdiscussion} were added

\item Eq.~(\ref{eq:energyvel}) was added

\item Fig.~\ref{fig:slabve} was added

\item Fig.~\ref{fig:cylve} was added

\item Fig.~\ref{fig:slabvemetal} was added

\item Fig.~\ref{fig:cylvemetal} was added

\end{itemize}

Chapter~\ref{ch:slowlight} is published as
\begin{itemize}
\item[]Benjamin R.\ Lavoie, Patrick M.\ Leung, and Barry C.\ Sanders. ``Slow light with three-level atoms in a metamaterial waveguide'' \emph{Phys.\ Rev.\ A,} 88:023860, 2013. ``Copyright (2013) by the American Physical Society''
\end{itemize}

The major changes to Chapter~\ref{ch:slowlight} from the original are as follows:
\begin{itemize}\label{Ch4 citation start}
\item The first paragraph of Sec.~I in original was split into two paragraphs:\ Now first and second paragraphs of Sec.~\ref{ch:slowlight-introduction}

\item Fig. 2 in original moved: Now Fig.~(\ref{fig:EITsys})

\item The second paragraph of Sec.~I in original was split into three paragraphs:\ Now third, fourth and fifth paragraphs of Sec.~\ref{ch:slowlight-introduction}

\item The second last sentence of the sixth paragraph of Sec.~\ref{ch:slowlight-introduction} was added

\item The first paragraph of Sec.~II in original was split into two paragraphs:\ Now first and second paragraphs of Sec.~\ref{ch:slowlight-theory}

\item Eq.~(1) in original moved: Now Eq.~(\ref{epsilonMM})

\item Eq.~(2) in original moved: Now Eq.~(\ref{muMM})

\item Eq.~(4) in original moved: Now Eq.~(\ref{eq:besselderiv})

\item Eq.~(5) in original moved: Now Eq.~(\ref{eq:assocbessderiv})

\item Eq.~(9) in original moved and modified: Now Eq.~(\ref{EITchi})

\item Eq.~(10) in original moved: Now Eq.~(\ref{eq:ndensity})

\item Eq.~(11) in original moved: Now Eq.~(\ref{rabifreq})

\item The fifth and sixth paragraphs of Sec.~II in original modified and merged into one:\ Now fifth paragraph of Sec.~\ref{ch:slowlight-theory}

\item The seventh and eighth paragraphs of Sec.~II in original moved and modified:\ Now eighth and ninth paragraphs of Sec.~\ref{ch:slowlight-theory}

\item The third paragraph of Sec.~III of original was split into two paragraphs:\ Now third and fourth paragraphs of Sec.~\ref{ch:slowlight-results}

\item The fourth paragraph of Sec.~III of original was split into two paragraphs:\ Now fifth and sixth paragraphs of Sec.~\ref{ch:slowlight-results}

\item The second and third sentences of the fifth paragraph of Sec.\ref{ch:slowlight-results} were added

\item Fig.~\ref{fig:dopedbeta} was added

\item The last paragraph of Sec.~III in original was removed.

\item Sec.\ref{ch:slowlight-AnApprox} was added

\item First four paragraphs of Sec.~\ref{ch:slowlight-discussion} were added

\item Table~\ref{gammaeffect} was added

\item The third paragraph of Sec.~IV in original was split into two paragraphs:\ Now seventh and eighth paragraphs of Sec.~\ref{ch:slowlight-discussion}

\item Last four paragraphs of Sec.~\ref{ch:slowlight-discussion} were added

\end{itemize}

%% file: Introduction.tex
\chapter{Introduction}

\section{Waveguides}

A waveguide is a physical device used to confine and control the path taken by energy travelling as a propagating wave. For example, electromagnetic waves can be guided by metal tubes or dielectric fibres~\cite{Yeh:2008}, acoustic waves can be guided by hollow tubes such as a flute or those in a pipe organ, and even waves in water can be guided by a canal. Waveguides are used as a means of containing and directing wave energy, usually for a specific application. Electromagnetic waveguides have a variety of applications ranging from long wavelength, such as radar, to short wavelength, such as fibre optics. Herein, the term waveguide will be used exclusively to refer to a waveguide for electromagnetic waves.

\begin{figure}[t,b] 
       \centering
	\includegraphics[width=.6\textwidth]{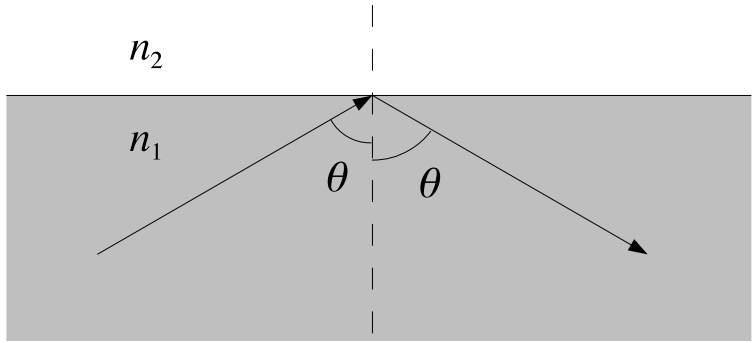}
	\caption[An illustration of total internal reflection.]{An illustration of the phenomenon of total internal reflection. For materials such that $n_{1}>n_{2}$, when the incident angle $\theta$ is above a certain threshold, called the critical angle, all of the light is reflected back from the interface. If the incident angle is less than the critical angle, a portion of the light travels across the interface and is refracted with the remaining light reflected.}
     \label{fig:totintref}   
\end{figure}

Strictly speaking, Maxwell's equations must be solved to obtain the exact solution of a waveguide, especially for those with a core radius on the order of a wavelength or less. However, a good way to illustrate how electromagnetic fields are guided is to use the principle of total internal reflection~\cite{Yeh:2008}. Total internal reflection, as the name implies, is the complete reflection of travelling electromagnetic fields off an interface between two media. Figure~\ref{fig:totintref} illustrates the phenomenon of total internal reflection.
This reflection requires two conditions to be met. 

The first is that the index of refraction of the medium in which the electromagnetic field is travelling be greater than that of the adjacent material. The second is that the angle at which the electromagnetic field is incident upon the interface must be greater than a critical value (the value depends on the refractive indices of the two materials). Otherwise a significant amount of energy will not be reflected and will continue to propagate into the second medium. An optical waveguide, then, is simply an outer material, the cladding, surrounding and inner material, the core, both having properties such that a propagating electromagnetic field in the inner material can experience total internal reflection at the interface.

It is not enough for an electromagnetic field simply to undergo total internal reflection in the core of the waveguide for it to be guided. Another condition that must be met is that the electromagnetic field reflect in such a way that the incident waves interfere constructively with the reflected waves.
When this condition is met, the electromagnetic field is able to propagate along the waveguide and is a guided mode. If the fields do not interfere constructively, however, they will undergo destructive interference upon reflection from the core-cladding interface. This will cause the electromagnetic energy to dissipate into the core and not propagate along the guide. 

Waveguides are capable of guiding electromagnetic fields in a number of different guided modes. The particular modes supported by a waveguide for an electromagnetic field of a given frequency depend on the composition and the geometry of the waveguide~\cite{Yeh:2008}. In general, waveguides with larger cores and with a larger ratio between the refractive index of the core to that of the cladding can support a greater number of modes. Different geometries and materials can affect the shape of the modes.

Electromagnetic waves propagating along a metamaterial-clad waveguide can do so in two ways. The most well known method of energy transport in a waveguide is the ordinary mode. Ordinary modes transport energy in the form of propagating electromagnetic fields that are partially confined to the core of the waveguide. These fields are confined by total internal reflection, while constructive interferences between incident and reflected wavefronts create discrete energy distributions called modes. 

The other method of energy transport in a metamaterial-clad guide is through surface waves~\cite{Novotny:1994}. Surface plasmon-polaritons are the only surface wave that will be considered in this thesis. However, it is important to distinguish surface plasmon-polaritons from other types of surface waves. 

For instance, when light undergoes total internal reflection at an interface between two dielectrics, the refracted wave vector is oriented parallel to the interface~\cite{Axelrod:1984}. This allows evanescent fields to propagate along the interface for a short distance as a surface wave. A surface wave can also be excited at interfaces between linear materials and certain nonlinear materials~\cite{Tomlinson:1980}. In this case, the self focussing effect of the nonlinear material traps the wave at the interface. From this point, the term surface mode will be used to refer exclusively to a surface plasmon-polariton mode.

\begin{figure}[t,b] 
     \centering
      \subfloat{\label{fig:SPP-schem}\includegraphics[width=0.45\textwidth]{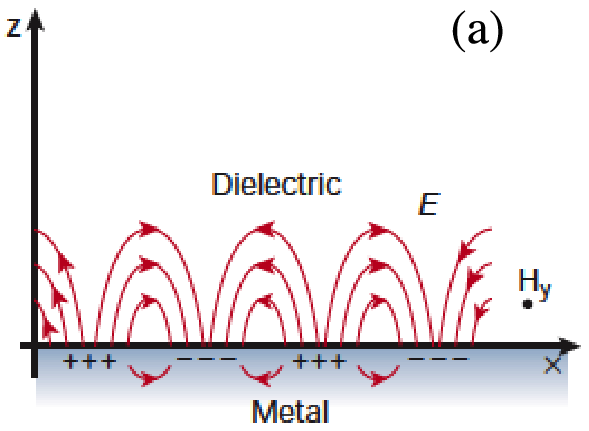}}\hspace{2cm}
      \subfloat{\label{fig:SPP-field}\includegraphics[width=0.3\textwidth]{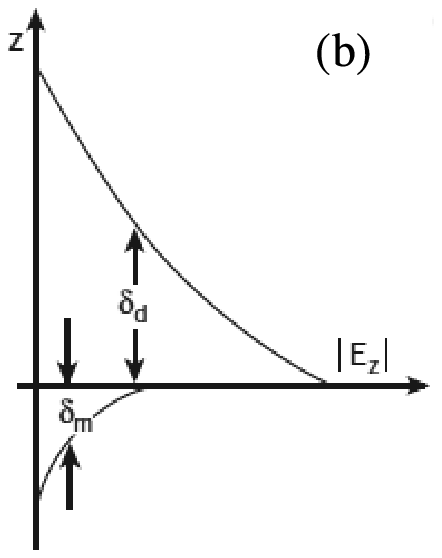}}
     \caption[Field of a surface plasmon-polariton.]{Diagrams showing a surface plasmon-polariton. The coherent oscillations of electrons near the surface \protect\subref{fig:SPP-schem} provide a medium for energy transport and give rise to evanescent fields \protect\subref{fig:SPP-field} that penetrate into the surrounding media. $\delta_{\text d}$ and $\delta_{\text m}$ show the skin depths of the fields on the dielectric and metal/metamaterial sides respectively. \textsc{Source:} \emph{Box 1 in W.\ L.\ Barnes et al.\ \emph{Nature,\ }424:824-830, 2003.}}\label{fig:SPP}
\end{figure}

Surface plasmon-polaritons are coherent oscillations of electrons coupled to electromagnetic fields at the interface between two materials. The condition that the permittivity or permeability must have opposite signs on either side of the interface must be met for these modes to exist. This condition will be explained in Sec.~\ref{subs:flat-int_theory}. The oscillating electrons provide a medium for energy transport in these modes, and the electromagnetic field propagates along the guide as evanescent tails (see Fig.~\ref{fig:SPP}). In these modes, no energy is transported by the evanescent fields.

Electromagnetic waveguides are used for a variety of applications. For example optical fibres play a vital role in the telecommunication industry. The ability of optical fibres to carry information at extremely high speeds, relative to metal cables such as coaxial cable, is attractive for data transfer and communications purposes. Optical fibres also have the advantage of supporting high bandwidth pulses, allowing a high information density per pulse. Techniques, such as multiplexing, have helped to increase the information carrying capacity even more~\cite{Brackett:1990}.

In addition to classical communication and information purposes, photons are well suited for quantum information processing, making optical networks extremely useful for quantum communication~\cite{NielsenChuang:2000}. With an optical fibre infrastructure already largely in place, it is likely that quantum communication protocols will take advantage of the suitability of photons for these types of tasks, if not entirely at least in part. For this reason, optical waveguide devices capable of handling and processing quantum signals will need to be designed and implemented.

\section{Metamaterials}

Metamaterials are composite materials that have been constructed out of two or more different materials and are designed to have properties that are not possessed by the constituent materials alone~\cite{Balmain:2005,Engheta:2006}. Consider metal alloys; by simply combining two or more metals together in solid solution the resultant mixture, the alloy, has properties that differ from the constituent metals. These properties can be controlled by adjusting the proportions of the constituents. In an analogous way electromagnetic metamaterials respond to electromagnetic fields in a manner that is not possible with just the constituent materials alone. Within this thesis the term metamaterial will be used to refer exclusively to electromagnetic metamaterials.

\begin{figure}[t,b] 
     \centering
      \subfloat{\label{fig:wire-pair-nimm}\includegraphics[width=0.45\textwidth]{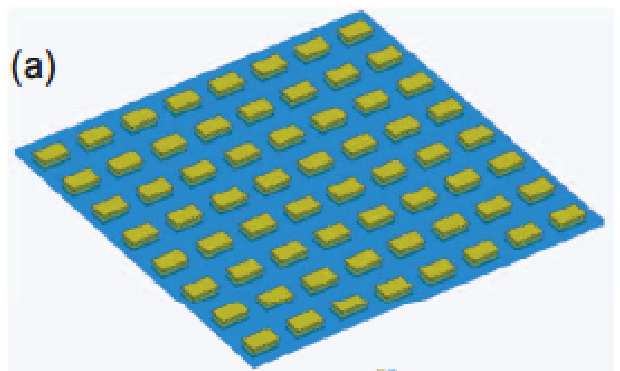}}\hfill
      \subfloat{\label{fig:fishnet-nimm}\includegraphics[width=0.45\textwidth]{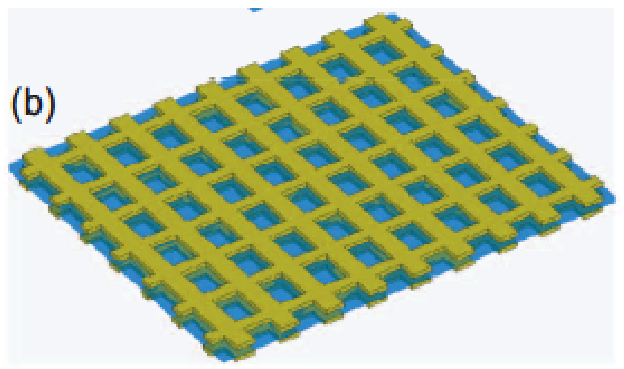}}
     \caption[Examples of metamaterial structure.]{Examples of possible sub-wavelength structures, \protect\subref{fig:wire-pair-nimm} a wire-pair construction and \protect\subref{fig:fishnet-nimm} a fishnet construction, used to construct a metamaterial. The yellow(lighter) areas indicate a metal and the blue(darker) areas indicate a dielectric. \textsc{Source:} \emph{Figure~1 in J. Zhou et al.\ \emph{Opt.\ Express, }16:11147-11152, 2008.}}\label{fig:nimmstructure}
\end{figure}

Electromagnetic metamaterials work using a different principle than metal alloys. Simply combining two materials does not elicit the response desired. An additional ingredient, specifically periodic nano-structuring at a scale smaller than the wavelength of the incident light, is required. It is the combination of different materials and nano-structuring that allows metamaterials to behave in ways not seen in naturally occurring materials. This behaviour can be tailored by the choice of constituent materials and the geometry (shape, size, spacing, etc.) of the nano-structures~\cite{Balmain:2005,Engheta:2006}. Figure~\ref{fig:nimmstructure} shows two examples of the possible sub-wavelength structures that can be used for constructing optical metamaterials.

The nano-structures in a metamaterial act to disrupt the typical flow of currents that form in the presence of a propagating electromagnetic field. In standard metals, there is nothing, aside from individual ions in the bulk, to prevent electrons from moving about. In metamaterials, the nano-structures are put in place to force the electrons to form oscillating rings of current. These oscillating currents interact with the electromagnetic field in a type of diamagnetic response, i.e.\ generating magnetic fields in response to the incident electromagnetic field~\cite{Smith:2000}. 

Some metamaterials are called negative-index metamaterials because they can have a negative refractive index for some frequencies. The implications of having a negative refractive index were first predicted in 1967 (an English translation was printed in 1968) by Victor Veselago~\cite{Veselago:1968}. He correctly predicted that simultaneously negative permittivity and permeability implies a negative refractive index and leads to effects like reversed Snell's law, negative phase velocity and reversed Doppler effect.

At the time, however, the technology was not in place to fabricate these materials and so, as they do not appear naturally, Veselago's work went largely unnoticed. That was until the early 2000s when technology finally caught up and metamaterials could be constructed~\cite{Boardman:2005,Ramakrishna:2005}. Now that it is possible to construct metamaterials, research into their potential applications has uncovered some interesting possibilities~\cite{Pendry:2000,Grbic:2004,Pendry:2006,Cai:2007,Shalaev:2008,Nielsen:2010}

Waveguides must be constructed using materials that have the requisite optical properties for the proposed application, and metamaterials are materials that display heretofore unseen optical properties. Thus it is interesting to ask: Given the unusual optical properties of metamaterials, what will be the effect of constructing a waveguide using metamaterials? Will waveguides constructed with metamaterials display new phenomena or offer any advantages over existing technologies?

These are questions that have been considered by a number of authors to varying degrees~\cite{Shadrivov:2003,Halterman:2003,Qing:2005,He:2005,D'Aguanno:2005,ZJWang:2006,Kim:2007,ZHWang:2008,He:2008,Kim:2009} but largely for lossless metamaterials. It is important to consider the effects of losses in these systems as real metamaterials exhibit losses~\cite{Zhou:2008,Xiao:2009,Tassin:2012}. Additionally, none of these previous studies has been a comprehensive characterization of waveguides constructed with lossy metamaterials. We address these questions in a more detailed study and find that there are new phenomena, as well as potential advantages, for certain applications.

\section{Motivation}\label{ch:intro-motivation}

This project is based on the foundation laid by recent work, in which surface plasmon-polaritons (SPPs) are excited along a flat metamaterial-dielectric interface~\cite{Kamli:2008}. It was previously found that a SPP excited on a flat metamaterial-dielectric interface will experience near-zero losses at and around a particular frequency, termed the abyss frequency. When three-level atoms are embedded into the dielectric near the interface, electromagnetically induced transparency (EIT) can be achieved. It was shown that this interface can be used to generate strong nonlinear interactions, via EIT, leading to large mutual phase shifts between two weak pulses~\cite{Moiseev:2010}. As the optical signals that are interacting are low intensity, or weak fields, confinement is necessary to concentrate the field energy, thereby increasing the nonlinear response. 

The increased nonlinear response has the dual effect of increasing the spectral width of the EIT transparency window, thereby allowing shorter pulses, as well as increasing the mutual phase shift experienced by two co-propagating pulses. The transverse confinement of the SPPs travelling along the flat interface was not explicitly considered in the previous work, but assumed to be provided by focussing. The possible effects of transverse confinement on the behaviour of the SPP was, therefore, not taken into account. Additionally, the function used to model the permeability is not a good representation of the response of actual metamaterials. 

The motivation behind this project has two major facets: the first is a fundamental understanding of the behaviour of metamaterials, specifically with respect to their ability to guide electromagnetic waves. Metamaterials are still in their infancy with much of their behaviour still an unknown, and the behaviour of metamaterial waveguides is interesting in its own right. Metamaterials bring new optical properties to the table, which could give rise to new waveguide properties and behaviours. 

The unusual optical properties of metamaterials lead to the second facet behind the motivation of this project, which is potential applications. Metamaterial waveguides display unusual phenomena that could potentially be useful for more than simply a fundamental understanding of metamaterials. For instance, in the telecommunications sector it is understood that the current infrastructure for delivering data to the end user, coaxial cables and twisted pair wire, will soon be unable to handle increasing demands. This is known as ``the last mile problem''~\cite{Green:2004}. One way to overcome this challenge and reliably deliver the needed data volume is with optical fibre networks. Optical fibres have a much larger capacity than either coaxial cable or twisted pair wire, while offering rapid signal transit times.

Optical networks, however, are not without drawbacks. One such drawback is the ability to effectively route optical signals. The current technology for routing optical signals is to convert them to electronic signals, then perform the required routing, then convert the electronic signals back into optical signals~\cite{Kivshar:2008,Saleh:2012}. This is a time consuming process when compared to the speed of optical networks, and will act as a bottleneck. There are other logistical advantages to a so-called ``transparent'' network as well~\cite{Saleh:2012}. 

If the switching can be done optically, it would bypass the need for signal conversion and eliminate the routing bottleneck. Metamaterial-dielectric waveguides are of interest in this respect, as they open up the possibility of strong nonlinear interactions between weak fields with low-losses. Strong nonlinear interactions create large phase shifts, which can be used to perform optical switching. In addition to switching, slow light can be used in controllable optical delay lines and optical buffers~\cite{Boyd:2006}, which would also be useful for fast optical networks.

Optical networks are also likely to play a major role in quantum computing, and are already used in existing quantum communication technology. This is due to the high data rate of optical networks, along with the compatibility of quantum communication technology and techniques with the properties of photons~\cite{NielsenChuang:2000}. The ability to do fast optical switching, along with other optical controls, within these networks will help quantum computing technology perform high speed computations unhindered by routing bottlenecks. 

Another potential application that motivated this project is deterministic quantum controlled-not gates. These gates could be realized by a device with the ability to impart a $\pi$ phase shift to a single photon. Additionally, metamaterial-clad waveguides can provide strong confinement with reduced losses for light stopped with electromagnetically induced transparency using rare-earth ions embedded in the waveguide core. The scheme proposed herein shows promise for quantum memories for weak fields with on-demand retrieval.

We propose a scheme for a metamaterial-clad waveguide device that would enable all-optical control of signals. Such a device would mitigate the bottleneck caused by conventional switching methods, thereby speeding up optical networks. This device is not intended to replace existing fibres but to replace conventional switching and routing devices already used in fibre networks.


This project has three goals. The first goal is to develop a theoretical framework to facilitate the study of metamaterial-clad waveguides. The second goal is to show, through a characterization, that metamaterial-clad waveguides display unusual and novel properties. The third goal is to show that metamaterial-dielectric waveguides have advantages over metal-clad waveguides for polaritonics, particularly for all-optical control via highly-confined slow light. Polaritonics is the design and construction of devices to transport and manipulate signals using polaritons.

To achieve these goals, previous results are extended by using a more realistic function for the metamaterial permeability, as well as explicitly considering the effects of transverse confinement. To explicitly introduce transverse confinement a cylindrical guiding structure is used in place of a flat interface, thereby providing confinement in both transverse directions. Confinement in the propagation direction is provided by the natural pulse compression that arises when slowing an optical signal.

\section{Metamaterial Waveguides: Claims}

\begin{figure}[h,t,b] 
     \centering
     \subfloat{\label{fig:EzAmpOrdinary}\includegraphics[width=0.3\textwidth]{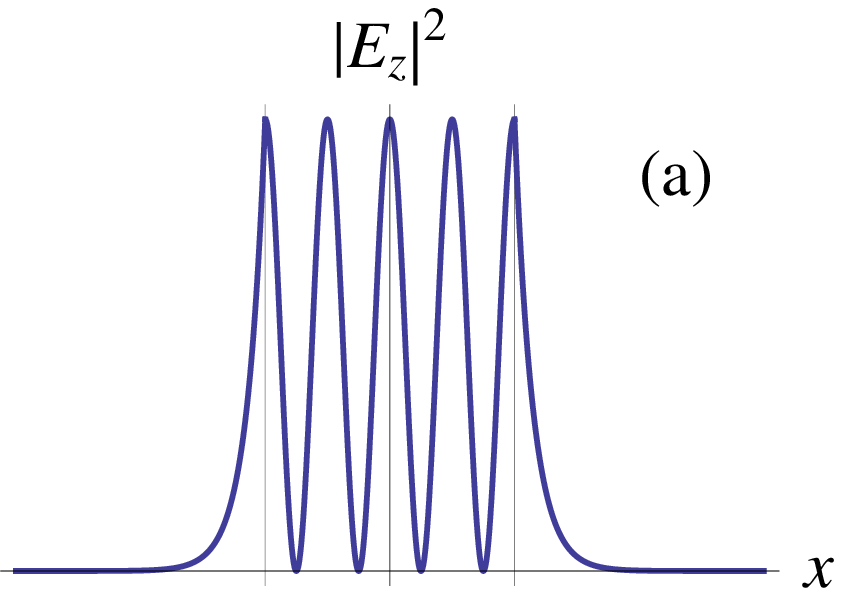}}
     \hfill
      \subfloat{\label{fig:EzAmpSurface}\includegraphics[width=0.3\textwidth]{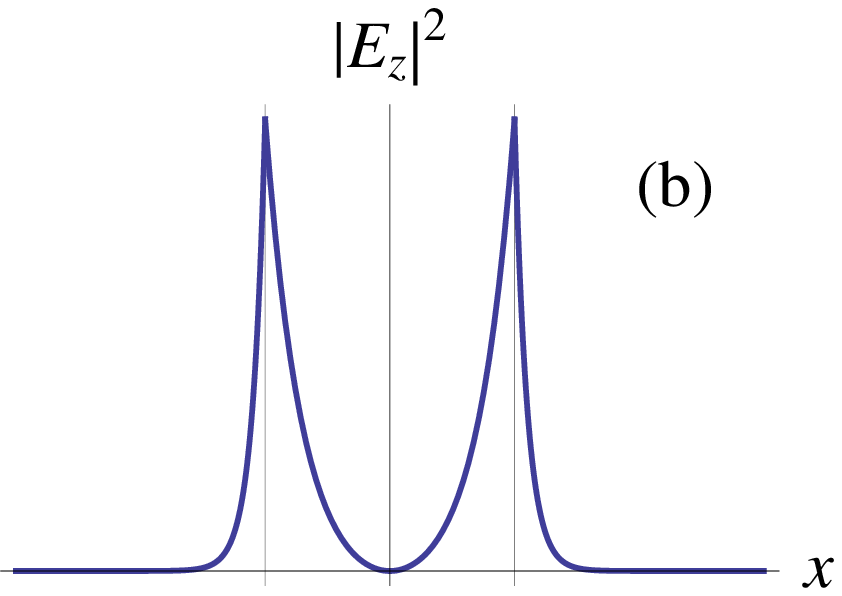}} 
         \hfill
         \subfloat{\label{fig:EzAmpHybrid}\includegraphics[width=0.3\textwidth]{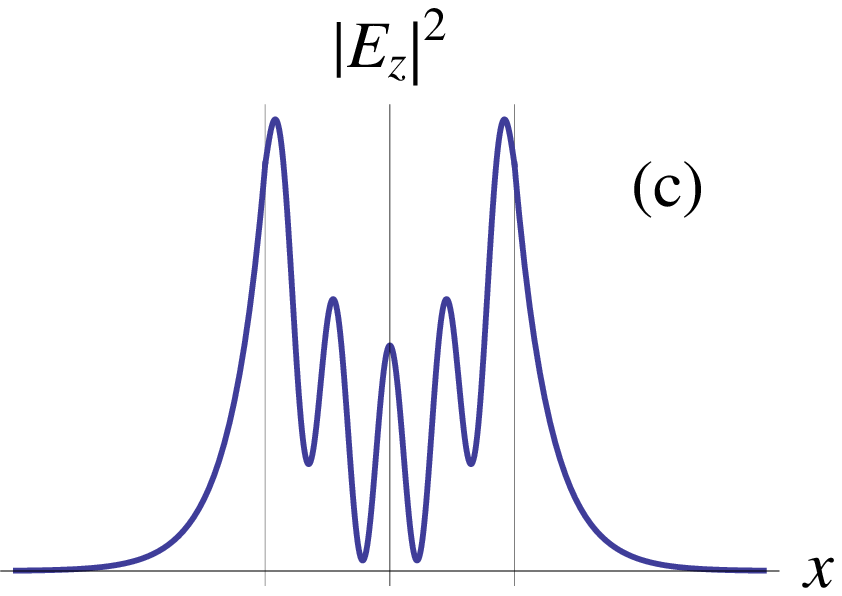}} 
         \caption[Diagram contrasting the three mode types.]{Plots of $|E_{z}|^{2}$ versus $x$ (direction perpendicular to the interface between the core and cladding) for the ordinary \protect{\subref{fig:EzAmpOrdinary}}, surface \protect{\subref{fig:EzAmpSurface}} and hybrid \protect{\subref{fig:EzAmpHybrid}} modes in a slab guide. The vertical lines on the left and right in each plot indicate the boundary between the core and cladding. The shape of the hybrid mode indicates it is a superposition of an ordinary and surface mode. A similar intensity pattern can be observed for the different mode types of a cylindrical guide.} \label{fig:EzAmp}   
\end{figure}

We have found that metamaterial-dielectric waveguides support a heretofore unobserved mode, in which both an ordinary and a surface mode coexist and travel at the same speed along the guide. These new modes, being a superposition of a plasmon-polariton and a guided mode (see Fig.~\ref{fig:EzAmp}), have a fraction of their energy carried by the electromagnetic field and some by electron oscillations at the core-cladding interface. The superposition of these two modes points to the importance of a general treatment of losses in waveguide theory. The existence of these mode types is, in some sense, surprising, as it is not obvious that an ordinary and surface mode should travel at the same speed along the guide.

As the superposition of the surface and ordinary modes behaves like a single mode, we label this a hybrid ordinary-surface mode, or simply a hybrid mode. These are not to be confused with the hybrid TE-TM modes, sometimes labelled HE and EH, supported by cylindrical guides. The HE and EH modes are simply modes having no component of the electric or magnetic field everywhere equal to zero.

As such there can be TM, TE, EH and HE hybrid ordinary-surface modes that are supported by a cylindrical metamaterial-dielectric waveguide. The language overlap is unfortunate, but the term hybrid is the only term which adequately describes the behaviour and energy distribution of these new modes. Therefore, the term hybrid mode is reserved in this text to refer exclusively to the hybrid ordinary-surface modes, unless explicitly indicated otherwise.

The underlying mechanisms of metamaterial behaviour are only partially understood. The fact that certain structures, split ring resonators for instance, can respond to electromagnetic fields in such a way so as to produce a negative permeability is well known~\cite{Balmain:2005,Penciu:2010}. However, other effects have not been studied. For instance, introducing a frequency dependent permeability, which includes an additional mechanism for energy dissipation, can reduce the attenuation in a metamaterial when compared to that in a metal. That is, for a materials with a complex permittivity, the addition of a complex permeability can reduce, rather than increase, the energy loss experienced by a propagating electromagnetic field. This seems counter intuitive, as adding a second mechanism for energy loss would suggest an increase in the loss rate.

We have shown that the reduced loss, due to the introduction of a complex function for the permeability, is also present in some of the guided modes of metamaterial-clad waveguides. When compared to a metal-clad waveguide with an identical permittivity, a metamaterial-clad waveguide outperforms in terms of the inherent loss. This is true when the real part of the metamaterial permeability is less than unity and the magnetic damping constant is less than that of the electric damping constant. Under these conditions, within a finite frequency range the attenuation of a surface mode in a metamaterial-clad waveguide is reduced by 20\% to 40\% from that in a metal-clad guide. 

We use a metal-clad waveguide for comparison because a dielectric-clad waveguide cannot support surface plasmon-polariton modes. The confinement provided by these modes is desirable for increasing the local intensity of the electromagnetic fields. As mentioned in Sec.~\ref{ch:intro-motivation}, having both strong confinement and low losses is desirable for all-optical control. This can be achieved by taking advantage of the low-loss surface plasmon-polariton mode of a metamaterial-clad waveguide. 

The effective width of a waveguide is the actual width of the core plus the skin depth (the distance that the mode penetrates into the cladding). Comparable confinement means that the effective width of the metamaterial guide is similar to that of the metal guide; the ratio of effective widths of the metamaterial guide to the metal guide considered in this thesis is 1.04 at the frequency of the low-loss surface mode.

To enable all-optical control of weak fields, three-level atoms are to be embedded in the dielectric core of the metamaterial-dielectric waveguide to facilitate the slowing of light through EIT. The three-level atoms change the susceptibility of the core, altering the dispersion and attenuation of the modes, near a resonance frequency. As one of the goals of this project is the slowing of weak fields, we assume that any nonlinear response of the metamaterial cladding is negligible. The only nonlinear response, then, is in the core and due only to the three-level atoms embedded there. We show that when light is slowed via EIT in a waveguide having a metamaterial cladding, it exhibits reduced propagation loss compared to a metal cladding. Additionally, the reduced loss does not come at the expense of either the confinement or group velocity reduction.

\section{Outcomes and Impact}

The first outcome of this work is the establishment of an analytical extension to waveguide theory. This extension provides a basis on which attenuation in waveguides may be treated in a general way. Partial analytical extensions, in the sense that only the permittivity is complex whereas the permeability remains real, have been done for both the slab~\cite{Kaminow:1974} and cylindrical~\cite{Novotny:1994} geometries, as well as methods for approximating the modes ~\cite{Reisinger:1973}.

The analytical extension provided herein is the first comprehensive consideration of waveguide theory allowing both the permittivity and permeability to take complex values. This is important because it highlights the effects of energy dissipation on the dispersion and attenuation of modes in a waveguide. Allowing the permeability to be complex, as well as the permittivity, has a drastic effect on the modes at certain frequencies. The most surprising is the coupling of surface plasmon-polariton modes with ordinary modes to form a new hybrid mode.

The second outcome of this work is the creation of a theoretical framework that can be used to design metamaterial waveguides and exploit the unusual properties of their guided modes. The analytical extension of waveguide theory leads directly to this, as metamaterials are characterized by their permittivity and permeability both having a frequency dependence and taking complex values. The framework for studying metamaterial waveguides is retrieved by choosing functions that reproduce the response of the metamaterial in question. 

Distinguished within the framework are three separate frequency regions in which the refractive index of a metamaterial displays distinct properties. We identify three frequency regions in the refractive index, each with different optical properties. The frequencies correspond to: A negative-index region, in which the real part of the refractive index is negative. A metal-like region, in which the real part of the refractive index is near zero and the imaginary part is large in comparison, and a dielectric-like region, in which the real part of the refractive index is large and the imaginary part can be either small or large. 

The impacts of this research are not immediate, but are to follow upon the construction of metamaterials with appropriate optical responses. Our work provides a platform on which future studies of metamaterial-clad waveguides can be based. These studies will expand the theoretical framework we have built, and determine how metamaterial-clad waveguides can be used in future technologies. For example, if an optical network is fitted with all-optical routing devices, rather than the current technology, the result would be higher throughput of data. The effect this would have on users of this network is faster data transfer speeds. Additionally, the existence of hybrid ordinary-surface modes in metamaterial-clad waveguides provides a novel tool, which could be useful for certain applications such as filtering.



\section{Methods}

We extensively characterize metamaterial-cladded waveguides in order gain a better understanding of their behaviour, as this has not been done before. Our characterization involves determining the dispersion and attenuation, as functions of frequency, for various modes of the waveguides. These quantities are then used to compute various other quantities, such as energy velocity and effective guide width, for each mode. To determine the modes of the undoped metamaterial-clad waveguides, I employ standard waveguide optics techniques. 



Metamaterials inherently dissipate energy, so any investigation of the properties of metamaterial-dielectric waveguides must incorporate this loss. Normally, a waveguide with large losses is not considered to be useful, and losses are assumed to be small if they are considered at all. When losses are considered they are typically included simply as an attenuation term to simulate the dissipation of energy. As such, these small losses are assumed not to affect the shape or behaviour of the modes. This is a good assumption if losses are small, but it is not valid if the imaginary part of the propagation constant is of similar magnitude to the real part.

The attenuation of fields propagating in metamaterials can be significant. Thus, in this work we do not assume that the attenuation of a metamaterial waveguide is small. For this reason various quantities, such as the propagation constant, that would be real under the assumption of small losses become complex.

I solve Maxwell's equations in their source free form within each region of the waveguide. I then match the solutions at the interfaces between regions by ensuring that they satisfy a set of boundary conditions. As the real parts of the permittivity and permeability can take negative values for metamaterials, it is not immediately clear how this might affect the boundary conditions. For instance, the wave vector of a beam reverses direction upon entering a metamaterial with a negative index of refraction. Due to the fact that the boundary conditions are derived with no requirements on the sign of either the permittivity or permeability, no modifications are made for the case of a negative-index metamaterial. Thus, I assume the boundary conditions are independent of material properties.

Ensuring the fields satisfy the boundary conditions yields an equation, called the dispersion relation, that can be solved to obtain the complex propagation constant. For the slab and cylindrical geometries, the dispersion relation is transcendental and must be solved numerically for a complex propagation constant. I use Mathematica\textsuperscript{\textregistered} to numerically solve the dispersion relation, as it has the built-in root-finding function {\tt FindRoot[]}. The {\tt FindRoot[]} function uses the Newton method to numerically search for the real or complex root of a function, given an initial guess. As the waveguide is capable of supporting multiple modes at some frequencies, the propagation constant for different modes can be obtained by an appropriate choice of the initial guess.

To generate data for plots of the dispersion and attenuation as functions of frequency, I solve the dispersion relation for a number of different frequencies. I wrote a routine in Mathematica\textsuperscript{\textregistered} to increment the frequency and employ {\tt FindRoot[]} at each frequency. I used Mathematica\textsuperscript{\textregistered} to plot the data obtained from this routine, as well as other quantities that are calculated from this data.

 
To slow light in a waveguide, three-level $\Lambda$ atoms are embedded in the dielectric core. To simulate this, we modified the expression for the electric susceptibility of the waveguide core by including an appropriate term for the EIT interaction~\cite{Lambropoulos:2007}. Two parameters are needed for this expression, the decay rates from an excited state to the ground state and from a hyperfine state to the ground state of the three-level atoms. These two parameters are taken from the literature, and they are consistent with those observed for Pr$^{3+}$.

The signal field interacts nonlinearly with the atoms when the pump field is on, resulting in EIT. We assume that the intensities of the fields are weak enough that any nonlinear interaction with the metamaterial cladding or the dielectric in the core can be neglected. Thus, the only source of nonlinearity is the three-level atoms, which are confined to the core. 

Though the pump field is assumed to be weak enough not to elicit a nonlinear response from the metamaterial and dielectric, we assume it is strong enough that it will not be significantly attenuated within the region of interest. This is to ensure that the pump field does not change appreciably in the propagation direction and can be treated as a parameter. Assuming a non-depleting pump field means the system behaves as a linear system with a spatially dependent refractive index, and is called a pseudo-linear approximation. The system is linear in the sense that the pump field is independent of the signal field and, hence, simply modifies the refractive index, which allows the use of linear equations.

We restrict our analysis to TM modes for both the signal and pump fields, as it simplifies the calculations due to the azimuthal invariance of these modes. A similar analysis can be performed using any choice of modes. However, using modes with a dependence on the azimuthal angle would require solving two coupled differential equations. 

We use Maxwell's equations to determine the modes of the doped waveguide, but unlike the undoped guide, an exact analytical solution cannot be found for the signal field. This is due to the fact that the pump field introduces a spatial dependence in the refractive index of the core for the signal field. As a result, we employ a numerical solution that is based on the shooting method for boundary-value problems. The details of the numerical solution are given in Sec.~\ref{ch:slowlight-AnApprox}, and the code is given in Appx.~\ref{ap:1}.



The wave equation, which is obtained from Maxwell's equations, is a partial differential equation that can be solved for the electric or magnetic field. When the material that electromagnetic fields are propagating within is invariant in the propagation direction, the wave equation can be simplified. This simplified form depends on the transverse directions and the propagation constant of the electromagnetic field. However, the value of the propagation constants must be solved for as well. Thus, for an electromagnetic field propagating along a waveguide, the wave equation must be solved for two unknowns, the electric field and the propagation constant.

To deal with the fact that the wave equation has two unknowns, I modify the shooting method to treat the propagation constant as the value to be guessed. This is possible due to the fact that the boundary conditions are known from the geometry and the electric and magnetic fields in the cladding region. By guessing a value for the propagation constant, the wave equation can be solved and the boundary conditions compared against the known values.

We verify the numerical solutions by comparing them with solutions that we obtain using an analytical approximation. We choose a control field frequency such that the presence of the pump field imposes a minimal spatial dependence on the refractive index in the core. Then, we make an analytical approximation to the wave equation at this frequency so that the solutions can be compared to those of the numerical method with the same pump field. The numerical solutions are found to match well with the analytical approximation, lending confidence to the numerical method.

\section{Outline of Thesis}

The electromagnetic theory that underpins the theoretical framework needed to investigate slow light in metamaterial waveguides is introduced and briefly discussed in Ch.~\ref{ch:background}. Maxwell's equations, which form the basis for classical electromagnetism and describe electromagnetic fields, are introduced both with and without sources. Electromagnetic fields in linear materials are discussed, along with the boundary conditions for fields at an interface between two different materials. The theory of waveguides is discussed, and the relevant equations are given for the flat interface, the slab, and cylindrical geometries. The Kramers-Kronig relations and their implications are introduced here, as well.

The nonlinear interaction of light with matter is also discussed, along with how these nonlinear interactions give rise to effects that cannot be explained with a linear theory. To this end, the polarization of the material is expanded into increasing powers of the electric field~\cite{Armstrong:1962}. The phenomenon of electromagnetically induced transparency and how it is a nonlinear process is covered, along with methods for calculating the electric susceptibility of a material. Plasmas and their connection to metals are also briefly discussed. Lastly, an overview of optical quantum memories, specifically focussing on EIT, is given.

Electromagnetic metamaterials are discussed in Ch.~\ref{ch:MM}. An overview of metamaterial classification is given and current fabrication techniques are discussed. Some considerations for calculating the stored energy in a metamaterial are covered, and the electromagnetic response of metamaterials is discussed.

The fourth and fifth chapters contain the details of the original work done for this thesis. A characterization of waveguides with a dielectric core and metamaterial cladding (metamaterial-dielectric waveguides) is given in Ch.~\ref{ch:mmwgs}. This chapter contains plots of dispersion, attenuation and effective guide width for selected modes of slab and cylindrical metamaterial-dielectric waveguides. The purpose of the characterization is to understand the behaviour of metamaterial-dielectric waveguides and determine if they display any novel phenomena. Chapter~\ref{ch:mmwgs} is reproduced from previously published work~\cite{Lavoie:2012}. Some changes to the notation are made for consistency, and a list of major changes is given on p.~\pageref{Ch3 citation start}.

A scheme for producing highly-confined slow light with reduced losses for weak fields is outlined in Ch.~\ref{ch:slowlight}. In this chapter the cylindrical metamaterial-dielectric guide is investigated as a platform for controllable, highly-confined, slow light with reduced losses. The feasibility of fabricating such a device is discussed along with insights gained into the behaviour of metamaterial-dielectric waveguides. Additionally, an outline of the numerical routine used to solve the wave equation is given, along with details of the analytical approximation used to verify the results of the numerical program. Chapter~\ref{ch:slowlight} is reproduced from published work~\cite{Lavoie:2013}. Some changes to the notation are made for consistency, and a list of major changes is given on p.~\pageref{Ch4 citation start}.

Practical considerations for constructing and testing a working device are discussed in Ch.~\ref{ch:pract-consid}. A discussion of other possible choices for the signal and pump modes is also given here. A summary of the thesis and concluding remarks are contained in Ch.~\ref{ch:summary}. Appendix~\ref{ap:1} contains the code used to numerically solve the wave equation for the metamaterial-clad guide with three-level atoms doped in the core.

%% file: Background.tex
\chapter{Background}\label{ch:background}

\section{Introduction}

This project focuses on the characterization of an optical waveguide that consists of a metamaterial cladding and a dielectric core. The waveguide is first characterized for a dielectric core having a constant, real permittivity and permeability. We characterize the waveguide over a range of frequencies by calculating the propagation and attenuation constants. We then compare the results to those we calculate for a similar guide but having a metal, rather than a metamaterial, cladding.

The core is then assumed to be doped with three-level $\Lambda$ atoms in order to enable optical control of signals via electromagnetically induced transparency. We again characterize the waveguide by calculating the propagation and attenuation constants. In this case, however, the permittivity of the core is no longer constant but both frequency and spatially dependent. 

To perform the required calculations, we have drawn on previous research and innovations. This chapter outlines the background material related to the fields of electromagnetism, particularly linear and nonlinear optics. Specifically, Maxwell's equations are introduced in Sec.~\ref{sec:Maxwell}, the relevant aspects of waveguide theory are covered in Sec.~\ref{sec:waveguides}, the Kramers-Kronig relations are discussed in Sec.~\ref{sec:KKrels}, a discussion of nonlinear optics is given in Sec.~\ref{sec:NLoptics}, two methods for calculating the electric susceptibility are given in Sec.~\ref{chiecalc}, electromagnetically induced transparency is discussed in Sec.~\ref{sec:EIT}, and an overview of optical quantum memories is given in Sec.~\ref{sec:QuantMem}. As the topic of metamaterials is quite large, it is covered separately in Ch.~\ref{ch:MM}.

\section{Maxwell's Equations}\label{sec:Maxwell}

Classical electromagnetic theory is rooted in Maxwell's equations, which describe the relation between electric and magnetic fields and how they arise due to charges and currents. As will be discussed in Sec.~\ref{sec:MMs}, the language of magnetic monopoles is useful when discussing the behaviour of metamaterials. Maxwell's equations can be written to accommodate sources of ``magnetic charge'', and in this form the four Maxwell's equations are~\cite{Jackson:1999}
\begin{subequations}\label{Maxwelleqns}
\begin{align}
\nabla\cdot\bm D &=\rho_{\text{e}}\\
\nabla\cdot\bm B &=\rho_{\text m},\\
-\nabla\times{\bm E} &= \frac{\partial{\bm B}}{\partial t}+\bm J_{\text m},\\
\nabla\times{\bm H} &=\frac{\partial{\bm D}}{\partial t}+{\bm J}_{\text e},
\end{align}
\end{subequations}
with $\bm E$ the electric field, $\bm D$ the electric displacement field, $\bm B$ the magnetic induction field, $\bm H$ the auxiliary magnetic field, $\rho_{\text e}$ and $\rho_{\text m}$ the free electric and magnetic charge densities, respectively, and ${\bm J}_{\text e}$ and ${\bm J}_{\text m}$ the free electric and magnetic current densities, respectively.

The electric displacement field $\bm D$ and the auxiliary magnetic field $\bm H$ result from a material response to the electric and magnetic fields, respectively. The two constitutive relations are
\begin{equation}\label{eq:genconstitutive}
{\bm D}=\epsilon_{0}{\bm E}+\bm P,\quad{\bm B}=\mu_{0}\left({\bm H}+\bm M\right),
\end{equation}
with $\epsilon_{0}$ and $\mu_{0}$ the permittivity and permeability of free space, respectively, and $\bm P$ and $\bm M$ the polarization and magnetization of the host material, respectively~\cite{Jackson:1999}. The expressions of Eq.~\ref{eq:genconstitutive} are general relations for electric and magnetic fields in a material. For a linear material response, the polarization becomes $\bm P=\epsilon_{0}\chi^{(1)}_{\text e}\bm E$ and the magnetization becomes $\bm M=\chi^{(1)}_{\text m}\bm B$, with $\chi^{(1)}_{\text e}$ and $\chi^{(1)}_{\text m}$ the linear electric and magnetic susceptibilities of the material, respectively. Thus, the constitutive relations for a linear material response are
\begin{equation}\label{eq:linconstitutive}
{\bm D}=\epsilon{\bm E},\quad{\bm B}=\mu{\bm H},
\end{equation}
with $\epsilon=\epsilon_{0}(1+\chi^{(1)}_{\text e})$ and $\mu=\mu_{0}(1+\chi^{(1)}_{\text m})$.

The magnetic charge density $\rho_{\text m}$ and the magnetic current density ${\bm J}_{\text m}$ are hypothetical magnetic analogues of their electric counterparts, and are predicated on the existence of magnetic monopoles. The existence of both the magnetic charge and current densities is a natural consequence of the existence of magnetic monopoles. These two quantities are included here for the sake of completeness, as magnetic monopoles are discussed later.

Some of the consequences of magnetic monopoles can be seen by examining Eqs.~(\ref{Maxwelleqns}). Excess magnetic charge, represented by the quantity $\rho_{\text m}$, would provide a single poled source of magnetic fields leading to a locally diverging magnetic field. This is in contrast to the case of no magnetic monopoles, where magnetic fields do not diverge but form closed loops. Also, a changing magnetic current, which is a flow of magnetic monopoles, could be used as a generator of electric fields. This is analogous to the generation of magnetic fields via a changing electric current.

The source-free form of Maxwell's equations are the resulting equations for the case $\rho_{\text{e}}=\rho_{\text m}=0$ and ${\bm J}_{\text{e}}=\bm J_{\text m}=0$. These equations are valid when there are no free charges, and lead to the electromagnetic wave equations, which have the form~\cite{Jackson:1999}
\begin{subequations}
\begin{align}
\nabla^2 {\bm E}(\bm r,\omega)&=-\omega^{2}\epsilon(\omega)\mu(\omega){\bm E}(\bm r,\omega),\label{generalwaveeqCh2}\\
\nabla^2 {\bm B}(\bm r,\omega)&=-\omega^{2}\epsilon(\omega)\mu(\omega){\bm B}(\bm r,\omega), \label{generalwaveeqCh2-B}
\end{align}
\end{subequations}
for spatially-invariant permittivity and permeability. When the permittivity or permeability is not spatially independent, the form of the wave equation changes slightly. For instance, in a material that has a spatially-dependent permittivity the wave equation for the $\bm E$ field has the form
\begin{equation}
{\nabla}^{2}{\bm E}+{\bm \nabla}\left(\frac{1}{\epsilon(\bm r,\omega)}{\bm E}\cdot{\bm \nabla}\epsilon(\bm r,\omega)\right)=\frac{\omega^{2}}{c^{2}}\mu(\omega)\epsilon(\bm r,\omega){\bm E(\bm r,\omega)}.\label{eq:genvectorwave}
\end{equation}
The wave equations are second-order partial differential equations that can be solved for the electric and magnetic fields. These equations are used to find the forms of propagating fields.

\section{Waveguides}\label{sec:waveguides}
\subsection{Overview}

A waveguide is any structure that can partially confine and direct the propagation of energy travelling in the form of waves. Within the context of this work, a waveguide will refer to any structure that can guide electromagnetic radiation. Such a structure consists of an inner core, which is transparent to electromagnetic fields, and an outer cladding layer, or layers, with a refractive index that is less than that of the core. The cladding material can be transparent to the electromagnetic fields, or it can be a reflective material, such as a metal.

The waveguide of interest in this thesis is a metamaterial-clad waveguide, and is composed of a dielectric core surrounded by a metamaterial cladding. We contrast the properties of the metamaterial-clad waveguide with a metal-clad waveguide that is composed of a dielectric core surrounded by a metal cladding. Both of these structures are capable of guiding electromagnetic waves at optical frequencies.

When electromagnetic waves travel along a waveguide, they do so by reflecting off the interface created by the boundary between the core and cladding regions. When the wavefronts reflect, they interact with wavefronts that are still travelling toward the boundary. If the wavefronts interfere constructively, the waves can continue to propagate along the guide. The resulting propagating field is called a mode.  If the wavefronts interfere destructively then the waves are not guided and the energy simply dissipates.

A mode of a waveguide refers to a particular cross-sectional energy distribution within the core of the waveguide. Constructive interference between the incident and reflected fields at the core-cladding boundary forms the shape of the mode. The number of modes that can propagate along a waveguide depends on a number of factors, including the frequency of the electromagnetic waves, and the waveguide geometry and material~\cite{Yeh:2008}. 

This is analogous to a vibrating string or membrane. In these two systems a superposition occurs for specific frequencies of vibration, sometimes called a standing wave pattern. A mode in a waveguide can be thought of as a standing wave pattern for the electromagnetic field.

Waveguides also support modes, called leaky modes, that are not strictly guided modes. These modes are radiation modes but, as they radiate energy very slowly, can be thought of as almost guided. Consider a ray-optics picture of a waveguide. If the propagating fields are incident on the core-cladding interface at an angle greater than the critical angle, they undergo total internal reflection. The electromagnetic energy is confined to the core as it travels along the waveguide. This is called a guided mode. 

For an electromagnetic field that is incident on the core-cladding boundary at an angle smaller than the critical angle, there will be a refracted field. This refracted field is energy that has escaped the waveguide and radiates away, hence the term radiation modes. If the angle of incidence is close to the critical angle, the amount of energy that is radiated away from the guide will be small. This is called a leaky mode, as the energy slowly leaks out of the guide, but the mode can travel a significant distance along the guide before it is undetectable~\cite{Snyder:1974}.

Mathematically, the distinction between guided and leaky modes in a waveguide with material losses is the sign of the real part of the transverse component of the wave vector. For guided modes, the real part of the transverse wave vector is positive, ensuring that the fields decay to zero far from the core-cladding interface. For leaky modes, the real part of the transverse wave vector is negative~\cite{Burke:1970}. 

The analysis performed in this thesis restricts the real part of the transverse wave vector to positive values. This ensures that the solutions found correspond to guided modes. We chose to restrict our analysis to guided modes, as leaky modes would provide another avenue for energy loss, which we are attempting to minimize. The term mode will be used in this thesis to refer to guided modes.

The modes of a waveguide can be one of a number of different types.  Transverse electric (TE) modes are characterized by the electric field only having components in the transverse (perpendicular to propagation) direction. Transverse magnetic (TM) modes have a magnetic field that only has transverse components. For transverse electromagnetic (TEM) modes, both electric and magnetic fields have only transverse components. Additionally, cylindrical guides support modes called HE or EH, which have components of both the electric and magnetic fields in the propagation direction as well as in the transverse direction.

Three different waveguide geometries, flat interface, parallel plate (or slab), and cylindrical, are discussed throughout this thesis. The flat interface geometry consists of two different materials joined together to form a single planar interface. This is the geometry used in the foundational work~\cite{Kamli:2008,Moiseev:2010}, and it is the simplest waveguide structure. The flat interface only supports surface plasmon-polariton modes. 

We also consider the slab geometry as it has features in common with both a flat interface (no transverse confinement) and a cylindrical waveguide (supports multiple types of modes). Additionally, the slab waveguide has been studied for various materials~\cite{Yeh:2008,Kaminow:1974}. This allows us to check our methods against accepted results. 

\begin{figure}[t,b,p] 
      \centering
      	\includegraphics[width=0.5\textwidth]{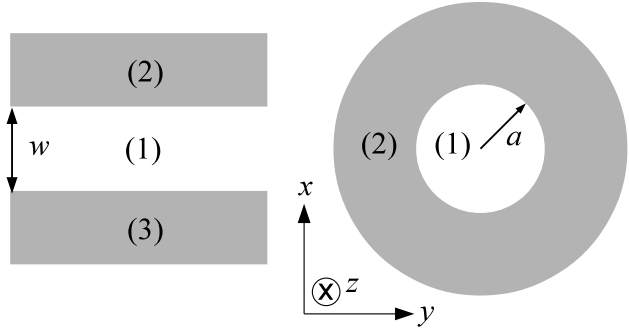}
	\caption[Diagram of the slab and cylindrical waveguides.]{A diagram of the structure of the slab and cylindrical waveguides. Region (1) in both guides is the core and is a lossless dielectric, whereas regions (2), and region (3) of the slab guide, are the cladding regions and are metamaterial. The width of the slab guide is $w$ and the radius of the cylindrical guide is $a$. Fields propagate in the $z$ direction (into the page), whereas the $x$ and $y$ directions are the transverse directions. \textsc{Source:} \emph{Figure~1 in B. R. Lavoie et al.\ \emph{Photon.\ Nanostruct:\ Fundam.\ Appl.,\ }10:602-614, 2012.}}
     \label{fig:guidediagramCh2}   
\end{figure}

As is seen in Fig.~\ref{fig:guidediagramCh2}, the slab waveguide consists of three distinct sections.
The centre section is the waveguide core, with $w$ the thickness of the core, and is surrounded on both sides by cladding. The cladding regions are assumed to be sufficiently thick so as to substantially contain the evanescent fields. This assumption is valid, as the effective width of the guide (the width of the core plus the skin depth at the interface) is typically only slightly larger than $w$. 

The cylindrical geometry is the focus of this thesis, however, as it provides confinement in both transverse directions, which is necessary for achieving highly confined fields. The geometry of the cylindrical guide (see Fig.~\ref{fig:guidediagramCh2}) is that of a central cylinder surrounded by a cladding layer. As with the slab guide, the cladding is assumed to be sufficiently thick so as to substantially contain the evanescent fields therein.

To determine the field configurations in a waveguide, Maxwell's equations must be solved in each region for both the electric and magnetic fields. The wave equation, Eq.~(\ref{generalwaveeqCh2}) or Eq.~(\ref{generalwaveeqCh2-B}), is a convenient form for this. As the fields will be travelling along the guide, which we can assume is oriented along the $z$ direction without loss of generality, they have the form
\begin{equation}
\bm E(x,y,z;\omega)=\bm E(x,y)\e^{i\left(\tilde{\beta}z-\omega t\right)}+ \text{c.c.},\label{fieldpropform}
\end{equation}
where $\tilde{\beta}=\beta+i\alpha$ is the complex propagation constant, with $\beta$ the dispersion and $\alpha$ the attenuation, both real, and c.c.\ indicates the complex conjugate.

The speed of light in materials depends on the refractive index of the material. In materials that have a frequency-dependent refractive index, light of different frequencies travels at different speeds. This is called chromatic dispersion. There are two sources of dispersion in waveguides~\cite{Yeh:2008}. The first is chromatic dispersion, and the second is waveguide dispersion, which is due to the interaction of the light with the waveguide boundaries. 

In this thesis, the term dispersion refers to the combined contributions from both chromatic and waveguide dispersion. For a continuous monochromatic field, the dispersion dictates the phase velocity of the fields. Pulses propagating in a waveguide, however, can be deformed by dispersion. 

A pulse travelling along a waveguide contains a range of frequencies, with short pulses containing more frequency components than long pulses. Dispersion causes these different frequency components to travel at different speeds along the guide. As a result, a pulse will spread out, or broaden, as it travels through the waveguide. 

Waveguide dispersion in multimode guides causes different modes to travel at different speeds along the waveguide. A pulse launched at one end of a waveguide will generally be coupled into multiple modes~\cite{Yeh:2008}. These different modes will travel along the waveguide at different speeds causing the pulse to be broadened.

To determine the dispersion of a waveguide, we ensure that the solutions to Eq.~(\ref{generalwaveeqCh2}) satisfy the electromagnetic boundary conditions at the interface between two materials, which we will label 1 and 2~\cite{Yeh:2008}:
\begin{subequations}\label{boundaryconditions}
\begin{align}
\bm n\times\left(\bm E_{1}-\bm E_{2}\right)&=0,\\
\bm n\times\left(\bm H_{1}-\bm H_{2}\right)&=\bm J_{\text s},\\
\left(\bm B_{1}-\bm B_{2}\right)\cdot\bm n&=0,\\
\left(\bm D_{1}-\bm D_{2}\right)\cdot\bm n&=\rho_{\text s},
\end{align}
\end{subequations}
with the subscripts 1 and 2 indicating in which material the field is located, $\bm n$ the direction normal to the interface, $\bm J_{\text s}$ the surface current density and $\rho_{\text s}$ the surface charge density. As we are assuming that there is zero net charge at the interface, we take $\rho_{\text s}=0$ and $\bm j_{\text s}=0$. Applying the boundary conditions results in an expression for the propagation constant $\tilde{\beta}$. The propagation constant of a waveguide can be thought of as an effective refractive index of the system, which depends on the core and cladding materials, and the configuration of the electromagnetic field (the mode). As the propagation constant depends on the propagating mode, there can be more than one valid value of $\tilde\beta$ at a given field frequency.
 
The dispersion relation of a waveguide, except for the flat interface geometry, is transcendental and must be solved numerically. Though the flat interface is not employed directly in this work, it is germane to this thesis as it is the structure on which the foundational work is done~\cite{Kamli:2008, Moiseev:2010}. For this reason an overview of the flat interface as a waveguide structure is given.

\subsection{Flat Interface}\label{subs:flat-int_theory}

The flat interface guide is different from other waveguides in the sense that the electromagnetic fields are not confined within a central core but are only confined to the interface due to evanescent decay of the fields. Due to this guiding mechanism, the flat interface is restricted in the type of modes that it can support. The flat interface can only support so called surface plasmon-polariton modes, which are a coupling between collectively oscillating charge carriers at the material interface and the electromagnetic fields~\cite{Nkoma:1974}. The oscillating charge carriers provide a medium for energy transport that allows the surface plasmon-polaritons to propagate along the interface.  

To determine the allowed modes of a waveguide in cartesian coordinates, Eq.~(\ref{generalwaveeqCh2}) or Eq.~(\ref{generalwaveeqCh2-B}) must be solved in each region of the guide. It is not necessary, however, to solve for each component of both the electric and magnetic fields. It is sufficient to solve for only the $\hat{\bm z}$ components, as these can be related, through Maxwell's equations, to the other four components. The relations have the form~\cite{Yeh:2008}
\begin{subequations}\label{slabfieldrels}
\begin{align}
\displaystyle E_{x}&=-\frac{i}{\tilde{\beta}^{2}-\omega^{2}\epsilon\mu}\left(\tilde{\beta}\frac{\partial E_{z}}{\partial x}
	+\omega\mu\frac{\partial H_{z}}{\partial y}\right),\\
\displaystyle E_{y}&=-\frac{i}{\tilde{\beta}^{2}-\omega^{2}\epsilon\mu}\left(\tilde{\beta}\frac{\partial E_{z}}{\partial y}
	-\omega\mu\frac{\partial H_{z}}{\partial x}\right),\\
\displaystyle H_{x}&=-\frac{i}{\tilde{\beta}^{2}-\omega^{2}\epsilon\mu}\left(\tilde{\beta}\frac{\partial H_{z}}{\partial x}
	-\omega\epsilon\frac{\partial E_{z}}{\partial y}\right),\\
\displaystyle H_{y}&=-\frac{i}{\tilde{\beta}^{2}-\omega^{2}\epsilon\mu}\left(\tilde{\beta}\frac{\partial H_{z}}{\partial y}
	+\omega\epsilon\frac{\partial E_{z}}{\partial x}\right).
\end{align}
\end{subequations}

\begin{figure}[t,b,p] 
      \centering
      	\includegraphics[width=0.6\textwidth]{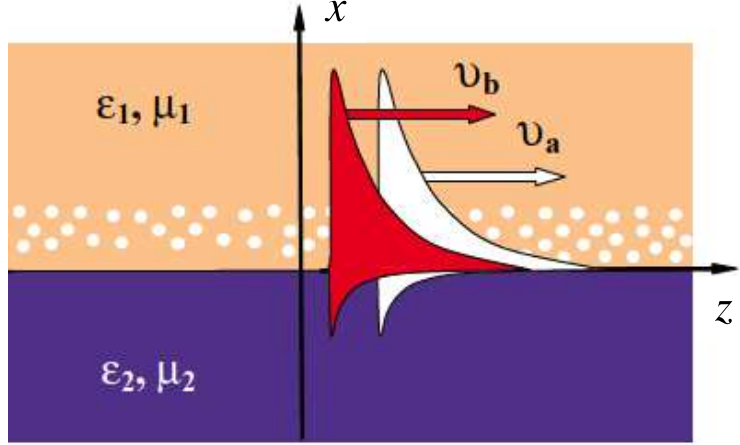}
	\caption[Diagram of the flat interface waveguide.]{A diagram of the structure of the flat interface waveguide. The region corresponding to $x>0$ is dielectric whereas $x<0$ is a metamaterial. The evanescent decay of the fields is indicated by the shape of the modes. \textsc{Source:} \emph{Figure~1 in S.\ A.\ Moiseev et al.\ \emph{Phys.\ Rev.\ A,\ }81:033839, 2010.}}
     \label{fig:flat-int}   
\end{figure}

The geometry of the flat interface is two materials with a planar interface, at $x=0$, between the two so that medium 1 is located at $x>0$ and medium 2 is at $x<0$. Figure~\ref{fig:flat-int} shows a schematic of the flat interface with two plasmon-polariton modes travelling along the interface.
Due to the fact that the flat interface is invariant in the $y$ direction, the derivative $\partial/\partial y=0$ everywhere. This implies that the fields themselves have no $y$ dependence. Thus, the wave equation for the $\hat{\bm z}$ component of the electric field is reduced to~\cite{Yeh:2008}
\begin{equation}
\left(\frac{\text d^{2}}{\text d x^{2}}+\omega^{2}\epsilon\mu-\tilde{\beta}^{2}\right)E_{z}(x)=0,\label{cartesianwaveeq}
\end{equation}
An equivalent equation for the magnetic field can be obtained as well.

Solving Eq.~(\ref{cartesianwaveeq}) yields $E_{z}$. Using Eqs.~(\ref{slabfieldrels}), the remaining field components of the TM modes of the flat interface are determined.
Applying the first two conditions from Eqs.~(\ref{boundaryconditions}) to the fields yields the dispersion relation for the flat interface:
\begin{equation}
\frac{\epsilon_{1}}{\gamma_{1}}=-\frac{\epsilon_{2}}{\gamma_{2}},\label{eq:flatdisprel}
\end{equation}
with $\gamma_{i}$ the complex wave number perpendicular to the interface in region $i$ and given by
\begin{equation}
\gamma_j:=\sqrt{\tilde{\beta}^2-\omega^2\epsilon_j\mu_j}.\label{transversewavenumberCh2}
\end{equation}

The negative sign in Eq.~(\ref{eq:flatdisprel}) means that for a guided mode to exist on a single interface the permittivity on each side of the interface must have opposite signs. That is, the permittivity on one side of the interface must be negative, whereas the permittivity on the other side must be positive. The implication is that guided modes cannot exist on a flat interface between two dielectrics or any two materials with strictly positive permittivity. The relation of Eq.~(\ref{eq:flatdisprel}) can be solved for $\tilde{\beta}$,
\begin{equation}
\tilde{\beta}=\frac{\omega}{c}\sqrt{\frac{\epsilon_{1}\epsilon_{2}(\epsilon_{2}\mu_{1}-\epsilon_{1}\mu_{2})}{\epsilon_{2}^{2}-\epsilon_{1}^{2}}},
\end{equation}
giving expressions for the dispersion and attenuation as functions of the frequency of the field and the material properties.

\subsection{Slab Geometry}

A waveguide with slab geometry, see Fig.~\ref{fig:guidediagramCh2}, has three regions; the core (region 1) is located at $0\leq x \leq w$, whereas the two cladding regions (regions 2 and 3) are located at $x>w$ and $x<0$. The geometry of the slab waveguide, like the flat interface, is such that it is invariant in the $y$ direction, meaning that the modes will have no $y$ dependence and $\partial \bm E/\partial y=\partial \bm B/\partial y=0$. As the modes of the slab guide are either TM or TE, the $z$ component of the magnetic or electric field, respectively is zero. This, makesi it necessary to solve only for the remaining non-zero $z$ component. Using the TM mode as an example, the non-zero components of the electric and magnetic fields of the modes are obtained by inserting the solution to Eq.~(\ref{cartesianwaveeq}) into the relations of Eq.~(\ref{slabfieldrels}). 

Ensuring that the boundary conditions are satisfied, by applying the conditions of Eq.~(\ref{boundaryconditions}), results in the dispersion relation for the slab geometry, which has the form~\cite{Yeh:2008}
\begin{equation}
\frac{\epsilon_1}{\gamma_1}\left(\frac{\epsilon_2}{\gamma_2}+\frac{\epsilon_3}{\gamma_3}\right)=-\left(\frac{\epsilon_1^2}{\gamma_1^2}+\frac{\epsilon_2 \epsilon_3}{\gamma_2\gamma_3}\right)\tanh\gamma_1 w.\label{slabdisprelCh2}
\end{equation}
To obtain the dispersion relation for the TE modes, simply replace all of the $\epsilon_j$ with $\mu_{j}$ and $\mu_{j}$ with $\epsilon_j$. The dispersion relation for the slab geometry is transcendental, unlike that for the flat interface, and must be solved numerically for $\tilde{\beta}$. The values of $\tilde{\beta}$ that are solutions to the dispersion relation correspond to the allowed modes of the waveguide. The number of solutions depends on the dimensions and materials of the guide, as well as the frequency of the light. 

\subsection{Cylindrical Geometry}

The cylindrical guide has two regions, an inner region of radius $a$, or core, and an outer region, or cladding. Unlike the two geometries discussed previously, the cylindrical geometry does not have a transverse direction that is invariant. As with the slab guide, a cylindrical guide supports TE and TM modes, but it also supports other modes, denoted HE and EH. 

To determine the allowed modes of the cylindrical waveguide, Eq.~(\ref{generalwaveeqCh2}) must be solved for the $\hat{\bm z}$ components of the electric and magnetic fields. Equation~(\ref{generalwaveeqCh2}) in cylindrical coordinates has the form 
\begin{equation}
\left(\frac{1}{r}\frac{\partial}{\partial r}r\frac{\partial}{\partial r}+\frac{1}{r^{2}}\frac{\partial^{2}}{\partial\phi^{2}}+\frac{\partial^{2}}{\partial z^{2}}+\omega^{2}\epsilon\mu\right)E_{z}(r,\phi,z,\omega)=0,\label{eq:cylwave}
\end{equation}
along with an equivalent equation for the magnetic field. The remaining components can be obtained from Maxwell's equations through the following relations~\cite{Yeh:2008}:
\begin{subequations}\label{eq:cylfieldrels}
\begin{align}
\displaystyle E_{r}&=-\frac{i}{\tilde{\beta}^{2}-\omega^{2}\epsilon\mu}\left(\tilde{\beta}\frac{\partial E_{z}}{\partial r}
	+\frac{\omega\mu}{r}\frac{\partial H_{z}}{\partial \phi}\right),\\
\displaystyle E_{\phi}&=-\frac{i}{\tilde{\beta}^{2}-\omega^{2}\epsilon\mu}\left(\frac{\tilde{\beta}}{r}\frac{\partial E_{z}}{\partial \phi}
	-\omega\mu\frac{\partial H_{z}}{\partial r}\right),\\
\displaystyle H_{r}&=-\frac{i}{\tilde{\beta}^{2}-\omega^{2}\epsilon\mu}\left(\tilde{\beta}\frac{\partial H_{z}}{\partial r}
	-\frac{\omega\epsilon}{r}\frac{\partial E_{z}}{\partial \phi}\right),\\
\displaystyle H_{\phi}&=-\frac{i}{\tilde{\beta}^{2}-\omega^{2}\epsilon\mu}\left(\frac{\tilde{\beta}}{r}\frac{\partial H_{z}}{\partial \phi}
	+\omega\epsilon\frac{\partial E_{z}}{\partial r}\right).
\end{align}
\end{subequations}

The solutions to Eq.~(\ref{eq:cylwave}), combined with the relations in Eq.~(\ref{eq:cylfieldrels}), yield the expressions for the fields in each region of the cylindrical guide. Using the definitions
\begin{equation}
J'_m\left(a\kappa_{1}\right):=\left.\frac{\text d J_m\left(r \kappa_{1}\right)}{\text d r}\right|_{r=a}\label{eq:besselderiv}
\end{equation}
and
\begin{equation}
K'_m\left(a\gamma_{1}\right):=\left.\frac{\text d K_m\left(r \gamma_{1}\right)}{\text d r}\right|_{r=a}{\textstyle ,}\label{eq:assocbessderiv}
\end{equation}
with $ J_m$ and $K_{m}$ the Bessel function and modified Bessel function, respectively, 
the dispersion relation for the cylindrical guide is obtained by applying the boundary conditions of Eq.~(\ref{boundaryconditions}). The dispersion relation has the form
\begin{equation}\label{cyldisprel}
\frac{\tilde{\beta}^2m^2}{a^2\omega^2}\left(\frac{1}{\kappa_{2}^{2}}+\frac{1}{\gamma_{2}^{2}}\right)^2
=\left(\frac{\mu_1}{\kappa_{1}^{2}}\frac{J'_m\left(a \kappa_1\right)}{J_{m}\left(a\kappa_1\right)}+\frac{\mu_2}{\gamma_{2}^{2}}\frac{K'_m\left(a\gamma_2\right)}{K_m\left(a\gamma_2\right)}\right)\left(\frac{\epsilon_1}{\kappa_{1}^{2}}\frac{J'_m\left(a \kappa_1\right)}{J_{m}\left(a\kappa_1\right)}+\frac{\epsilon_2}{\gamma_{2}^{2}}\frac{K'_m\left(a\gamma_2\right)}{K_m\left(a\gamma_2\right)}\right),
\end{equation}
with $\kappa_{1}=i\gamma_{1}$ and $m$ an integer characterizing the azimuthal symmetry. As with the dispersion relation for the slab geometry, this equation must be solved numerically for $\tilde{\beta}$.

\section{The Kramers-Kronig Relations}\label{sec:KKrels}

The complex permittivity of a material consists of two parts, the real part, which quantifies dispersion, and the imaginary part, which quantifies the attenuation rate. It is easy to assume that these two quantities are independent, as there is no obvious mechanism linking them. Because they describe physical quantities, however, they must satisfy causality (i.e.\ an effect cannot happen earlier in time than the underlying cause). As a consequence of satisfying causality, the complex permittivity takes the form~\cite{LandauLifshitz:1960,Jackson:1999}
\begin{equation}
\frac{\epsilon(\omega)}{\epsilon_{0}}=1+\int_{0}^{\infty}G(\tau)\text{e}^{i\omega\tau}\,\text{d}\tau,
\end{equation}
with $G(\tau)$ the Fourier transform of $\epsilon(\omega)/\epsilon_{0}-1$ and relates the time-dependent forms of $\bm E$ and $\bm D$.

Due to the fact that $\bm E$ and $\bm D$ are real quantities, $G(\tau)$ must also be real. If we also require that $G(\tau)$ is finite for all $\tau$ and $G(\tau)\rightarrow 0$ for $\tau\rightarrow\infty$, then $\epsilon(\omega)$ is an analytic function of $\omega$ in the upper half plane~\cite{Jackson:1999}. That is, if we allow $\omega$ to take complex values, then  $\epsilon(\omega)$ is infinitely differentiable for $\text{Im}(\omega)\geq0$. This implies that there are restrictions on the allowed values of the permittivity of a material. 

The real and imaginary parts of any complex function that is analytic in the upper half plane are related via the Kramers-Kronig relations~\cite{LandauLifshitz:1960}, which have the form:
\begin{subequations}\label{eq:kramers-kronig}
\begin{equation}
\zeta_{\text r}(\omega)=\frac{2}{\pi}\;\mathcal{P}\!\!\int_{0}^{\infty}\frac{\omega'\zeta_{\text i}(\omega')}{\omega'^{2}-\omega^{2}}\text d\omega',
\end{equation}
and
\begin{equation}
\zeta_{\text i}(\omega)=-\frac{2\omega}{\pi}\;\mathcal{P}\!\!\int_{0}^{\infty}\frac{\zeta_{\text r}(\omega')}{\omega'^{2}-\omega^{2}}\text d\omega',
\end{equation}
\end{subequations}
with $\zeta_{\text r}$ and $\zeta_{\text i}$ real such that $\zeta_{\text r}+i\zeta_{\text i}$ is an analytic function of $\omega$ in the upper half plane, and $\mathcal{P}$ denotes the Cauchy principal value. The Kramers-Kronig relations are valid for electromagnetic fields propagating through vacuum or any material (or combination of materials) as they are a direct consequence of causality. As an analytical, complex valued function of frequency, the values of the real and imaginary parts of the propagation constant (i.e.\ the dispersion and attenuation) are restricted by the Kramers-Kronig relations. The same is true for the permeability of a material and the complex propagation constant, both of which must also satisfy causality.

In addition to connecting the real and imaginary parts of these quantities, two sum rules can be obtained. For a material with a permittivity described by Eq.~(\ref{genpermittivity}), it can be shown that~\cite{Jackson:1999}
\begin{equation}
\omega_{\text e}^{2}=\frac{2}{\pi}\int_{0}^{\infty}\omega\text{Im}(\epsilon(\omega))/\epsilon_{0}\,\text{d}\omega,
\end{equation}
with $\omega_{\text e}$ the plasma frequency, which will be defined later. This equation is sometimes called the sum rule for oscillator strengths. With the assumption that $\text{Re}(\epsilon(\omega'))/\epsilon_{0}-1=-\omega_{\text e}^{2}/\omega'^{2}+O(1/\omega'^{4})$ for all $\omega'>N$, the second sum rule has the form~\cite{Jackson:1999}
\begin{equation}
\frac{1}{N}\int_{0}^{N}\text{Re}(\epsilon(\omega))/\epsilon_{0}\,\text{d}\omega=1+\frac{\omega_{\text e}^{2}}{N^{2}},
\end{equation}
with $N$ the number of molecules per unit volume. This second sum rule states that when $N\rightarrow\infty$ the average value of $\text{Re}(\epsilon(\omega))/\epsilon_{0}$ is unity.

\section{Nonlinear Optics}\label{sec:NLoptics}

This section follows Boyd~\cite{Boyd:2008}. When an atom is exposed to an electric field the electrons that are bound to the atom are displaced to one side and, along with the positively charged nucleus of the atom, create an electric dipole. Such an electric dipole has a non-zero electric field. In the case of a dipole induced by an applied electric field, the direction of the field generated by the dipole is reversed from that of the applied field. In a material such as a dielectric, which contains numerous atoms, an applied electric field causes each individual atom to become a dipole. The result of this large number of dipoles aligned in the same direction is an electric field created within the material.

A material that has a number of dipoles induced is said to be polarized, and the dipole moment per unit volume is called the polarization. In other words, a polarized material is one that is generating an internal electric field in response to an external field. The resulting field, called the displacement field, was introduced in Sec.~\ref{sec:Maxwell} and is given by
\begin{equation}
\bm D=\epsilon_{0}\bm E+\bm P,\label{eq:Ddef}
\end{equation}
with $\bm E$ the external, or applied, electric field and $\bm P$ the polarization of the material .

The degree of polarization in a material is related to the size of the dipole moment induced in each atom. For low intensity light the polarization response $\bm P$ is linear with respect to the electric field. A material's polarization response is not instantaneous, however. It takes a nonzero amount of time for the individual dipoles to form and become aligned.

Thus, the polarization depends not only on the electric field at the present time, but also on the electric field at all past times as well. The linear polarization can be written as
\begin{equation}
\bm P(\bm r,t)=\epsilon_{0}\int_{0}^{\infty}\text d\tau R^{(1)}(\bm r,\tau)\bm E(\bm r,t-\tau),\label{eq:linearPtime}
\end {equation}
with $R^{(1)}(\bm r,\tau)$ the linear response function, which is the contribution of the electric field to the polarization at time $\tau$. This is the linear polarization at time $t$ of a material in response to an applied electric field. 

Using the the Fourier transform and its inverse, defined as
\begin{equation}
\bm F(\bm r,\omega)=\int_{-\infty}^{\infty}\text dt\bm F(\bm r,t)\e^{i\omega t},
\end{equation}
and
\begin{equation}
\bm F(\bm r,t)=\frac{1}{2\pi}\int_{-\infty}^{\infty}\text d\omega\bm F(\bm r,\omega)\e^{-i\omega t},
\end{equation}
Eq.~(\ref{eq:linearPtime}) can be expressed as a function of the frequency of the incident light, rather than time. This is a useful representation, as often the electromagnetic properties of materials are given as functions of frequency. Replacing $\bm E(\bm r,t-\tau)$ with its Fourier transform yields
\begin{equation}
\bm P(\bm r,t)=\frac{\epsilon_{0}}{2\pi}\int_{0}^{\infty}\text d\tau R^{(1)}(\bm r,\tau)\int_{-\infty}^{\infty}\text d\omega\bm E(\bm r,\omega)\e^{-i\omega (t-\tau)}.
\end{equation}
By defining the linear susceptibility as
\begin{equation}
\chi^{(1)}(\bm r,\omega):=\int_{0}^{\infty}\text d\tau R^{(1)}(\bm r,\tau)\e^{i\omega\tau},\label{eq:chi1def}
\end{equation}
and replacing $\bm P(\bm r,t)$ with its Fourier transform, Eq.~(\ref{eq:linearPtime}) becomes
\begin{equation}
\frac{1}{2\pi}\int_{-\infty}^{\infty}\text d\omega\bm P(\bm r,\omega)\e^{-i\omega t}=\frac{\epsilon_{0}}{2\pi}\int_{-\infty}^{\infty}\text d\omega\chi^{(1)}(\bm r,\omega)\bm E(\bm r,\omega)\e^{-i\omega t},
\end{equation}
leading directly to the relation
\begin{equation}
P(\bm r,\omega)=\epsilon_{0}\chi^{(1)}(\bm r,\omega)\bm E(\bm r,\omega),\label{eq:linearPfreq}
\end{equation}
which describes the linear polarization of a material as a function of the frequency of the incident light.

By inserting Eq.~(\ref{eq:linearPfreq}) into Eq.~(\ref{eq:Ddef}) and comparing it to the constitutive equation ${\bm D}=\epsilon{\bm E}$ the permittivity of a linear material is simply
\begin{equation}
\epsilon(\bm r,\omega)=\epsilon_{0}\left(1+\chi^{(1)}(\bm r,\omega)\right),
\end{equation}
with the relative permittivity $\epsilon/\epsilon_{0}=1+\chi^{(1)}$ . The refractive index of a material is also defined through the susceptibility and for the case of a linear material is
\begin{equation}
n=\sqrt{\frac{\epsilon}{\epsilon_{0}}}=\sqrt{1+\chi^{(1)}}.\label{eq:linearrefindex}
\end{equation}
For many materials, especially dielectrics, $\chi^{(1)}$ is spatially invariant and, for some frequency windows, is nearly frequency invariant. For this reason the refractive index of many materials is quoted as a constant, sometimes with a frequency range where the value is applicable.

Electric fields with a large enough intensity can elicit a polarization with a nonlinear response to the external field. In this case, Eq.~(\ref{eq:linearPtime}) is not sufficient to describe the polarization response of the material. The polarization can, however, be expanded in powers of $\bm E$, such that~\cite{Armstrong:1962},
\begin{equation}\label{eq:nonlinearPtime}
\begin{split}
\bm P(\bm r,t)&=\bm P^{(1)}(\bm r,t)+\bm P^{(2)}(\bm r,t)+\bm P^{(3)}(\bm r,t)+\cdots\\
	&=\epsilon_{0}\int_{0}^{\infty}\text d\tau_{1}\, R^{(1)}(\bm r,\tau_{1}):\bm E(\bm r,t-\tau_{1})\\
		&\quad+\epsilon_{0}\int_{0}^{\infty}\int_{0}^{\infty}\text d\tau_{1}\text d\tau_{2}\,R^{(2)}(\bm r,\tau_{1},\tau_{2})
			:\bm E(\bm r,t-\tau_{1})\bm E(\bm r,t-\tau_{2})\\
		&\quad+\epsilon_{0}\int_{0}^{\infty}\int_{0}^{\infty}\int_{0}^{\infty}\text d\tau_{1}\text d\tau_{2}\text d\tau_{3}\,
			R^{(3)}(\bm r,\tau_{1},\tau_{2},\tau_{3}):\bm E(\bm r,t-\tau_{1})\\
			&\qquad\times\bm E(\bm r,t-\tau_{2})\bm E(\bm r,t-\tau_{3})\\
		&\quad+\cdots.
\end{split}
\end{equation}

Often it is more convenient to work with a frequency dependent form, rather than a time dependent one. To this end the time-dependent electric fields in Eq.~(\ref{eq:nonlinearPtime}) are replaced by their Fourier transforms. Defining the $n$th order nonlinear susceptibility as
\begin{equation}
\chi^{(n)}(\bm r,\omega_{1},\cdots,\omega_{n})=\int_{0}^{\infty}\cdots\int_{0}^{\infty}\text d\tau_{1}\cdots \text d\tau_{n}\,
	R^{(n)}(\bm r,\tau_{1},\cdots,\tau_{n})\e^{i(\omega_{1}\tau_{1}+\cdots+\omega_{n}\tau_{n})},
\end{equation}
for $n\geq2$, Eq.~(\ref{eq:nonlinearPtime}) can be written in terms of the frequency of the light:
\begin{equation}\label{eq:nonlinearPfreq}
\begin{split}
\bm P(\bm r,\omega)&=\epsilon_{0}\chi^{(1)}(\bm r,\omega):\bm E(\bm r,\omega)\\
		&\quad+\epsilon_{0}\int_{-\infty}^{\infty}\int_{-\infty}^{\infty}\text d\omega_{1}\text d\omega_{2}\,
			\chi^{(2)}(\bm r,\omega_{1},\omega_{2}):\bm E(\bm r,\omega_{1})\bm E(\bm r,\omega_{2})
			\delta(\omega-\omega_{1}-\omega_{2})\\
		&\quad+\epsilon_{0}\int_{-\infty}^{\infty}\int_{-\infty}^{\infty}\int_{-\infty}^{\infty}\text d\omega_{1}\text d\omega_{2}\text d\omega_			{3}\,\chi^{(3)}(\bm r,\omega_{1},\omega_{2},\omega_{3}):\bm E(\bm r,\omega_{1})\\
		&\quad\qquad\times\bm E(\bm r,\omega_{2})\bm E(\bm r,\omega_{3})\delta(\omega-\omega_{1}-\omega_{2}-\omega_{3})\\
		&\quad+\cdots,
\end{split}
\end{equation}
which must be satisfied for all $\omega$.

Equation~(\ref{eq:nonlinearPfreq}) accounts for the contributions from all frequencies. If, however, the fields are monochromatic, the contribution from the $n$th order polarization term reduces to
\begin{equation}
\bm P^{(n)}(\bm r,\omega=\omega_{1}+\cdots+\omega_{n})=\frac{\epsilon_{0}}{(2\pi)^{n}}
		\chi^{(n)}(\bm r,\omega_{1},\cdots,\omega_{n}):\bm E(\bm r,\omega_{1})\cdots\bm E(\bm r,\omega_{n}).
\end{equation}
In general, $\chi^{(n)}$ is a tensor of rank $n+1$. Thus, the $i$th component of the second-order polarization term, for example, would be
\begin{equation}\label{eq:p2freqcomp}
P_{i}^{(2)}(\bm r,\omega=\omega_{n}+\omega_{m})=\sum_{jk}\sum_{(nm)}\frac{\epsilon_{0}}{(2\pi)^{2}}\chi_{ijk}^{(2)}(\bm r,\omega_{n},\omega_{m})E_{j}(\bm r,\omega_{n})E_{k}(\bm r,\omega_{m}),
\end{equation}
with the indices $ijk$ representing the Cartesian components of the fields, and the sum over $(nm)$ taken such that $\omega=\omega_{n}+\omega_{m}$ always. 

As an example, consider an electric field with two frequency components,
\begin{equation}\label{eq:2ndordexampleE}
\bm E(\bm r,t)=\bm E_{1}(\bm r)\e^{-i\omega_{1}t}+\bm E_{2}(\bm r)\e^{-i\omega_{2}t}+\bm E^{*}_{1}(\bm r)\e^{i\omega_{1}t}
	+\bm E^{*}_{2}(\bm r)\e^{i\omega_{2}t}.
\end{equation}
The second order nonlinear polarization term, due to the presence of the field, becomes
\begin{equation}\label{2ndordnonlin}
\begin{split}
\bm P(\bm r,t)&=\frac{\epsilon_{0}}{4\pi^{2}}\left(\chi^{(2)}(\bm r,\omega_{1},\omega_{1})
		:\bm E_{1}(\bm r)\bm E_{1}(\bm r)\e^{-2i\omega_{1}t}\right.\\
	&\quad+\chi^{(2)}(\bm r,\omega_{2},\omega_{2}):\bm E_{2}(\bm r)\bm E_{2}(\bm r)\e^{-2i\omega_{2}t}\\
	&\quad+2\chi^{(2)}(\bm r,\omega_{1},\omega_{2}):\bm E_{1}(\bm r)\bm E_{2}(\bm r)\e^{-2i(\omega_{1}+\omega_{2})t}\\
	&\quad+2\chi^{(2)}(\bm r,\omega_{1},-\omega_{2}):\bm E_{1}(\bm r)\bm E_{2}^{*}(\bm r)\e^{-2i(\omega_{1}-\omega_{2})t}\\
	&\quad+\chi^{(2)}(\bm r,\omega_{1},-\omega_{1}) :\bm E_{1}(\bm r)\bm E_{1}^{*}(\bm r)\\
	&\quad+\left.\chi^{(2)}(\bm r,\omega_{2},-\omega_{2}) :\bm E_{2}(\bm r)\bm E_{2}^{*}(\bm r)+ \text{c.c.}\right),
\end{split}
\end{equation}
with c.c.\ indicating the complex conjugate. 

Here we have used complex field notation, as we are assuming a harmonic field. This notation makes clearer how the different frequency components in the polarization expansion yield new frequency components. To recover the real field from the complex notation simply take the real part:
\begin{equation}
\bm E(\bm r,t)=\text{Re}\left(\bm E(\bm r)\e^{-i\omega t}\right)=\frac{1}{2}\left(\bm E(\bm r)\e^{-i\omega t}+\bm E^{*}(\bm r)\e^{i\omega t}\right).
\end{equation}
Complex field notation will also be used later in this thesis.

The first two terms on the right hand side of Eq.~(\ref{2ndordnonlin}) correspond to fields having frequencies of $2\omega_{1}$ and $2\omega_{2}$ respectively, and the phenomenon is called second-harmonic generation. To generate these fields, two photons of frequency $\omega_{1}$ or $\omega_{2}$ interact in a nonlinear manner, facilitated by the surrounding material, and combine to form a new single photon of frequency $2\omega_{1}$ or $2\omega_{2}$, respectively. In this process two photons are destroyed while a new photon of higher frequency is simultaneously generated. The last two terms of Eq.~(\ref{2ndordnonlin}) correspond to what is called optical rectification, which is the generation of a static electric field across a nonlinear material. 

The third and fourth terms of Eq.~(\ref{2ndordnonlin}) are known as the sum- and difference-frequency generation terms and correspond to fields with frequencies $\omega_{1}+\omega_{2}$ and $\omega_{1}-\omega_{2}$, respectively. The sum-frequency generation process is similar to second-harmonic generation; two photons, one of each frequency $\omega_{1}$ and $\omega_{2}$, are destroyed to form a new higher frequency photon. The difference-frequency generation, however, works in a different way. To generate a lower-frequency photon, a single photon is destroyed and two new photons are created at lower frequencies of $\omega_{2}$ and $\omega_{1}-\omega_{2}$. The presence of the lower frequency field stimulates the splitting of the higher frequency field, thereby selecting the frequency of the generated field.

It is through these aforementioned processes that a material with a second-order nonlinearity can facilitate the generation of frequency components in a field that are absent from the input fields. Higher-order nonlinearities exist as well, with the third-order being the next order of nonlinear response. An expansion of the polarization, similar to Eq.~(\ref{2ndordnonlin}), can be done for the third-order nonlinear response. In a similar manner to the second-order nonlinear response, the third-order nonlinear response leads to third-harmonic generation and various sum and difference combinations of the three contributing frequency components.

A third order nonlinear process exists that does not produce a field of a different frequency. Instead, this particular nonlinear interaction changes the way in which the material responds to the applied electromagnetic field. This nonlinear process is called the Kerr nonlinearity or Kerr effect, and is described by the third-order polarization expansion term
\begin{equation}
P^{(3)}_{i}(\bm r,\omega_{1})=\frac{3\epsilon_{0}}{8\pi^{3}}\sum_{jkl}\chi^{(3)}_{ijkl}(\bm r,\omega_{1},\omega_{1},-\omega_{1})E_{j}(\bm r,\omega_{1})E_{k}(\bm r,\omega_{1})E_{l}(\bm r,-\omega_{1}),\label{eq:Kerromega}
\end{equation}
with the subscript $i$ indicating a vector component. To show how the material response is affected by the Kerr nonlinearity, we will consider an isotropic material, such as glass.

The third-order nonlinear susceptibility $\chi^{(3)}$ is a rank-four tensor, which means it has 81 individual elements. The number of nonzero and independent elements for an isotropic material, however, is greatly reduced due to symmetry considerations. As isotropic materials respond the same regardless of direction, the only nonzero elements of $\chi^{(3)}$ are of the form
\begin{equation}\label{eq:chi3iso}
\begin{split}
\chi_{ijkl}^{(3)}(\omega_{1},\omega_{2},\omega_{3})&=\chi_{1122}^{(3)}(\omega_{1},\omega_{2},\omega_{3})\delta_{ij}\delta_{kl}+\chi_{1212}^{(3)}(\omega_{1},\omega_{2},\omega_{3})\delta_{ik}\delta_{jl}\\
	&\quad+\chi_{1221}^{(3)}(\omega_{1},\omega_{2},\omega_{3})\delta_{il}\delta_{jk},
\end{split}	
\end{equation}
which is true for arbitrary field frequencies. Restricting to a single frequency, which is the case for the Kerr effect, reduces the independent elements even further due to an intrinsic permutation symmetry. The remaining independent elements of $\chi^{(3)}$ for the Kerr nonlinearity have the form
\begin{equation}\label{eq:chi3isoKerr}
\chi_{ijkl}^{(3)}(\omega_{1},\omega_{1},-\omega_{1})=\chi_{1122}^{(3)}(\omega_{1},\omega_{1},-\omega_{1})\left(\delta_{ij}\delta_{kl}+\delta_{ik}\delta_{jl}\right)+\chi_{1221}^{(3)}(\omega_{1},\omega_{1},-\omega_{1})\delta_{il}\delta_{jk}.
\end{equation}

Combining Eqs.~(\ref{eq:Kerromega}) and (\ref{eq:chi3isoKerr}) yields
\begin{equation}\label{eq:PKerr}
\begin{split}
\bm P_{\text{Kerr}}^{(3)}(\bm r,\omega_{1})&=\frac{6\epsilon_{0}}{8\pi^{3}}\chi_{1122}^{(3)}(\omega_{1},\omega_{1},-\omega_{1})(\bm E(\bm r,\omega_{1})\cdot\bm E^{*}(\bm r,\omega_{1}))\bm E(\bm r,\omega_{1})\\
	&\quad+\frac{3\epsilon_{0}}{8\pi^{3}}\chi_{1221}^{(3)}(\omega_{1},\omega_{1},-\omega_{1})(\bm E(\bm r,\omega_{1})\cdot\bm E(\bm r,\omega_{1}))\bm E^{*}(\bm r,\omega_{1}),
\end{split}
\end{equation}
which is the nonlinear polarization due to the Kerr effect. It can immediately be seen that there are two separate contributions to the nonlinear polarization. The term proportional to $\bm E$ produces a polarization with the same handedness as the electric field, and the term proportional to $\bm E^{*}$ produces a polarization with the opposite handedness. The effect of these separate contributions is to rotate the polarization of elliptically polarized light. There is no effect on linear or circularly polarized light, however, so we will not discuss the separate contributions any further.

The total polarization of an isotropic material is the sum of Eqs.~(\ref{eq:PKerr}) and (\ref{eq:linearPfreq}), which can be inserted into Eq.~(\ref{eq:Ddef}) to yield the displacement field
\begin{equation}\label{eq:DKerrspm}
\bm D(\bm r,\omega_{1})=\epsilon_{0}\left(1+\chi^{(1)}(\omega_{1})\right)\bm E(\bm r,\omega_{1})+\bm P_{\text{Kerr}}^{(3)}(\bm r,\omega_{1}),
\end{equation}
This is the effective electric field inside the material. It can be shown that for linearly polarized fields, the refractive index due to the polarization depends on the intensity of the electric field and has the form
\begin{equation}
n_{\text{Kerr}}(\bm r,\omega_{1})=n_{0}(\omega_{1})+n_{2}(\omega_{1})\left|E(\bm r,\omega_{1})\right|^{2}
\end{equation}
with
\begin{equation}
n_{0}(\omega_{1})=\sqrt{1+\chi^{(1)}(\omega_{1})}
\end{equation}
the linear refractive index of the material, and
\begin{equation}
n_{2}(\omega_{1})=\frac{3\chi^{(3)}_{1111}(\omega_{1},\omega_{1},-\omega_{1})}{2n_{0}(\omega_{1})}.
\end{equation}

The effect of the refractive index of a material is to modify the speed of light in that material to a new speed given by $v=c/n$. The resulting, usually slower, speed causes the field to be phase shifted. Having an intensity-dependent refractive index means that the intensity of the field itself affects the refractive index, thereby modifying the resulting phase shift. This phenomenon is known as self-phase modulation.

The intensity-dependent polarization due to the Kerr nonlinearity also modifies the refractive index for other fields propagating through the material at the same location and at the same time. This effect, known as cross-phase modulation, is similar to self-phase modulation, but is a phase shift due to the presence of a second electromagnetic field, rather than a self-induced phase shift. The two fields must be distinguishable from one another, so they must have different frequencies or different propagation directions. The term in the polarization expansion that corresponds to cross-phase modulation of a field with frequency $\omega_{1}$ due to the presence of a field with frequency $\omega_{2}$ is
\begin{equation}\label{eq:PXPM}
\begin{split}
\bm P_{\text{XPM}}^{(3)}(\bm r,\omega_{1})&=\frac{12\epsilon_{0}}{8\pi^{3}}\chi_{1122}^{(3)}(\omega_{2},-\omega_{2},\omega_{1})(\bm E(\bm r,\omega_{2})\cdot\bm E^{*}(\bm r,\omega_{2}))\bm E(\bm r,\omega_{1})\\
	&\quad+\frac{6\epsilon_{0}}{8\pi^{3}}\chi_{1221}^{(3)}(\omega_{2},-\omega_{2},\omega_{1})(\bm E(\bm r,\omega_{2})\cdot\bm E(\bm r,\omega_{2}))\bm E^{*}(\bm r,\omega_{1}).
\end{split}
\end{equation}
This expression shows that the effect of cross-phase modulation is twice as strong as self-phase modulation.

The refractive index of a material due to cross-phase modulation is similar to that of self-phase modulation but with the nonlinear term increased by a factor of 2:
\begin{equation}
n_{\text{XPM}}(\omega_{1},\omega_{2})=n_{0}(\omega_{1})+2n_{2}(\omega_{1})\left|E(\bm r,\omega_{2})\right|^{2}.\label{eq:nXPM}
\end{equation}
From Eq.~(\ref{eq:nXPM}) it can be seen that the two-fold increase in the polarization response for cross-phase modulation translates directly to a two-fold increase in the nonlinear term of the refractive index. Thus, cross-phase modulation is twice as effective at modifying the phase of a field as self-phase modulation.

The third-order susceptibility $\chi^{(3)}$ is generally small and on the order of $10^{-22}\,\text m^{2}/\text V^{2}$~\cite{Boyd:2008}, meaning the nonlinear interaction is weak, and the nonlinear contribution to the refractive index $n_{2}$ is very small. The result of a small change in the refractive index is that any phase shift acquired due to self-phase modulation or cross-phase modulation will also be small. Increasing the nonlinear response will, in turn, increase the acquired phase shift. 

Increasing the intensity of the fields increases the nonlinear response of the refractive index by increasing $\left|E(\bm r,\omega)\right|^{2}$. The intensity can be increased simply by increasing the total power of the field. The drawback to this is more energy must be imparted to the field, which may not be an option for some systems, or already existing fields. Decreasing the volume of the field, without changing the total energy, effectively concentrates the energy and increases the local intensity, thereby strengthening the nonlinear interaction. 

Another way to increase the phase shift due to self- or cross-phase modulation is to increase the length of the nonlinear interaction. As the phase shift is the result of the modified speed of light in the material, allowing the interaction to take place over a longer propagation distance will allow the phase to be shifted more. As the propagation distance is not tied to the intensity of the fields, not accounting for energy losses, the two can be combined to increase the resulting phase shift even further.

\section{Calculating the electric susceptibility $\chi^{(n)}$}\label{chiecalc}

The electric susceptibility of a material is a measure of its ability to respond to an externally applied electric field by polarizing, thereby generating an internal electric field. An expression for the electric susceptibility of an isotropic material can be derived classically by solving the equation of motion for each electron in the material. The equation of motion for each individual electron has the form~\cite{Jackson:1999}
\begin{equation}
\frac{\text d^{2}}{\text d t^{2}}\bm x+\Gamma_{\text e}\frac{\text d}{\text d t}\bm x+\omega_{0}^{2}\bm x=-\frac{e}{m_{\text e}}\bm E(\bm x,t)\label{permiteqmot}
\end{equation}
with $\bm x$ the position of the electron, $\omega_{0}$ a binding frequency that corresponds to a binding energy of the electron, $\Gamma_{\text e}$ a phenomenological damping term, $-e$ the electron charge and $m_{\text e}$ the electron mass.

Using the solution to Eq.~(\ref{permiteqmot}) along with Eq.~(\ref{eq:linearPfreq}), the electron dipole moment $\bm p=-e\bm x$, and the fact that the material polarization is the sum of all the individual dipoles, an expression can be obtained for the electric susceptibility. For materials with a single atom type, every electron will have the same binding energy. The permittivity will then have the form
\begin{equation}
\frac{\epsilon(\omega)}{\epsilon_{0}}=1+\frac{N_{\text e}e^{2}}{\epsilon_{0}m_{\text e}}\left(\omega_{0}^{2}-\omega(\omega+i\Gamma_{\text e})\right)^{-1},\label{genpermittivity}
\end{equation}
with $\epsilon_{0}$ the permittivity of free space, $N_{\text e}$ the number of electrons per unit volume, $\omega$ the frequency of the electromagnetic field, and $\Gamma_{\text{e}}$ the electric damping term. The electric susceptibility is related to the permittivity through the expression
\begin{equation}\label{eq:epsdef}
\chi^{(1)}(\omega)=\frac{\epsilon(\omega)}{\epsilon_{0}}-1.
\end{equation}

In most materials, the electrons are bound to the atoms that make up the material. In metals, however, an electron from each atom is free to move about the material resulting in an electron `gas'. These free, or conduction, electrons in a metal can be considered a plasma, as they are not bound to the atoms that form the metal, but rather are free to move about the metal. A plasma is a collection of disassociated ions, whether ionized atoms, single electrons or both. As such, the ions are free to move independently of each other. The plasma of free electrons in a metal is responsible for the electromagnetic response of the metal, such as conductivity and permittivity.

The permittivity of a metal can be described by the Drude model~\cite{Jackson:1999,Sernelius:2001}, which takes the form
\begin{equation}
\frac{\epsilon(\omega)}{\epsilon_0}=\epsilon_{\text b}-\frac{\omega_{\textrm{e}}^2}{\omega\left(\omega+i\Gamma_{\textrm{e}}\right)},\label{epsilonMM}
\end{equation}
with $\epsilon_{\text{b}}$ a background permittivity, and the plasma frequency of the metal given by
\begin{equation}
\omega_{\text{e}}=\sqrt{\frac{N_{\text e}e^{2}}{\epsilon_{0}m_{\text e}}}.
\end{equation}
The Drude model can be obtained from Eq.~(\ref{genpermittivity}) by setting the binding frequency $\omega_{0}$ to zero (equivalent to a zero binding energy), which is in agreement with having free electrons in a metal. The background permittivity $\epsilon_{b}$ is generally equal to one, but could be different due to contributions from bound electrons in the metal.

As was discussed in the previous section, the susceptibility can depend nonlinearly on the external electric field. By using a quantum mechanical density matrix approach, the linear and nonlinear susceptibilities can be calculated. The Hamiltonian of the system can be written as (the remainder of this section follows Boyd~\cite{Boyd:2008}),
\begin{equation}
\hat H=\hat H_{0}-\hat{\bm\mu}\cdot\bm E(t),
\end{equation}
with $\hat H_{0}$ the hamiltonian for the free atom, $\hat{\bm\mu}=e\hat{\bm r}$ the electric dipole moment operator of the atom and $\bm E(t)$ the applied electric field. The unperturbed hamiltonian $\hat H_{0}$ satisfies the relation
\begin{equation}
\hat H_{0}\varphi_{n}=E_{n}\varphi_{n},
\end{equation}
with $\varphi_{n}$ the energy eigenfunctions of the free atom and $E_{n}$ the corresponding energies.

The atom-field interaction is usually assumed to be weak in the sense that the expectation value of $\hat{\bm\mu}\cdot \bm E(t)$ is small relative to $\hat H_{0}$. With this assumption, the system is described by the following quantum mechanical equation of motion:
\begin{equation}
\dot\rho_{nm}=\frac{-i}{\hbar}\left[\hat H,\hat\rho \right]_{nm}-\frac{\Gamma_{nm}}{2}\left(\rho_{nm}-\rho_{nm}^{(\text{eq})}\right),\label{quanteqmotionelec}
\end{equation}
with $\hat\rho$ the density matrix of the system, $\hat\rho^{(\text{eq})}$ the state of the system at equilibrium and $\Gamma_{nm}$ phenomenological relaxation (decay) terms. Equation~(\ref{quanteqmotionelec}) can be solved using perturbation theory to determine the susceptibility. 

To obtain a solution using perturbation theory, the atom-field interaction term $\hat{\bm\mu}\cdot \bm E(t)$ is modified to the form $\lambda\hat{\bm\mu}\cdot \bm E(t)$, with $\lambda$ a parameter for the strength of the perturbation, which has a value between 0 and 1. The parameter $\lambda$ is set to 1 afterwards to obtain the physical solution. We now seek a solution to Eq.~(\ref{quanteqmotionelec}) that has the form
\begin{equation}\label{eq:perturbform}
\rho_{nm}=\rho^{(0)}_{nm}+\lambda\rho^{(1)}_{nm}+\lambda^{2}\rho^{(2)}_{nm}+\ldots,
\end{equation}
with terms having increasing powers of $\lambda$ corresponding to smaller perturbations of the steady state solution. To find the perturbation solution, we require Eq.~(\ref{eq:perturbform}) to be a solution to the equation of motion for any value of $\lambda$.

Inserting Eq.~(\ref{eq:perturbform}) into the equation of motion and collecting terms according to the powers of lambda yields the following set of equations for consecutive orders of the $\rho_{nm}$ expansion:
\begin{subequations}\label{eq:perturbeqs}
\begin{align}
\dot\rho^{(0)}_{nm}&=-i\omega_{nm}\rho^{(0)}_{nm}-\frac{1}{2}\Gamma_{nm}\left(\rho^{(0)}_{nm}-\rho^{(\text{eq})}_{nm}\right),\label{eq:rho0}\\
\dot\rho^{(1)}_{nm}&=-\left(i\omega_{nm}+\frac{1}{2}\Gamma_{nm}\right)\rho^{(1)}_{nm}-\frac{i}{\hbar}\left[\hat{\bm\mu}\cdot \bm E(t),\rho^{(0)}\right]_{nm},\label{eq:rho1}\\
\dot\rho^{(2)}_{nm}&=-\left(i\omega_{nm}+\frac{1}{2}\Gamma_{nm}\right)\rho^{(2)}_{nm}-\frac{i}{\hbar}\left[\hat{\bm\mu}\cdot \bm E(t),\rho^{(1)}\right]_{nm},\label{eq:rho2}
\end{align}
\end{subequations}
and so on for higher expansion orders. The transition frequency $\omega_{nm}$ is defined as
\begin{equation}
\omega_{nm}=\frac{E_{n}-E_{m}}{\hbar}.
\end{equation}
The system of equations for the $\rho^{(i)}_{nm}$ can be solved with direct integration, starting with $\rho^{(0)}_{nm}$ up to the highest order perturbation desired. To obtain expressions for the different orders of the susceptibility, the same order perturbation solution is required. For example, to determine the expression for $\chi^{(2)}$, the equation of motion must be solved up to second order for the density matrix, which has elements denoted by $\rho^{(2)}_{nm}$. 

The solution to Eq.~(\ref{eq:rho0}) is the state of the system in the absence of any external fields, therefore the equilibrium state $\hat\rho^{(\text{eq})}$ is taken as the solution, i.e.\ $\hat\rho^{(0)}=\hat\rho^{(\text{eq})}$. We assume that the atoms are cooled to cryogenic temperatures, so the population of the excited states will be negligible. Thus, we make the assumption that the density matrix for $\hat\rho^{(0)}$ is diagonal, i.e.\
\begin{equation}
\rho^{(0)}_{nm}=0\quad\text{for}\quad m\neq n.
\end{equation}

The polarization of a material was defined in Sec.~\ref{sec:NLoptics} to be the average dipole moment per unit volume, which can be expressed in terms of the first-order perturbation solution $\hat\rho^{(1)}$ as
\begin{equation}\label{eq:avgdipolemt}
\overline{\left<\tilde{\bm \mu}(t)\right>}=\sum_{nm}\rho^{(1)}_{nm}(t)\bm\mu_{mn}.
\end{equation}
This expression is both a quantum and ensemble average of the dipole moments generated by the atoms. To obtain an explicit expression for the average dipole moment, we solve Eq.~(\ref{eq:rho1}) for $\hat\rho^{(1)}$ using that fact that the form of $\hat\rho^{(0)}$ is known. The first-order perturbation solution is then
\begin{equation}
\rho^{(1)}_{nm}(t)=\frac{1}{\hbar}\left(\rho^{(0)}_{mm}-\rho^{(0)}_{nn}\right)\sum_{p}\frac{\bm \mu_{nm}\cdot\bm E(\omega_{p})\text e^{-i\omega_{p}t}}{\left(\omega_{nm}-\omega_{p}\right)-i\Gamma_{nm}/2}.
\end{equation}
The average dipole moment per unit volume can now be written as
\begin{equation}
\overline{\left<\tilde{\bm \mu}(t)\right>}=\sum_{nm}\frac{1}{\hbar}\left(\rho^{(0)}_{mm}-\rho^{(0)}_{nn}\right)\sum_{p}\frac{\bm\mu_{mn}\left[\bm \mu_{nm}\cdot\bm E(\omega_{p})\right]\text e^{-i\omega_{p}t}}{\left(\omega_{nm}-\omega_{p}\right)-i\Gamma_{nm}/2}
\end{equation}

From Eq.~(\ref{eq:linearPfreq}) we can see that the linear polarization is related to the linear electric susceptibility. Writing the average dipole moment per unit volume in terms of its frequency components,
\begin{equation}\label{eq:dipolemtfreq}
\overline{\left<\tilde{\bm\mu}(t)\right>}=\sum_{p}\overline{\left<\hat{\bm\mu}(\omega_{p})\right>}\text e^{-i\omega_{p}t},
\end{equation}
and using it to describe the linear polarzation along with Eq.~(\ref{eq:linearPfreq}) leads to
\begin{equation}
\bm P(\omega_{p})=\epsilon_{0}\chi_{(1)}(\omega_{p})\cdot\bm E(\omega_{p})=N\overline{\left<\hat{\bm\mu}(\omega_{p})\right>}.
\end{equation}
This relation shows how the linear susceptibility is related to the average dipole moment per unit volume. The linear susceptibility can now be written as
\begin{equation}
\chi^{(1)}(\omega_{p})=\frac{N}{\epsilon_{0}\hbar}\sum_{nm}\left(\rho^{(0)}_{mm}-\rho^{(0)}_{nn}\right)\frac{\bm\mu_{mn}\bm \mu_{nm}}{\left(\omega_{nm}-\omega_{p}\right)-i\Gamma_{nm}/2}.
\end{equation}
The expression for the susceptibility depends on the dipole moments of the atoms as well as the difference in population between excited states. When only one energy level of an atom is populated the first order susceptibility has the form~\cite{Boyd:2008}
\begin{equation}
\chi^{(1)}_{ij}(\omega)=\frac{N}{\epsilon_{0}\hbar}\sum_{n}\left[\frac{\mu_{an}^{i}\mu_{na}^{j}}{(\omega_{na}-\omega)-i\Gamma_{na}/2}
	+\frac{\mu_{na}^{i}\mu_{an}^{j}}{(\omega_{na}+\omega)+i\Gamma_{na}/2}\right]\label{quantelecsuscept}
\end{equation}
with $N$ the atomic number density, $\mu_{nm}^{i}$ is the $i$th cartesian component of the dipole moment for the $\left|n\right>\rightarrow\left|m\right>$ transition and $\omega_{nm}$ is the frequency of the transition. For an isotropic material with only one resonance frequency Eq.~(\ref{quantelecsuscept}) reduces to Eq.~(\ref{eq:epsdef}), in agreement with the classical model. 

Expressions for higher order susceptibilities can be calculated in a similar manner, though they quickly become very compilcated. For instance, the second order polarization can be obtained through the relation
\begin{equation}\label{eq:p2avgdipole}
\bm P^{(2)}(\omega=\omega_{p}+\omega_{q})=N\overline{\left<\hat{\bm\mu}(\omega_{p}+\omega_{q})\right>},
\end{equation}
with the average dipole moment per unit volume given by
\begin{equation}
\overline{\left<\tilde{\bm \mu}(t)\right>}=\sum_{nm}\rho^{(2)}_{nm}(t)\bm\mu_{mn}=\sum_{r}\overline{\left<\hat{\bm\mu}(\omega_{r})\right>}\text e^{-i\omega_{r}t}.
\end{equation}
The second-order term in the perturbation expansion is the solution to Eq.~(\ref{eq:rho2}) and has the form
\begin{align}
\rho^{(2)}_{nm}(t)&=\frac{1}{\hbar^{2}}\sum_{\nu}\sum_{pq}\text e^{-i(\omega_{p}+\omega_{q})t}\\
	&\quad\times\left(\frac{\left[\rho^{(0)}_{mm}-\rho^{(0)}_{\nu\nu}\right]\left[\bm\mu_{n\nu}\cdot\bm E(\omega_{q})\right]\left[\bm\mu_{\nu m}\cdot\bm E(\omega_{p})\right]}{\left[\left(\omega_{nm}-\omega_{p}-\omega_{q}\right)-i\Gamma_{nm}/2\right]\left[\left(\omega_{\nu m}-\omega_{q}\right)-i\Gamma_{\nu m}/2\right]}\right.\\
	&\quad-\left.\frac{\left[\rho^{(0)}_{\nu\nu}-\rho^{(0)}_{nn}\right]\left[\bm\mu_{n\nu}\cdot\bm E(\omega_{p})\right]\left[\bm\mu_{\nu m}\cdot\bm E(\omega_{q})\right]}{\left[\left(\omega_{nm}-\omega_{p}-\omega_{q}\right)-i\Gamma_{nm}/2\right]\left[\left(\omega_{n\nu}-\omega_{p}\right)-i\Gamma_{n\nu}/2\right]}\right)
\end{align}
Inserting this expression into the one for the average dipole moment per unit volume and then comparing Eqs.~(\ref{eq:p2avgdipole}) and (\ref{eq:p2freqcomp}), yields the second-order nonlinear susceptibility (with a slight modification to ensure it possesses permutation symmetry):
\begin{equation}
\begin{split}
\chi^{(2)}_{ijk}(\omega_{p},\omega_{q})&=\frac{N}{2\epsilon_{0}\hbar^{2}}\sum_{lmn}\left[\rho^{(0)}_{ll}-\rho^{(0)}_{mm}\right]\\
	&\quad\times\left(\frac{\mu^{i}_{ln}\mu^{j}_{nm}\mu^{k}_{ml}}{\left[\left(\omega_{nl}-\omega_{p}-\omega_{q}\right)-i\Gamma_{nl}/2\right]\left[\left(\omega_{ml}-\omega_{p}\right)-i\Gamma_{ml}/2\right]}\right.\\
	&\quad+\frac{\mu^{i}_{ln}\mu^{k}_{nm}\mu^{j}_{ml}}{\left[\left(\omega_{nl}-\omega_{p}-\omega_{q}\right)-i\Gamma_{nl}/2\right]\left[\left(\omega_{ml}-\omega_{q}\right)-i\Gamma_{ml}/2\right]}\\
	&\quad+\frac{\mu^{j}_{ln}\mu^{i}_{nm}\mu^{k}_{ml}}{\left[\left(\omega_{nm}+\omega_{p}+\omega_{q}\right)-i\Gamma_{nm}/2\right]\left[\left(\omega_{ml}-\omega_{p}\right)-i\Gamma_{ml}/2\right]}\\
	&\quad+\left.\frac{\mu^{k}_{ln}\mu^{i}_{nm}\mu^{j}_{ml}}{\left[\left(\omega_{nm}+\omega_{p}+\omega_{q}\right)-i\Gamma_{nm}/2\right]\left[\left(\omega_{ml}-\omega_{q}\right)-i\Gamma_{ml}/2\right]}\right).
\end{split}
\end{equation}
Similar calculations can be done for higher order nonlinear susceptibilities as well, but the expressions become increasingly large.

\section{Electromagnetically Induced Transparency}\label{sec:EIT}

Electromagnetically induced transparency (EIT) is an optical effect in which matter that is normally opaque is made to be transparent to light of a particular frequency. It is a nonlinear phenomenon in the sense that the linear susceptibility $\chi^{(1)}$ is a function of the pump field amplitude~\cite{Lambropoulos:2007}. Though the expression for the susceptibility of an EIT system cannot be obtained from the solutions of Eq.~(\ref{permiteqmot}), it can be determined by solving a system of Maxwell-Bloch equations for a three level atomic system~\cite{Lambropoulos:2007}.

\begin{figure}[t,b] 
       \centering
	\includegraphics[width=.4\textwidth]{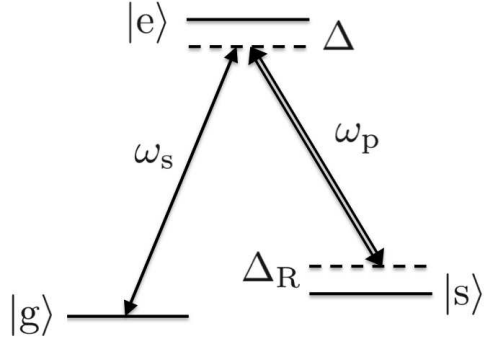}
	\caption[A schematic of the three-level $\Lambda$ atom]{A schematic of the three-level $\Lambda$ atom. The ground state is $\left|\text g\right>$, the excited state is $\left|\text e\right>$ and $\left|\text s\right>$ is a metastable state. The frequency of the pump field is $\omega_{\text p}$ and of the signal field is $\omega_{\text s}$, and $\Delta$ and $\Delta_{\text R}$ are detunings. \textsc{Source:} \emph{Figure 2 in B.\ R.\ Lavoie et al.,\ \emph{Phys.\ Rev.\ A,} 88:023860, 2013.}}
     \label{fig:EITsys}   
\end{figure}

To achieve EIT, atoms with a certain energy level configuration are needed. There are two three-level configurations that can produce EIT, namely lambda ($\Lambda$) and the ladder, or cascade, configuration ($\Xi$). Of these, the $\Lambda$ configuration results in the best transparency so it is the configuration considered for this thesis. A diagram of the energy levels of a $\Lambda$ atom is given in Fig.~\ref{fig:EITsys},
with $\left|\text g\right>$, $\left|\text e\right>$ and $\left|\text s\right>$ the ground state, excited state and hyperfine state respectively, $\omega_{\text p}$ the frequency of the pump field and $\omega_{\text s}$ the frequency of the signal field.

The $\Lambda$ atoms are initially prepared for EIT with a strong pump field that is tuned to the $\left|\text s\right>\rightarrow\left|\text e\right>$ transition. Thus, any electrons in the state $\left|\text s\right>$ are excited to the state $\left|\text e\right>$ and subsequently decay into the ground state $\left|\text g\right>$. This is done to ensure that any excited atoms are put into the ground state. The intensity of the pump field must then be decreased, but not to zero, in order to achieve EIT. The presence of the pump serves to split the excited state into two hyperfine states.

The susceptibility of the prepared $\Lambda$ atoms due to the presence of the pump field is~\cite{Lambropoulos:2007}
\begin{equation}
\chi^{(1)}_{\Lambda}(\omega)=\frac{2a_{0}c}{n\omega}\frac{i\gamma_{\text{eg}}}{\gamma_{\text{eg}}-i\Delta
+\left|\Omega_{\text p}\right|^{2}\left(\gamma_{\text{sg}}-i\Delta_{\text R}\right)^{-1}},\label{EITchi}
\end{equation}
with $c$ the speed of light in vacuum, $n$ the refractive index of the host material, $\gamma_{ij}$ the decay rate from level $i$ to level $j$, and
\begin{equation}
a_{0}:=\frac{3\pi c^{2}}{n^{2}\omega^{2}}\rho_{\text a}\approx\frac{3\pi c^{2}}{n^{2}\omega_{\text{eg}}^{2}}\rho_{\text a},
\label{eq:ndensity}
\end{equation}
with $\rho_{\text a}$ the number density of three-level atoms. The approximation in Eq.~(\ref{eq:ndensity}) is valid near the EIT resonance frequency. 
Defining $\omega_{ij}$ as the transition frequency for the $\left|i\right>\leftrightarrow\left|j\right>$ transition, the detunings are $\Delta=\omega_{\text{eg}}-\omega_{\text s}$, $\Delta_{\text R}:=\Delta-\omega_{\text{es}}+\omega_{\text p}$. The pump is set on resonance with the $\left|\text{s}\right>\leftrightarrow\left|\text{e}\right>$ transition such that $\Delta_{\text R}=\Delta$. The Rabi frequency of the pump is
\begin{equation}
\Omega_{\text p}=\frac{1}{\hbar}{\bm d}\cdot {\bm E}_{\text{p}},\label{rabifreq}
\end{equation} 
with $\bm d$ the dipole moment of the $\left|\text{s}\right>\leftrightarrow\left|\text{e}\right>$ transition, and $\bm E_{\text p}$ the pump field.

The strength of the pump field is an important consideration for achieving EIT, as one can easily mistake the qualitatively similar, yet fundamentally different, Autler-Townes splitting for EIT~\cite{Abi-Salloum:2010,Anisimov:2011}. Autler-Townes splitting is a phenomenon that bears the hallmark of EIT, namely an absorption dip located between two resonances that is accompanied by a steep dispersion curve. However, Autler-Townes splitting and EIT are the results of different physical processes. 

The separation of the split states created by the pump field is directly related to the pump intensity. For a low-intensity pump the split states are close together, and for a high-intensity pump they are further apart. Autler-Townes splitting occurs for high-intensity pumps ($\Omega_{\text p}>\gamma_{31}-\gamma_{21}$), such that the split states have a large separation and the overlap between them is small. As a result, there is no destructive interference caused when electrons are excited into the split states. The absorption dip is simply the result of the edges of the resonances from the split states overlapping. Thus, the absorption dip between the resonance peaks can never reach zero.

\begin{figure}[t,b] 
      \centering
	\subfloat{\label{fig:genEITatten}\includegraphics[width=0.6\textwidth]{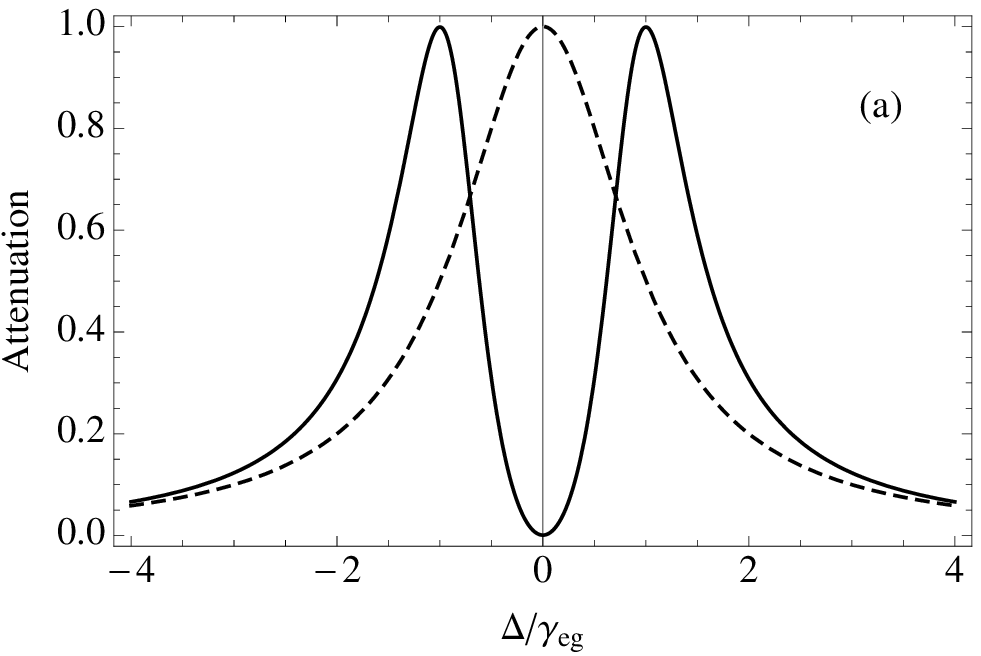}}\\
	\subfloat{\label{fig:genEITdisp}\includegraphics[width=0.6\textwidth]{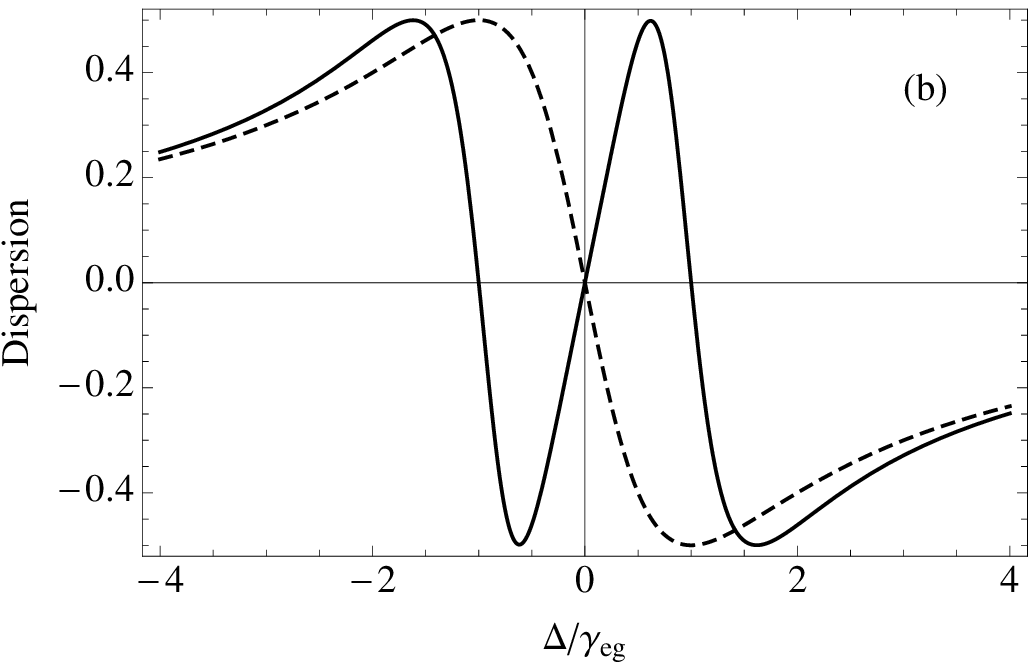}}
	\caption[Attenuation and dispersion due to EIT.]{Attenuation \protect\subref{fig:genEITatten} and dispersion \protect\subref{fig:genEITdisp} of a collection of $\Lambda$ atoms for a signal field in units of $a_{0}$, with the pump on resonance ($\Delta_{\text R}=\Delta$) and $\gamma_{\text{sg}}/\gamma_{\text{eg}}=10^{-3}$. The solid lines are for $\Omega_{\text p}/\gamma_{\text{eg}}=1$ and the dashed lines are for $\Omega_{\text p}=0$ (two-level atom).}
     \label{fig:genEITplots}   
\end{figure}

For pump intensities that are low enough ($\Omega_{\text p}<\gamma_{31}-\gamma_{21}$) the two split states overlap significantly. Coherent, destructive interference is exhibited for electrons that are excited into one of these overlapping states. The result is that electrons cannot be excited into either of the split states, and light on resonance with the $\left|\text g\right>\rightarrow\left|\text e\right>$ transition is not absorbed (see Fig.~\ref{fig:genEITplots}\subref{fig:genEITatten}).  This is the phenomenon known as EIT.

The dispersion and attenuation of an electromagnetic field propagating through a material are connected via the Kramers-Kronig relation~\cite{LandauLifshitz:1960}. The implication of this connection is that accompanying the drop in attenuation due to EIT is a dispersion that changes rapidly with frequency (see Fig.~\ref{fig:genEITplots}\subref{fig:genEITdisp}). The group velocity of a pulse is defined by the gradient of the dispersion~\cite{Lambropoulos:2007},
\begin{equation}
v_{\text g}(\omega)=\left[\frac{\text d}{\text d \omega}\left(n(\omega)\frac{\omega}{c}\right)\right]^{-1}=c\left[1+\frac{\omega}{2}\frac{\text d}{\text d \omega}\chi^{(1)}_{\Lambda}(\omega)\right]^{-1},\label{vgdef}
\end{equation}
with
\begin{equation}
n(\omega)=\sqrt{1+\chi^{(1)}_{\Lambda}(\omega)},
\end{equation}
the refractive index of the pumped $\Lambda$ atoms. Thus, the dispersion curve generated by EIT drastically affects the group velocity of a pulse propagating through an EIT medium.

The steep slope of the dispersion (Fig.~\ref{fig:genEITplots}\subref{fig:genEITdisp}) due to EIT causes a reduction in the group velocity of a pulse that is inversely proportional to the intensity of the pump field~\cite{Lambropoulos:2007}. If the pump field is turned off entirely while the signal pulse is inside the EIT medium, the signal pulse is coherently stored within the $\Lambda$ atoms. Light that is stopped with EIT retains information about the frequency, amplitude, and phase of the pulse, which is restored to the pulse when it is recovered from the EIT medium~\cite{Liu:2001}. Thus, the phenomenon of EIT is an effective way of achieving both slow and stopped light~\cite{Turukhin:2001}.

\section{Optical Quantum Memories}\label{sec:QuantMem}

For any computing system or means of information communication, it is extremely useful to be able to store information and retrieve it at a later time. With the advent of quantum information, and the subsequent growth of this field, the ability to store quantum information in a retrievable way has become very important. This is the primary motivation for developing quantum memories.

The usefulness of photons for quantum information processing and quantum communication has made optical quantum memories an attractive method for storing quantum states. The most obvious use of a quantum memory is simply storing the results of quantum computations for use later, akin to how memory is used in a classical computer. There are, however, a number of other useful ways that quantum memories can be employed besides simply information storage~\cite{Simon:2010}.

For example, single photon sources are used for a number of applications in quantum information science, such as for quantum cryptography~\cite{Gisin:2002}. There are two main methods of generating single photons, faint laser pulses and entangled photon generation via parametric downconversion. The faint laser pulse method relies on generating pulses with an extremely low probability of having more than one photon. The drawback to such a low intensity is that most of the pulses will contain zero photons, and the probability of a pulse having one photon is low.

The parametric downconversion method relies on exciting a second-order nonlinear response in a crystal. The nonlinear interaction generates two entangled photons that are emitted in different directions. One photon, called the signal, is used in the quantum information process and the other, called the idler, can simply be discarded, but is often used to herald the generation of the signal photon. As with the faint pulse method, the fields must be weak enough to ensure that the probability of generating more than one signal photon at a time is small. Although the parametric downconversion method is heralded, it is still probabilistic in nature and cannot generate a single photon on demand.

If it is known that a single photon will be needed in the future for a quantum information process, a quantum memory can be employed. A single photon is loaded into the quantum memory beforehand, and is recalled at the time it is needed. This eliminates the need to wait for a probabilistic process to generate a single photon, significantly speeding up some quantum information processes.

Another use of quantum memories is to aid in long-distance secure quantum communication, which requires sharing of entangled pairs, and transmitting quantum states between separate parties. These physical systems must be sent through a channel of some sort. In the case of optical quantum communication, the channel is most likely an optical-fibre network. Because a channel is a physical system, it interacts with anything that it is transporting. When quantum states interact with the environment, in this case the channel, noise is introduced that causes decoherence of the state. When the state is being used to transport information or entanglement, decoherence degrades the amount of information or entanglement that the state contains.

To prevent the quantum state from becoming so far degraded that the information or entanglement it is carrying is lost, a series of quantum repeaters can be set up to ``boost'' the signal. For the case of entanglement, a pair of entangled photons is created and shared between neighbouring repeaters. The entanglement is then shared between the two communicating parties by means of entanglement swapping at each repeater.

The objective is to take an entangled pair of photons created by one party, or a third party, and share the entanglement of these photons along the quantum channels. This is done until both parties are in possession an entangled photon. In this way, quantum memories facilitate long-distance entanglement sharing. The entangled photons can then be used to perform various quantum communication and computation operations.

There are various schemes by which an optical quantum memory can be realized~\cite{Lvovsky:2009}, such as EIT, controlled reversible inhomogeneous broadening, atomic frequency combs, and the so-called DLCZ protocol, all of which have their own benefits and drawbacks. The scheme presented herein employs EIT to produce slow light, and an optical memory is suggested as a possible application. For this reason an overview of the operation of an EIT-based quantum memory is given.

As described in Sec.~\ref{sec:EIT}, electromagnetically induced transparency arises from two electromagnetic fields creating coherent, destructive interference via a particular interaction with three-level atoms, in this case $\Lambda$ atoms. The destructive interference prevents the signal field from causing an electronic transition in the $\Lambda$ atoms. Thus, the atoms are transparent to electromagnetic fields having a frequency within a narrow window around the atomic transition resonance. Any pulse that fits within this spectral window will pass through the EIT medium without being absorbed.

The existence of a transparency window, by means of the Kramers-Kronig relation, causes a steep change in the dispersion, as a function of frequency, causing the transmitted pulse to slow down significantly. The reduced group velocity allows a spatially long pulse to be compressed, as the leading edge slows down when entering the EIT material, while the remainder of the pulse continues at its original group velocity until entering the EIT region. In this way a pulse can be contained in a region that is small compared to the pulse length.

\begin{figure}[t,b] 
       \centering
	\includegraphics[width=0.6\textwidth]{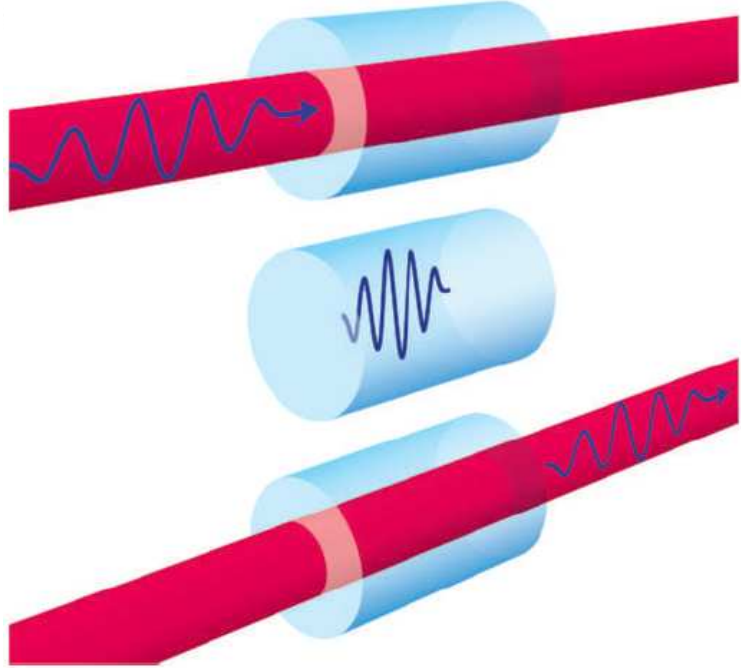}
	\caption[Depiction of the operation of an EIT-based quantum memory.]{The operation of an EIT-based quantum memory is depicted. \textsc{Top:} The signal pulse is entering the memory, and pump field is turned on making the EIT medium transparent to the signal pulse. \textsc{Middle:} The pump field is turned off, trapping, or storing, the signal pulse in the memory. The pulse compression is depicted in this stage as well, and can be seen by comparing the wavelength to the incoming and outgoing pulses. \textsc{Bottom:} By turning the pump field back on, the medium is again made transparent to the signal pulse and it resumes its propagation and exits the medium. \textsc{Source:} \emph{Figure~2 in A.\ Lvovsky, B.\ C.\ Sanders and W.\ Tittel,\ \emph{Nature Photon.,\ }3:706--714, 2009.}}
     \label{fig:EITmem}   
\end{figure}

While the pulse is contained within the EIT medium the pump field can be turned off, effectively turning off the EIT. With the pump turned off the EIT atoms are no longer transparent to the signal field, and the field is subsequently absorbed by the atoms. This absorption puts the atoms in excited states in such a way so as to preserve information about not only the frequency of the signal but also the amplitude and phase~\cite{Liu:2001,Appel:2008}. The stored pulse can easily be retrieved simply by turning the pump back on. A depiction of a pulse entering an EIT medium, being stored, and then retrieved is shown in Fig.~\ref{fig:EITmem}.

%% file: metamaterials.tex
\chapter{Electromagnetic Metamaterials}\label{ch:MM}

\section{Introduction}

When electromagnetic fields interact with a material, they cause the electrons of the atoms that make up the material to move~\cite{Jackson:1999}. This motion generates electric, and sometimes magnetic, fields as a result, which augment the original fields. The way in which the electric and magnetic fields are augmented is described by the permittivity and permeability, respectively. The permittivity of most materials, which is discussed in Sec.~\ref{chiecalc}, is different from the permittivity of free space $\epsilon_{0}$, and frequency dependent. The permeability of most materials, on the other hand is usually equal to $\mu_{0}$, the permeability of free space.

The purpose of metamaterials is to create a desired electromagnetic response by altering the permeability and/or permittivity of the constituent materials. Depending on the electromagnetic response of the metamaterial, they can be classified into one of four categories: negative-index, double-negative, single-negative, and double-positive. The electromagnetic response of metamaterials is frequency-dependent, however, so a single metamaterial can fall into different categories, depending on the frequency of the incident fields. In this chapter, the different categories of metamaterials will be discussed.

This chapter is divided into three sections. Section~\ref{MMclassification} defines the different metamaterial types, how they are distinct from other materials, and where metamaterial properties overlap with those of other materials. Section~\ref{ch:intro-MMconstruction} is an overview of the different fabrication methods for various metamaterials. Lastly, Sec.~\ref{sec:MMs} is a discussion of the electromagnetic response of metamaterials and considerations for calculating the stored energy in a metamaterial.

\section{Material and Metamaterial Classification}\label{MMclassification}

Metamaterials are a class of man-made materials that have electromagnetic properties not reproducible with the constituent materials alone. However, metamaterials are not the first, nor the only artificial material. Artificial dielectrics are a class of artificial material that has sub-wavelength structures, and are typically designed as lightweight alternatives to existing materials for long-wavelength applications (e.g.\ radar). Photonic crystals are an artificial material with wavelength scale structures that are designed to control photon propagation~\cite{Russell:2003}.

Metamaterials are distinguished from other artificial materials by the combination of artificial sub-wavelength structuring with a novel electromagnetic response~\cite{Balmain:2005}. The electromagnetic response of a metamaterial is usually due to a frequency-dependent permeability that can have negative values for some frequencies. This important distinction is the reason that artificial dielectrics, though they employ sub-wavelength structures, are not considered metamaterials. Artificial dielectrics are designed to mimic the response of existing materials and are considered predecessors to metamaterials.

\begin{figure}[t,b] 
       \centering
	\includegraphics[width=0.5\textwidth]{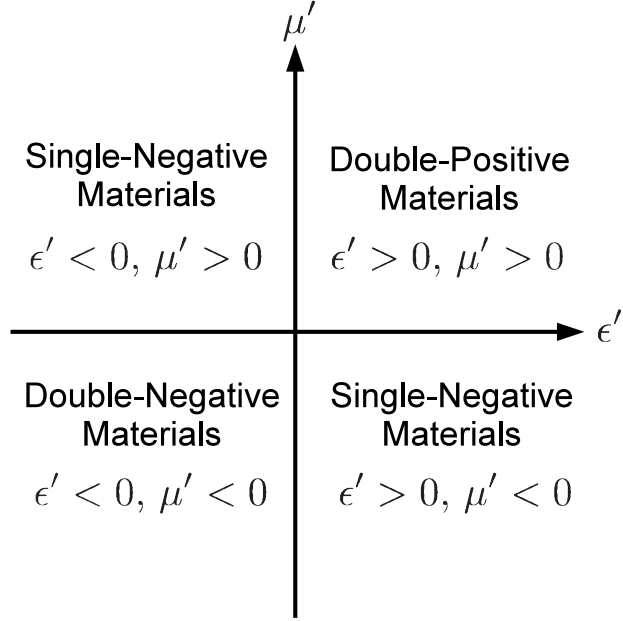}
	\caption[Metamaterial types]{A graphic showing the different metamaterial classifications, and how they relate to the real parts of the permittivity and permeability.}
     \label{fig:MMtypes}   
\end{figure}

The electromagnetic response of a metamaterial can be categorized as either double-negative, single-negative, or double-positive. As the response of a metamaterial is frequency dependent, the way a metamaterial is classified can depend on how it is being employed. The different classifications refer to the sign of the real parts of the permittivity and permeability (see Fig.~\ref{fig:MMtypes}). 
The imaginary parts of the permeability and permittivity are always positive for passive materials. Negative values would imply that the material is imparting energy to the fields, which is called gain, and would require the material to have an energy source. We are concerned with only passive metamaterials for this work.

Double-positive materials are characterized by $\text{Re}(\epsilon)>0$ and $\text{Re}(\mu)>0$. Most materials fall into this category, including most naturally occurring and synthetic materials. Due to the prevalence of double-positive materials in nature, there is generally no need to go to the effort of fabricating a metamaterial specifically for this purpose. Double-positive materials, however, are necessary to observe some of the effects of single-negative, and double-negative materials. This is because their unusual properties are observed when fields either cross, or propagate along, an interface between two materials of different classes~\cite{Engheta:2006}. 

Single negative materials are those with either $\text{Re}(\epsilon)<0$ (epsilon-negative) or $\text{Re}(\mu)<0$ (mu-negative) but not both. There are some instances of epsilon-negative materials that are naturally occurring. In metals, for example, the real part of Eq.(\ref{epsilonMM}) can be negative for frequencies less than the plasma frequency~\cite{Jackson:1999}. However, when the frequency becomes comparable to the damping term $\Gamma_{\text e}$, the imaginary part of the permittivity dominates~\cite{Pendry:1996}. Materials with a negative permeability, rather than a negative permittivity, are not naturally occurring. However, it is possible to achieve a negative permeability in some materials by applying a magnetic field~\cite{Dechant:2006}.

By writing $\epsilon=\epsilon'+i\epsilon''$ and $\mu=\mu'+i\mu''$ the refractive index of a metamaterial (or any material described by both a complex permittivity and permeability) can be expanded into the form
\begin{equation}
n_{\text{MM}}=\frac{c}{\sqrt{2}}\left(\text{sgn}\left(\Im{\epsilon\mu}\right)
	\sqrt{\left|\epsilon\mu\right|+\epsilon'\mu'-\epsilon''\mu''}+ i\sqrt{\left|\epsilon\mu\right|-\epsilon'\mu'+\epsilon''\mu''}\right),\label{expandedn}
\end{equation}
which shows how a negative permittivity or permeability will affect the refractive index of a material. The most striking consequence of either a negative permittivity or permeability is that it drastically reduces the size of the real part of the refractive index, and increases the imaginary part. This effect is more pronounced when the imaginary parts of the permittivity and permeability are small. For this reason, single-negative materials are not generally transparent to electromagnetic fields.

Double-negative materials are those with $\text{Re}(\epsilon)<0$ and $\text{Re}(\mu)<0$. From Eq.~(\ref{expandedn}), it can be seen that the condition for a material having a negative index of refraction is
\begin{equation}
\text{Im}(\epsilon\mu)=\epsilon'\mu''+\mu'\epsilon''<0.\label{NIMcondition}
\end{equation}
This relation shows that if the real part of both the permittivity and permeability are negative, then so too is the real part of the refractive index. Thus, passive double-negative materials are necessarily negative-index materials. Unlike with single-negative materials, the real part of the refractive index of double-negative materials can be large, but negative. In that case, fields are able to propagate through the double-negative material.

A negative-index material is a material that has the real part of its refractive index less than zero.  The real part of Eq.~(\ref{expandedn}) is negative for single-negative materials if $\text{Im}(\epsilon\mu)<0$. In this way, a single-negative material can be a negative-index material, though the real part of the refractive index would likely be small.

It is also possible to have negative-index materials that are not metamaterials. Some photonic crystals, for instance, can exhibit a negative index of refraction~\cite{Parimi:2004}. Photonic crystals are not considered metamaterials, however, as the size of the artificial structures is on the order of a wavelength.

There is one class of material, the zero-index material, that does not fit nicely into any one category. A zero-index material is characterized by having the real part of its refractive index equal to zero. This can be achieved for lossless materials by making one or both the permittivity and permeability equal to zero~\cite{Litchinitser:2008}. For lossy materials, having only one of  $\text{Re}(\epsilon)=0$ or $\text{Re}(\mu)=0$ only yields a zero-index material for frequencies where $\left|\epsilon\mu\right|+\epsilon'\mu'-\epsilon''\mu''=0$. However, like the lossless material, a lossy material with both $\text{Re}(\epsilon)=0$ and $\text{Re}(\mu)=0$ yields a zero-index material for all frequencies. Note that the imaginary part of the refractive index is not zero for lossy zero-index materials.

Some of the equations in this thesis, such as Eq.(\ref{vectorwave}), can break down for $\epsilon=0$, $\mu=0$ or both. The metamaterial considered in this thesis has $\text{Re}(\epsilon(\omega))=0$ at $\omega=\omega_{\text e}$, $\text{Re}(\mu(\omega))=0$ at $\omega=0.2\omega_{\text e}$ and near $\omega=0.28\omega_{\text e}$, and has $\text{Re}(n_{\text MM})=0$ near $\omega=0.16\omega_{\text e}$ and near $\omega=0.68\omega_{\text e}$. However, the imaginary parts of $\epsilon(\omega)$, $\mu(\omega)$, and $n_{\text MM}$ are never equal to zero for the range of frequencies considered in this thesis. For this reason, there is no issue with equations breaking down when any of the permittivity, permeability or refractive index is equal to zero.

\section{Metamaterial Construction}\label{ch:intro-MMconstruction}

Metamaterial construction is not simple, particularly for optical frequencies, as it requires a periodic structure on a sub-wavelength scale. The structures must be sub-wavelength so that the interaction is effectively averaged over a wavelength making the metamaterial behave as if it is homogeneous, rather than having discrete structures. If the structures were wavelength scale or larger, individual structures would interact with one or more wavelengths of the field altering the behaviour.
\begin{figure}[t,b] 
       \centering
	\includegraphics[width=0.4\textwidth]{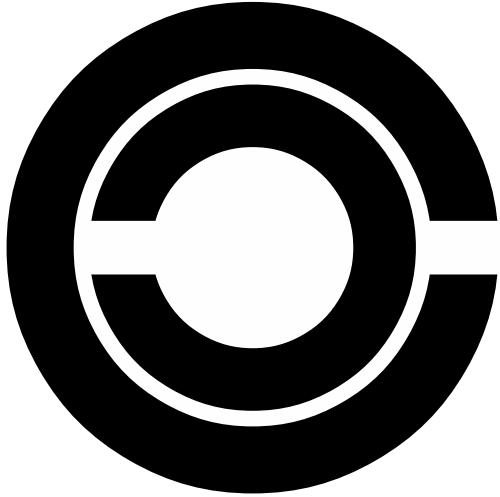}
	\caption[Diagram of a split-ring resonator]{A diagram of a split-ring resonator showing the concentric rings with the breaks, or splits, on opposite sides.}
     \label{fig:SRR}   
\end{figure}

The first metamaterials that were constructed were effective at microwave frequencies. This was because the periodic structure could be easily assembled by depositing metal wires on dielectric substrates. The periodic structure in these early metamaterials was the split-ring resonator, which consists of two concentric circular wires each having a break on opposite sides. The result is two concentric `C' shaped wires facing in opposite directions, as can be seen in Fig.~\ref{fig:SRR}.
The splits in each wire allow the structure to be resonant for wavelengths larger than the ring diameters~\cite{Smith:2000}. This makes the split-ring resonator a sub-wavelength structure, an array of which constitutes a metamaterial.

Due to the fact that these first metamaterials consisted solely of split-ring resonators, they only had a frequency dependent permeability and were able to achieve negative values. However, they did not have a frequency dependent permittivity. Including thin metal wires alongside the split-ring resonators generates a frequency dependent permittivity as well, leading to what are now known as negative-index metamaterials. The use of metal wires, rather than bulk metal, is to reduce the attenuation of the microwaves as the attenuation rate of microwaves in metal is large.

While there are a number of possible uses for metamaterials that operate at microwave frequencies and below, it is also desirable to have metamaterials that can operate at infrared, optical, and frequencies even higher. Part of the challenge in building optical metamaterials is designing structures that can be implemented at nanometer scales. A number of techniques have been adapted to fabricating optical metamaterials with varying degrees of success~\cite{Boltasseva:2008}.

Early optical metamaterials were effectively two-dimensional metamaterials in the sense that they were thin layers and only worked when light was incident normal to the material plane. Techniques similar to photolithography are used to construct these metamaterials. Fabrication of optical metamaterials is not possible with standard lithography techniques, as the light used has a larger wavelength than the size of the structures needed. To reduce the wavelength, electron-beam lithography can be used to construct sub-optical-wavelength structures. The beam in electron beam lithography can have a width on the order of nanometers.

\begin{figure}[t,b,p] 
       \centering
	\includegraphics[width=\textwidth]{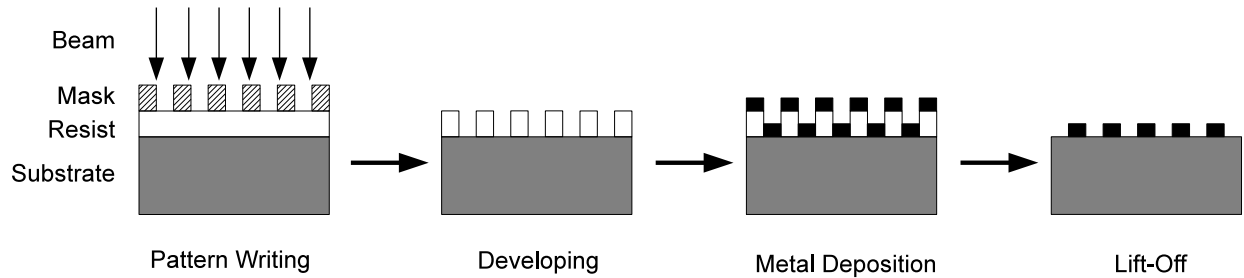}
	\caption[Overview of the lithography process.]{An overview of the basic lithography process. The resist is first etched using either an optical or electron beam. The mask is required for standard lithography to form a pattern on the resist. Electron-beam and interference lithography do not need a mask. A layer of metal is then deposited over the entire surface. When the resist is removed afterward, the metal that was deposited on the resist is also removed, leaving behind the desired pattern.}
     \label{fig:e-beamlitho}   
\end{figure}

\begin{figure}[t,b] 
     \centering
      \subfloat{\label{fig:NIM-eb1}\includegraphics[width=0.45\textwidth]{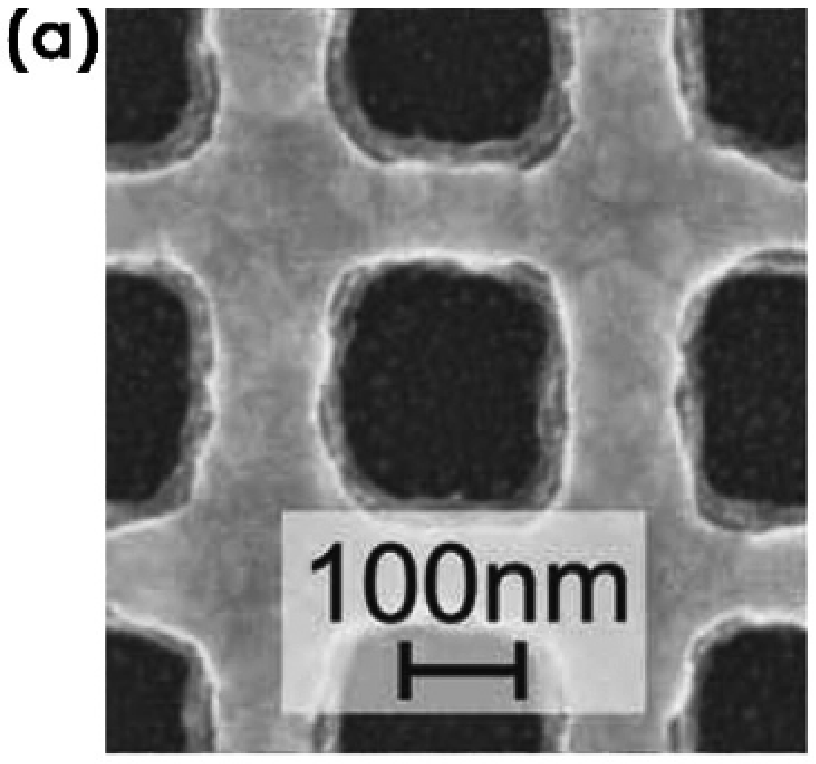}}\hfill
      \subfloat{\label{fig:NIM-eb2}\includegraphics[width=0.45\textwidth]{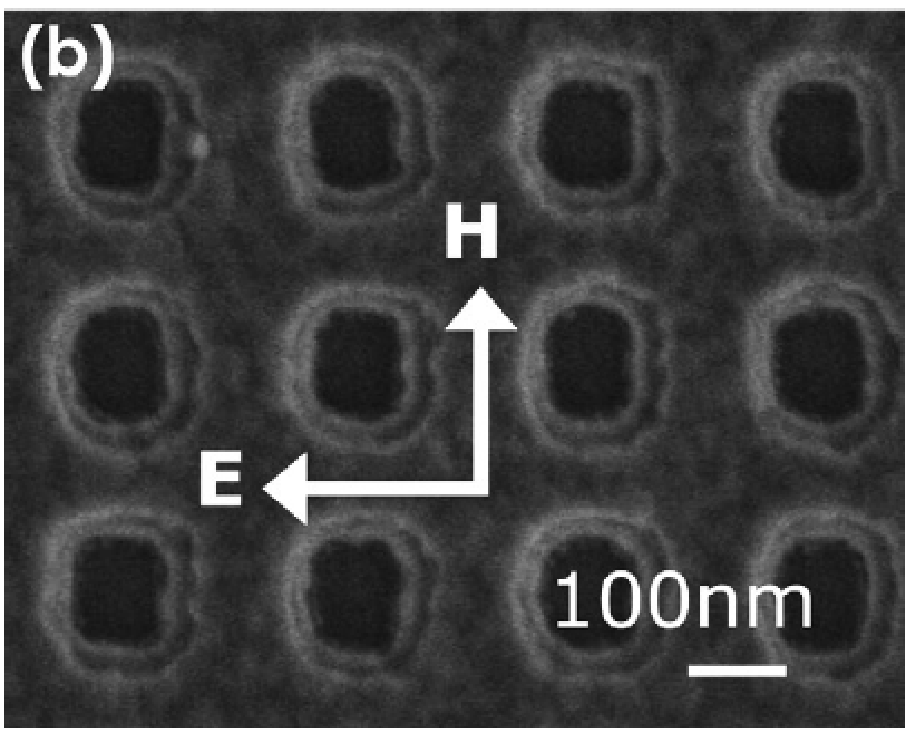}}
     \caption[Metamaterials fabricated using electron-beam lithography.]{Scanning electron microscope images of metamaterials fabricated using electron-beam lithography. \textsc{Source:} \emph{Figure~2 in A. Boltasseva and V. M. Shalaev.\ \emph{Metamaterials, }2:1--17, 2008.}}\label{fig:NIM-eb}
\end{figure}

Electron-beam lithography, similar to photolithography, is used to print a pattern onto a layer called a photoresist (Fig.~\ref{fig:e-beamlitho} outlines the process of electron-beam lithography).
Depending on the photoresist used, either the printed pattern or the negative image is then removed. With some of the photoresist removed, the layer underneath is now exposed in the desired pattern. At this point, the metal, or other material, can be deposited over the entire surface. Now, when the resist is removed it takes with it the material deposited on top of it. The material remaining on the substrate forms the desired pattern. Figure~\ref{fig:NIM-eb} shows scanning electron microscope images of two metamaterials fabricated using electron-beam lithography.

Electron-beam lithography, however, can be a very time consuming process as the electron beam must be scanned across the surface to generate the required pattern. By contrast, photolithography uses a filter, called a mask, to project the entire pattern onto the photoresist at once. A method to take advantage of the quicker photolithography, and bring the resolution to that required for constructing metamaterials, is interference lithography. 

The basic process is the same as for photolithography or electron-beam lithography. The only difference is how the pattern is printed on to the photoresist. Rather than use a mask to filter out the unwanted light or scanning a beam across the surface, a standing wave pattern is created by interfering two or more coherent beams of light. The peaks of constructive interference in the standing wave pattern print on to the photoresist, allowing it to be removed at these points for etching. Where the interference is not constructive the intensity is not enough to affect the photoresist.

This process is inherently well suited to constructing metamaterials as the interference pattern is periodic, and so must be the pattern of nano-structures in the metamaterial. In addition, due to the fact that an interference pattern is used, interference lithography is effective at patterning large areas, meaning larger metamaterials can be constructed. The drawback, however, is that the resolution is not as fine as with electron-beam lithography.

Another possible way to produce optical metamaterials is with what is called nano-imprint lithography. In this process a physical stamp, rather than light or electrons, is used to print a pattern on the photoresist. With nano-imprint lithography, the resolution of the structures is limited only by the fabrication process, and because a stamp is being used it can cover a large area.

These processes, however, are inherently two-dimensional. They involve creating a pattern on a thin flat surface and etching away the unwanted pieces. This is problematic when we consider that, ultimately, optical metamaterials will be used in practical applications that may require more than a thin material, or that metamaterials be insensitive to the direction and/or orientation of the fields.

\begin{figure}[t,b] 
       \centering
	\includegraphics[width=0.6\textwidth]{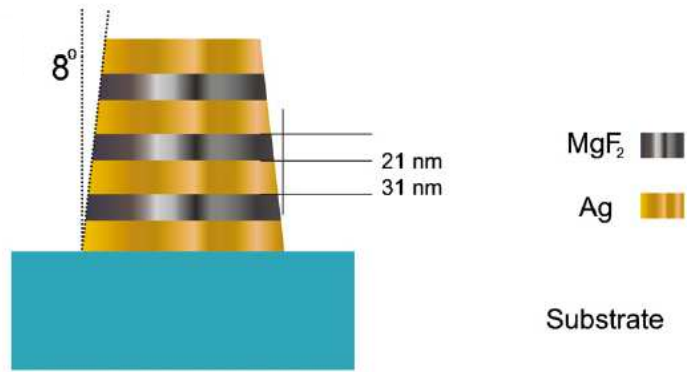}
	\caption[Picture of a 3D layered metamaterial.]{A cross section of part of a 3D metamaterial made by layering 2D structures. The thinning of the structures caused by the limited resolution of the lithography techniques is shown. \textsc{Source:} \emph{Figure~5(a) in A. Boltasseva and V. M. Shalaev.\ \emph{Metamaterials, }2:1--17, 2008.}}
     \label{fig:3dlayernim}   
\end{figure}

Three-dimensional metamaterials can be created with lithography processes simply by repeating the process a number of times on subsequent layers. In this way the metamaterial can be built up from individual layers. The problem with layering the 2-D structures is that the resulting metamaterial is not truly 3-D, as it will only function for light incident on two opposing faces (those that are the top and bottom layer). The remaining faces are made up of the edges of the stacked layers and incident light will not elicit the desired response. An additional problem with stacking the layers is that, due to the lithography process, the structures on each subsequent layer become thinner resulting in tapered, rather than straight, vertical structures (see Fig.~\ref{fig:3dlayernim}). 

True 3-D metamaterials can be constructed by other means. A promising candidate is a process called two-photon photopolymerization, wherein a two-photon nonlinear interaction with a material causes the material to polymerize (small molecules bond together creating large molecules). This process only occurs at the focal point of a highly focused laser and is thereby capable of producing structures at wavelength resolutions. The polymerization process can be parallelized, using an array of mirrors, to create many polymer structures at once. This technique can be adapted to creating metallic structures either by depositing a metallic film on the polymer structures, or by creating the polymer structures next to a metal layer. This is inherently a layering method to create 3-D metamaterials. Due to the added flexibility of the polymerization process, however, two-photon photopolymerization allows for the fabrication of true 3-D metamaterials.

Recently, a 3-D optical metamaterial was created using a self-assembly technique~\cite{Vignolini:2012}. Self assembly is the process by which certain compounds, when brought together, arrange themselves in a consistent way through local interactions. In this way, large intricate structures can be formed
simply by allowing the reaction to happen. 

\begin{figure}[t,b] 
     \centering
      \subfloat{\label{fig:3dNIM-SA}\includegraphics[width=0.45\textwidth]{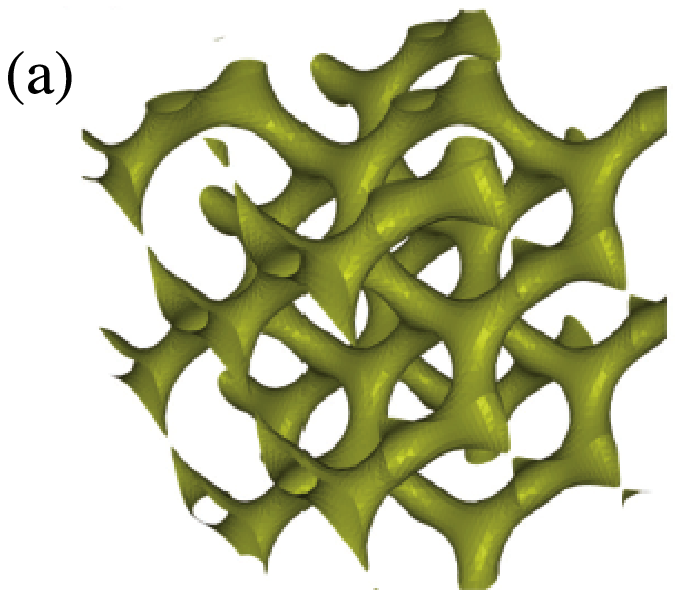}}\hfill
      \subfloat{\label{fig:3dNIM-SA-SEM}\includegraphics[width=0.45\textwidth]{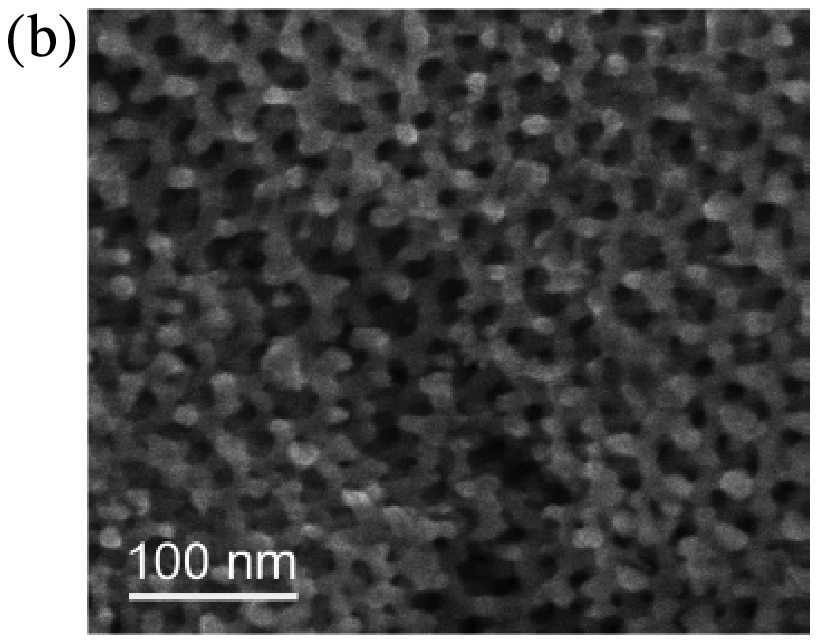}}
     \caption[Metamaterial fabricated with self-assembly.]{A metamaterial created by self-assembly, \protect\subref{fig:3dNIM-SA} shows a computer simulation of the metamaterial structure and \protect\subref{fig:3dNIM-SA-SEM} shows an image of the actual metamaterial taken using a scanning electron microscope. \textsc{Source:} \emph{Figures~2(d) and 1(b) in S.\ Vignolini et al.\ \emph{Adv.\ Mater.,\ }24:OP23-OP27, 2012.}}\label{fig:SA-NIM}
\end{figure}

The metamaterial, see Fig.~\ref{fig:SA-NIM},
was made by combining three substances into what is called a block copolymer, which is two or more individual polymers joined by a covalent bond. To fabricate the metamaterial, the block copolymer is allowed to self assemble. In this process, two of the three blocks of the copolymer form interwoven structures, called interpenetrating gyroid networks, within a bulk framework of the third block.

One of the interwoven networks is then removed through UV and chemical etching, leaving a void. This void is then filled with a metal, in this case gold, and the other two blocks are removed leaving a metal structure. Due to the periodic nature, and small unit cell, of the remaining metal structure it acts as a metamaterial. The structure made by Vignolini et al. was polarization sensitive, but it is not unreasonable to imagine a self assembly structure that leads to a polarization independent metamaterial.

Techniques for fabricating metamaterials will continue to improve in the future, and will no-doubt reach a level of reliability and scalability that optical metamaterials, and perhaps metamaterials for higher frequency electromagnetic radiation, will be available for commercial use. What is not clear at the moment is what benefits to existing technology and new metamaterial inspired technology will emerge.

\section{Electromagnetic Response and Energy of Metamaterials}\label{sec:MMs}

This work uses the the so-called fishnet metamaterial, as this design is well studied~\cite{Boltasseva:2008,Zhou:2008,Xiao:2009,Penciu:2010}, can be tailored for optical wavelengths, and also can be fabricated with current techniques. The fishnet metamaterial is a two-dimensional structure that consists of two metal layers separated by a third layer, which may be dielectric or vacuum. A periodic structure of holes with sub-wavelength dimensions is introduced onto these layers. The holes are sub-wavelength to ensure that their effect is averaged out over a wavelength, creating an effectively homogeneous electromagnetic response. 

As the electromagnetic permittivity response of metals at optical frequencies is very strong it is the dominant contributor to the permittivity of metamaterials. Thus, for fishnet metamaterials the permittivity is the same as it is for the constituent metal and given by Eq.~(\ref{epsilonMM}). The permeability, on the other hand, is due to the sub-wavelength structure of the metamaterial. 

The magnetic response of the split-ring resonator geometry is due to currents flowing at the edges of the rings~\cite{Koschny:2007,Economou:2008}. These currents are driven by the oscillations of the incident electromagnetic field. These oscillating currents generate a magnetic field that opposes the incident magnetic field, thereby reducing the net magnetic field within the metamaterial. In this way, a metamaterial responds diamagnetically to incident electromagnetic fields.

In a standard metal, the path that electrons travel in response to an electric field is interrupted only by the presence of atoms, which cause the electrons to scatter. In a metamaterial, the path of some of the electrons is interrupted by the sub-wavelength structure, and the electrons are forced to travel around the holes. This creates small oscillating current loops, which generate magnetic fields. The resulting permeability has the form~\cite{Penciu:2010}
\begin{equation}
\frac{\mu(\omega)}{\mu_{0}}=\mu_{\text{b}}+\frac{F \omega^2}{\omega_0^2-\omega\left(\omega+i\Gamma_{\text{m}}\right)},\label{muMM}
\end{equation}
with $\mu_{0}$ the permeability of free space, $\mu_{\text{b}}$ the background permeability, $F$ a geometric parameter, $\omega_0$ the resonance frequency, and $\Gamma_{\text{m}}$ the magnetic damping term. The combination of permittivity from the metal layers and permeability resulting from the sub-wavelength structures gives metamaterials unusual electromagnetic properties that are not possible with just the constituent materials alone.

The correct method for calculating the energy in an electromagnetic field must take into account the material in which it is propagating. The electromagnetic energy density $u$ in vacuum or a dispersion-free dielectric is given by~\cite{LandauLifshitz:1960}
\begin{equation}
u=\frac{1}{2}\left(\epsilon |\bm E|^{2}+\mu|\bm H|^{2}\right).\label{vacEMU}
\end{equation}
For electromagnetic fields propagating in a metamaterial with a negative index of refraction (both $\epsilon<0$ and $\mu<0$), Eq.~(\ref{vacEMU}) is not valid as it implies a negative energy. In fact, Eq.~(\ref{vacEMU}) cannot be used to calculate the energy of a field propagating in any dispersive material~\cite{LandauLifshitz:1960}.

The correct form of the electromagnetic energy density in dispersive materials is
\begin{equation}
u=\frac{1}{4}\left(\frac{\text d\left[\text{Re}(\epsilon(\omega))\omega\right]}{\text d \omega} |\bm E|^{2}+\frac{\text d\left[\text{Re}(\mu(\omega))\omega\right]}{\text d \omega}|\bm H|^{2}\right),\label{dispEMU}
\end{equation}
which reduces to Eq.~(\ref{vacEMU}) for constant permittivity and permeability. This more general form, however, is still not sufficiently general to accommodate metamaterials~\cite{Cui:2004,Cao:2006,Luan:2009}. Equation~(\ref{dispEMU}) can still be negative for frequencies where the refractive index is negative. For metamaterials with a permittivity and permeability given by Eqs.~(\ref{epsilonMM}) and (\ref{muMM}), respectively, and $\epsilon_{\text b}=\mu_{\text b}=1$, the following form for the electromagnetic energy density has been proposed~\cite{Luan:2009}
\begin{equation}
u=\frac{\epsilon_{0}|\bm E|^{2}}{4}\left(1+\frac{\omega_{\text e}^{2}}{\omega^{2}+\Gamma_{\text e}^{2}}\right)+\frac{\mu_{0}|\bm H|^{2}}{4}\left(1+F\frac{\omega^{2}(3\omega_{0}^{2}-\omega^{2})}{(\omega_{0}^{2}-\omega^{2})^{2}+\omega^{2}\Gamma_{\text m}^{2}}\right).\label{nimenergy}
\end{equation}
This form for the electromagnetic energy density is derived by considering the mechanism for power loss in a metamaterial, namely $I^{2}R$ heating, to ensure that it is treated properly. Equation~(\ref{nimenergy}) remains positive even for negative index metamaterials, so this form of the electromagnetic energy is used for all energy calculations.

As the behaviour of metamaterials is not intuitive, it has been helpful and enlightening to use the language of magnetic monopoles when discussing metamaterials. Magnetic monopoles are not known to exist, but the ideas and terminology are good tools to highlight the uniqueness of metamaterials.  Discussing the magnetic response of metamaterials in terms of magnetic charge carriers, both free and bound, is more intuitive than talking about split-ring resonators and oscillating current loops.

The use of magnetic monopole terminology is not without motivation; consider the quantization of electric charges. It was shown by Paul Dirac that if any magnetic monopoles exist, then electric charges must be quantized~\cite{Dirac:1931}. That is not to say that quantized charge necessitates the existence of magnetic monopoles, but it is an interesting idea that, when used to discuss metamaterials, can help to elicit clear discussions and explanations. In fact, Veselago himself used this language in an attempt to clarify the nature and possible construction of metamaterials~\cite{Veselago:1968}.

When discussing the response of materials to electromagnetic fields, we talk about the movement of charge carriers, which are electrons. In a metal the electrons form a plasma and are free to move about. Electromagnetic fields cause the electrons to oscillate, resulting in fields created inside the material. Electrons in metamaterials respond in much the same way, but the artificial structures alter their behaviour somewhat, which causes a magnetic response as well as an electric response.

Some of the modes that are supported by a metamaterial-clad guide are surface plasmon-polaritons, which arise through coherent oscillations of electrons near the surface of the metamaterial. At some frequencies, a portion of the energy of the surface plasmon-polariton modes is carried by the magnetic response of the metamaterial, i.e.\ oscillating magnetic dipoles. The magnetic dipoles are created by oscillating currents in the metamaterial, which are induced by the incident electromagnetic field. 

To simplify discussions, we have found that it is helpful to think of the magnetic response in an analogous way to the electric response. In this way, a surface plasmon-polariton travelling on a metamaterial-dielectric interface can be thought of as arising due to coherent oscillations of both electric and magnetic charges in the metamaterial. To this end we can imagine a material with free magnetic charges, rather than electrons, which we will call a ``magnetic metal''.

A ``magnetic metal'' would have a magnetic response that mirrors the electric response of a standard metal. By combining the two types of metals, a metal with two coexisting plasmas would result, one electric, consisting of electrons, and one magnetic, consisting of magnetic monopoles. Incident electromagnetic fields would cause both the electric and magnetic plasmas to oscillate giving a metamaterial its unusual electromagnetic response. In this way, then, a metamaterial can be thought of as an alloy of two metals, one a standard metal with an electric plasma, and one a ``magnetic metal'' with a magnetic plasma. 

The permeability of the type of metamaterial that is considered here arises from current loops that are in fixed locations, and so the alloy analogy is not quite applicable. It is more correct to imagine a magnetic analogue of an atom, with magnetic charges of one polarity in the centre, forming a nucleus, and the opposite polarity magnetic charges orbiting this nucleus. By doping these magnetic atoms into a metal, the magnetic charges bound to the magnetic atoms would elicit a magnetic response resulting in a permeability similar to Eq.~(\ref{muMM}). The resulting metamaterial would behave much like actual metamaterials.

%% file: LLmodesMMwgs.tex
\chapter{Low-Loss Surface Modes and Lossy Hybrid Modes in Metamaterial Waveguides}\label{ch:mmwgs}
\section{Introduction}

Electromagnetic metamaterials are materials designed to produce an electromagnetic response that is not possible with just the constituent materials alone~\cite{Balmain:2005}. Within the past decade metamaterials have been of interest~\cite{Boardman:2005,Ramakrishna:2005}, with metamaterials that operate at optical frequencies under current development~\cite{Shalaev:2007,Boltasseva:2008,Xiao:2009}. Potential uses include cloaking devices~\cite{Cai:2007}, and perfect lenses~\cite{Pendry:2000}. Given that the electromagnetic response of a metamaterial is, in principle, tailorable, there has been considerable interest in developing metamaterial waveguides~\cite{Shadrivov:2003,Halterman:2003,Qing:2005, He:2008,D'Aguanno:2005}. The modes supported by metamaterial waveguides exhibit a wide range of properties with applications to improving the design of optical and near ultra-violet systems. We investigate the properties of modes in both slab and cylindrical waveguides with a lossy metamaterial cladding and show how both waveguide geometries are able to support low-loss surface modes.

By exploiting effects that arise from surface plasmon-polaritons propagating along a single interface~\cite{Kamli:2008,Moiseev:2010}, a metamaterial-dielectric waveguide is able to support low-loss surface modes. The low-loss modes in a metamaterial-dielectric waveguide display a lower attenuation than is possible for the same mode in a metal-dielectric waveguide, thereby opening up the possibility of low-loss plasmonic waveguides. The low-loss surface modes we observe arise due to the same interference effects that cause the near-zero-loss surface mode observed on a single metamaterial-dielectric interface~\cite{Kamli:2008}. The single-interface waveguide is used to confine the fields in order to enhance optical nonlinearity for all-optical control at a single-photon level~\cite{Moiseev:2010}. Rather than using a single-interface waveguide for all-optical control, employing a low-loss surface mode of a cylindrical waveguide, which confines the fields in the transverse direction, could enhance the optical nonlinearity that is required for all-optical control of pulses.

Energy loss, which is a major obstacle in the implementation of metamaterial waveguides, is present in metamaterials due to scattering and heating effects in the materials used to construct them~\cite{Tassin:2012}.  The artificial structure of metamaterials alters the scattering and heating effects of the materials~\cite{Penciu:2010}. To fully understand the properties of the modes in a metamaterial waveguide, it is crucial to include loss in the model. So far little consideration has been paid to the effects of loss~\cite{D'Aguanno:2005,He:2005}. In order to model energy loss, we allow both the permeability and permittivity of the metamaterial to be complex-valued and frequency-dependent. For the permeability we use a Lorentz-like model that describes the behaviour of metamaterials designed for optical frequencies~\cite{Penciu:2010}, whereas the standard Drude model is used for the permittivity of the metamaterial. 

We compute dispersion, attenuation, and effective guide width for various modes of a cylindrical guide and, for comparison purposes, a slab guide. We show that a metamaterial-dielectric waveguide supports both surface and ordinary TM modes as well as hybrid ordinary-surface TM modes. Hybrid ordinary-surface modes are supported in metal-dielectric parallel-plate waveguides~\cite{Barlow:1973}. Hybrid modes are not predicted by lossless models. We find that a metamaterial-dielectric guide is able to support hybrid modes over a much wider frequency range than is seen in metal-dielectric guides. Some modes have attenuation curves with sharp changes over narrow frequency regions. The frequency at which this change occurs depends on the structural parameters, thereby enabling a metamaterial-dielectric guide to act as a frequency filter.


\section{Theory}

To characterize light in a waveguide, we solve the wave equation, given by Eqs.~(\ref{generalwaveeqCh2}) and (\ref{generalwaveeqCh2-B}), in each region (core and cladding) for both the electric field, ${\bm E}(x,y,z;t)$, and the magnetic field, ${\bm H}(x,y,z;t)$. We are interested in traveling plane-wave solutions, which have the form of Eq.~(\ref{fieldpropform}).

As the permittivity and permeability of the claddings are complex functions, so too is the propagation constant $\tilde{\beta}:=\beta+i\alpha$ with $\alpha$ the attenuation per unit length for a propagating field and $\beta$ the propagation wavenumber. Both $\beta$ and $\alpha$ are real, and $\alpha>0$. A negative value for $\alpha$ implies the material is active (exhibits gain) whereas negative $\beta$ corresponds to fields propagating in the negative $z$ direction. Once the solutions to Eq.~(\ref{generalwaveeqCh2}) in each region are found, we apply the appropriate boundary conditions at each interface. Applying the boundary conditions yields the dispersion relation for the waveguide, which is geometry-specific.

As we are interested in lossy waveguides, the permittivity and permeability of the metamaterial must both be complex-valued and are denoted $\epsilon=\epsilon'+i\epsilon''$ and $\mu=\mu'+i\mu''$, respectively, for $\epsilon'$, $\epsilon''$, $\mu'$, and $\mu''$ all real, and $\epsilon''>0$ and $\mu''>0$. The expressions for $\epsilon$ and $\mu$ are determined by how metamaterials are constructed~\cite{Penciu:2010}. A common approach used in metamaterial design for optical frequencies is to exploit the naturally occurring electric response of metals to achieve a negative permittivity and design the structure of the metamaterial to generate a magnetic response that gives a negative permeability~\cite{Boltasseva:2008}. 

The natural electric response of metals to an electromagnetic field is described by a frequency-dependent expression for the permittivity given by the Drude model as defined in Eq.~(\ref{epsilonMM}). 
The structurally induced magnetic response of a metamaterial takes the form of Eq.~(\ref{muMM}). 
The magnetic response is generated by resonant currents that are driven by the electromagnetic field. For parameters that are real and positive, the real part of both Eqs.~(\ref{epsilonMM}) and (\ref{muMM}) can be either positive or negative whereas the imaginary parts are strictly positive.

The index of refraction of any lossy material is in general complex and given by
\begin{equation}
n=\pm\sqrt{\frac{\epsilon\mu}{\epsilon_0\mu_0}}=n_{\rm{r}}+i n_{\rm{i}},\label{complexn}
\end{equation}
where the sign is chosen such that $n_{\rm{i}}\geq 0$ to ensure the material is passive (gain-free). The other case, $n_{\rm{i}}< 0$, corresponds to a material that exhibits gain; i.e., the intensity of the fields increases as the fields propagate through the material. The choice of the sign in Eq.~(\ref{complexn}) to ensure $n_{\rm{i}}>0$, means that at frequencies where both $\epsilon'<0$ and $\mu'<0$, the metamaterial behaves as a negative-index material with $n_{\rm{r}}<0$. 

The frequency region where $\epsilon'$ and $\mu'$ have opposite signs (one negative and one positive) correspond to metamaterials that do not allow propagating waves. Many metals at optical frequencies act as electric plasmas and are characterized by $\epsilon'<0$ and $\mu'>0$; hence, for metamaterials that are similarly characterized, we use the term metal-like to describe them. The metal-like region is what D'Aguanno et al.\ call the opaque region~\cite{D'Aguanno:2005}. For most frequencies in the metal-like region $n_{\rm{i}}>\left|n_{\rm{r}}\right|$. In contrast, metamaterials with $\epsilon'>0$ and $\mu'<0$ are not metal-like per se but would correspond to materials that display electromagnetic responses consistent with an effective magnetic plasma~\cite{Ramakrishna:2005}. For frequency regions with both $\epsilon'>0$ and $\mu'>0$, the metamaterial simply behaves as a dielectric with a frequency-dependent refractive index.

We begin by studying the slab geometry as it has features in common with both a single interface (no transverse confinement) and cylindrical waveguides (supports multiple types of modes). Additionally, the slab waveguide has been studied for various materials~\cite{Yeh:2008,Kaminow:1974}, which allows us to check our methods against accepted results. The slab waveguide consists of three layers as seen in Fig.~\ref{fig:guidediagramCh2}.
The centre layer is the core, and the width of the waveguide, $w$, is the thickness of the core. We choose the width of the slab guide to be $w=4\pi c/\omega_{\rm{e}}$, such that the frequency at which $w$ is equal to the wavelength of the field is near the middle of the frequency range in which the metamaterial is metal-like.

The core is surrounded on both sides by cladding. The cladding layers are sufficiently thick so as to substantially contain the evanescent fields, as the effective width of the guide (the width of the core plus the skin depth at the interface) is typically only slightly larger than $w$. The permittivity and permeability of the cladding material is described by Eqs.~(\ref{epsilonMM}) and (\ref{muMM}), whereas they are constant in the core.

The slab guide supports two distinct classes of modes, transverse magnetic (TM) and transverse electric (TE). The TM modes have non-zero magnetic field components only in the transverse direction ($H_{z,x}=0$) whereas the TE modes have electric field components which are non-zero only in the transverse direction ($E_{z,x}=0$). The existence of TM surface modes in our system is partially dependent on the cladding having $\epsilon'<0$ and $\mu'>0$, whereas the TE modes require $\epsilon'>0$ and $\mu'<0$ in the cladding. As the expressions for $\epsilon$ and $\mu$, Eqs.~(\ref{epsilonMM}) and (\ref{muMM}) respectively, are different, we cannot expect TE surface modes to simply be a generalization of TM surface modes.  

We choose to study the TM modes, as the forms of the response functions allow for the existence of TM surface modes for a larger set of parameters than the TE surface modes. Applying the boundary conditions to the solutions of Eq.~(\ref{generalwaveeqCh2}) for the TM modes leads to the dispersion relation given by Eq.~(\ref{slabdisprelCh2}),
which must be satisfied for the existence of a propagating mode~\cite{Yeh:2008}. Here
\begin{equation}
\gamma_j:=\sqrt{\tilde{\beta}^2-\omega^2\epsilon_j\mu_j}\label{transversewavenumber}
\end{equation}
are the complex wave numbers for the transverse components of the fields for $j=1,\,2,\,3$ referring to the core and two cladding layers respectively. As Eq.~(\ref{slabdisprelCh2}) is transcendental, we solve numerically for the complex propagation constant $\tilde{\beta}$.

Cylindrical guides enable transverse confinement of the modes but are of interest to us for more than this reason. The cylindrical geometry is well studied in other contexts and is used in a number of applications, e.g.,\ fibre optics. Additionally, as for the slab geometry, the dispersion relation for a guide with cylindrical geometry can be obtained analytically. The cylindrical waveguide, shown in Fig.~\ref{fig:guidediagramCh2}, consists of a dielectric cylindrical core of circular cross-section with radius $a$. A cladding layer, which has a frequency dependence described by Eqs.~(\ref{epsilonMM}) and (\ref{muMM}), surrounds the core. As with the slab geometry, the cladding layer is assumed to be arbitrarily thick.

We again assume traveling-wave solutions, but we must include all components of both electric and magnetic fields. In general the modes of a cylindrical guide are not TE or TM but in-between modes that contain all of the field components and are called HE or EH~\cite{Yeh:2008}. The names HE and EH allude to the fact that all components of both electric and magnetic fields are non-zero. The TM and TE modes of the cylindrical guide are special cases of the general solution, which is given by Eq.~(\ref{cyldisprel}). With the dispersion relations for both geometries, we are able to determine the propagation constants for the allowed modes.




\section{Modes}\label{sec:LLWGmodes}

The metamaterial-dielectric slab waveguide supports a number of TM modes with three distinct mode behaviours, namely ordinary, surface and hybrid modes. Each mode may display more than one behaviour, with different behaviours supported for different frequency regions. The mode behaviour is determined by the relative sizes of the real and imaginary parts of $\gamma_{1}$, the wavenumber perpendicular to the interface inside the core. To simplify the discussion we use ordinary wave, surface wave, and hybrid wave when referring to the character, or behaviour, of a mode and use the term mode or explicitly name the mode (e.g.\ TM$_{j}$) when we are discussing a specific solution to the dispersion relation. 

We label the modes in a manner consistent with lossless waveguides~\cite{Yeh:2008}:\ A numeric subscript $j$ refers to ordinary waves and indicates how many \emph{zeros} (how many times the $H_{y}$ field changes sign) the mode has inside the core. The subscripts ``s'' and ``a'' indicate the symmetric and antisymmetric surface modes, respectively. Symmetric and anti-symmetric refers to the symmetry present in the $H_{y}$ field. We extend the labelling scheme to the modes of lossy waveguides even though the defining features may not be present for all frequencies.

Using the number of zeros in the $H_{y}$ field inside the core as the definition of the TM$_{j}$ modes is not the best way for the modes in a lossy waveguide to be distinguished. As the modes have both real and imaginary field components, it is not always clear when using the components individually which mode is being considered. However, by plotting the magnitude of the magnetic field $|H_{y}|^{2}=H_{y}H_{y}^{*}$ as a function of $x$, the modes can readily be distinguished from one another. 

\begin{figure}[t,b] 
      \centering
	\subfloat{\label{fig:TM0def}\includegraphics[width=0.5\textwidth]{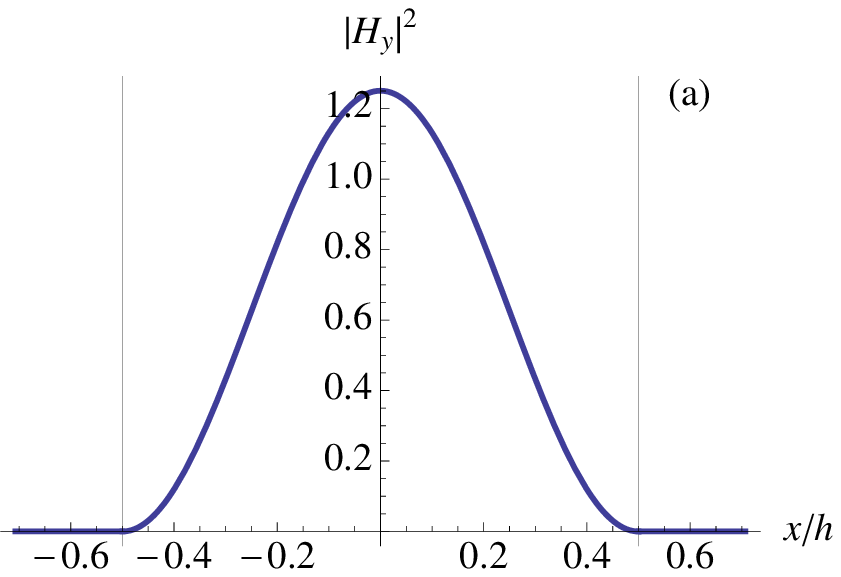}}\\
	\subfloat{\label{fig:TM1def}\includegraphics[width=0.5\textwidth]{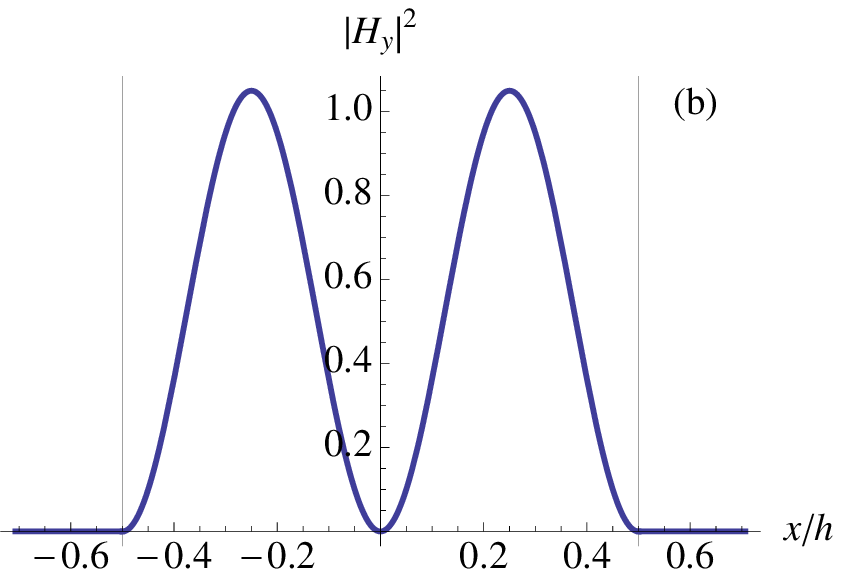}}
	\subfloat{\label{fig:TM2def}\includegraphics[width=0.5\textwidth]{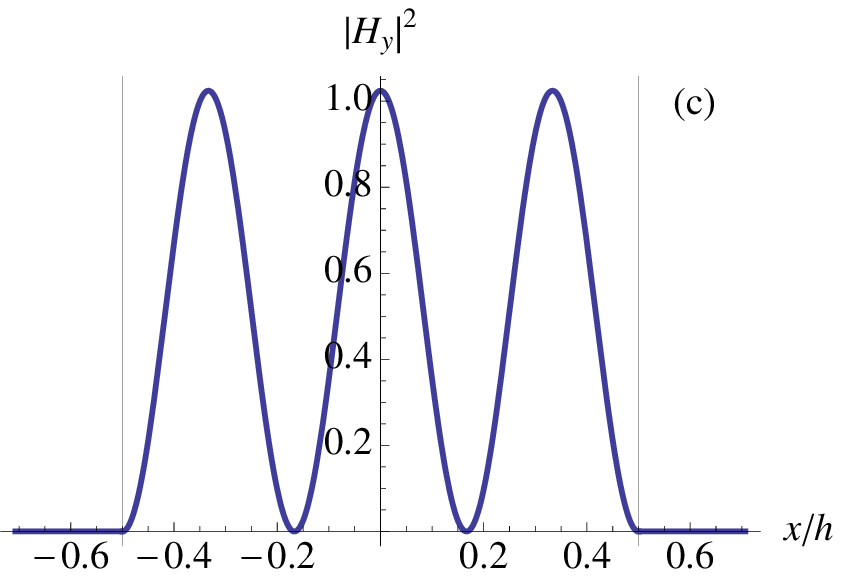}}
	\caption[Plots showing how the labels TM$_{j}$ are defined.]{Plots showing $|H_{y}|^{2}$ for the TM$_{0}$ \protect\subref{fig:TM0def}, TM$_{1}$ \protect\subref{fig:TM1def} and TM$_{2}$ \protect\subref{fig:TM2def} modes of the slab guide. The index $j$ in the label TM$_{j}$ can be defined by either $j+1$ equals the number of local maxima within the core, or $j$ equals the number of local minima, not including those at the core-cladding boundaries. The thin vertical lines indicate the core-cladding boundaries.}
     \label{fig:TMjdef}   
\end{figure}

The definition of $j$ in the mode labels can be done in one of two equivalent ways. Either, $j+1$ equals the number of local maxima that $|H_{y}|^{2}$ has inside the core, or $j$ equals the number of local minima that $|H_{y}|^{2}$ has inside the core, discounting those at the core-cladding boundaries. Figure~\ref{fig:TMjdef} shows three examples of $|H_{y}|^{2}$ for TM$_{j}$ modes to show their relation to the value of $j$.

Ordinary waves are caused by traveling electromagnetic fields inside the core reflecting from the core-cladding interface(s). The reflected fields overlap with the incident fields causing an interference effect, with constructive interference corresponding to the allowed modes. The condition for ordinary-wave behaviour is $\rm{Im}\left(\gamma_{1}\right)>> \rm{Re}\left(\gamma_{1}\right)$, which ensures that, in the slab guide, the fields have an oscillating wave pattern in the transverse direction of the core.

Surface waves are the result of electrons at the core-cladding interface carrying energy along the waveguide. The energy from the fields is transferred to the electrons causing them to oscillate, providing a medium for energy transport. This type of collective oscillation of electrons at a surface coupled with electromagnetic waves is called a surface polariton or surface plasmon-polariton~\cite{Nkoma:1974}.

\begin{figure}[t,b] 
      \centering
	\subfloat{\label{fig:TMsdef}\includegraphics[width=0.5\textwidth]{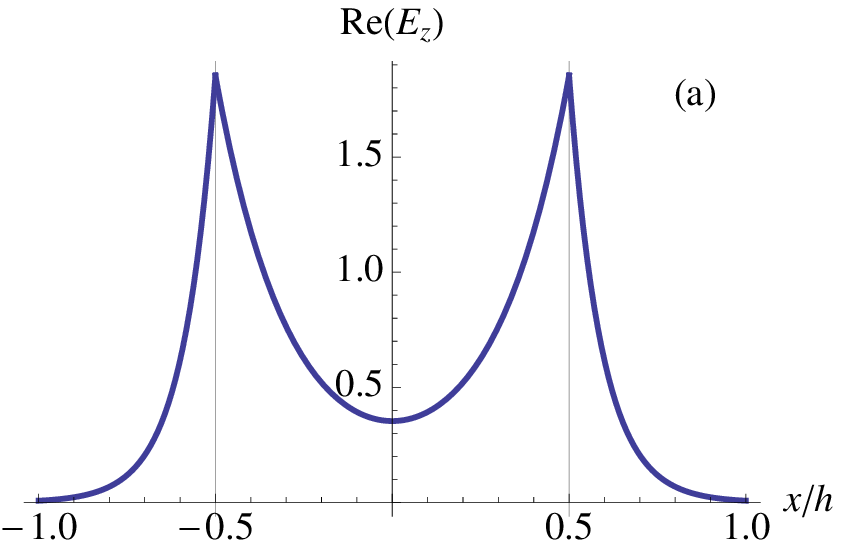}}
	\subfloat{\label{fig:TMadef}\includegraphics[width=0.5\textwidth]{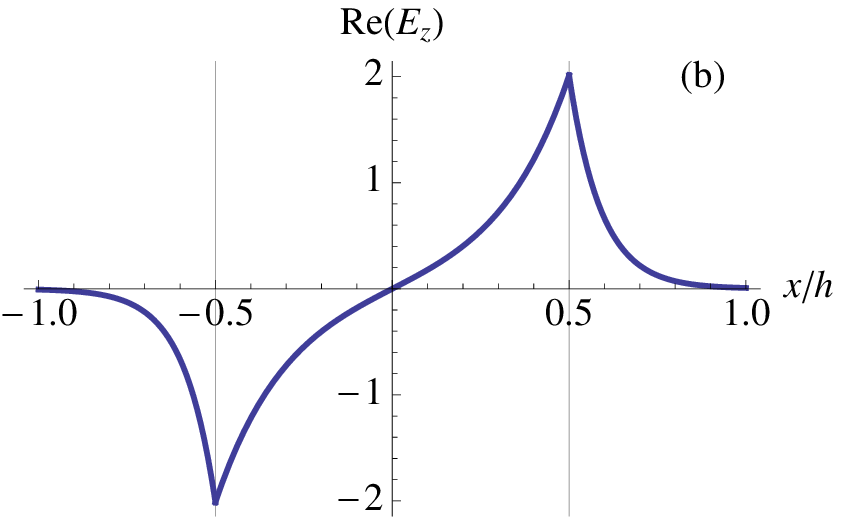}}
	\caption[Plots showing examples of the TM$_{\text s}$ and TM$_{\text a}$ modes.]{Plots showing $\text{Re}(E_{z})$ for the TM$_{\text s}$ \protect\subref{fig:TMsdef} and TM$_{\text a}$ \protect\subref{fig:TMadef} modes of the slab guide. These modes have only evanescent fields generated by coherent oscillations of electrons near the surface. The thin vertical lines indicate the core-cladding boundaries.}
     \label{fig:surfTMdef}   
\end{figure}

The polariton field of the surface mode decays exponentially as a function of the distance from the interface. Mathematically the condition for surface waves is $\rm{Re}\left(\gamma_{1}\right)>> \rm{Im}\left(\gamma_{1}\right)$, which causes the intensity to be concentrated around the interface(s) between the core and cladding and a relatively small intensity through most of the core. In slab guides only two surface wave TM modes are supported, namely symmetric (TM$_{\rm{s}}$) and anti-symmetric (TM$_{\rm{a}}$). Figure~\ref{fig:surfTMdef} shows $\text{Re}(E_{z})$ as a function of $x$ for the symmetric and anti-symmetric surface modes. Unlike ordinary modes, the two surface modes are readily distinguishable using $\text{Re}(E_{z})$. In fact, the surface modes are difficult to tell apart using $|H_{y}|^{2}$ because it is positive everywhere.

\begin{figure}[t,b] 
      \centering
	\subfloat{\label{fig:TM2hyb}\includegraphics[width=0.5\textwidth]{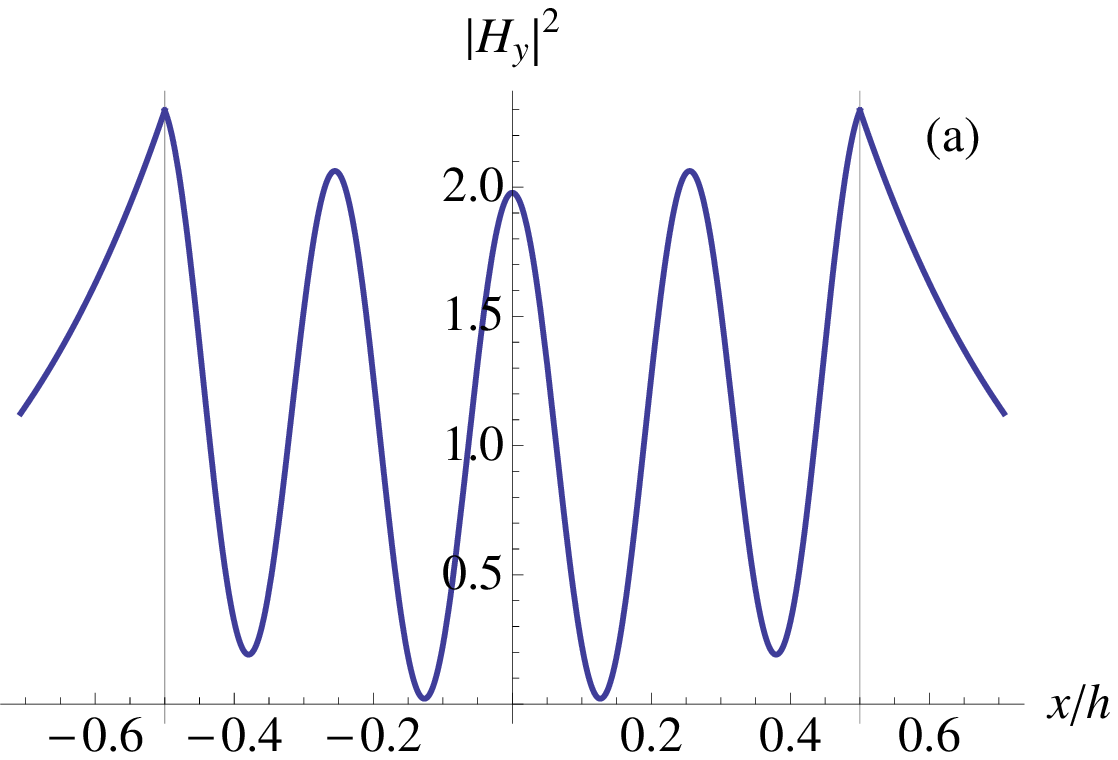}}
	\subfloat{\label{fig:TM3hyb}\includegraphics[width=0.5\textwidth]{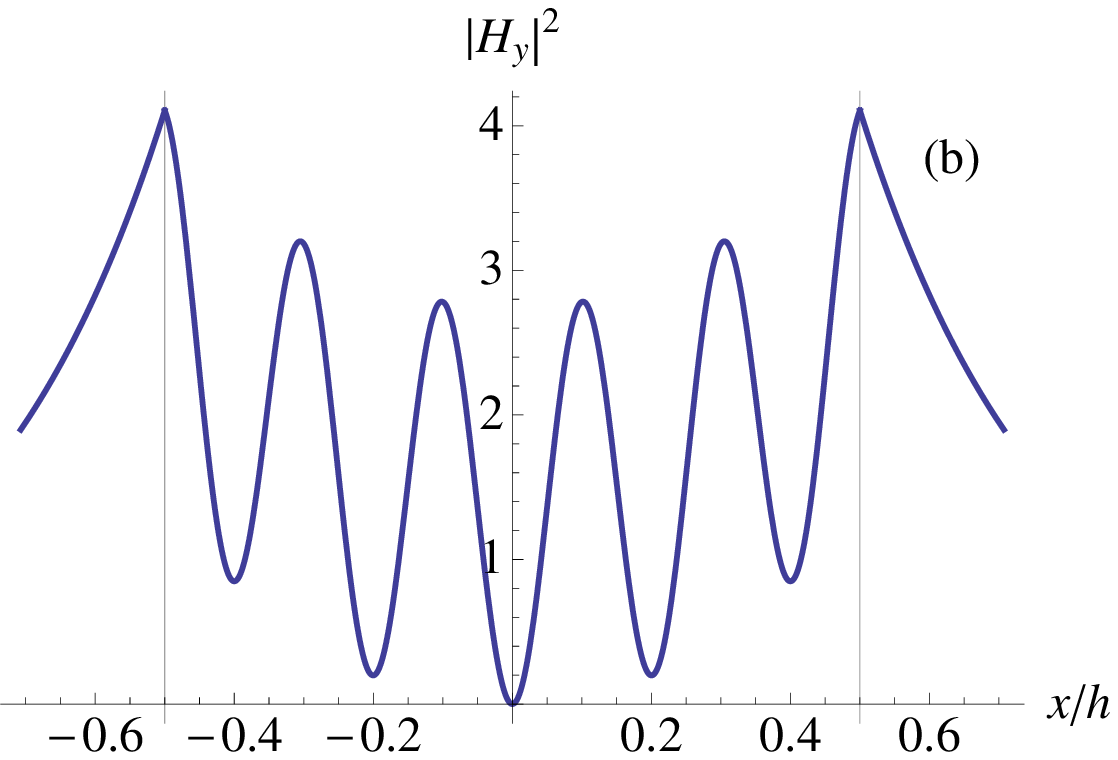}}
	\caption[Plots showing examples of the TM$_{2}$ and TM$_{3}$ hybrid modes.]{Plots showing $|H_{y}|^{2}$ for the TM$_{2}$ \protect\subref{fig:TM2hyb} and TM$_{3}$ \protect\subref{fig:TM3hyb} modes of the slab guide at frequencies where they are hybrid ordinary-surface modes. The same definition of the label $j$ cannot be used as with the ordinary TM$_{j}$ modes, due to the fact that $|H_{y}|^{2}$ for hybrid waves is not always consistent with the ordinary waves. The labels for the hybrid waves are chosen due to the fact that the dispersion and attenuation curves for these modes are continuations of those for the ordinary waves. The thin vertical lines indicate the core-cladding boundaries.}
     \label{fig:TMhyb}   
\end{figure}

In frequency regions where both $\rm{Re}\left(\gamma_{1}\right)$ and $\rm{Im}\left(\gamma_{1}\right)$ are comparable, the mode is a hybrid wave. Hybrid waves can be understood as a product of two features in the transverse direction, namely an evanescent feature and an oscillatory feature. To see how the two features combine, consider the components of the magnetic field in the core of the the slab waveguide, which have the form
\begin{align}
H&=H_{0}\e^{\gamma_{1}x}\e^{i(\tilde{\beta}z-\omega t)}\nonumber\\
&=H_{0}\e^{\gamma'_{1}x}\e^{i\gamma''_{1}x}\e^{i(\tilde{\beta}z-\omega t)},\label{hybridfield}
\end{align}
where $H_{0}$ is a constant, the complex wavenumber $\gamma_{1}=\gamma'_{1}+i\gamma''_{1}$, with $\gamma'_{1}$ and $\gamma''_{1}$ both real. The terms $\e^{\gamma'_{1}x}$ and $\e^{i\gamma''_{1}x}$ represent surface-wave and ordinary-wave behaviour, respectively. The hybrid waves are then a combination of surface waves, i.e., fields being carried by electron oscillations, and ordinary waves due to fields traveling in the core and reflecting off the interfaces, thereby resulting in interference effects. Figure~\ref{fig:TMhyb} shows plots of $|H_{y}|^{2}$ for the TM$_{2}$ and TM$_{3}$ modes of the metamaterial-clad slab guide at frequencies where they are hybrid modes.

The labelling scheme for the hybrid modes is not based on the number of local maxima or minima in a particular field component, as this may be different for hybrid modes than for ordinary modes. Instead, the labelling scheme is chosen based on the dispersion and attenuation curves for the waveguide. The dispersion and attenuation curves for the hybrid modes are extensions of those for some of the ordinary and surface modes. Thus, if a hybrid mode is found on a dispersion curve that transitions into one for an ordinary mode, TM$_{2}$ for instance, the hybrid mode is given the same label. The labelling scheme was chosen to reflect the fact that the transition between ordinary and hybrid, or surface and hybrid, modes is continuous. The choice of labels also allows a single dispersion or attenuation curve to be labelled with a single label, such as TM$_{2}$, which serves to avoid confusion.

For hybrid waves, a substantial portion of the energy is transferred along the interfaces as well as in the core, so hybrid waves are like a combination of ordinary waves and surface waves. Hybrid waves are not seen in dielectric-dielectric guides and only for a very narrow frequency range of the TM$_{\rm{a}}$ mode in metal-dielectric guides, where the mode changes from a surface wave to an ordinary wave.

The metamaterial-dielectric cylindrical guide supports the same three mode types as the slab guide, though they differ slightly as the cross section of the cylindrical guide has a circular symmetry. The fields of the ordinary wave TM modes in the cylindrical guide have the oscillating wave pattern of a vibrating circular membrane.  The fields in the cladding decay exponentially, as $\rm{Re}\left(\gamma_{2}\right)>> \rm{Im}\left(\gamma_{2}\right)$. The modes of the cylindrical guide are labeled similarly to the slab guide, with the transverse magnetic modes labeled as TM$_{j}$. For the cylindrical guide, however, a numeric value of $j$ refers to the number of \emph{zeros} in $H_{\phi}$ that lie between the centre of the core and the core-cladding boundary. 

As with the slab guide, the number of zeros is not always the best way to differentiate between the ordinary modes of a cylindrical guide that dissipates energy. Again, this is due to the fact that all of the field components have both real and imaginary parts and which mode they correspond to is not always clear from these individually. The ordinary modes of a cylindrical guide can be differentiated by comparing the number of local minima or maxima in plots of $|H_{\phi}|^{2}$ as a function of $r$.

\begin{figure}[t,b] 
      \centering
	\subfloat{\label{fig:TM0cyldef}\includegraphics[width=0.5\textwidth]{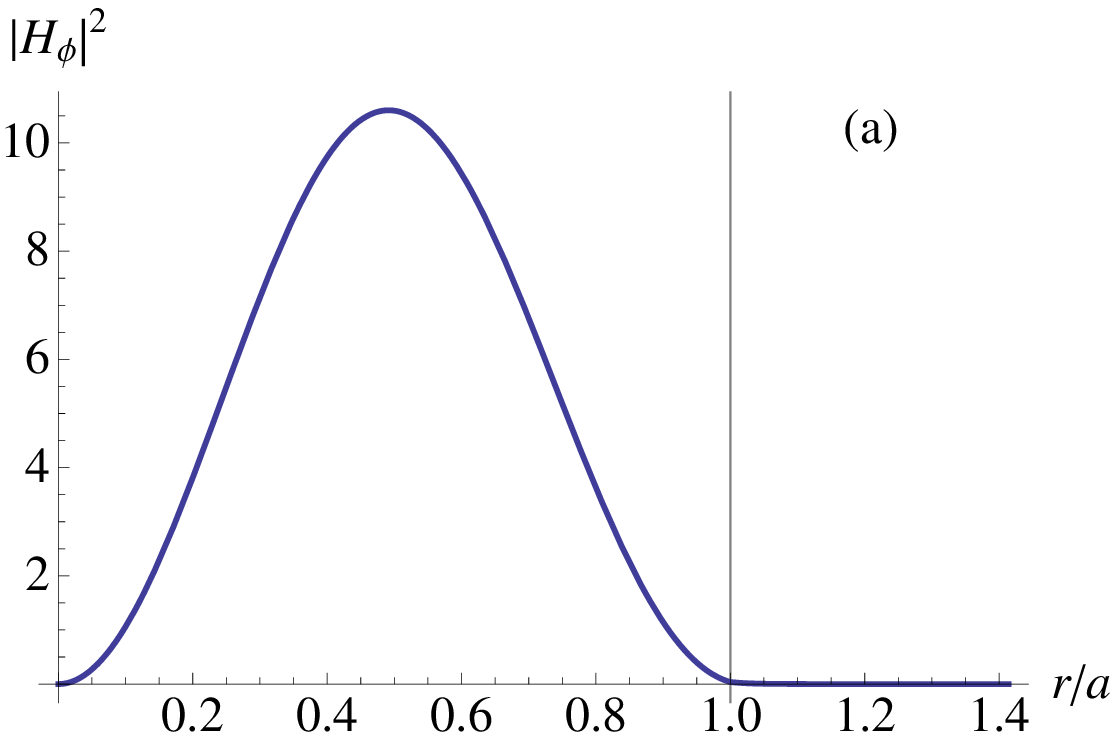}}\\
	\subfloat{\label{fig:TM1cyldef}\includegraphics[width=0.5\textwidth]{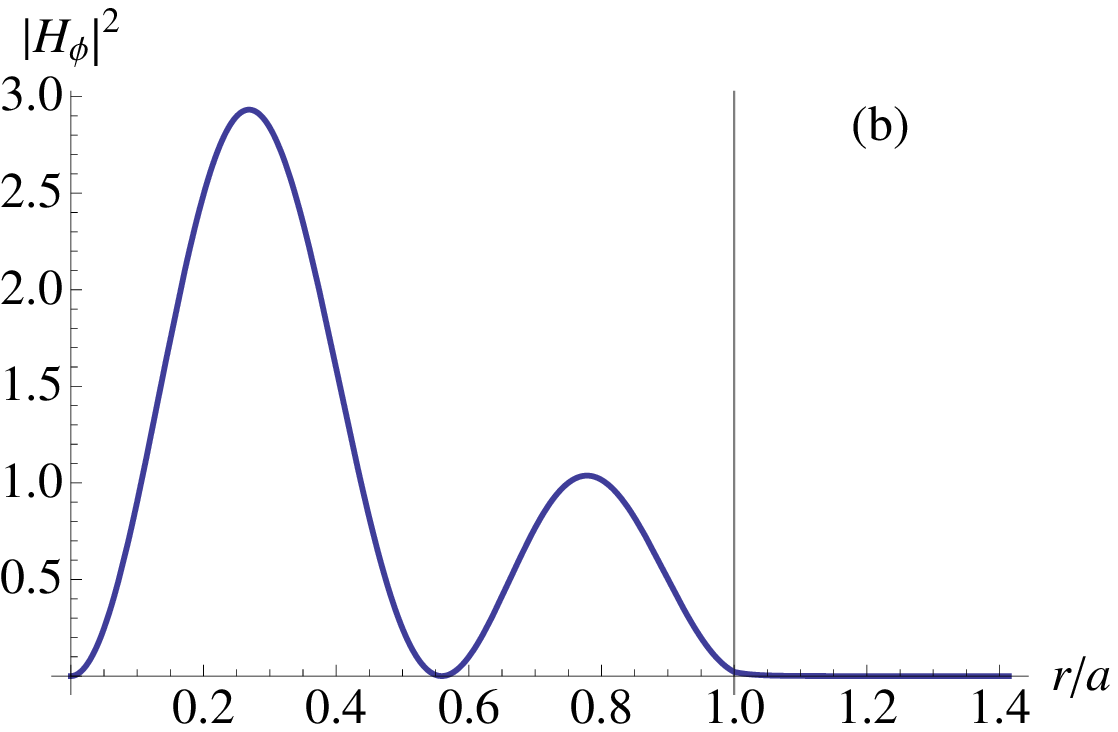}}
	\subfloat{\label{fig:TM2cyldef}\includegraphics[width=0.5\textwidth]{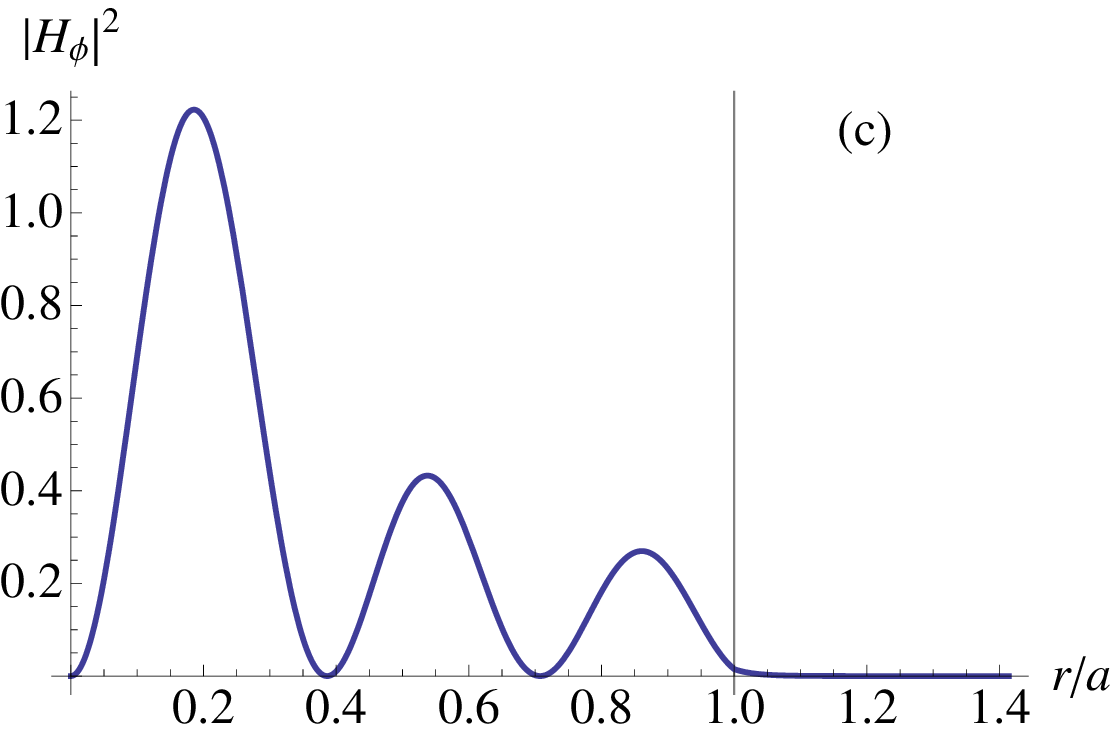}}
	\caption[Plots showing how the labels TM$_{j}$ are defined for the cylindrical guide.]{Plots showing $|H_{\phi}|^{2}$ for the TM$_{0}$ \protect\subref{fig:TM0cyldef}, TM$_{1}$ \protect\subref{fig:TM1cyldef} and TM$_{2}$ \protect\subref{fig:TM2cyldef} modes of the cylindrical guide. The index $j$ in the label TM$_{j}$ can be defined by either $j+1$ equals the number of local maxima the range $0<r<a$, or $j$ equals the number of local minima the range $0<r<a$, not including those at $r=0$, or $r=a$. The thin vertical line indicates the core-cladding boundary.}
     \label{fig:TMjcyldef}   
\end{figure}

To label the modes, a similar approach was used to the one for the slab guide; $j+1$ is equal to the number of local maxima in the range $0<r<a$, or equivalently $j$ equals the number of local minima in that range, not including any at $r=0$ or $r=a$. Figure~\ref{fig:TMjcyldef} shows $|H_{\phi}|^{2}$ as a function of $r$ for the TM$_{0}$, TM$_{1}$ and TM$_{2}$ ordinary modes of the metamaterial-clad cylindrical waveguide. As with the plots for the slab guide, the modes are clearly distinguishable by the number of local maxima and minima, which are used to define $j$ as described above.

The HE modes have two indices (HE$_{mj}$), where the azimuthal symmetry is characterized by the integer $m$, which determines the number of oscillations in the fields over an angle of $2\pi$. The index $m$ here is the same as that in Eq.~(\ref{cyldisprel}).  The index $j$ plays the same role for HE modes as it does for TM modes. Due to the circular symmetry of the cylindrical guide, it supports only one surface wave TM mode. However, the cylindrical guide supports a number of surface wave HE modes of different orders with the symmetry determined, as with the ordinary wave HE modes, by the azimuthal parameter $m$. Modes of the cylindrical guide with surface-wave behaviour are denoted by the index ``s'' for both TM and HE (e.g. TM$_{\text s}$, HE$_{1\text s}$).

The hybrid modes supported by the metamaterial-clad cylindrical waveguide are defined in the same way as those of the slab guide. The hybrid modes of the cylindrical guide, however, are harder to distinguish from ordinary modes than their slab guide counterparts. This is due to the fact that the ordinary modes of the cylindrical guide naturally have a decaying amplitude as a function of $r$.

\begin{figure}[t,b] 
      \centering
	\subfloat{\label{fig:TMscyl}\includegraphics[width=0.5\textwidth]{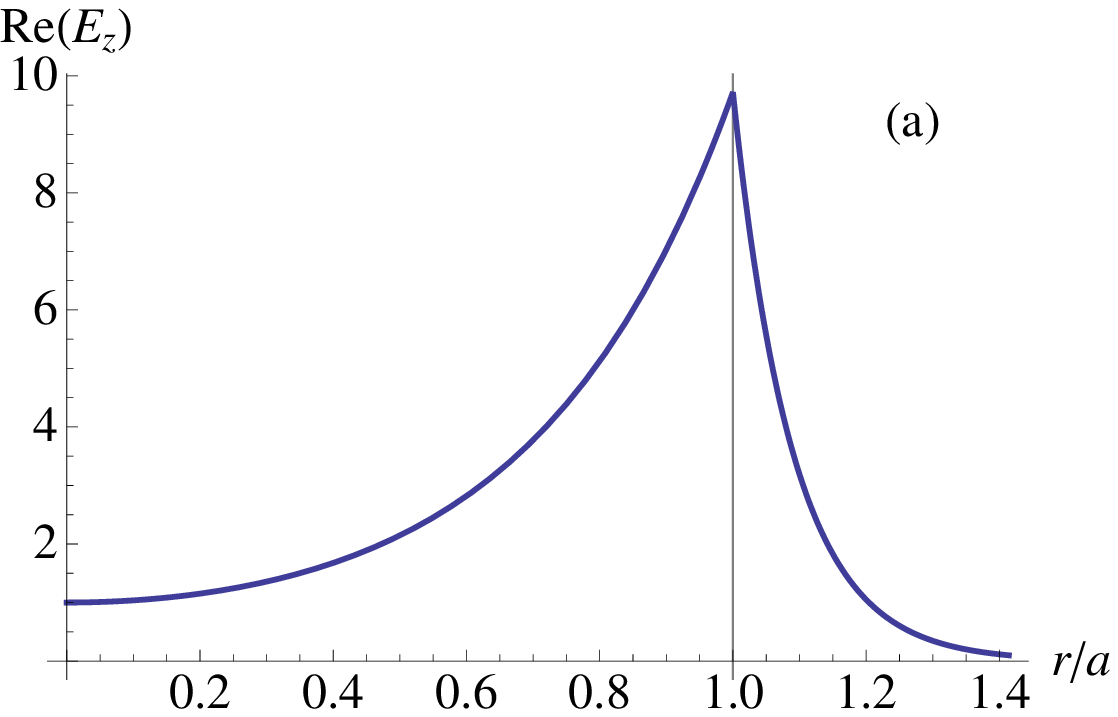}}\\
	\subfloat{\label{fig:TM1cylhyb}\includegraphics[width=0.5\textwidth]{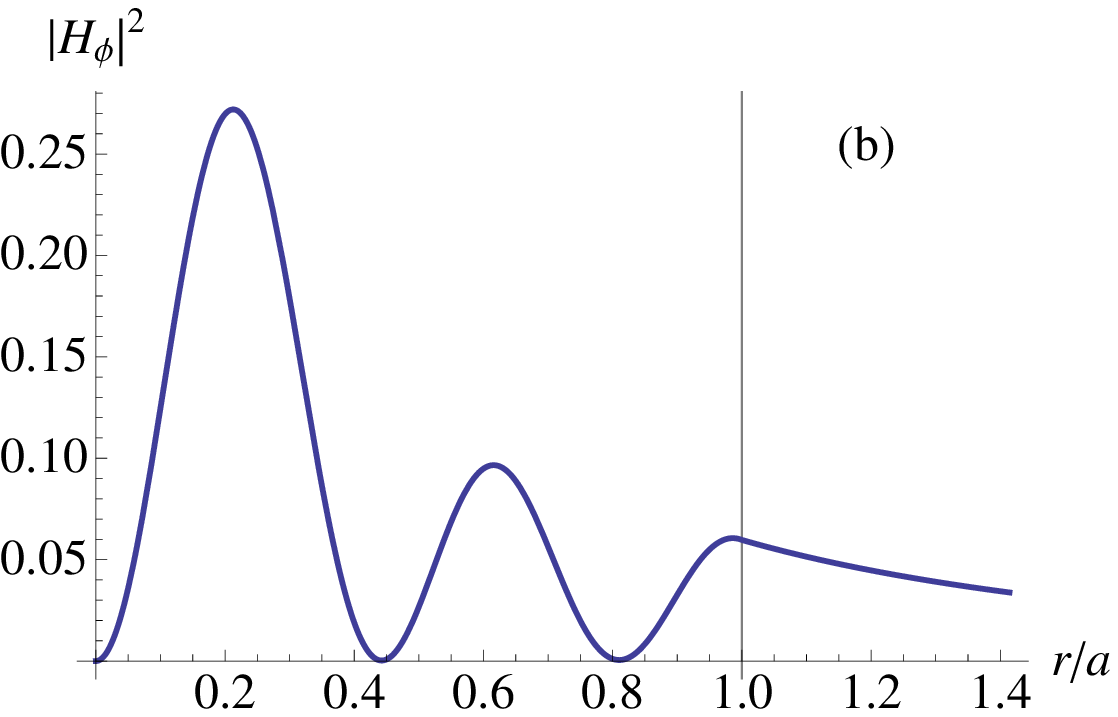}}
	\subfloat{\label{fig:TM3cylhyb}\includegraphics[width=0.5\textwidth]{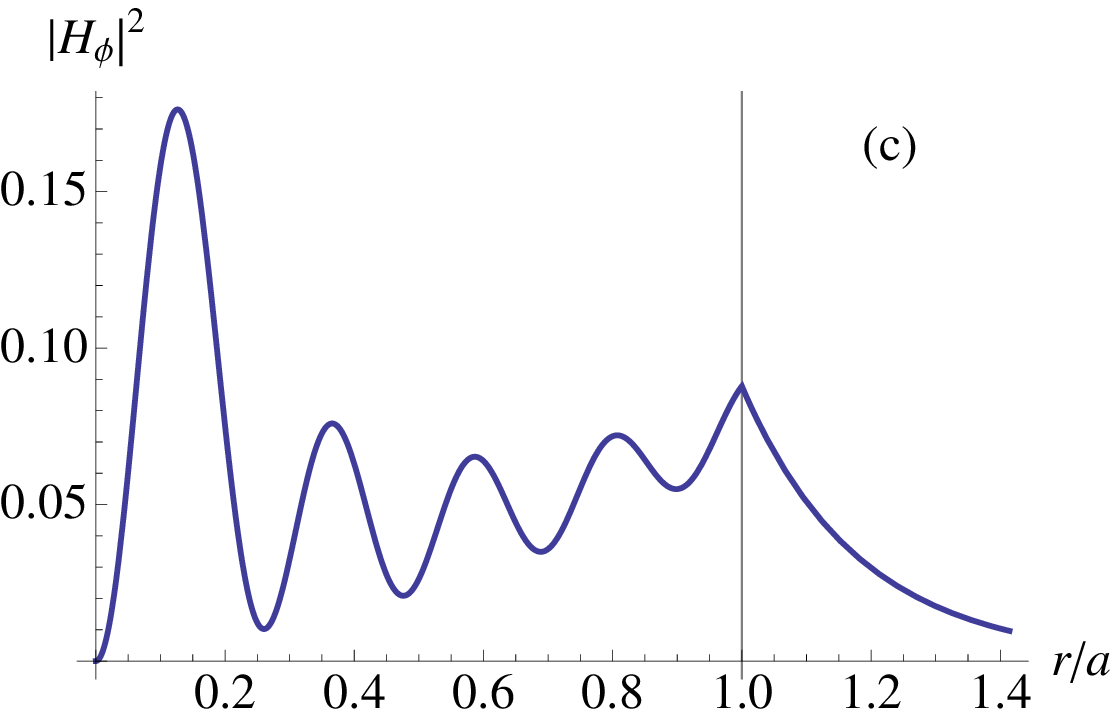}}
	\caption[Plots showing profiles of surface and hybrid modes for the cylindrical guide.]{Plots showing $\text{Re}(E_{z})$ for the TM$_{s}$ mode \protect\subref{fig:TMscyl}, and $|H_{\phi}|^{2}$ for the TM$_{1}$ \protect\subref{fig:TM1cylhyb} and TM$_{3}$ \protect\subref{fig:TM3cylhyb} hybrid modes of the metamaterial-clad cylindrical guide. As with the slab guide, surface modes are generally easily distinguished from ordinary modes. However, the hybrid modes can be hard to discern from ordinary modes. Here, the TM$_{3}$ mode has clear hybrid mode properties with the field amplitude increasing away from the centre of the core, whereas the TM$_{1}$ mode is not obviously a hybrid mode. One sign, however, that the TM$_{1}$ mode is hybrid is the slow decay of the field on in the cladding. The thin vertical line indicates the core-cladding boundary.}
     \label{fig:TMcylsurf-hyb}   
\end{figure}

Ordinary modes in a cylindrical guide are described by the Bessel functions $J_{\text m}(r)$, which decay with increasing values of $r$. The result is that it is often difficult to discern whether the mode is an ordinary mode or hybrid mode. One telltale sign of a hybrid mode is slow decay of the fields into the cladding. Figure~\ref{fig:TMcylsurf-hyb} shows a plot of $\text{Re}(E_{z})$ as a function of $r$ for the TM$_{\text s}$ mode, and $|H_{\phi}|^{2}$ as a function of $r$ for the TM$_{1}$ and TM$_{3}$ hybrid modes, for the metamaterial-clad cylindrical guide.


\section{Characterization}\label{sec:LLWGcharacter}

With the physical nature of the three wave types supported by a metamaterial-dielectric guide understood, we now characterize the waveguides. To characterize the two guide geometries, we examine how the effective refractive index, attenuation along the propagation direction, and effective guide width for a variety of supported modes vary with frequency. The following values are used for the metamaterial parameters~\cite{Kamli:2008}: $\omega_{\rm e}=1.37\times10^{16}{\rm s}^{-1}$, $\Gamma_{\rm{m}}=\Gamma_{\rm e}=2.73\times10^{13}{\rm s}^{-1}$, $\omega_0=0.2\omega_{\rm e}$, $F=0.5$ whereas the core is described by $\epsilon_1=1.3\epsilon_{0}$ and $\mu_1=\mu_{0}$.

\begin{figure}[t,b] 
      \centering
	\subfloat{\label{fig:epsmu}\includegraphics[width=0.5\textwidth]{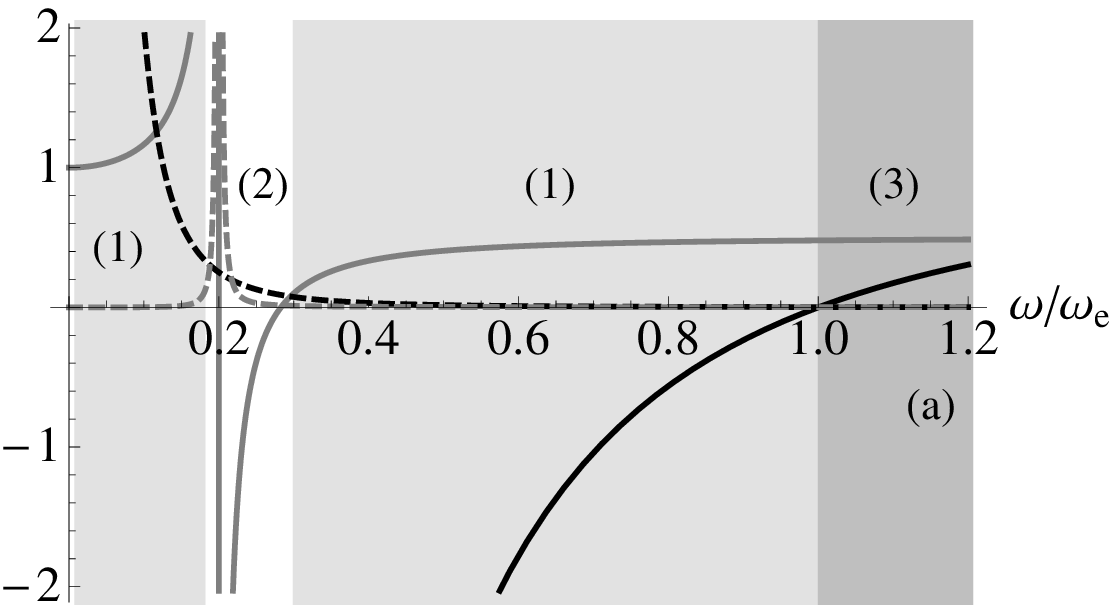}}
	\subfloat{\label{fig:nimmindex}\includegraphics[width=0.5\textwidth]{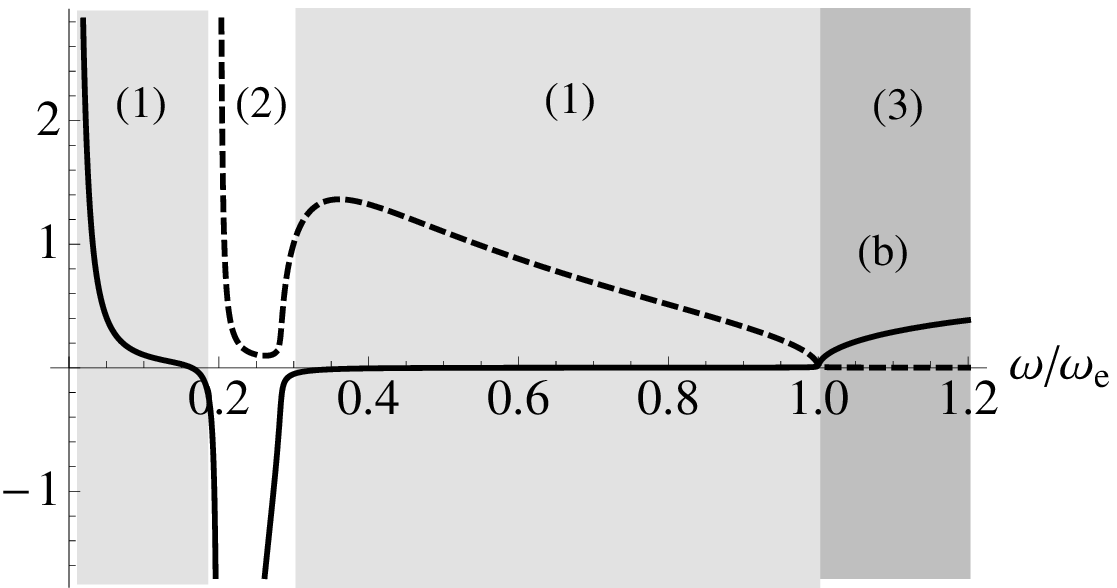}}
	\caption[Electromagnetic characteristics of a metamaterial.]{Plots of the real parts (solid lines) and imaginary parts (dashed lines) of \protect\subref{fig:epsmu} $\epsilon/\epsilon_{0}$ (black) and $\mu/\mu_{0}$ (grey), and \protect\subref{fig:nimmindex} the refractive index of the metamaterial as a function of frequency. Region (1) in each graph is the metal-like region, region (2) is the negative-index region and region (3) is the dielectric-like region.}
     \label{fig:nimmregions}   
\end{figure}

Frequency regions with differing refractive index characteristics for the metamaterial are identifiable (Fig.~\ref{fig:nimmregions})
when the above parameters are used along with Eqs.~(\ref{epsilonMM}) and (\ref{muMM}). Above $\omega_{\rm{e}}$ the metamaterial behaves as a lossy dielectric with both $\epsilon'>0$ and $\mu'>0$. There are two \emph{metal-like} regions, one for $0.3\omega_{\rm{e}}\lesssim\omega<\omega_{\rm{e}}$ and one for $\omega\lesssim0.2\omega_{\rm{e}}$. We say \emph{metal-like} because the regions are characterized by $\epsilon'<0$ and $\mu'>0$, which is typical for metals at optical frequencies but not seen in dielectrics. In the frequency range $0.2\omega_{\rm{e}}\lesssim\omega\lesssim 0.3\omega_{\rm{e}}$ the metamaterial has a negative-index. The binding frequency, $\omega_{0}$, in Eq.~(\ref{muMM}) causes rapid changes in the refractive index (Fig.~\ref{fig:nimmregions}\subref{fig:nimmindex}) around $\omega=0.2\omega_{\rm e}$. 

We characterize the slab geometry first, and to provide a basis for comparison, we also include results for a metal-dielectric waveguide. We assume the slab guide is symmetric, meaning all of the parameters are identical for both cladding layers. The parameters we use to describe the metal are the same as for the metamaterial, with the exception that $\mu_{2,\,3}=\mu_{0}$. To characterize the behaviour of the modes we compare the real and imaginary parts of $\gamma_{1}$.
\begin{figure}[t,b] 
     \centering
      \subfloat{\label{fig:relgammaSlabOrd}\includegraphics[width=0.5\textwidth]{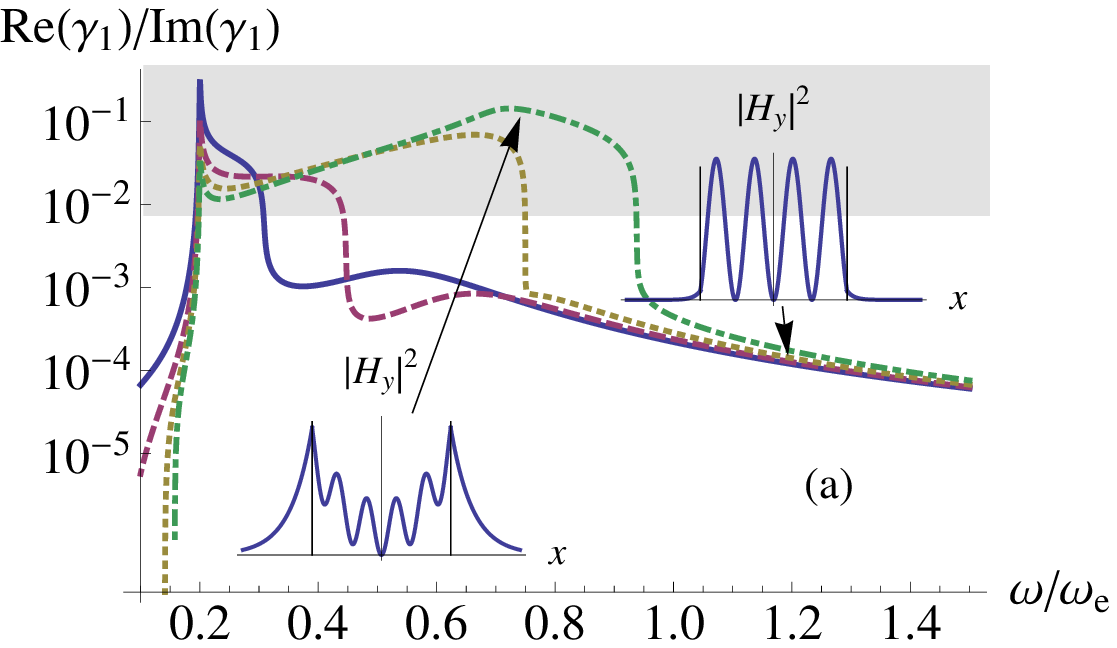}}
      \subfloat{\label{fig:relgammaSlabSurf}\includegraphics[width=0.5\textwidth]{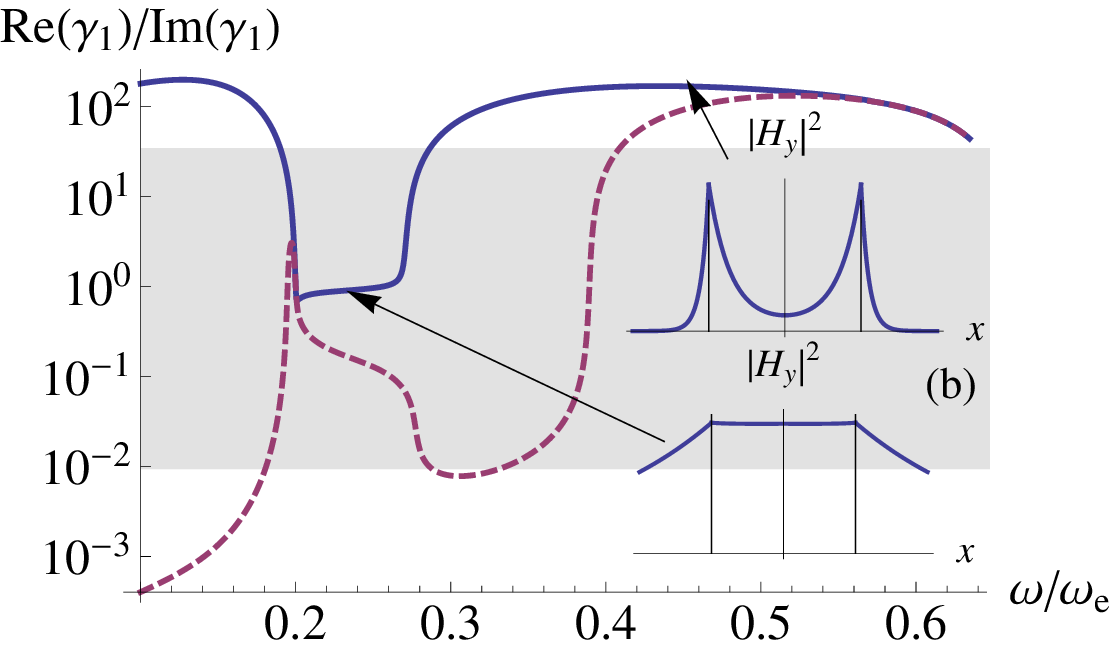}}
     \caption[Plots of mode behaviour as a function of frequency for the slab waveguide.]{Plots of $\rm{Re}\left(\gamma_{1}\right)/\rm{Im}\left(\gamma_{1}\right)$ for \protect\subref {fig:relgammaSlabOrd} the TM$_{0}$ (solid), TM$_{1}$ (dashed), TM$_{2}$ (dotted), and TM$_{3}$, (dot-dashed) and \protect\subref{fig:relgammaSlabSurf} the TM$_{\rm{s}}$ (solid) and TM$_{\rm{a}}$ (dashed) modes of the metamaterial-dielectric slab guide. The shaded region indicates the nominal range of hybrid waves, as there is no well-defined boundary. The insets are plots of $|H_{y}|^{2}$, which show the transverse profile of the mode at select frequencies. The thin vertical bars in the insets indicate the core-cladding boundary.}\label{fig:relgammaSlab}   
\end{figure}
A plot of $\rm{Re}\left(\gamma_{1}\right)/\rm{Im}\left(\gamma_{1}\right)$ (see Fig.~\ref{fig:relgammaSlab}) shows where the real and imaginary parts of $\gamma_{1}$ are of comparable magnitude, indicated by shaded regions, which correspond to hybrid waves. Outside the shaded regions, either the real or imaginary part is dominant and the mode is a surface or ordinary wave, respectively. The insets of the plots also show the transverse profile of select modes at frequencies that lie in regions of different mode behaviour. 

The transverse mode profile does not always present a clear distinction between ordinary, surface, and hybrid modes. As noted earlier, the fields of ordinary waves are generally distributed evenly across the core and have an oscillatory pattern. Surface modes, however, are characterized by the fields being concentrated at the core-cladding boundary and decaying toward the centre of the core. Hybrid waves have characteristics of both wave types. One distinguishing feature of hybrid waves is a large amount of the field is present in the cladding layer (fat tails). The two insets for hybrid waves in Fig.~\ref{fig:relgammaSlab} clearly show the fat tails indicative of hybrid waves.

Four modes are plotted in Fig.~\ref{fig:relgammaSlab}\subref{fig:relgammaSlabOrd} showing the regions of ordinary and hybrid-wave behaviour for each. Hybrid-wave behaviour is present for $\rm{Re}\left(\gamma_{1}\right)/\rm{Im}\left(\gamma_{1}\right)\gtrsim10^{-2}$, which translates to the TM$_{0}$ mode for $0.2\omega_{\rm e}\lesssim\omega\lesssim0.3\omega_{\rm{e}}$, the TM$_{1}$ mode for $0.2\omega_{\rm e}\lesssim\omega\lesssim0.45\omega_{\rm{e}}$, the TM$_{2}$ mode for $0.2\omega_{\rm e}\lesssim\omega\lesssim0.75\omega_{\rm{e}}$ and the TM$_{3}$ mode for $0.2\omega_{\rm e}\lesssim\omega\lesssim0.95\omega_{\rm{e}}$. The insets are plots of $|H_{y}|^{2}$ for the TM$_{3}$ mode and show the difference in the transverse profile between the ordinary and hybrid waves. The inset for $\omega\approx1.2\omega_{\rm{e}}$ is an ordinary wave ($\rm{Re}\left(\gamma_{1}\right)/\rm{Im}\left(\gamma_{1}\right)\approx10^{-3}$), whereas the inset for $\omega\approx0.8\omega_{\rm{e}}$ is a hybrid wave ($\rm{Re}\left(\gamma_{1}\right)/\rm{Im}\rm\left(\gamma_{1}\right)\approx10^{-1}$).

Figure~\ref{fig:relgammaSlab}\subref{fig:relgammaSlabSurf} shows the TM$_{\rm{s}}$ and TM$_{\rm{a}}$ modes of the metamaterial-dielectric slab guide. Hybrid-wave behaviour is present for $10^{-2}\lesssim\rm{Re}\left(\gamma_{1}\right)/\rm{Im}\left(\gamma_{1}\right)\lesssim50$, which translates to the TM$_{\rm s}$ mode for  $0.2\omega_{\rm e}\lesssim\omega\lesssim0.3\omega_{\rm{e}}$ and the TM$_{\rm{a}}$ mode for $0.18\omega_{\rm e}\lesssim\omega\lesssim0.4\omega_{\rm{e}}$. From the plots, we see that the TM$_{\rm s}$ and TM$_{\rm a}$ modes are surface waves for frequencies corresponding to $\rm{Re}\left(\gamma_{1}\right)/\rm{Im}\left(\gamma_{1}\right)\gtrsim50$. The insets in this figure are plots of $|H_{y}|^{2}$ for the TM$_{\rm{s}}$ mode and show that when $\rm{Re}\left(\gamma_{1}\right)/\rm{Im}\left(\gamma_{1}\right)\gtrsim50$ the mode is a surface wave (fields concentrated at core-cladding boundary), whereas it is a hybrid wave when $\rm{Re}\left(\gamma_{1}\right)/\rm{Im}\left(\gamma_{1}\right)\lesssim50$.

\begin{figure}[t,b] 
       \centering
       \subfloat{\label{fig:TMSlabDispBOUND}\includegraphics[width=0.5\textwidth]{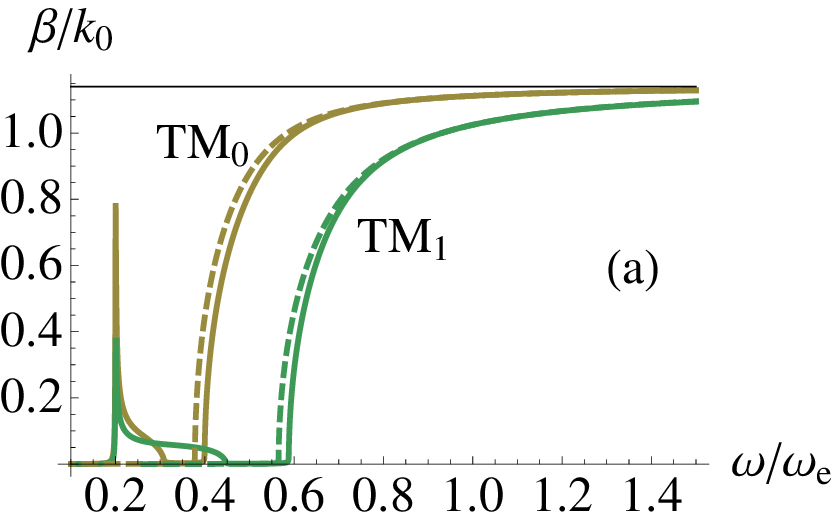}}
       \subfloat{\label{fig:TMSlabDispMIXED}\includegraphics[width=0.5\textwidth]{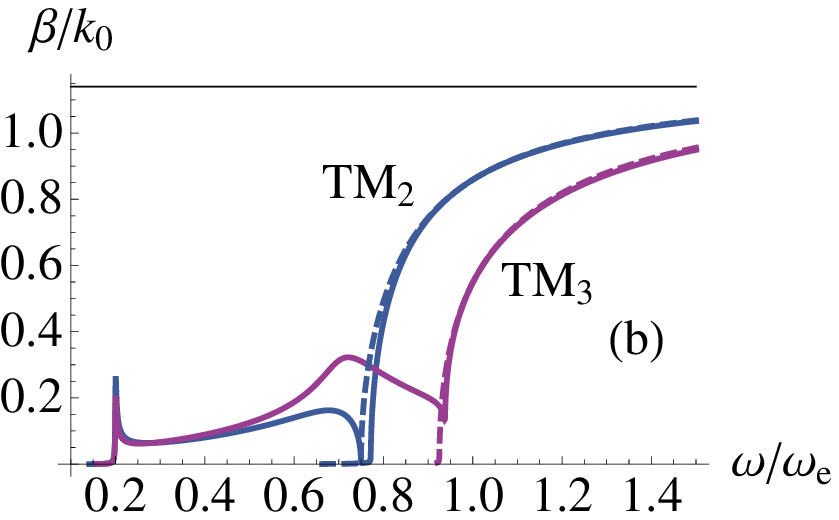}}\\
	\subfloat{\label{fig:TMSlabDispSurface}\includegraphics[width=0.5\textwidth]{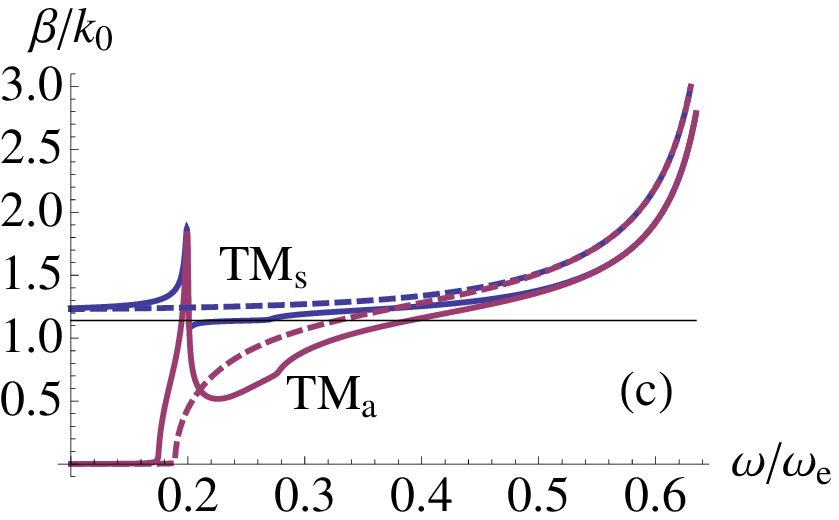}}
\caption[Plots of effective refractive index for selected modes of the slab waveguide.]{Effective refractive index for \protect\subref{fig:TMSlabDispBOUND} the TM$_{0}$ and TM$_{1}$, \protect\subref{fig:TMSlabDispMIXED} the TM$_{2}$ and TM$_{3}$, and \protect\subref{fig:TMSlabDispSurface} the TM$_{\rm{s}}$ and TM$_{\rm{a}}$ modes modes of the slab guide. The solid lines are for the modes of the metamaterial-dielectric guide, and the dashed lines are for the modes of the metal-dielectric guide. The thin horizontal line in each is the refractive index of the core.}
\label{fig:TMSlabDisp}   
\end{figure}

All guided modes in a waveguide experience an effective refractive index along the propagation direction. The effective refractive index of a waveguide is related to the propagation constant, $\beta$, by the relation $n_{\rm{eff}}=\beta/k_{0}$ with $k_{0}=\omega/c$ the free-space wave number, and c is the speed of light in vacuum. The effective refractive index of ordinary waves in a metamaterial-dielectric guide is very similar to that of a dielectric-dielectric guide, increasing with the frequency of the wave and bound from above by the refractive index of the core, $n_{\rm{core}}=c\sqrt{\epsilon_{1}\mu_{1}}$. The effective refractive index of the TM$_{0}$ and TM$_{1}$ modes is plotted in Fig.~\ref{fig:TMSlabDisp}\subref{fig:TMSlabDispBOUND}. For frequencies above $0.3\omega_{\rm e}$ and $0.45\omega_{\rm e}$ the TM$_{0}$ and TM$_{1}$ modes, respectively, are ordinary waves.

Modes with surface-wave behaviour exist in metal-dielectric guides as well as in metamaterial-dielectric guides. The requirement of $\epsilon'$ or $\mu'$ to have opposite signs on either side of the interface means surface modes are not possible in dielectric-dielectric guides. This requirement is reflected in the dispersion relation for the flat interface, Eq.~(\ref{eq:flatdisprel}). Only two modes in the slab guide show surface-wave characteristics, TM$_{\rm{s}}$ and TM$_{\rm{a}}$, which is present at frequencies such that $n_{\rm{eff}}>n_{\rm{core}}$. The effective refractive index of surface waves in a metamaterial-dielectric guide behaves as for a metal-dielectric guide, increasing with the frequency, and is plotted for the slab-guide modes TM$_{\rm{s}}$ and TM$_{\rm{a}}$ in Fig.~\ref{fig:TMSlabDisp}\subref{fig:TMSlabDispSurface}.

Hybrid waves sometimes exhibit anomalous dispersion (decreasing $n_{\rm{eff}}$ with increasing frequency). We have already seen hybrid-mode behaviour in Fig.~\ref{fig:TMSlabDisp}\subref{fig:TMSlabDispBOUND}, where it is present at lower frequencies than the ordinary wave of the same mode. There are also regions of hybrid-wave behaviour in the TM$_{\rm{s}}$ and TM$_{\rm{a}}$ modes, shown in Fig.~\ref{fig:TMSlabDisp}\subref{fig:TMSlabDispSurface}, which occur for $0.2\omega_{\rm{e}}\lesssim\omega\lesssim0.3\omega_{\rm{e}}$ for the TM$_{\rm s}$ mode and for $0.2\omega_{\rm e}\lesssim\omega\lesssim0.4\omega_{\rm{e}}$ for the TM$_{\rm{a}}$ mode. 

Figure~\ref{fig:TMSlabDisp}\subref{fig:TMSlabDispMIXED} shows the effective refractive index for the TM$_{2}$ and TM$_{3}$ modes of the metamaterial-dielectric slab waveguide. The hybrid-wave behaviour occurs for $0.2\omega_{\rm e}\lesssim\omega\lesssim0.75\omega_{\rm e}$ in the TM$_{2}$ mode and $0.2\omega_{\rm e}\lesssim\omega\lesssim0.95\omega_{\rm e}$ in the TM$_{3}$ mode. Both modes have regions of anomalous dispersion with a particularly large frequency region of anomalous dispersion in the TM$_{3}$ mode. 

The effective guide width of a waveguide $w_{\rm{eff}}$ is the nominal width $w$ plus the skin-depth at each interface, where $w_{\rm{eff}}=w+1/\rm{Re}\left(\gamma_2\right)+1/\rm{Re}\left(\gamma_3\right)$ for the slab guide. Ordinary and surface waves in a metamaterial-dielectric guide have an effective width that is only slightly larger than the nominal width; this is consistent with ordinary and surface waves of a metal guide. Hybrid waves, however, display an increased effective width when compared to the ordinary and surface waves.

\begin{figure}[t,b] 
     \centering
       \subfloat{\label{fig:TMSlabHeightBOUND}\includegraphics[width=0.5\textwidth]{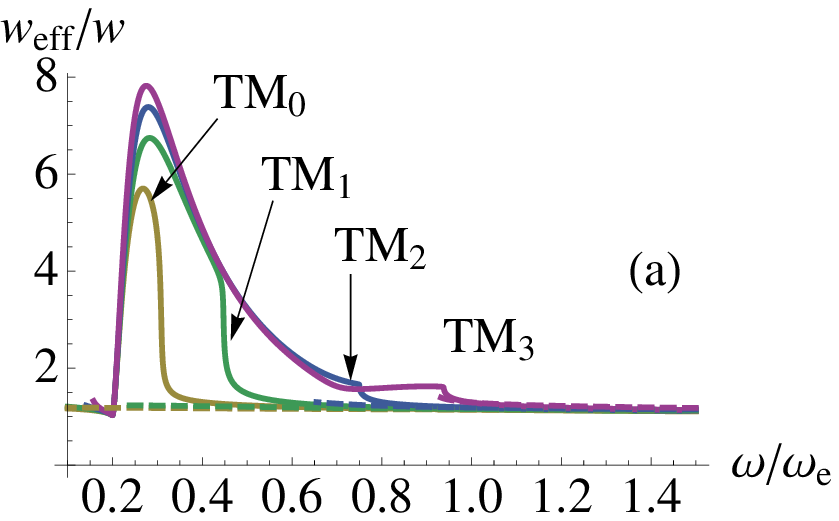}}
       \subfloat{\label{fig:TMSlabHeightSURFACE}\includegraphics[width=0.5\textwidth]{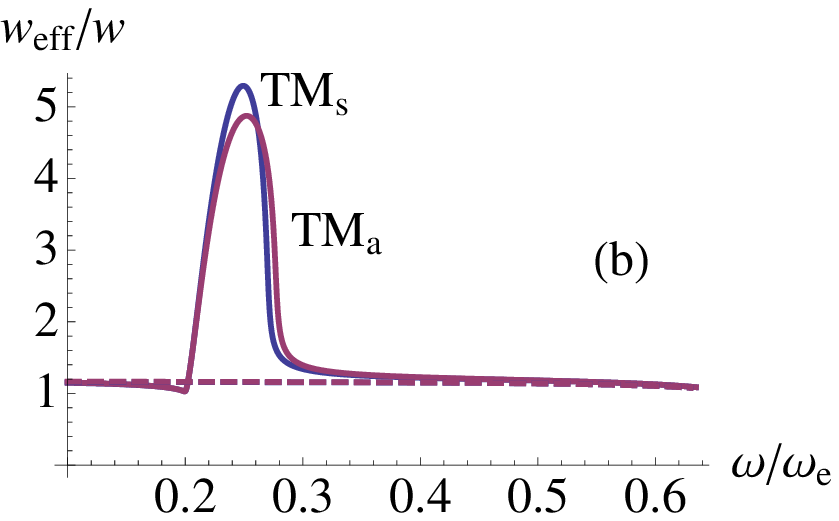}}
     \caption[Plots of effective guide width for selected modes of the slab waveguide.]{Effective guide width for \protect\subref {fig:TMSlabHeightBOUND} the TM$_{0}$ to TM$_{3}$, and \protect\subref{fig:TMSlabHeightSURFACE} the TM$_{\rm{s}}$ and TM$_{\rm{a}}$ modes of the slab guide. Solid lines are for the metamaterial-dielectric guide, and dashed lines are for the metal-dielectric guide.}\label{fig:TMSlabHeight}   
\end{figure}

Figures~\ref{fig:TMSlabHeight}\subref{fig:TMSlabHeightBOUND} and \ref{fig:TMSlabHeight}\subref{fig:TMSlabHeightSURFACE} are plots of the relative effective width $w_{\rm{eff}}/w$ and show the large effective width of the hybrid waves.
The largest effective width for each mode is seen in the negative index region, where all modes are hybrid waves. Significantly increased effective width is a feature unique to metamaterial guides. The effective width of metal-dielectric and dielectric-dielectric guides does not have such a large variation.

A large effective width is due to the field having a large extent in the cladding. In the case of hybrid modes, the large effective width is due to the fact that over half of the total energy of the mode is carried outside the core. As the cladding dissipates energy, there is a connection between effective width and losses for hybrid modes.

The surface waves, however, can have a small effective width even when there is a large fraction of the total energy in the cladding, which is due to the fact that the energy of a surface mode is concentrated around the interface with almost no energy in the centre of the core. Thus, even when there is little penetration of the fields into the cladding, there can be a comparably small penetration away from the interface into the core. The effect, then, is a relatively even distribution of energy about the interface and a small effective width. 

The attenuation of a mode in a waveguide is a measure of how much the intensity of the mode decreases as a function of the distance travelled. The imaginary part of $\tilde\beta$ is the attenuation constant, denoted $\alpha$, and characterizes the amount of attenuation per unit length that affects a propagating mode. All of the mode types in the metamaterial guide show some amount of attenuation, with some wave types having more than others. 

\begin{figure}[t,b] 
       \centering
   \subfloat{\label{fig:TMSlabAbsMIXED}\includegraphics[width=0.5\textwidth]{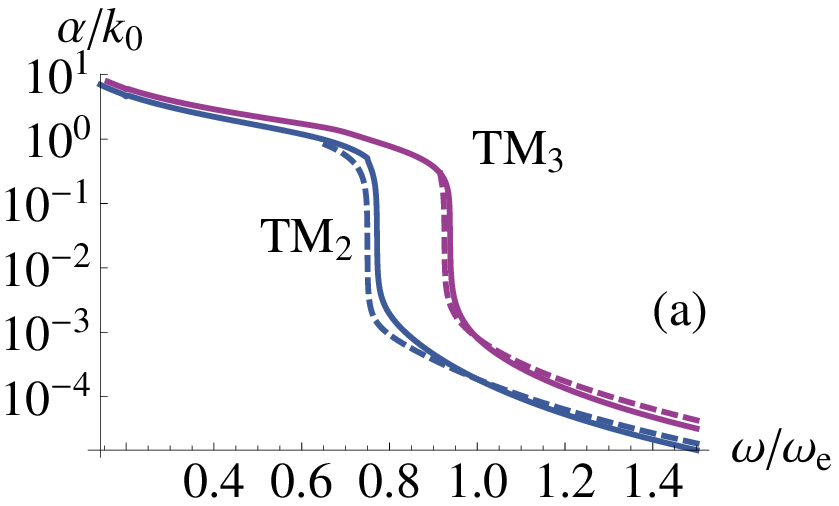}}
   \subfloat{\label{fig:TMSlabAbsSurface}\includegraphics[width=0.5\textwidth]{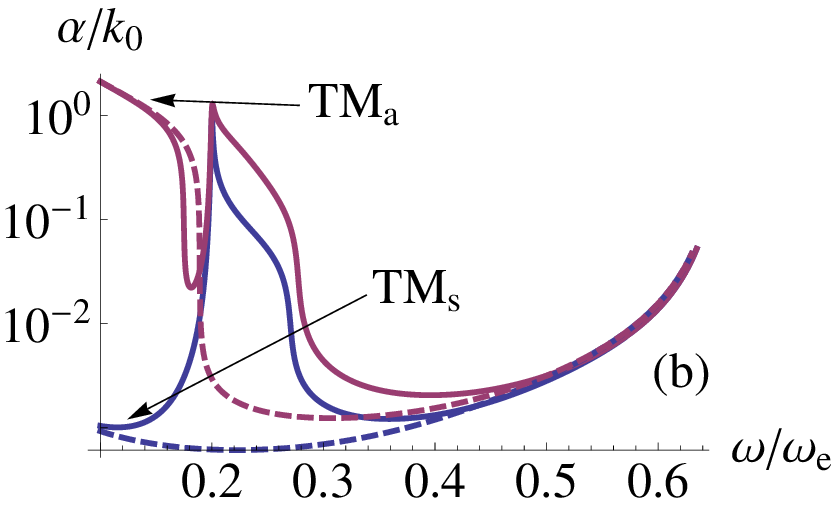}}
     \caption[Plots of the attenuation for selected modes of the slab waveguide.]{Attenuation for \protect\subref{fig:TMSlabAbsMIXED} the TM$_{2}$ and TM$_{3}$, and \protect\subref{fig:TMSlabAbsSurface} the TM$_{\rm{s}}$ and TM$_{\rm{a}}$ modes of the slab guide. Solid lines are for the metamaterial-dielectric guide, and dashed lines are for the metal-dielectric guide.}\label{fig:TMSlabAbs}
\end{figure}

Hybrid waves have the largest attenuation, which is partly due to the large fraction of energy that propagates in the cladding for these modes. The attenuation of the TM$_{2}$ and TM$_{3}$ modes of the slab guide is seen in Fig.~\ref{fig:TMSlabAbs}\subref{fig:TMSlabAbsMIXED}.
The regions of large attenuation occur where the modes display hybrid-wave behaviour. Hybrid waves also have larger attenuation than surface waves in the same mode, though the difference is much less pronounced. The attenuation of the TM$_{\rm{s}}$ and TM$_{\rm{a}}$ modes is shown in Fig.~\ref{fig:TMSlabAbs}\subref{fig:TMSlabAbsSurface}.

At frequencies where the mode displays hybrid-wave behaviour, the attenuation is considerably larger than for either ordinary-wave or surface-wave behaviour. The exception is the TM$_{\rm{a}}$ mode for $\omega\lesssim 0.2\omega_{\rm{e}}$ where it is an ordinary wave. Here the attenuation is large, but consistent with the TM$_{\rm{a}}$ mode in the metal guide (see the dashed line in Fig.~\ref{fig:TMSlabAbs}\subref{fig:TMSlabAbsSurface}), which is also an ordinary wave at low frequencies. The attenuation of both ordinary and hybrid waves decreases as the frequency increases, whereas the attenuation of surface waves generally increases with increasing frequency.

\begin{figure}[t,b] 
     \centering
      \subfloat{\label{fig:relgammaCylOrd1}\includegraphics[width=0.5\textwidth]{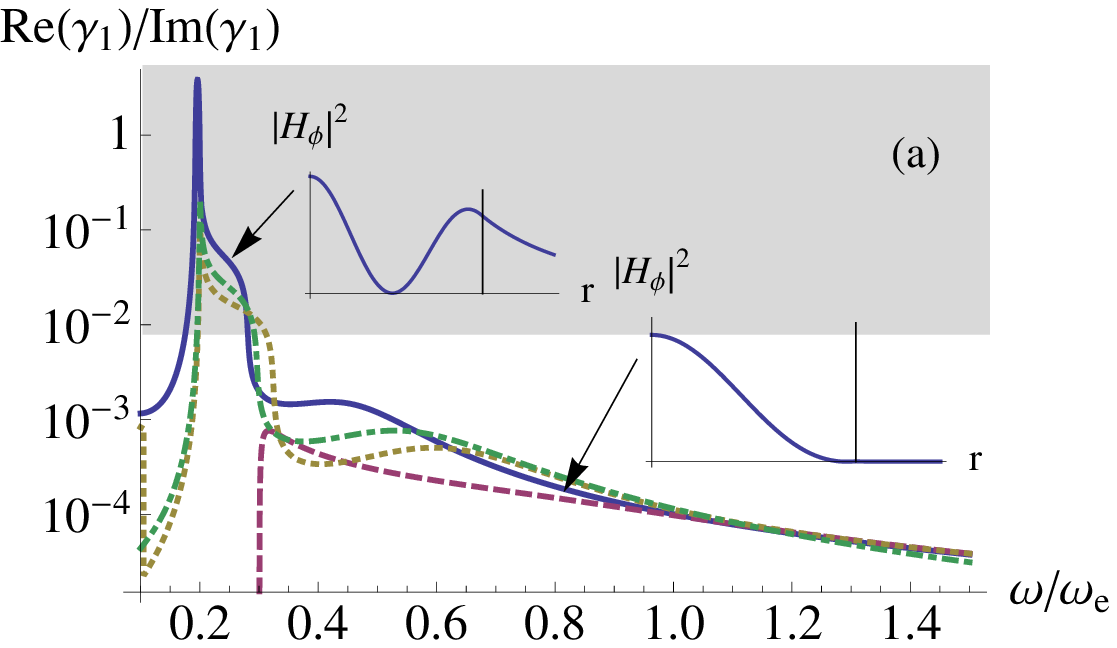}}
      \subfloat{\label{fig:relgammaCylOrd2}\includegraphics[width=0.5\textwidth]{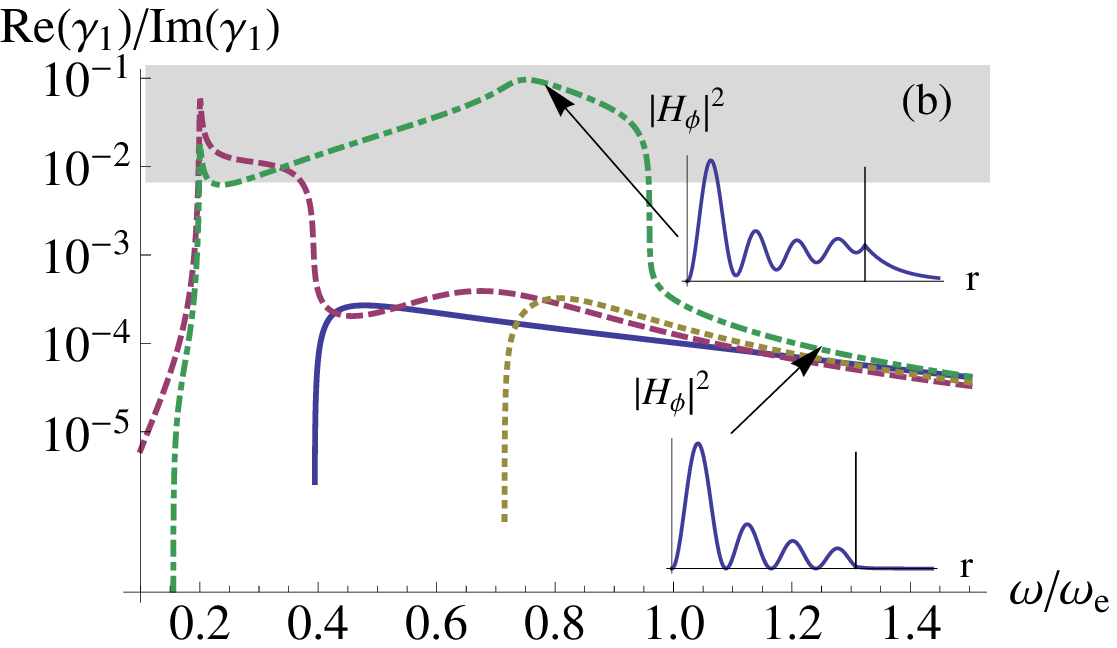}}\\
      \subfloat{\label{fig:relgammaCylSurf}\includegraphics[width=0.5\textwidth]{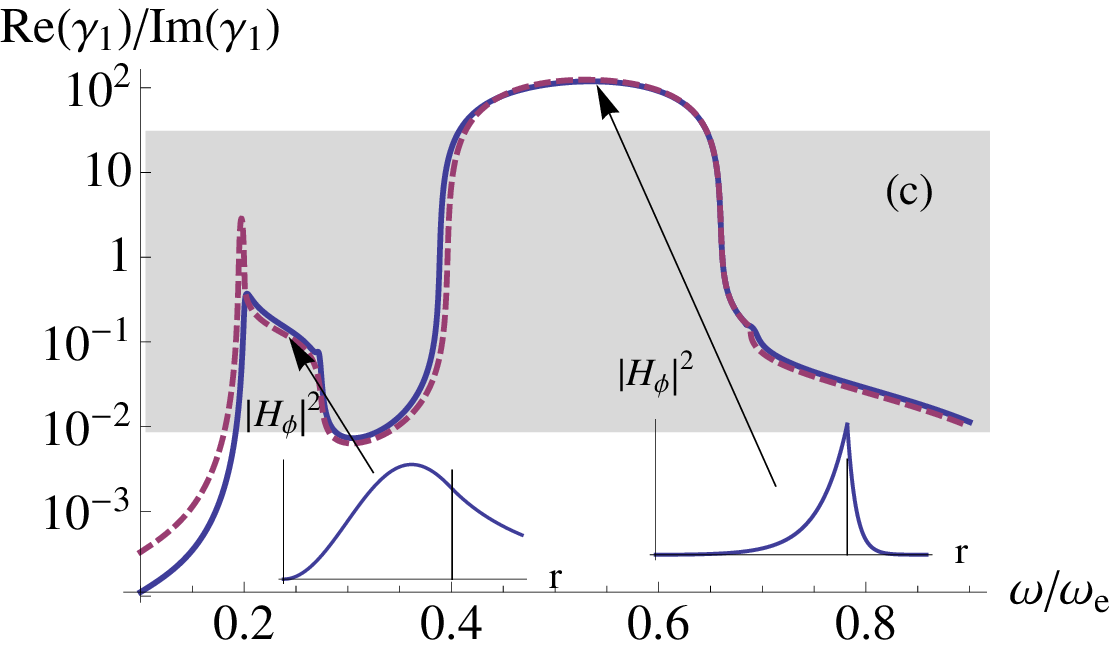}}
     \caption[Plots of mode behaviour as a function of frequency for the cylinrical guide.]{Plots of $\rm{Re}\left(\gamma_{1}\right)/\rm{Im}\left(\gamma_{1}\right)$ for \protect\subref {fig:relgammaCylOrd1} the HE$_{10}$ (solid), HE$_{11}$ (dashed), HE$_{12}$ (dotted), and TM$_{0}$, (dot-dashed), \protect\subref{fig:relgammaCylOrd2} the HE$_{13}$ (solid), TM$_{1}$ (dashed), TM$_{2}$ (dotted), and TM$_{3}$, (dot-dashed) and \protect\subref{fig:relgammaCylSurf} the HE$_{1\rm{s}}$ (solid) and TM$_{\rm{s}}$ (dashed) modes of the cylindrical metamaterial-dielectric guide. The shaded regions indicate the nominal range of hybrid waves, as there is no well-defined boundary. The insets are plots of $|H_{\phi}|^{2}$ for half of the guide at select frequencies and show the transverse profile of the mode at these frequencies. The thin vertical bar in the insets indicates the core-cladding boundary.}\label{fig:relgammaCyl}
\end{figure}
As with the slab guide, we provide plots of $\rm{Re}\left(\gamma_{1}\right)/\rm{Im}\left(\gamma_{1}\right)$ for the cylindrical metamaterial-dielectric guide in Fig.~\ref{fig:relgammaCyl} to clarify where the modes are hybrid waves. Hybrid waves are indicated by the shaded region. The fat tails indicative of hybrid waves in a slab guide are also present for hybrid waves in a cylindrical guide. The transverse profiles of the hybrid waves in Figs.~\ref{fig:relgammaSlab} and \ref{fig:relgammaCyl} show this increased field penetration, which is also reflected in the effective guide width for the modes.  The plots in Fig.~\ref{fig:relgammaCyl} show at which frequencies the modes of the cylindrical metamaterial-dielectric guide display which behaviour.

Figure~\ref{fig:relgammaCyl}\subref{fig:relgammaCylOrd1} shows plots of $\rm{Re}\left(\gamma_{1}\right)/\rm{Im}\left(\gamma_{1}\right)$ for the modes HE$_{10}$ to HE$_{12}$ and TM$_{0}$. From these plots the different frequency regions exhibit where the modes display hybrid and ordinary-wave behaviour. What is immediately clear is that there are no frequencies where the HE$_{11}$ mode is a hybrid wave. Instead it has a cut-off frequency near $0.3\omega_{\rm e}$. The three other modes (HE$_{10}$, HE$_{12}$ and TM$_{0}$) are hybrid waves for $0.2\omega_{\rm e}\lesssim\omega\lesssim 0.3\omega_{\rm{e}}$. Beyond about $0.3\omega_{\rm e}$ all four modes are ordinary waves. The insets in Fig.~\ref{fig:relgammaCyl}\subref{fig:relgammaCylOrd1} are $|H_{\phi}|^{2}$ for the HE$_{10}$ mode and show the transverse mode profile at two frequencies, $\omega\approx0.8\omega_{\rm e}$, where the mode is an ordinary wave, and $\omega\approx0.25\omega_{\rm e}$, where the mode is a hybrid wave. 

Similarly, Fig.~\ref{fig:relgammaCyl}\subref{fig:relgammaCylOrd2} shows plots of $\rm{Re}\left(\gamma_{1}\right)/\rm{Im}\left(\gamma_{1}\right)$ for the modes HE$_{13}$ and TM$_{1}$ to TM$_{3}$. Two of the modes (HE$_{13}$ and TM$_{2}$) do not show hybrid-wave behaviour, whereas the TM$_{1}$ and TM$_{3}$ are hybrid waves for $0.2\omega_{\rm e}\lesssim\omega\lesssim 0.4\omega_{\rm{e}}$ and $0.2\omega_{\rm e}\lesssim\omega\lesssim 0.95\omega_{\rm{e}}$, respectively. The insets here are plots of  $|H_{\phi}|^{2}$ for the TM$_{3}$ mode at two frequencies. Again, one of these plots represents an ordinary mode, $\omega\approx1.2\omega_{\rm e}$, and the other represents a hybrid mode, $\omega\approx0.8\omega_{\rm e}$. 

Modes with surface-wave behaviour are supported by the metamaterial-dielectric cylindrical guide as well. Plots of $\rm{Re}\left(\gamma_{1}\right)/\rm{Im}\left(\gamma_{1}\right)$ for the HE$_{1\rm{s}}$ and TM$_{\rm s}$ modes are shown in Fig.~\ref{fig:relgammaCyl}\subref{fig:relgammaCylSurf}. Both of these modes are surface waves for $0.4\omega_{\rm e}\lesssim\omega\lesssim 0.65\omega_{\rm{e}}$, ordinary waves for $\omega\lesssim0.2\omega_{\rm e}$, and hybrid waves otherwise. The insets in Fig.~\ref{fig:relgammaCyl}\subref{fig:relgammaCylSurf} show plots of $|H_{\phi}|^{2}$ for the TM$_{s}$ mode showing a surface wave, $\omega\approx0.5\omega_{\rm e}$, and a hybrid wave, $\omega\approx0.25\omega_{\rm e}$. 

\begin{figure}[t,b] 
     \centering
       \subfloat{\label{fig:TMCylDispBOUND}\includegraphics[width=0.5\textwidth]{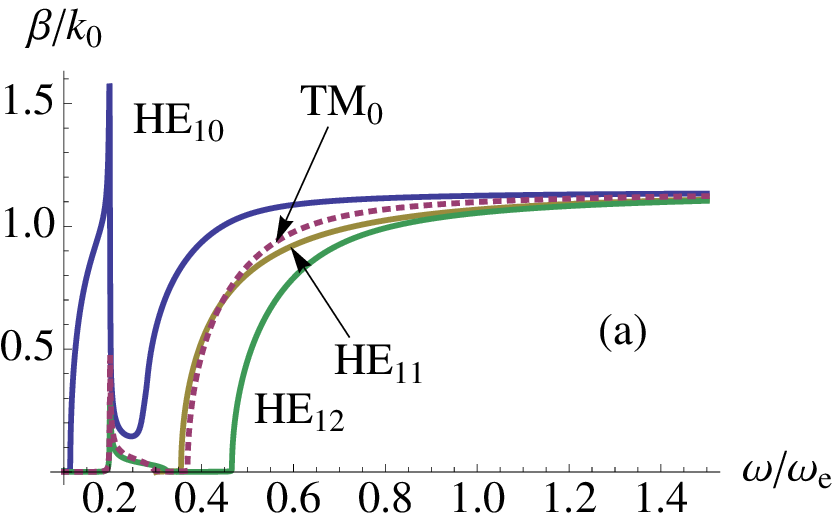}}
       \subfloat{\label{fig:CylDispOrdinary5-8}\includegraphics[width=0.5\textwidth]{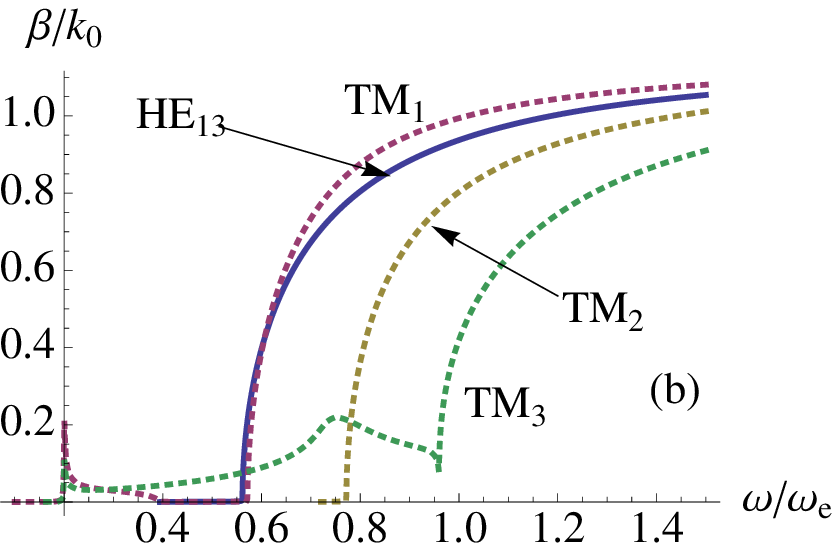}}\\
       \subfloat{\label{fig:HE1CylDispSURFACE}\includegraphics[width=0.5\textwidth]{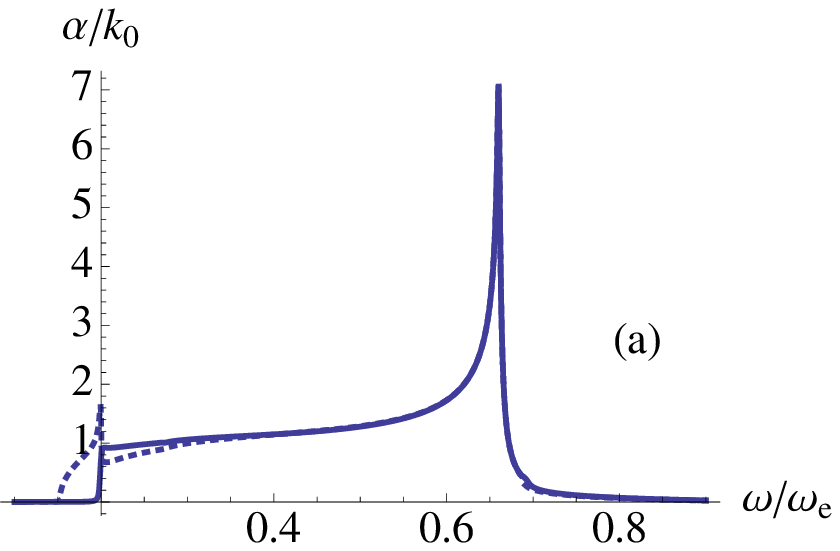}}
     \caption[Plots of effective refractive index for selected modes of the cylindrical waveguide.]{Effective refractive index for \protect\subref{fig:TMCylDispBOUND} the HE$_{10}$ to HE${12}$, and TM${0}$, \protect\subref{fig:CylDispOrdinary5-8}the  HE$_{13}$, and TM${1}$ to TM${3}$, and \protect\subref{fig:HE1CylDispSURFACE} the HE$_{1\rm{s}}$ and TM$_{\rm{s}}$ modes of the metamaterial-dielectric cylindrical guide. The solid lines indicate HE modes and the dotted lines indicate TM modes.}\label{fig:TMCylDisp}   
\end{figure}

The modes of a cylindrical guide display similar dispersion characteristics to those in a slab guide, including the appearance of hybrid waves at lower frequencies. Figures~\ref{fig:TMCylDisp}\subref{fig:TMCylDispBOUND} to \subref{fig:HE1CylDispSURFACE}
show $n_{\rm{eff}}$ for the modes HE$_{10}$ to HE$_{13}$, TM$_{0}$ to TM$_{3}$, as well as HE$_{\rm s}$, and TM$_{\rm s}$ of the metamaterial cylindrical guide. In contrast to the slab guide, which supports only two surface modes, the cylindrical guide supports a number of modes with surface-wave behaviour (one TM and the rest HE), all of different orders (number of oscillations in the electric field over an angle of $2\pi$). 

\begin{figure}[t,b] 
     \centering
     \subfloat{\label{fig:CylHeightOrdinary5-8}\includegraphics[width=0.5\textwidth]{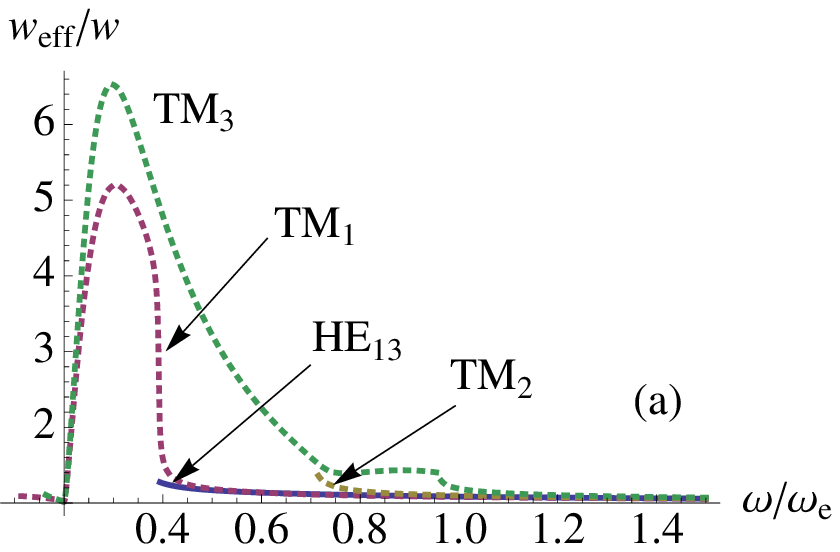}}
     \subfloat{\label{fig:CylHeightSurface}\includegraphics[width=0.5\textwidth]{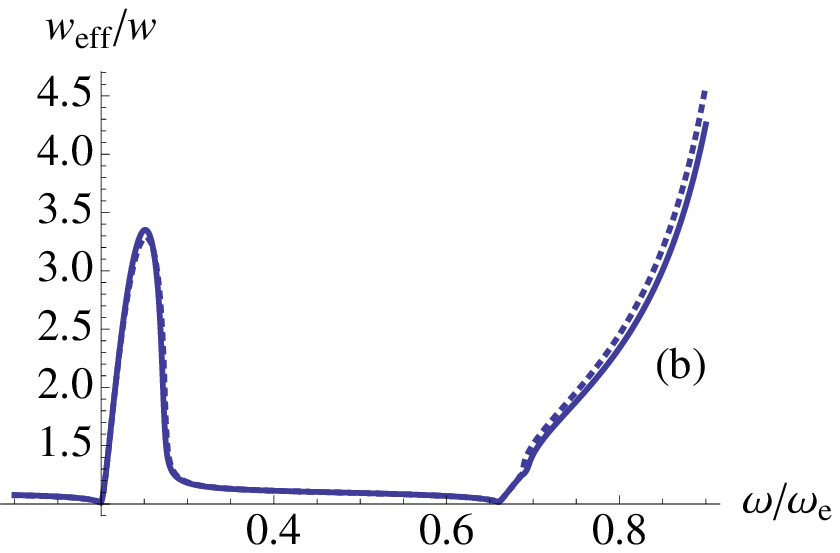}}
  \caption[Plots of effective guide width for selected modes of the cylindrical waveguide.]{Effective guide width for \protect\subref{fig:CylHeightOrdinary5-8} the HE$_{13}$ mode, and TM$_{1}$ to TM$_{3}$, and \protect\subref{fig:CylHeightSurface} the HE$_{1\rm{s}}$ and TM$_{\rm{s}}$ modes of the metamaterial-dielectric cylindrical guide. The solid lines indicate HE modes and the dotted lines indicate TM modes.}\label{fig:CylHeight}   
\end{figure}

The effective widths of the modes supported by the metamaterial-dielectric cylindrical guide show similar characteristics to the effective widths of the modes in a metamaterial-dielectric slab guide. Figure~\ref{fig:CylHeight}\subref{fig:CylHeightOrdinary5-8}
shows the effective width of the four modes HE$_{13}$, TM$_{1}$, TM$_{2}$, and TM$_{3}$ of the cylindrical guide, and Fig.~\ref{fig:CylHeight}\subref{fig:CylHeightSurface} shows the width for the HE$_{1\rm{s}}$ and TM$_{\rm{s}}$ modes. The effective widths of the ordinary and surface waves are only slightly larger than the nominal width of the guide, whereas the effective widths of the hybrid waves are large, when compared to the nominal guide width and the effective widths of the other wave types.

\begin{figure}[t,b] 
     \centering
     \subfloat{\label{fig:CylAbsOrdinary5-8}\includegraphics[width=0.5\textwidth]{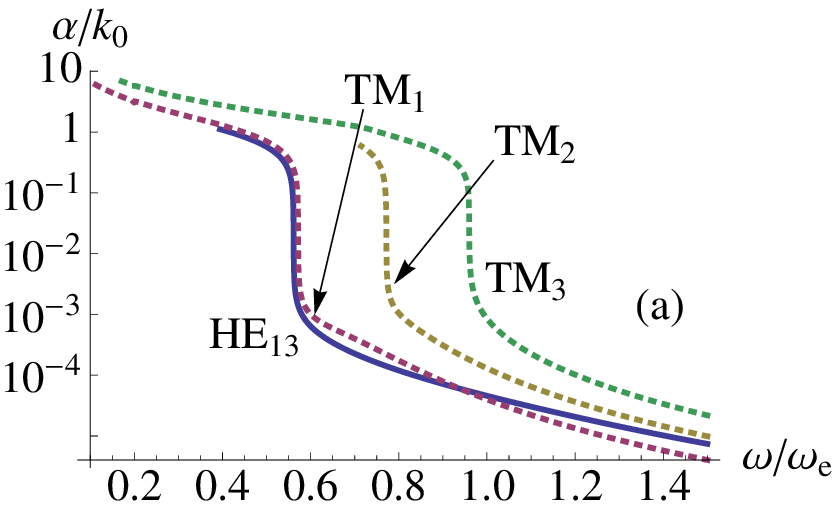}}
     \subfloat{\label{fig:CylAbsSurface}\includegraphics[width=0.5\textwidth]{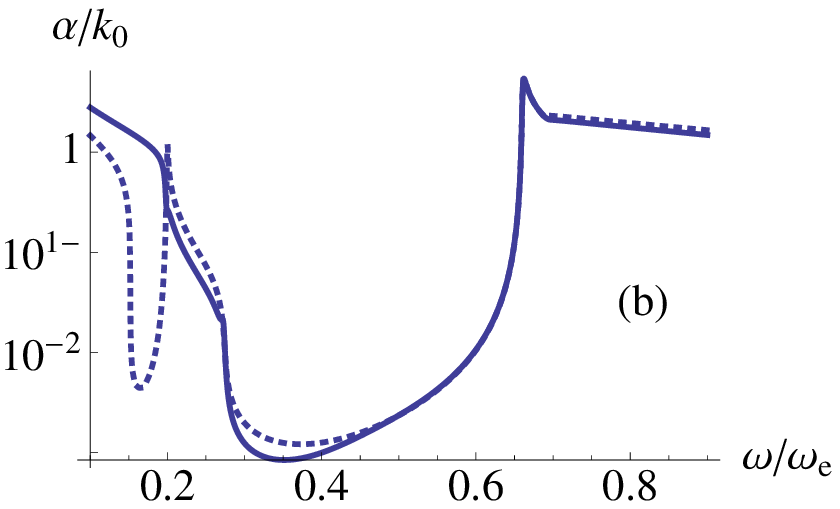}}
  \caption[Plots of attenuation for selected modes of the cylindrical waveguide.]{Attenuation for \protect\subref{fig:CylAbsOrdinary5-8} the HE$_{13}$ and TM$_{1}$ to TM$_{3}$, and \protect\subref{fig:CylAbsSurface} the HE$_{1\rm{s}}$ and TM$_{\rm{s}}$ modes of the metamaterial-dielectric cylindrical guide. The solid lines indicate HE modes and the dotted lines indicate TM modes.}\label{fig:CylAbs}   
\end{figure}

As with the effective widths, the attenuation of the modes of the metamaterial-dielectric cylindrical guide is similar to that of the slab guide. The attenuation for the HE$_{13}$, TM$_{1}$, TM$_{2}$, and TM$_{3}$ modes is shown in Fig.~\ref{fig:CylAbs}\subref{fig:CylAbsOrdinary5-8}.
In frequency regions of hybrid-wave behaviour the modes have large attenuation. The same trend of decreasing attenuation with increasing frequency is also present. The TM$_{\rm{s}}$ and HE$_{1\rm{s}}$ modes of the cylindrical guide, shown in Fig.~\ref{fig:CylAbs}\subref{fig:CylAbsSurface}, have attenuation characteristics similar to the TM$_{\rm{a}}$ mode of the slab guide. 

This chapter outlines the results of a theoretical characterization of a metamaterial-clad waveguide. However, it is important for theoretical predictions to be compatible with measurements that are realizable in an experimental setting. If this is not the case, theoretical predictions cannot be verified. Thus, I present here a short overview of experimental waveguide characterization.

The waveguide parameters sought in a waveguide characterization depend on the desired application. For instance, the dispersion of a waveguide is an important parameter, as it determines many of the waveguides propagation characteristics, and can be used to determine pulse spreading~\cite{Cohen:1985}. However, for some applications, such as telecommunication, there is also interest in the attenuation, mode field diameter, and the cut-off wavelength~\cite{Gisin:1993}. 

There are a number of different techniques for experimentally characterizing waveguides. The dispersion of a fibre can be determined in different ways, such as measuring the phase shift experienced by a beam passing through the guide, or inferred from the mode field diameter~\cite{Cohen:1985,Gisin:1993}. The mode field diameter can be acquired through simultaneous measurement of the transmitted and refracted near field~\cite{Gisin:1993}. For planar waveguides, it is possible to couple in to the side of the waveguide using a prism and measure the reflected power. This can in turn be used to calculate the propagation and attenuation constants for the guide~\cite{Lin:2012}.

We now understand how metamaterial-dielectric waveguides behave when the effect of absorption is considered. The subject of possible implications and implementations of such waveguides can now be approached.


\section{Discussion: Low-loss Surface Modes}\label{sec:LLWGdiscussion}

In a metamaterial-dielectric guide, attenuation of ordinary and surface waves is much less than for hybrid waves but is still comparable to attenuation in metal-dielectric waveguides. With current metamaterial technology, losses due to atomic interaction with the magnetic field are significant~\cite{Xiao:2009, Penciu:2010}. Previous results for optical metamaterials, however, show magnetic losses could be reduced through structural improvements~\cite{Zhou:2008}. Our results  show that lowering the magnetic losses opens up the possibility of low-loss metamaterial-dielectric waveguides, with the attenuation reduction due to interference effects in the metamaterial~\cite{Kamli:2008,Moiseev:2010}.

\begin{figure}[t,b] 
      \centering
       \subfloat{\label{fig:TMSlabAbsSurfacegm0}\includegraphics[width=0.5\textwidth]{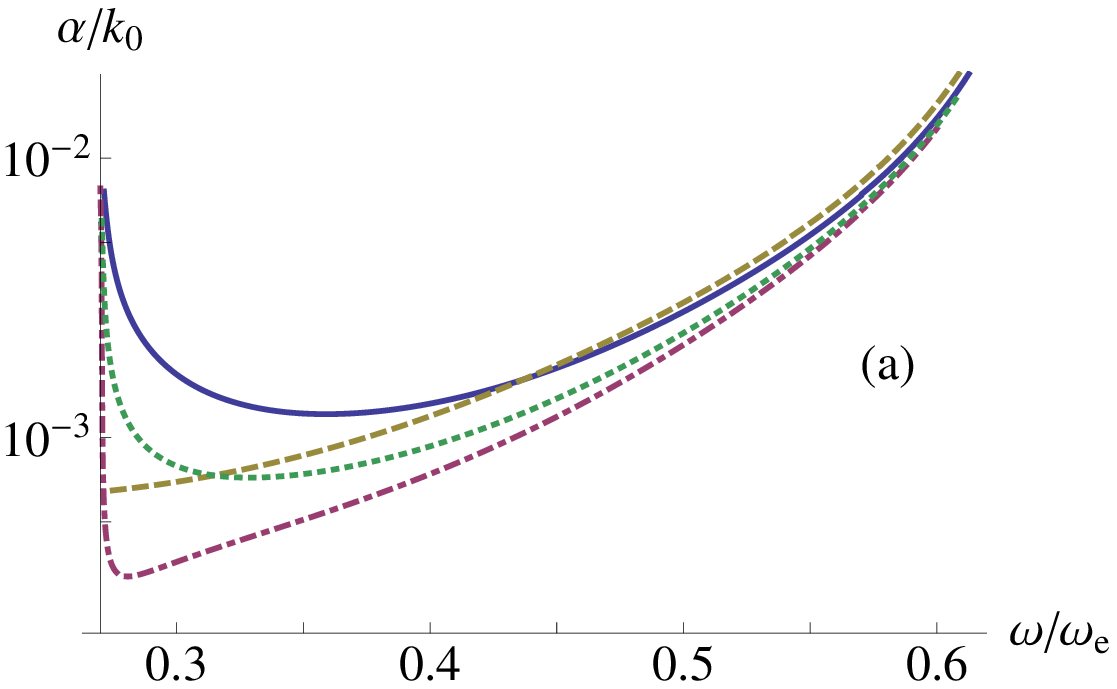}}\\
	\subfloat{\label{fig:HE1CylAbsSurfacegm0}\includegraphics[width=0.5\textwidth]{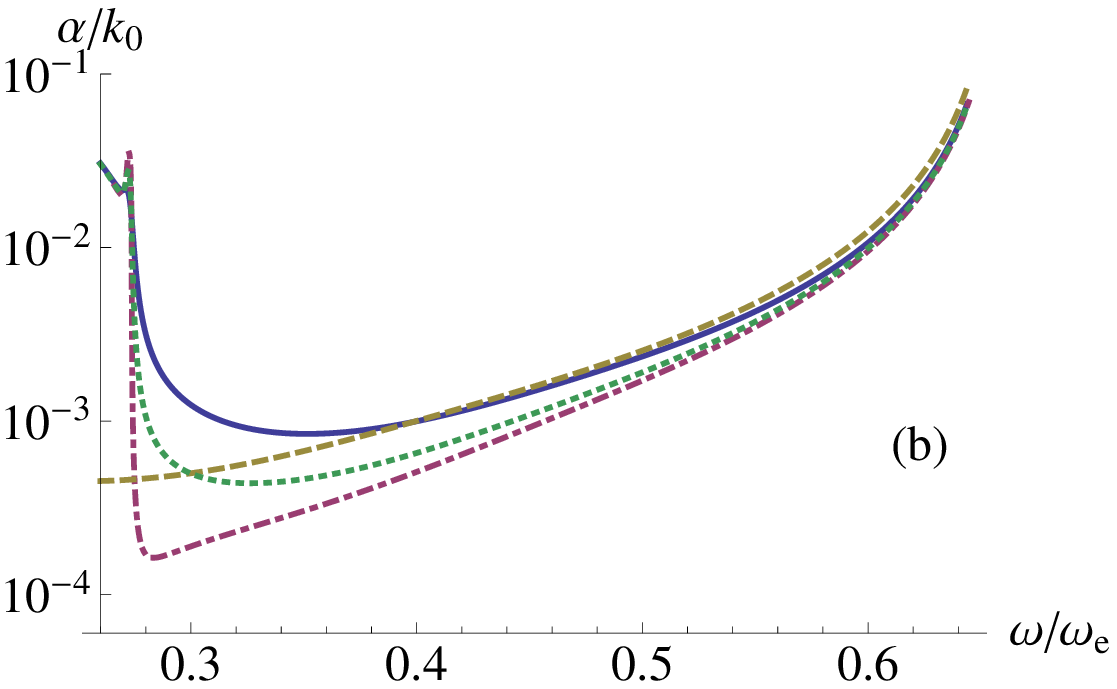}}
	\subfloat{\label{fig:HE1CylHeightSurfacegm0}\includegraphics[width=0.5\textwidth]{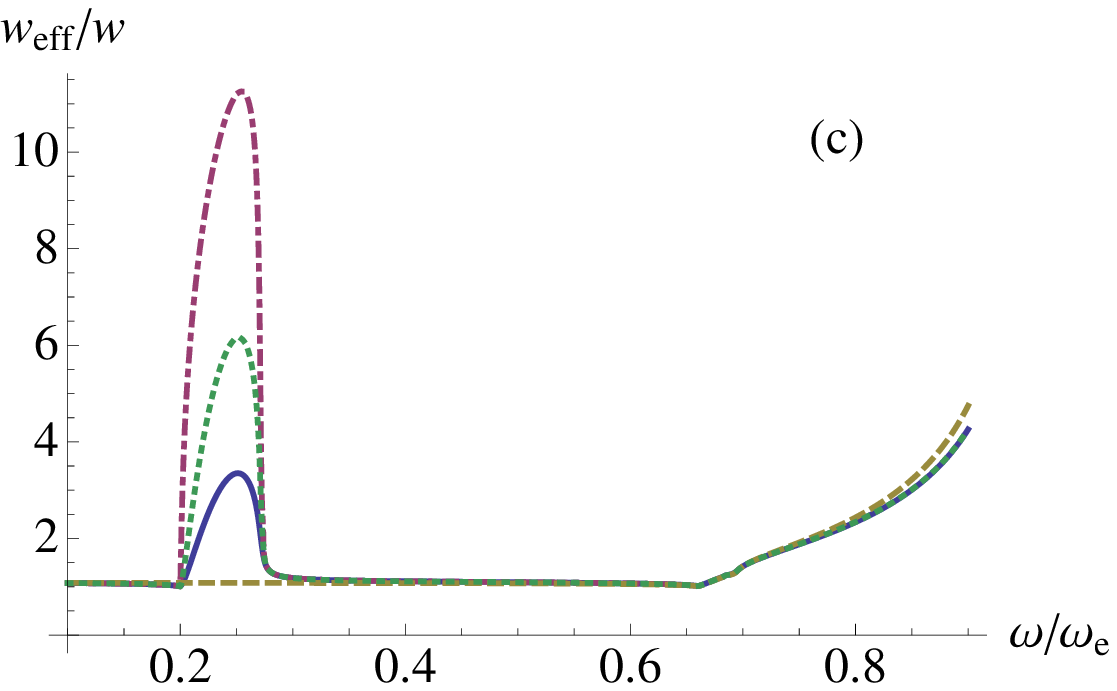}}
	\caption[Plots showing the advantage of the low-loss surface mode.]{Attenuation for \protect\subref{fig:TMSlabAbsSurfacegm0} the TM$_{\rm{s}}$ mode of the slab guide and \protect\subref{fig:HE1CylAbsSurfacegm0} the HE$_{\rm{1s}}$ mode of the cylindrical guide, and \protect\subref{fig:HE1CylHeightSurfacegm0} the effective width of the HE$_{\rm{1s}}$ mode of the cylindrical guide. The solid line is for the metamaterial-dielectric guide, the dashed line is for the metal-dielectric guide the dot-dashed line is for a metamaterial-dielectric guide with $\Gamma_{\rm{m}}=\Gamma_{\rm{e}}/100$ for the metamaterial and the dotted line is for a metamaterial-dielectric guide with $\Gamma_{\rm{m}}=\Gamma_{\rm{e}}/3$ for the metamaterial.}
     \label{fig:AbsSurfacegm0}   
\end{figure}

The losses associated with the magnetic interaction are included in the permeability through the parameter $\Gamma_{\rm{m}}$. Figure~\ref{fig:AbsSurfacegm0}\subref{fig:TMSlabAbsSurfacegm0}
shows a comparison of the TM$_{\rm{s}}$ modes for the metamaterial slab guide ($\Gamma_{\rm{m}}=\Gamma_{\rm{e}}=2.73\times10^{13}{\rm s}^{-1}$), the metal slab guide ($\Gamma_{\rm{m}}=0$), the metamaterial slab guide with reduced magnetic losses ($\Gamma_{\rm{m}}=\Gamma_{\rm e}/3$), as well as the metamaterial slab guide with greatly reduced magnetic losses ($\Gamma_{\rm{m}}=\Gamma_{\rm e}/100$). We choose $\Gamma_{\rm{m}}=\Gamma_{\rm{e}}/3$ to generate the plots as this should be practically achievable~\cite{Zhou:2008}.  The attenuation dip seen in metamaterial-dielectric waveguides is related to that seen for surface waves on a single metamaterial-dielectric interface~\cite{Kamli:2008,Moiseev:2010}. The solution for a single interface, as well as the attenuation dip, is recoverable from the metamaterial guides in the limit of large guide width.

Close inspection of Fig.~\ref{fig:AbsSurfacegm0}\subref{fig:TMSlabAbsSurfacegm0} reveals that for $\omega\gtrsim0.45\omega_{\rm e}$ the attenuation of a metamaterial-dielectric guide with $\Gamma_{\rm{m}}=\Gamma_{\rm e}$ is slightly less than that of a metal-dielectric guide. Reducing $\Gamma_{\rm m}$ has the dual effect of further reducing the attenuation of the metamaterial-dielectric guide as well as broadening the frequency window in which the attenuation is less than that of a metal-dielectric guide. Similarly, the attenuation of the HE$_{1\rm{s}}$ mode of the metamaterial-dielectric cylindrical guide has lower attenuation than that of the metal-dielectric guide for $\omega\gtrsim0.45\omega_{\rm e}$. Reducing $\Gamma_{\rm m}$ for the cylindrical guide has the effect of reducing modal attenuation, as it does for the slab guide. Also, when $\Gamma_{\rm{m}}=\Gamma_{\rm e}/100$ the attenuation of the the HE$_{1\rm{s}}$ mode of the metamaterial guide at $\omega\approx0.3\omega_{\rm e}$ is 0.36 times that for the metal guide.

Reducing the value of $\Gamma_{\rm{e}}$, which parameterizes loss due to atomic interactions with the electric field, also has the effect of reducing the total attenuation of the TM and HE modes in a metamaterial-dielectric waveguide. However, reducing $\Gamma_{\rm{e}}$ for a metal-dielectric waveguide also reduces the attenuation of its TM and HE modes. Thus for a metamaterial-dielectric waveguide, reducing $\Gamma_{\rm{e}}$ alone does not yield any benefit over a similar reduction for a metal-dielectric waveguide. The benefit is in combining the reduced $\Gamma_{\rm{e}}$ with a reduced $\Gamma_{\rm{m}}$ to compound the two effects, which significantly reduces attenuation in metamaterial-dielectric waveguides.

All-optical control of low-intensity pulses requires low attenuation and strong transverse field confinement for large cross-phase modulation. The cylindrical metamaterial-dielectric guide must meet the two requirements simultaneously to suffice for all-optical control. The first requirement, low attenuation, allows a low-intensity pulse to pass through the guide without being lost. By extending the results of Kamli, Moiseev and Sanders~\cite{Kamli:2008}, we have shown cylindrical metamaterial-dielectric waveguides can have lower attenuation than metal-dielectric waveguides when the magnetic losses of the metamaterial are reduced. 

The second requirement for all-optical control is strong confinement of the field in the transverse direction (normal to the core-cladding interface). Strong confinement allows the pulse to be confined to a smaller volume, thus increasing the local field intensity and enhancing the nonlinear response. As the low-loss modes of the cylindrical guide are due to field expulsion from the metamaterial cladding, the fields are almost entirely contained in the core. Thus, the metamaterial-dielectric cylindrical waveguide is a good candidate for enhancing cross-phase modulation between low-intensity pulses.

Metamaterial-dielectric waveguides support modes with three distinct regimes, ordinary, surface, and hybrid ordinary-surface waves. The existence of the hybrid wave regime is a direct result of the waveguide dissipating energy. Models for metamaterial waveguides that do not dissipate energy (lossless) do not predict hybrid modes~\cite{Shadrivov:2003,Qing:2005} due to the fact that the wavenumbers in the lossless model are allowed to be either real or imaginary, but not complex. Thus, the supported modes are either surface or ordinary waves, respectively.

Modelling the waveguide with a dissipative material allows the wavenumbers to be complex, meaning the modes can have characteristics of both surface and ordinary waves. Hybrid waves in metamaterial-dielectric guides are supported for a much broader frequency range than those in a metal-dielectric guide. The wider frequency range may make observing hybrid waves in practice easier for metamaterial guides than for metal guides.

Modes with hybrid-wave behaviour also have large attenuation relative to the other supported wave types. One possible use of hybrid modes is frequency filtering to remove unwanted frequencies from pulses. Pulses with frequencies outside of the hybrid wave region are allowed to propagate further, either as an ordinary wave or as a surface wave depending on the mode. Any pulse with frequencies that fall in the region of large attenuation propagate as a hybrid wave and are quickly extinguished.

\begin{figure}[t,b] 
      \centering
	\subfloat{\label{fig:slabvesurf}\includegraphics[width=0.5\textwidth]{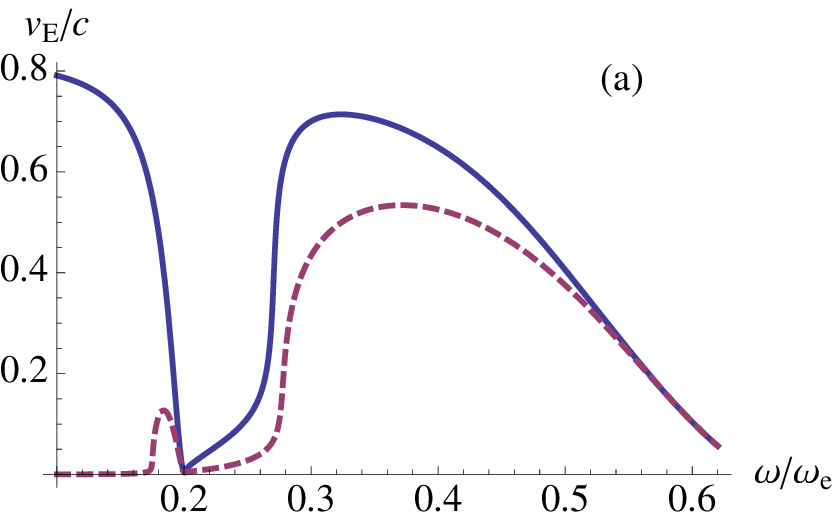}}\\
	\subfloat{\label{fig:slabveord}\includegraphics[width=0.5\textwidth]{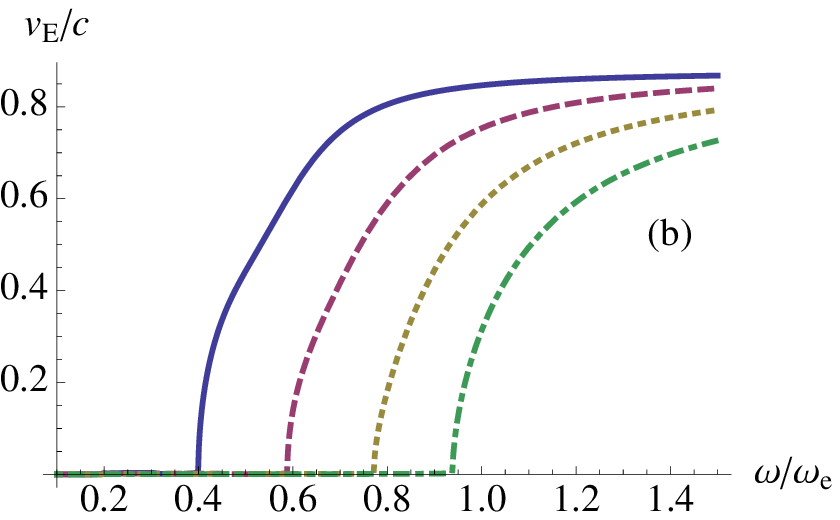}}	
	\subfloat{\label{fig:slabveord-zoom}\includegraphics[width=0.5\textwidth]{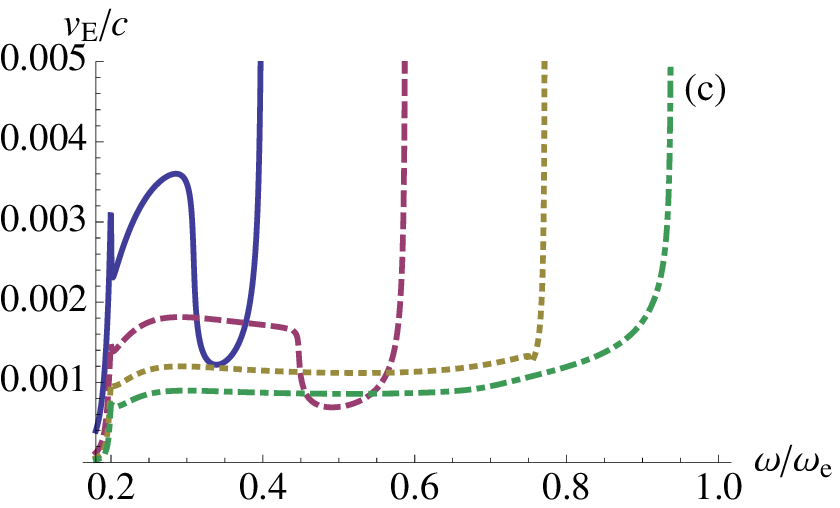}}
	\caption[Energy velocity of various modes in a metamaterial-clad slab waveguide.]{Plots showing the energy velocity as a function of frequency for \protect\subref{fig:slabvesurf} the TM$_{\text s}$ (solid) and TM$_{\text a}$ (dashed) modes, and \protect\subref{fig:slabveord} the TM$_{0}$ (solid), TM$_{1}$ (dashed) and TM$_{2}$ (dotted) and TM$_{3}$ (dot-dashed) modes of the metamaterial-clad slab guide. A close-up on the frequency range $0.18\omega_{\text e}$ to $\omega_{\text e}$ for the modes plotted in \protect\subref{fig:slabveord} is shown in \protect\subref{fig:slabveord-zoom} to better show the velocities for the hybrid waves.}
     \label{fig:slabve}   
\end{figure}

\begin{figure}[t,b] 
      \centering
	\subfloat{\label{fig:cylvesurf}\includegraphics[width=0.5\textwidth]{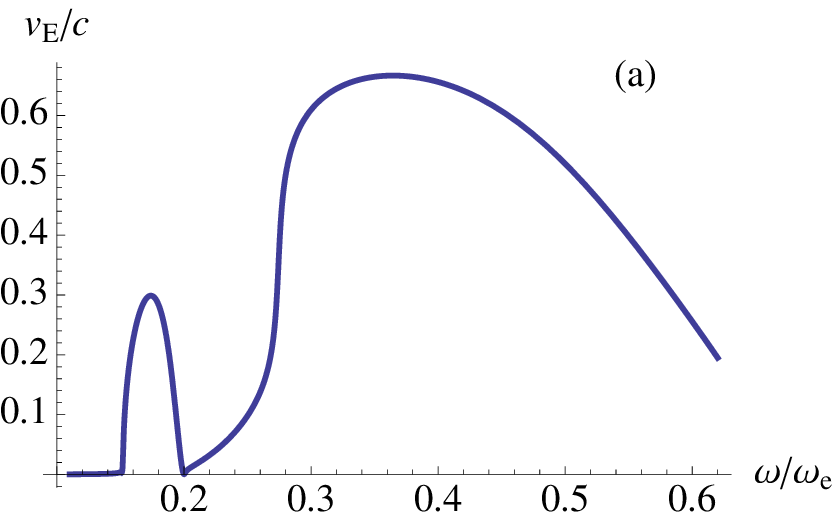}}\\
	\subfloat{\label{fig:cylveord}\includegraphics[width=0.5\textwidth]{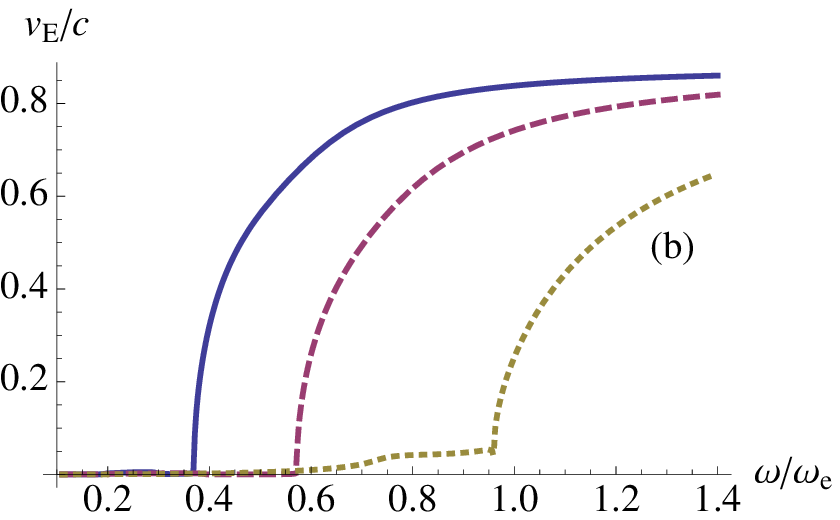}}	
	\subfloat{\label{fig:cylveord-zoom}\includegraphics[width=0.5\textwidth]{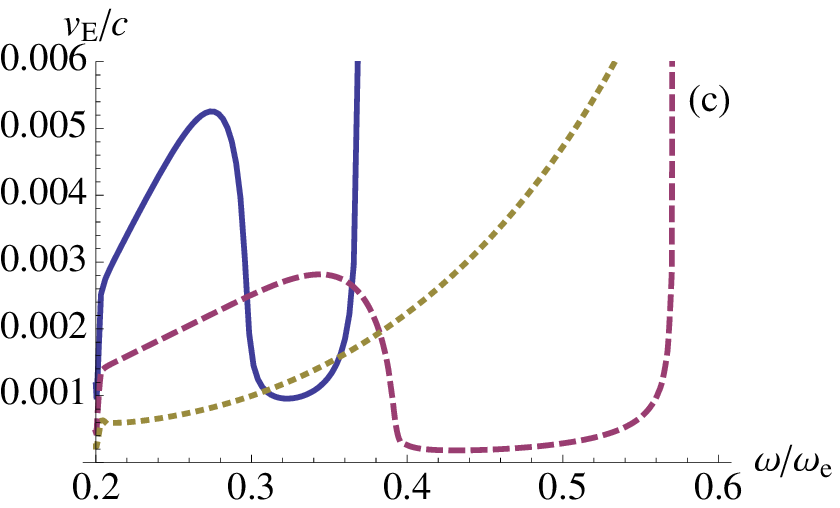}}
	\caption[Energy velocity of various modes in a metamaterial-clad cylindrical waveguide.]{Plots showing the energy velocity as a function of frequency for \protect\subref{fig:cylvesurf} the TM$_{\text s}$ and \protect\subref{fig:cylveord} the TM$_{0}$ (solid), TM$_{1}$ (dashed) and TM$_{3}$ (dotted) modes of the metamaterial-clad cylindrical guide. A close-up on the frequency range $0.2\omega_{\text e}$ to $0.6\omega_{\text e}$ for the modes plotted in \protect\subref{fig:cylveord} is shown in \protect\subref{fig:cylveord-zoom} to better show the velocities for the hybrid waves.}
     \label{fig:cylve}   
\end{figure}

Metamaterial-clad waveguides are also capable of slowing light due to the fact that the energy flux in the metamaterial cladding can be in the opposite direction to the energy flux in the core~\cite{Shadrivov:2003}. To determine the speed of propagation along the guide we use the energy velocity~\cite{Loudon:1970}, defined as
\begin{equation}
v_{\text E}=\frac{\int\bm S\cdot\bm{\text d A}}{\int u\;\text d A},\label{eq:energyvel}
\end{equation}
with $\bm S=\frac{1}{2}\Re{\bm E \times \bm H^{*}}$ the Poynting vector, $u$ the energy density given by Eq.~\ref{nimenergy}, $\bm{\text d A}$ the infinitesimal area element perpendicular to the direction of energy flow (the $\hat z$ direction in this case), and $\text d A$ the non-directional area element. The group velocity cannot be used as hybrid modes have large attenuation, and the group velocity is not a good measure for systems with large attenuation~\cite{Loudon:1970}. The energy velocity is plotted as a function of frequency in Fig.~\ref{fig:slabve} for the metamaterial-clad slab waveguide and in Fig.~\ref{fig:cylve} for the metamaterial clad cylindrical waveguide.

For the slab guide, by comparing Fig.~\ref{fig:slabve} with Figs.~\ref{fig:relgammaSlab}, \ref{fig:TMSlabDisp} and \ref{fig:TMSlabAbs}, and for the cylindrical guide, by comparing Fig.~\ref{fig:cylve} with Figs.~\ref{fig:relgammaCyl}, \ref{fig:TMCylDisp} and \ref{fig:CylAbs}, it can be seen that the modes are generally hybrid waves at the frequencies where they exhibit a low energy velocity. The exceptions are the modes that have a small energy velocity in a frequency region that has a near-zero value of $\beta/k_{0}$, but are still ordinary waves.

\begin{figure}[t,b] 
      \centering
	\subfloat{\label{fig:slabvesurfmetal}\includegraphics[width=0.5\textwidth]{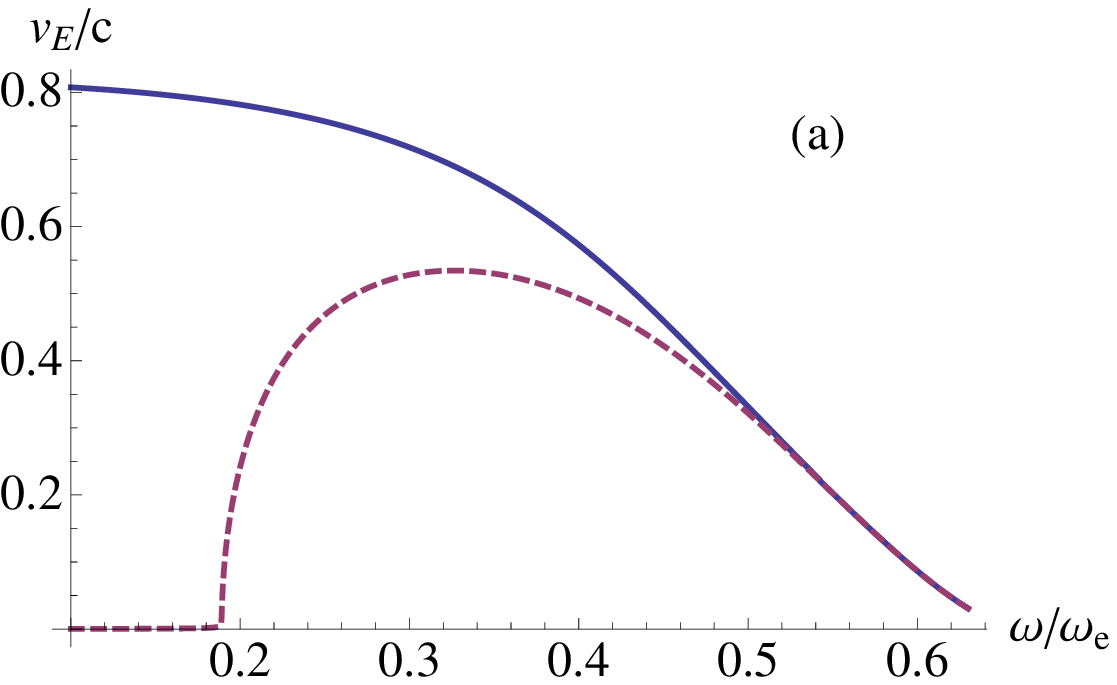}}\\
	\subfloat{\label{fig:slabveordmetal}\includegraphics[width=0.5\textwidth]{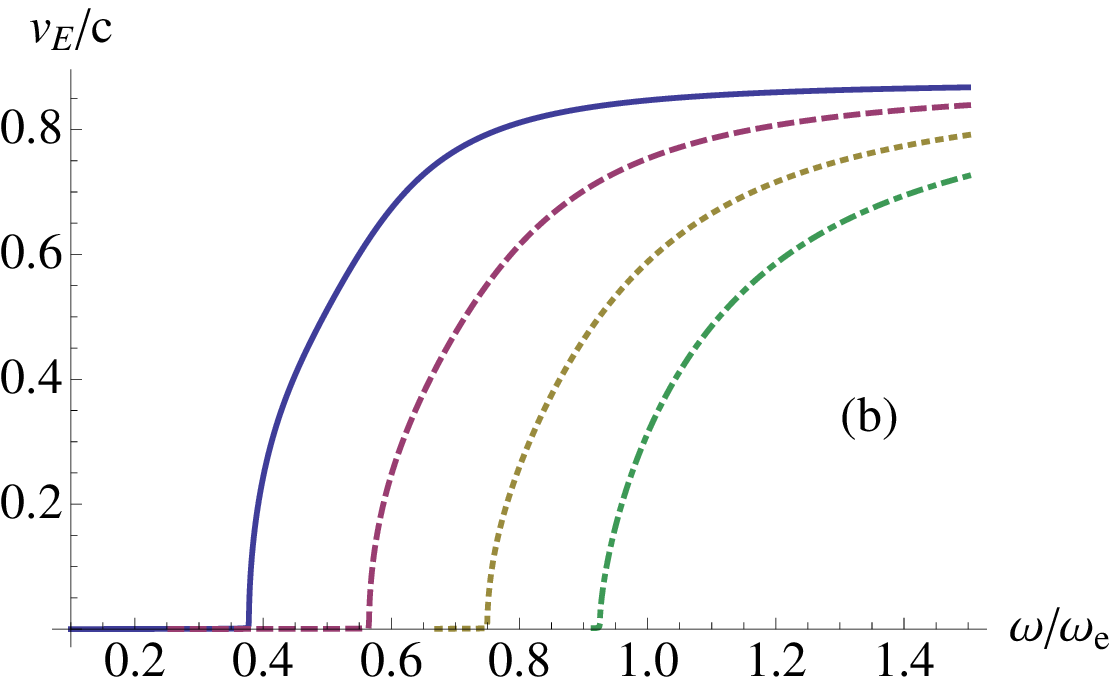}}	
	\subfloat{\label{fig:slabveordmetal-zoom}\includegraphics[width=0.5\textwidth]{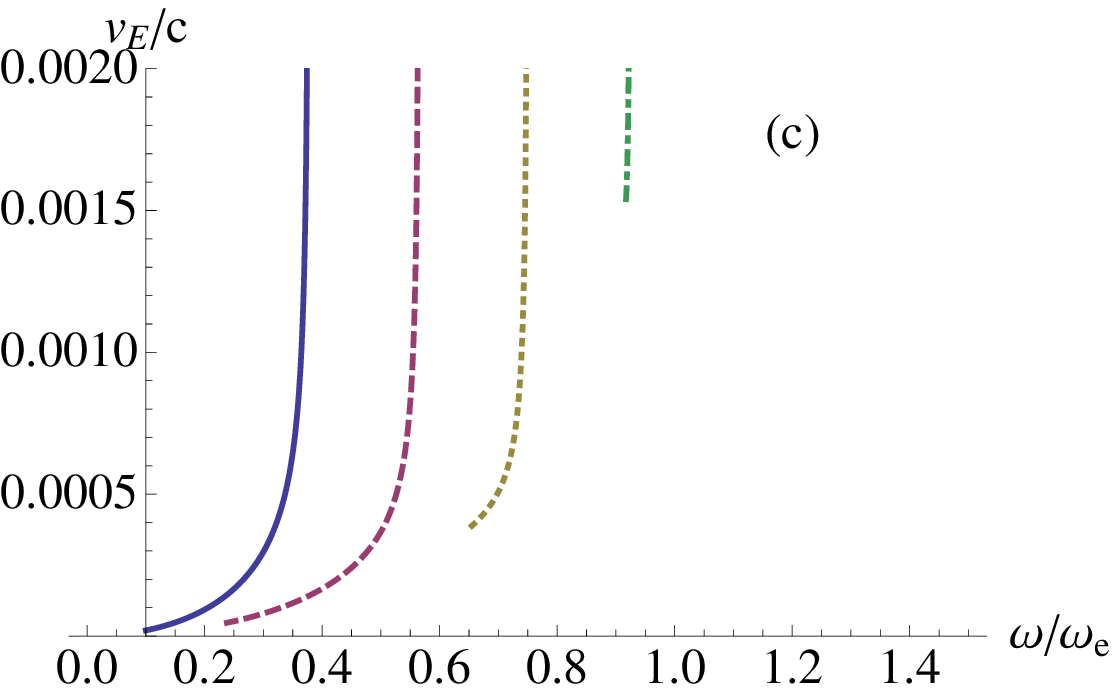}}
	\caption[Energy velocity of various modes in a metal-clad slab waveguide.]{Plots showing the energy velocity as a function of frequency for \protect\subref{fig:slabvesurfmetal} the TM$_{\text s}$ (solid) and TM$_{\text a}$ (dashed) modes, and \protect\subref{fig:slabveordmetal} the TM$_{0}$ (solid), TM$_{1}$ (dashed) and TM$_{2}$ (dotted) and TM$_{3}$ (dot-dashed) modes of the metal-clad slab guide. A close-up on the modes plotted in \protect\subref{fig:slabveordmetal} is shown in \protect\subref{fig:slabveordmetal-zoom} to better show the velocities near cut-off.}
     \label{fig:slabvemetal}   
\end{figure} 

\begin{figure}[t,b] 
      \centering
	\subfloat{\label{fig:cylvesurfmetal}\includegraphics[width=0.5\textwidth]{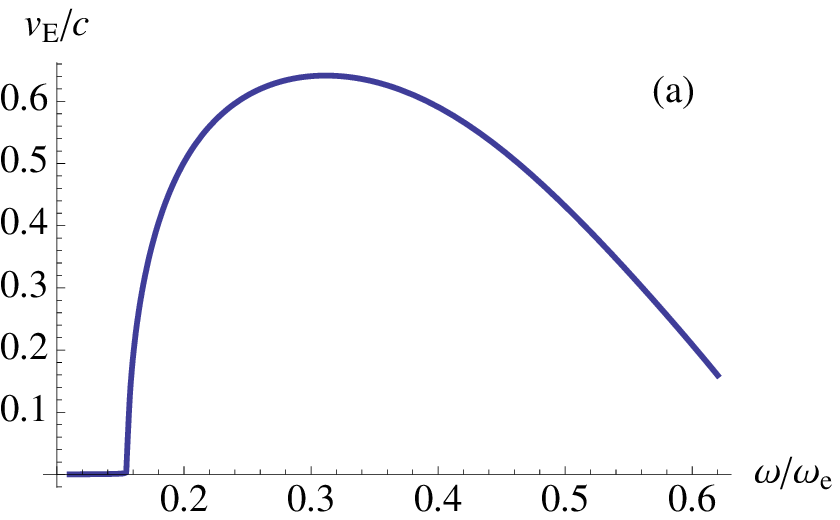}}\\
	\subfloat{\label{fig:cylveordmetal}\includegraphics[width=0.5\textwidth]{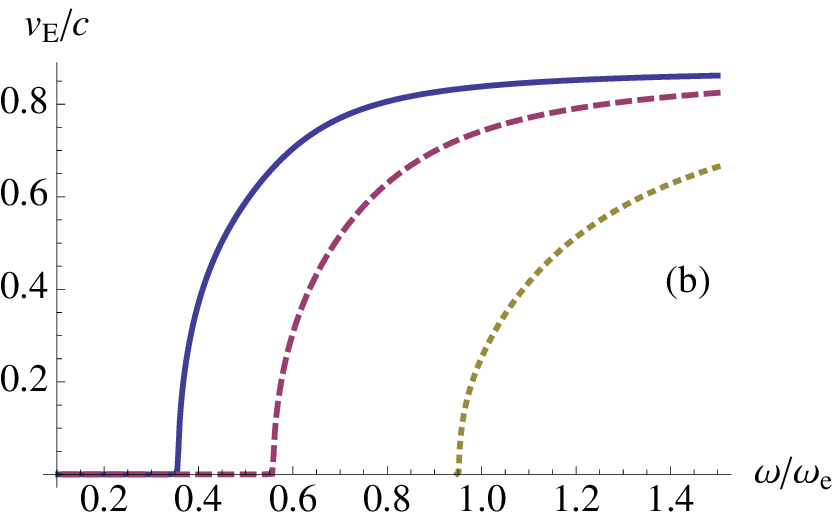}}	
	\subfloat{\label{fig:cylveordmetal-zoom}\includegraphics[width=0.5\textwidth]{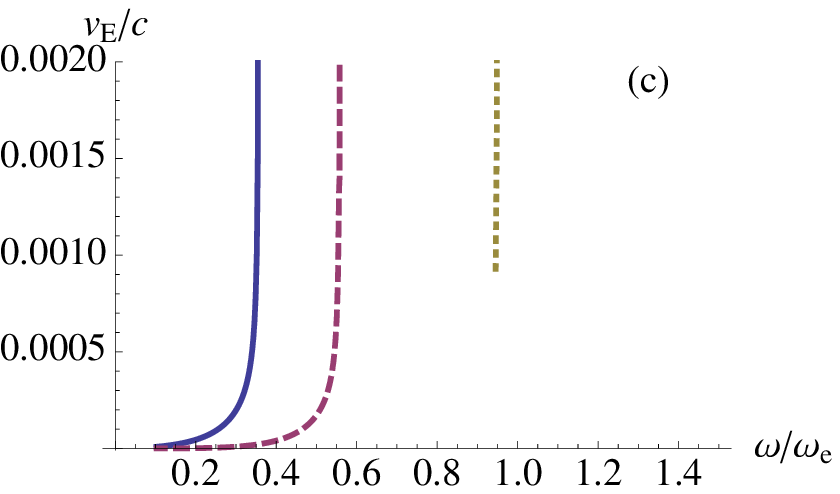}}
	\caption[Energy velocity of various modes in a metal-clad cylindrical waveguide.]{Plots showing the energy velocity as a function of frequency for \protect\subref{fig:cylvesurfmetal} the TM$_{\text s}$ and \protect\subref{fig:cylveordmetal} the TM$_{0}$ (solid), TM$_{1}$ (dashed) and TM$_{3}$ (dotted) modes of the metal-clad cylindrical guide. A close-up on the frequency range $0.2\omega_{\text e}$ to $0.6\omega_{\text e}$ for the modes plotted in \protect\subref{fig:cylveordmetal} is shown in \protect\subref{fig:cylveordmetal-zoom} to better show the velocities near cut-off.}
     \label{fig:cylvemetal}   
\end{figure}

The low energy velocity of hybrid waves may be part of the reason that they have such a large attenuation, which is a measure of energy dissipation per unit length. Having a slower energy velocity gives the hybrid waves more time to interact with the metamaterial, thereby dissipating more energy within the same length of waveguide.

The slow light achieved in this manner has the disadvantage that it is not directly controllable. The energy velocity ultimately depends on a number of factors, including the metamaterial parameters, signal frequency and propagation mode. However, these factors must be set either during device fabrication or at the time of pulse launch and cannot be tuned during operation.

The group velocity is, therefore, set for each particular pulse and cannot be changed during propagation through the device. Thus, for a device that is to controllably slow light, a scheme consisting of simply a dielectric core surrounded by a metamaterial cladding is insufficient. The low-loss surface mode supported by this scheme can be exploited, but a method for slowing the light controllably, electromagnetically induced transparency for instance, must be included.


\section{Summary}

Our results show that reducing the energy loss associated with the magnetic interactions reduces the attenuation of the surface modes in metamaterial-dielectric waveguides. For $\Gamma_{\rm{m}}\lesssim\Gamma_{\rm{e}}$ the attenuation is below that for a metal-dielectric guide at certain frequencies, despite the metal and metamaterials having identical expressions for the permittivity. Reducing $\Gamma_{\rm m}$ further reduces the attenuation in the metamaterial-dielectric guide as well as broadening the frequency window of reduced attenuation. For example, with our choice of parameters, the attenuation of the HE$_{1\rm{s}}$ mode of the metamaterial-clad cylindrical guide can be reduced to as low as about 0.36 times that of the HE$_{1\rm{s}}$ mode in the metal-clad guide.

With the use of structural improvements, or some other means of reducing $\Gamma_{\rm{m}}$, it is possible, in principle, to construct metamaterial-dielectric waveguide devices with low losses compared to a metal-dielectric guide. With the capability of transverse confinement while supporting low-loss modes, the metamaterial cylindrical guide is a good candidate for enhancing cross-phase modulation and ultimately all-optical control of low-intensity pulses.

We have examined the modal properties of a slab and cylindrical metamaterial-dielectric waveguide using a model that includes energy loss. We have shown that the modes of metamaterial-dielectric waveguides support three distinct regimes. They are ordinary, surface and hybrid, with hybrid modes being a feature unique to models that include energy loss. Ordinary and surface modes are well known and have been studied previously. Hybrid waves, however, have not been previously examined for metamaterial-dielectric waveguides. 

Ordinary waves have the bulk of their energy distributed throughout the core and are the result of internal reflections and interference effects. Surface waves, on the other hand, have their energy concentrated at the interface(s) of the guide, which is due to the fact that the energy is being transported along the guide through the oscillations of surface electrons. Hybrid waves, as their name suggests, show a combination of the two mechanisms and, therefore, a combination of their respective energy distributions.

Hybrid waves typically have a large effective width when compared to the other mode types, which implies a large amount of energy is in the cladding. When the energy of a mode is carried in a cladding that dissipates energy, the result is attenuation of the field. Our model incorporates a lossy cladding so the increased amount of energy in the cladding for hybrid modes is consistent with the increased attenuation. Within the negative-index frequency range of the metamaterial, all of the modes of the waveguides are hybrid waves and display the associated large attenuation and increased effective guide width.

The sharp change seen in the attenuation curves could be advantageous as a frequency filter. A pulse propagating through a metamaterial guide would have the frequency components in the high-attenuation regions removed through energy dissipation, while the remainder of the pulse continues to propagate. The frequencies at which the attenuation change occurs may be selected through the design of the metamaterial.

%% file: Slowlight.tex
\chapter{Slow Light With Three-Level Atoms in a Metamaterial Waveguide}\label{ch:slowlight}
\section{Introduction}\label{ch:slowlight-introduction}

Plasmonic devices show promise in increasing the speed of electronic devices and networks while still meeting the size requirements of modern electronic devices due to their ability to confine light to sub-wavelength scales~\cite{Brongersma:2010}. Combining such devices with optical phenomena, such as slow light, could yield an array of new optical devices to augment or replace existing technology. The feasibility of using nonlinear metamaterials for slow light has been considered~\cite{D'Aguanno:2008,Tassin:2009}, and a scheme for photon-echo quantum memory for surface plasmon-polaritons on a metamaterial interface has been proposed~\cite{Moiseev:2010book}.

Controllable slow light is useful for optical delay lines, optical buffers~\cite{Boyd:2006}, and enhanced nonlinear interactions~\cite{ZBWang:2006}; however, achieving highly-confined slow light with weak fields is impractical due to the high losses suffered as a result of the plasmonic confinement of light~\cite{Dionne:2006}. There exists a scheme~\cite{Kamli:2008,Moiseev:2010} using a flat interface metamaterial-dielectric waveguide for slowing highly-confined light through electromagnetically induced transparency (EIT)~\cite{Lambropoulos:2007}, while using metamaterial to minimize losses. EIT with weak fields is interesting in its own right for fundamental reasons, to distinguish it from Autler-Townes splitting for example, and also for applications to weak-field sensing~\cite{Anisimov:2011}. However, the scheme has the limitation that field confinement in the waveguide is only in one transverse spatial dimension, which allows propagating fields to diverge. 

\begin{figure}[t,b] 
      \centering
	\includegraphics[width=.6\textwidth]{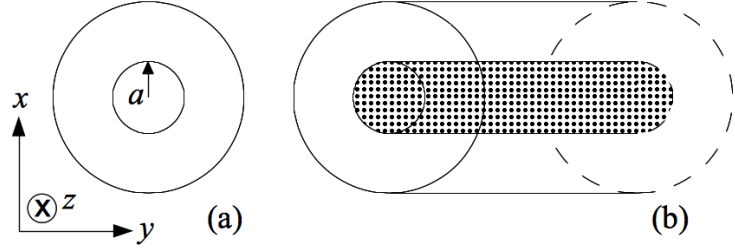}
	\caption[The cylindrical waveguide doped with three-level atoms.]{(a) The cylindrical waveguide cross section. (b) A schematic of the metamaterial-dielectric waveguide embedded with three-level $\Lambda$ atoms (black dots) in the core. The outer layer (cladding) is a metamaterial whereas the core is a dielectric.}
     \label{fig:guidediagram}   
\end{figure}

We propose using a cylindrical waveguide structure as shown in Fig.~\ref{fig:guidediagram}, rather than a flat interface, to inherently provide confinement in both transverse directions and prevent field divergence. The waveguide is composed of a dielectric core and a metamaterial cladding that surrounds the core. Three-level $\Lambda$ atoms (Fig.~\ref{fig:EITsys}) are homogeneously embedded throughout the core of the waveguide and are driven by a pump field in a low-loss surface mode~\cite{Moiseev:2010}, thereby enabling EIT for slowing the signal field in another low-loss surface mode.

To prepare EIT, the atoms are initially strongly pumped to depopulate the $\left|\text s\right>$ level and relax to the $\left|\text g\right>$ state. This process creates Fano interference in the $\left|\text{g}\right>\leftrightarrow\left|\text{e}\right>$ transition resulting in a transparency window at the $\left|\text{g}\right>\leftrightarrow\left|\text{e}\right>$ transition frequency. As a result of the refractive index change, the group velocity of the signal pulse is reduced when traveling through the transparent medium.

After preparing for the EIT, the intensity of the pump field is lowered because the weaker the pump field is, the greater the reduction of group velocity of the signal field. Hence, coherently controlled slow light is achievable by tuning the intensity of the pump field. Furthermore, by comparing the metamaterial-clad waveguide to a metal-clad one with the same permittivity, we show that a low-loss surface mode~\cite{Lavoie:2012} of the metamaterial-clad guide enables the same degree of slowing of light, but with reduced losses. Our analysis is important to fabricating coherently-controlled low-loss slow light devices with current metamaterial technology, and we provide conditions necessary for the existence of the low-loss surface mode for slowing light in the waveguide. 



In Sec.~\ref{ch:slowlight-theory}, we develop the theory for field propagation in the cylindrical metamaterial-clad waveguides without the doping of the three-level atoms. Then we extend the theory to allow for doping the atoms in the core of the waveguide. In Sec.~\ref{ch:slowlight-results}, we describe the numerical method we use to solve Maxwell's equations in the doped waveguide and present results obtained from the solutions. The numerical method used and the analytical approximation used to verify it are detailed in Sec.~\ref{ch:slowlight-AnApprox}. A discussion of our scheme, along with some considerations for implementation and possible applications, is given in Sec.~\ref{ch:slowlight-discussion}, and a summary of the work is given in Sec.~\ref{ch:slowlight-summary}.

\section{Theory}\label{ch:slowlight-theory}

Metamaterials are designed with artificial structure, giving them electromagnetic responses different from those of their constituent materials. We consider a fishnet-type metamaterial as they have been well studied, both experimentally and theoretically~\cite{Ramakrishna:2005,Boardman:2005,Boltasseva:2008,Xiao:2009}. Metamaterials of this design are inherently lossy, so the permittivity and permeability expressions must necessarily include dissipation terms.

The permittivity of the fishnet metamaterial is described by the Drude model given in Eq.~(\ref{epsilonMM}). 
The permeability, which is in practice achieved through the structure of the metamaterial, is described by Eq.~(\ref{muMM}). 
Equations (\ref{epsilonMM}) and (\ref{muMM}) describe the electromagnetic properties of the metamaterial cladding and are used, along with a constant permittivity for the core, to determine the modes of the undoped metamaterial-dielectric waveguide.

After preparation for EIT, the $\left|\text s\right>$ state is not populated and the atoms are no longer excited by the pump field, which means the pump field effectively travels in the modes of the undoped waveguide. On the other hand, the signal field can excite the atoms, so the signal field travels in the modes of the doped waveguide. In the following, we develop the theory for finding the modes in the undoped waveguide before finding the modes of the doped waveguide.


The guided modes in a cylindrical waveguide are transverse electric (TE), transverse magnetic (TM), or a combination of the two (denoted HE or EH)~\cite{Yeh:2008}. The TM and TE modes are those with field components $H_{z}=0$ and $E_{z}=0$, respectively, with $z$ the propagation direction. We restrict our analysis to TM modes, as they are invariant in azimuthal angle $\phi$, which allows us to simplify the calculations. In general, the analysis can be done with any choice of modes.

For the TM modes of the metamaterial-dielectric waveguide, the electric fields have the form
\begin{equation}
\tilde{\bm E}(r,\phi,z,t)={\bm E}(r,\phi)\e^{i(\tilde{\beta} z-\omega t)}+\text{c.c.},
\end{equation}
with c.c.\ indicating the complex conjugate, and $\tilde{\beta}=\beta+i\alpha$ the complex propagation constants for the allowed modes of the waveguide, which are obtained by solving the dispersion relation. Using the definitions given in Eqs.~(\ref{eq:besselderiv}) and (\ref{eq:assocbessderiv}),
the dispersion relation for TM modes in an undoped cylindrical metamaterial-dielectric waveguide is~\cite{Yeh:2008}
\begin{equation}
\frac{\epsilon_{\text{d}}}{\kappa^{2}}\frac{J'_m\left(a \kappa\right)}{J_{m}\left(a\kappa\right)}+\frac{\epsilon(\omega)}{\gamma^{2}}\frac{K'_m\left(a\gamma\right)}{K_m\left(a\gamma\right)}=0,
\end{equation}
with
\begin{equation}
\gamma:=\sqrt{\tilde{\beta}^{2}-\omega^{2}\epsilon(\omega)\mu(\omega)},\quad\kappa:=\sqrt{\omega^{2}\epsilon_{\text d}\mu_{\text{d}}-\tilde{\beta}^{2}},
\end{equation}
$\epsilon_{\text d}$ the permittivity of the dielectric core, $\epsilon(\omega)$ and $\mu(\omega)$ given by Eqs.~(\ref{epsilonMM}) and (\ref{muMM}), respectively, and $a$ the core radius. With this dispersion relation, we can find the low-loss TM surface mode for the pump field as detailed in~\cite{Lavoie:2012}.

The susceptibility of the undoped waveguide core is simply the dielectric susceptibility,
\begin{equation}
\chi_{\text{d}} =\epsilon_{\text{d}}/\epsilon_{0}-1.
\end{equation}
When the core is doped with the three-level atoms, the susceptibility of the core becomes the sum $\chi_{\text{d}}+\chi_{\Lambda}^{(1)}(\omega,r)$, which alters the guided modes near the EIT resonance frequency from those of the undoped guide. To find the low-loss TM surface mode of the signal field, we must solve the wave equation with the susceptibility of the doped waveguide core. The susceptibility of the pumped three-level $\Lambda$ atoms under EIT is given by Eq.~(\ref{EITchi}) with $n=n_{\text d}=c\sqrt{\epsilon_{\text d}\mu_{\text d}}$ the refractive index of the dielectric core, and $\mu_{\text d}$ the permeability of the dielectric core. 

Inside the waveguide core, the pump field expression is given by
\begin{equation}
{\bm E}_{\text{p}}(r)=A\left(J_{0}(\kappa_{\text{p}}r)\,\hat{\bm z}-i\frac{\tilde{\beta}_{\text{p}}}{\kappa_{\text{p}}^{2}}
J_{1}(\kappa_{\text{p}}r)\,\hat{\bm r}\right)\label{Epump},
\end{equation}
with $\hat{\bm z}$ and $\hat{\bm r}$ unit vectors in the propagation and radial directions, respectively; $A$ a constant that determines the amplitude of the pump, $\tilde{\beta}_{\text{p}}$ the complex propagation constant for the pump mode, and
\begin{equation}
\kappa_{\text{p}}:=\sqrt{\omega_{\text{es}}^{2}\epsilon_{\text d}\mu_{\text d}-\tilde{\beta}_{p}^{2}}.
\end{equation}
We define $\eta:=Ad$, which is a parameter adjusted by the amplitude of the pump, with $d=\bm d \cdot\hat{\bm z}=\bm d \cdot\hat{\bm r}$ the dipole moment of the $\left|\text{s}\right>\leftrightarrow\left|\text{e}\right>$ transition. This yields
\begin{equation}
\Omega_{\text p}(r)=\frac{\eta}{\hbar}\left(J_{0}(\kappa_{\text{p}}r)-i\frac{\tilde{\beta}_{\text{p}}}{\kappa_{\text{p}}^{2}}
J_{1}(\kappa_{\text{p}}r)\right),\label{rabifreqfunc}
\end{equation}
the spatially dependent Rabi frequency of the $\left|\text{s}\right>\leftrightarrow\left|\text{e}\right>$ transition.

As the pump field, given by Eq.~(\ref{Epump}), has an $r$ dependence but no $\phi$ or $z$ dependence, so do equations~(\ref{EITchi}) and (\ref{rabifreq}). This is because the pump is in a TM mode, which has a radial dependence but no azimuthal dependence, implying $\phi$ invariance. Furthermore, the pump is approximated as non-depleting, which implies $z$ invariance. This approximation is valid for short propagation lengths, such that the pump does not deplete considerably. The results presented in this paper are based on this approximation. 

For longer propagation lengths the pump does deplete, which reduces the Rabi frequency $\Omega_{\text p}(r)$ and leads to a further reduction in the group velocity. In this case, one can treat the whole waveguide as a series of concatenated short sections along the propagation direction, such that the pump intensity in each section is approximately constant. This treatment allows one to determine the group velocity of the signal in each section and calculate the overall delay of the signal in the whole waveguide.

\begin{figure}[t,b] 
       \centering
	\includegraphics[width=0.45\textwidth]{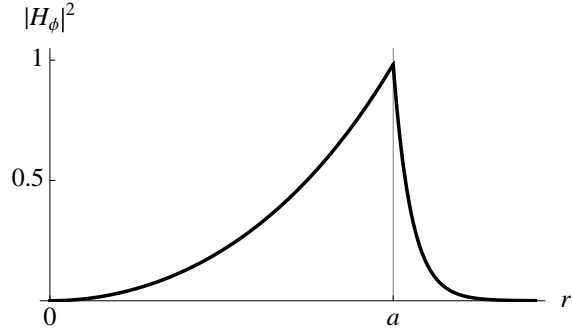}
	\caption[A plot of $|H_{\phi}|^{2}$ of a surface mode]{A plot of $|H_{\phi}|^{2}$ of a surface mode showing the transverse profile in arbitrary units. The thin vertical line indicates the interface between the core and cladding regions.}
     \label{fig:surfacemode}   
\end{figure}

The permittivity of the doped core depends on the pump field. The pump field is in a TM surface mode (see Fig.~\ref{fig:surfacemode}),
so the intensity varies steeply in the radial direction; thus, the signal field experiences a steeply graded refractive index in the core. As a result, within the core, the scalar wave equation used to solve for the modes of the undoped guide is no longer valid for the modes of the doped guide. In the cladding region, however, the permittivity and permeability are not spatially dependent, so we can still find the fields in this region using the scalar wave equation. 

To solve for the propagation constants and the fields in the doped guide, we must begin with the vector wave equation (obtained from Eq.~\ref{eq:genvectorwave})
\begin{equation}
{\nabla}^{2}{\bm E}+{\bm \nabla}\left(\frac{1}{\epsilon_{\text{eff}}(\omega,r)}{\bm E}\cdot{\bm \nabla}\epsilon_{\text{eff}}(\omega,r)\right)+\omega^{2}\mu_{0}\epsilon_{\text{eff}}(\omega,r){\bm E}=0,\label{vectorwave}
\end{equation}
with ${\nabla}^{2}$ the Laplacian and
\begin{equation}
\epsilon_{\text{eff}}(\omega,r)=\begin{cases}
\epsilon_{0}\left(1+\chi_{\text{d}}+\chi_{\Lambda}^{(1)}(\omega, r)\right)&r<a,\\
\epsilon(\omega)&r\geq a,
\end{cases}
\end{equation}
the effective permittivity of the waveguide with three-level $\Lambda$-atoms embedded in the core.
As we are restricting our analysis to TM modes for both the pump and signal fields, Eq.~(\ref{vectorwave}) can be decoupled into scalar equations with the $z$ component taking the form
\begin{equation}
\frac{\text{d}^{2}}{\text{d}r^{2}}E_{z}+\left(\frac{1}{r}-\frac{\tilde{\beta}^{2}}{\kappa_{\text{eff}}^{2}(\omega,r)\epsilon_{\text{eff}}(\omega,r)}\frac{\text{d}}{\text{d}r}\epsilon_{\text{eff}}(\omega,r)\right)\frac{\text{d}}{\text{d}r}E_{z}+\kappa_{\text{eff}}^{2}(\omega,r)E_{z}=0\label{TMvectwave},
\end{equation}
with
\begin{equation}
\kappa_{\text{eff}}(\omega,r):=\begin{cases}
\sqrt{\omega^{2}\mu_{0}\epsilon_{\text{eff}}(\omega,r)-\tilde{\beta}^{2}}&r<a,\\
i\gamma&r\geq a.
\end{cases}
\end{equation}
Equation~(\ref{TMvectwave}) is not generally analytically solvable, except at certain frequencies, so we numerically determine the dispersion and attenuation of the low-loss surface mode near the EIT resonance frequency.


\section{Method and Results}\label{ch:slowlight-results}

There are two main problems with solving Eq.~(\ref{TMvectwave}). First, $\tilde{\beta}$ is complex, meaning it has two free parameters so a standard shooting method cannot be employed~\cite{Press:2007}. The second problem is that the fitness function,
\begin{equation}
f=\left|\frac{H_{\phi,\text{core}}-H_{\phi,\text{clad}}}{H_{\phi,\text{clad}}}\right|,\label{fitness}
\end{equation}
with $H_{\phi,\text{core}}$ and $H_{\phi,\text{clad}}$ the $\phi$ component of the magnetic field in the core and cladding regions, respectively, is not convex. Therefore, implementing a hill descent algorithm is not sufficient to determine if a particular value of $\tilde{\beta}$ is a solution. To circumvent these problems we draw inspiration from the shooting method and employ a trial method to solve Eq.~(\ref{TMvectwave}) for $E_{z}(r,\omega)$. 

We need to find $E_{z}(r,\omega)$ for a range of frequencies of interest. We begin at a frequency where an exact solution for a TM surface mode is found. To find the solution at an adjacent frequency, we slightly deviate the known $\tilde{\beta}$ at the former frequency to obtain a trial value for $\tilde{\beta}$ and numerically solve Eq.~(\ref{TMvectwave}) for $E_{z}(r, \omega)$; other components of the field are calculated from this trial $E_{z}(r,\omega)$. To test whether the trial $\tilde{\beta}$ is acceptable, we check if the corresponding trial $E_{z}(r,\omega)$ satisfies the boundary condition at the core-cladding interface to within a tolerance. Mathematically, the acceptance criterion is $f<10^{-9}$. If the trial $E_{z}(r,\omega)$ does not satisfy the boundary condition, a different trial $\tilde{\beta}$ is chosen, by slightly deviating the known $\tilde{\beta}$ at the former frequency and the search is repeated. When an acceptable $\tilde{\beta}$ is found, we repeat the search for the next frequency until $E_{z}(r,\omega)$ of the TM surface mode is determined for a range of frequencies of interest.

We apply the same metamaterial parameters presented in our previous work~\cite{Lavoie:2012} to our calculations in this work. Hence, the TM surface mode of the undoped metamaterial-clad guide is the mode for the pump of our scheme~\cite{Lavoie:2012}. The metamaterial parameters are shown in Table~\ref{MMparams}. 
\begin{table}[t,b]
\centering
\caption{Parameter values for the metamaterial cladding.}
\begin{tabular}{cc}
\hline\hline Parameter & Value\\
\hline
$\epsilon_{\text b}$&1\\

$\mu_{\text b}$&1\\

$\omega_{\rm e}$&$1.37\times10^{16}{\rm s}^{-1}$\\

$\Gamma_{\rm e}$&$2.73\times10^{13}{\rm s}^{-1}$\\

$\Gamma_{\rm{m}}$&$\Gamma_{\rm e}$\\

$\omega_0$&$0.2\omega_{\rm e}$\\

$F$&$0.5$\\
\hline\hline
\end{tabular}
\label{MMparams}
\end{table}
The core parameters are $\epsilon_{\text d}=1.3\epsilon_{0}$, $\mu_{\text d}=\mu_{0}$ and the core radius $a=4\pi c/\omega_{\text e}$. For the three-level atoms embedded in the core, we set the number density $\rho_{\text a}=1.26\times10^{21}\text{m}^{-3}$ so that the resulting permittivity remains positive.

The decay rate of the excited state is $\gamma_{\text{eg}}=10^{5}\text{s}^{-1}$, and that of the hyperfine state is $\gamma_{\text{sg}}=10^{-2}\text{s}^{-1}$. These decay rates are consistent with observed values for $\text{Pr}^{3+}$ ions embedded in bulk Y$_{2}$SiO$_{5}$ crystal and under cryogenic conditions~\cite{Turukhin:2001}. The mean signal frequency is $\omega_\text{s}=\omega_\text{eg}= 0.409\omega_{\text e}$ and the mean pump frequency is $\omega_\text{p}=\omega_\text{es}=0.41\omega_{\text e}$. The group velocity of the pump mode $v_{\text p}=0.47c$.

\begin{figure}[t,b] 
      \centering
	\subfloat{\label{fig:dopedcyldisp1E4}\includegraphics[width=0.5\textwidth]{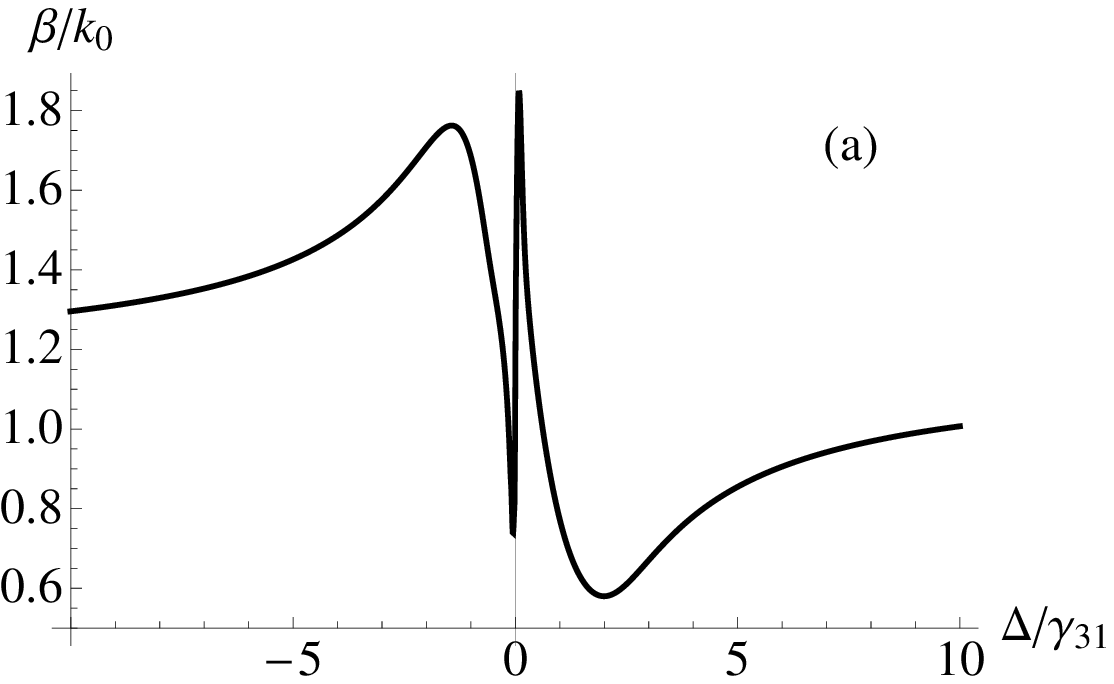}}
	\subfloat{\label{fig:dopedcyldisp2E4}\includegraphics[width=0.5\textwidth]{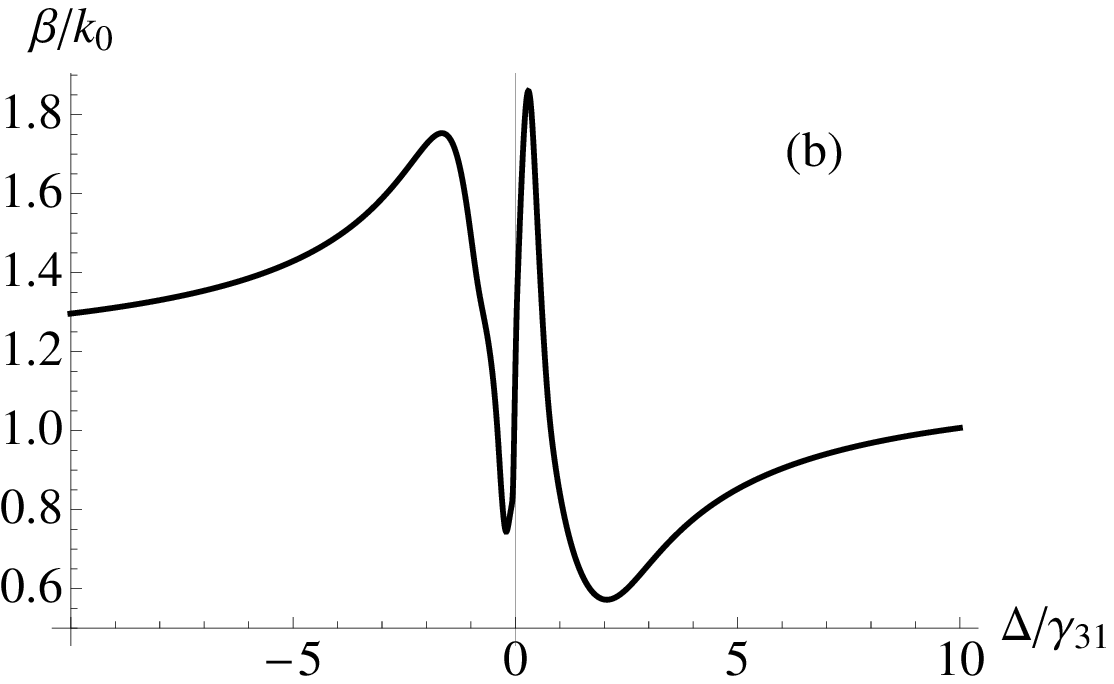}}\\
	\subfloat{\label{fig:dopedcylabs1E4}\includegraphics[width=0.5\textwidth]{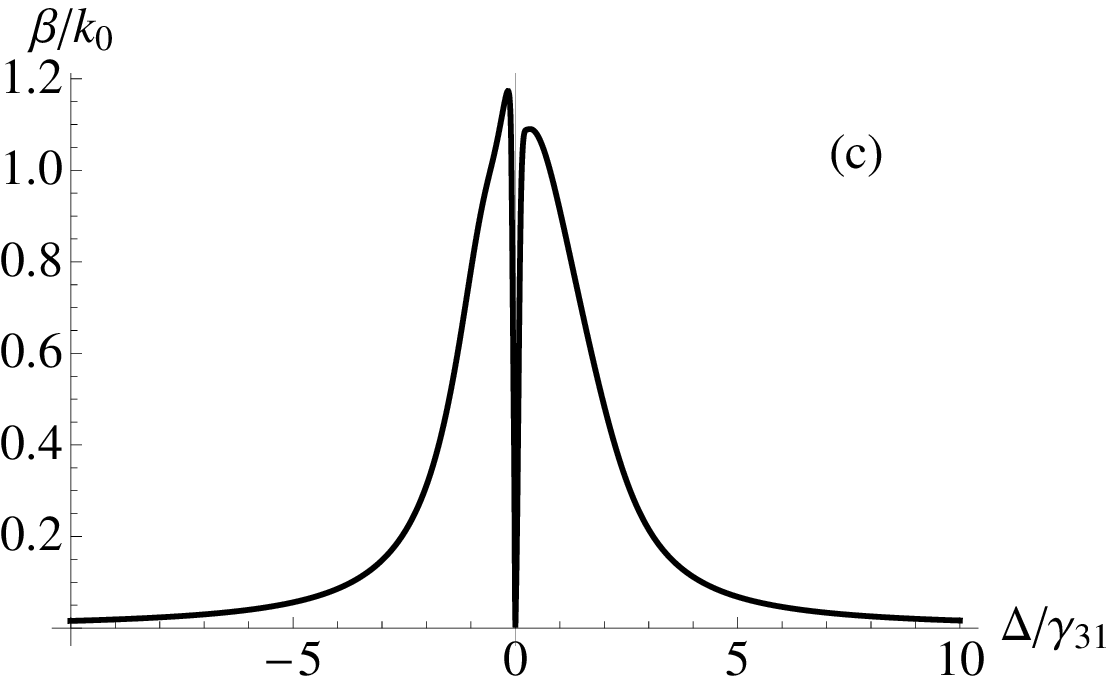}}
	\subfloat{\label{fig:dopedcylabs2E4}\includegraphics[width=0.5\textwidth]{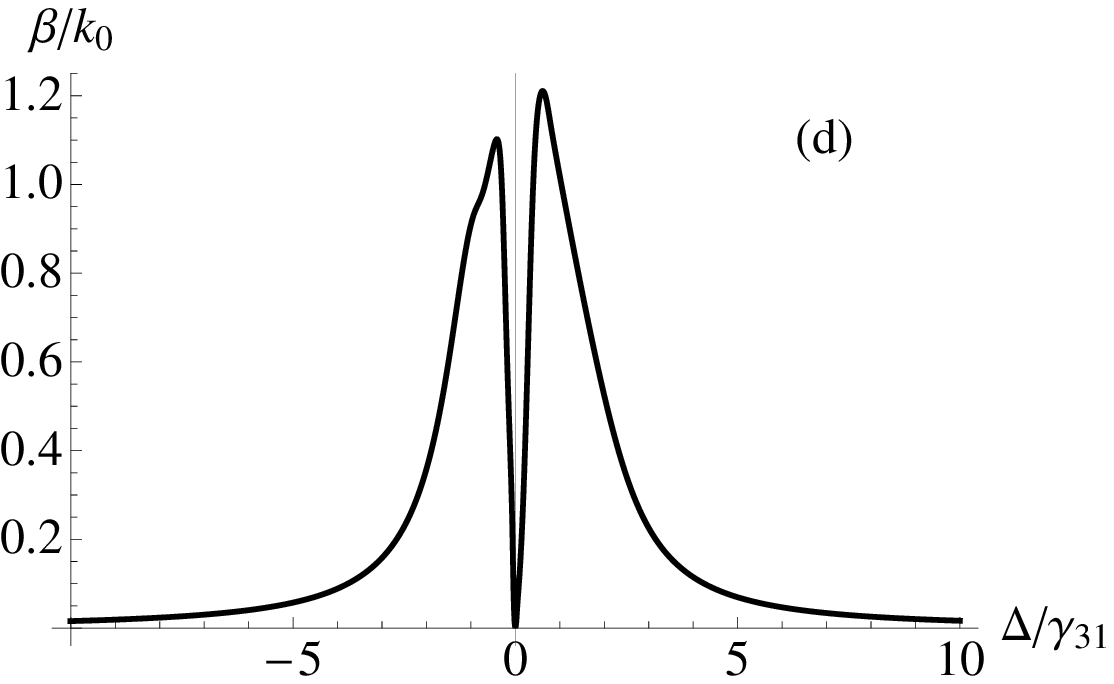}}
	\caption[Dispersion and absorption spectra for the signal field in the doped guide]{Plots of the dispersion \subref{fig:dopedcyldisp1E4} and \subref{fig:dopedcyldisp2E4}, and absorption \subref{fig:dopedcylabs1E4} and \subref{fig:dopedcylabs2E4} as functions of frequency for $\eta/\hbar=1\times10^{4}\text s^{-1}$ and $\eta/\hbar=2\times10^{4}\text s^{-1}$ respectively.}
     \label{fig:dopedbeta}   
\end{figure}

Using the method and parameters outlined above, we have calculated the complex propagation constants of the signal and pump TM surface modes of our scheme. The dispersion and absorption spectra of the signal field are characteristic of EIT near the resonance frequency, with the dispersion having a steep positive slope and the absorption having a deep well. Figure~\ref{fig:dopedbeta} shows the dispersion and absorption for the signal field as a function of frequency. These plots are for the two pump intensities corresponding to $\eta/\hbar=1\times10^{4}\text s^{-1}$ and $2\times10^{4}\text s^{-1}$, at frequencies near the EIT resonance.

\begin{figure}[t,b] 
      \centering
	\subfloat{\label{fig:vgplot}\includegraphics[width=0.5\textwidth]{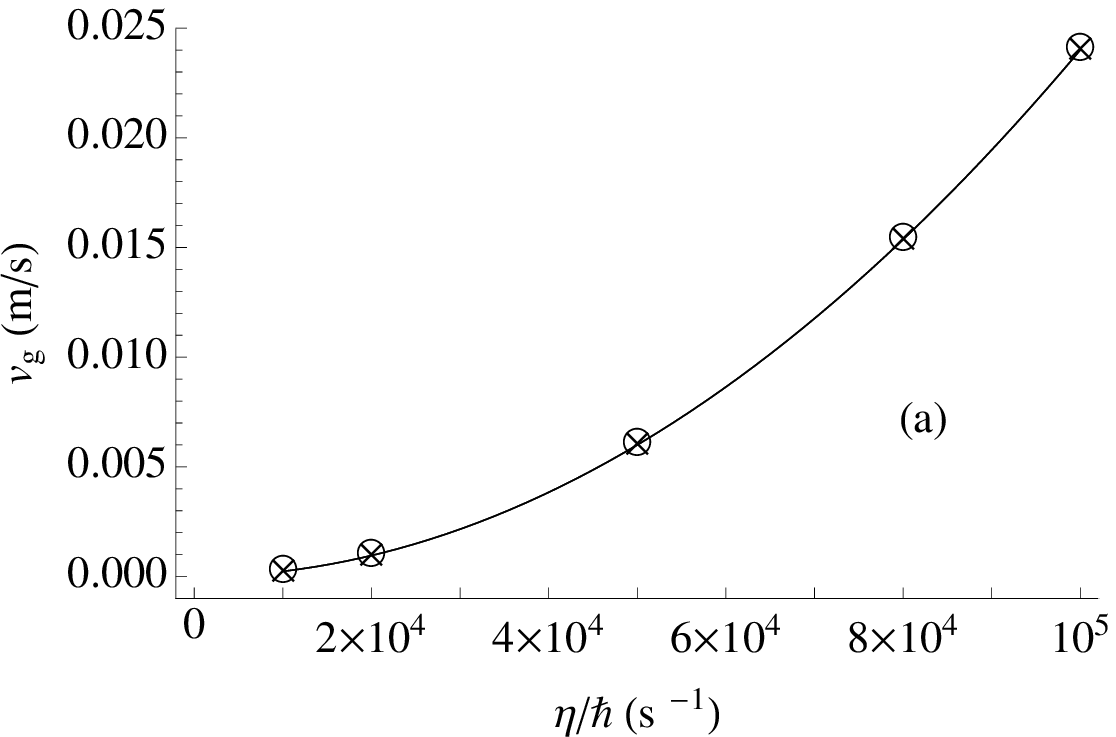}}
	\subfloat{\label{fig:vlogplot}\includegraphics[width=0.5\textwidth]{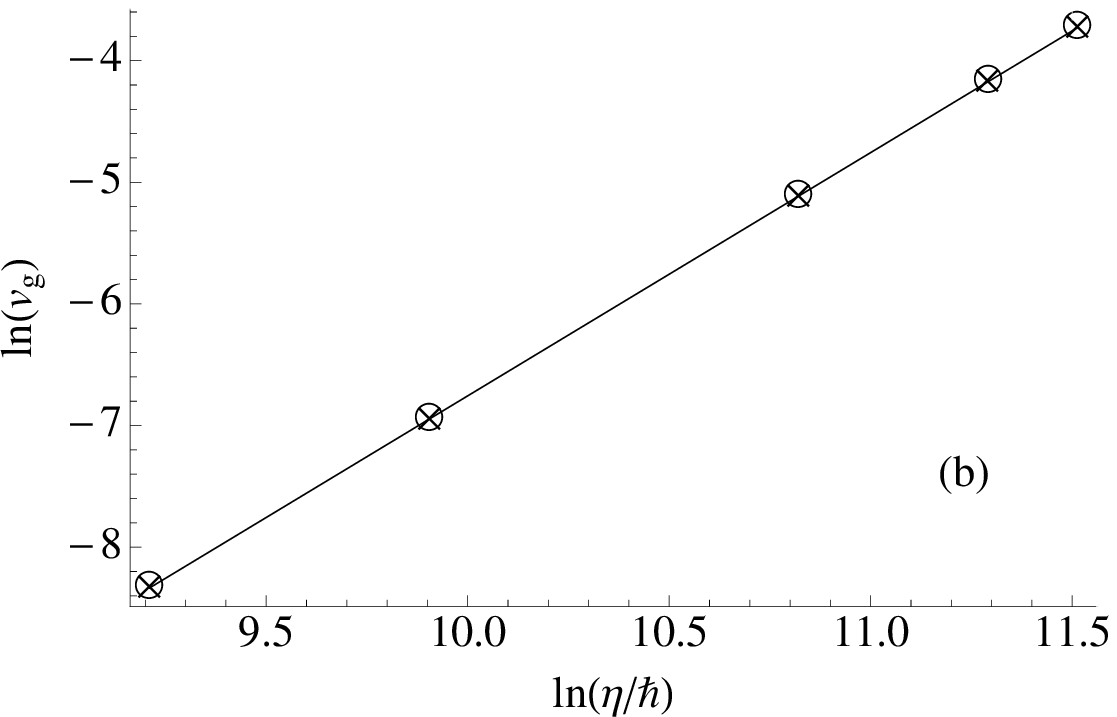}}
	\caption[Plots of the numerically calculated group velocity.]{The numerically calculated group velocity is plotted at $\Delta=0$ as a function of pump intensity for a metamaterial-clad guide ($\ocircle$) with $\Gamma_{\text m}=\Gamma_{\text e}/3$ and a metal-clad guide ($\times$). A linear plot \subref{fig:vgplot} shows the values of the group velocity in m/s, and a log-log plot \subref{fig:vlogplot} shows the relationship between the group velocity and $\eta$. The solid curve in \subref{fig:vgplot} is a least squares quadratic fit, and that in \subref{fig:vlogplot} is a linear regression fit with a slope of 2.0003.}
	 \label{fig:vgplots}   
\end{figure}

With the complex propagation constant $\tilde{\beta}=\beta+i\alpha$ for a range of frequencies, we now calculate the group velocity of the signal field using $v_{\text{g}}=(\text{d}\beta/\text{d}\omega)^{-1}$. Figure~\ref{fig:vgplot} shows the group velocity of the signal field surface mode in both the metal- and metamaterial-clad waveguides for a range of $\eta$. Figure~\ref{fig:vlogplot} shows that the group velocity of the signal for a metamaterial-clad guide is proportional to $\eta^{2}$, and as $\eta\propto|\Omega_{\text p}(r)|$, this agrees with EIT theory~\cite{Lambropoulos:2007}.

\begin{figure}[t,b] 
       \centering
	\includegraphics[width=0.6\textwidth]{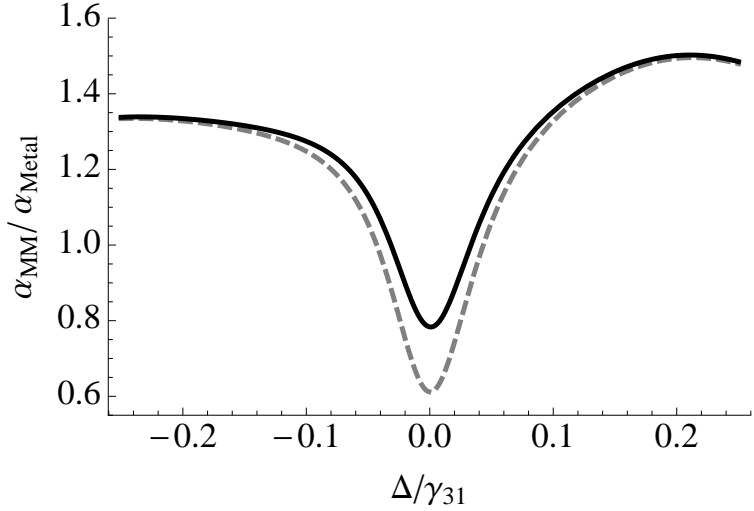}
	\caption[Plots of the relative attenuation of the surface mode]{Plots of the relative attenuation of the surface mode of the metamaterial guide to that of a metal guide ($\alpha_{\text{MM}}/\alpha_{\text{Metal}}$) with $\eta/\hbar=5\times10^{4}\text{s}^{-1}$, for $\Gamma_{\text{m}}=\Gamma_{\text{e}}/3$ (solid) and $\Gamma_{\text{m}}=\Gamma_{\text{e}}/100$ (dashed).}
     \label{fig:lowlosseit}   
\end{figure}

Using the imaginary part of the propagation constant, we obtain curves of the relative attenuation of the metamaterial-clad to the metal-clad waveguides as a function of detuning $\Delta$, as shown in Fig.~\ref{fig:lowlosseit}. At the EIT resonance (i.\e.\ zero detuning), the metamaterial-dielectric guide provides nearly a 20\% reduction in attenuation over a metal-dielectric guide for $\Gamma_{\text{m}}=\Gamma_{\text{e}}/3$ (solid line). For $\Gamma_{\text{m}}=\Gamma_{\text{e}}/100$ (dashed line), the attenuation reduction improves to near 40\% over a metal-dielectric guide. When $\Gamma_{\text{m}}=\Gamma_{\text{e}}$ the attenuation of a metamaterial-dielectric guide is comparable to a metal-dielectric guide. For layered structures, such as the fishnet design, previous work~\cite{Penciu:2010} has shown that $\Gamma_{\text{m}}\leq\Gamma_{\text{e}}$ and a value of $\Gamma_{\text{m}}=\Gamma_{\text{e}}/3$ should be feasible with current metamaterial technology~\cite{Zhou:2008}. Hence, our cylindrical metamaterial-dielectric waveguide is capable of supporting a low-loss TM surface mode.

\begin{figure}[t,b] 
       \centering
	\includegraphics[width=0.6\textwidth]{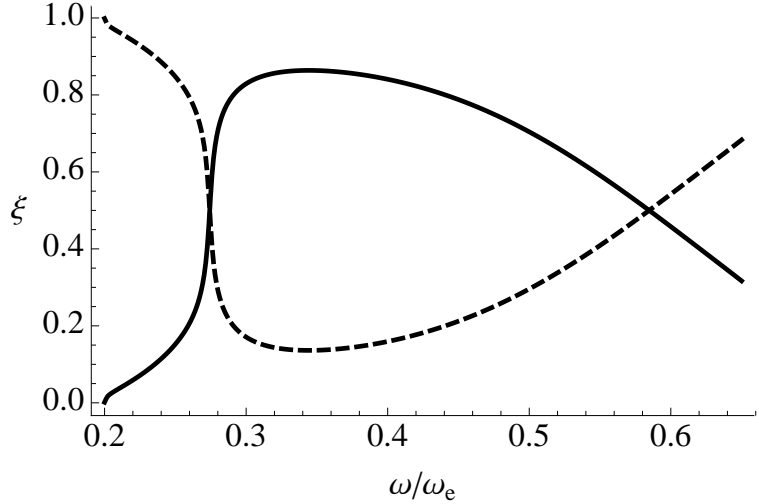}
	\caption[Plot of the fractional energy in a metamaterial guide]{Plots of the fraction of the total energy in the core (solid) and the cladding (dashed) as a function of frequency for the low-loss surface mode of the cylindrical metamaterial-dielectric waveguide.}
     \label{fig:fracen}   
\end{figure}

To understand why the attenuation of the low-loss surface mode does not approach zero near $\omega_{\text s}$, as it does for the flat interface~\cite{Kamli:2008}, we plot the fraction of the total energy of the mode that resides in both the core and cladding. Figure~\ref{fig:fracen} shows the fractional energy $\xi$ of the TM surface mode of the cylindrical metamaterial-dielectric guide as a function of frequency. For a given mode the fractional energy is the ratio of the energy in one part of the guide, the core for instance, to the total energy of the mode. The solid line is for the fractional energy in the core whereas the dashed line is for the cladding. The fraction of energy in the dielectric does not reach 1, and some of the signal field penetrates into the metamaterial, leading to a non-zero attenuation.

 

\section{Numerical Calculation and Analytical Approximation}\label{ch:slowlight-AnApprox}

\noindent\emph{This section is not included in the original publication.}

\begin{figure}
\centering
\begingroup
    \fontsize{10pt}{12pt}\selectfont
\begin{tikzpicture}[node distance = 3cm, auto]
    \node [block] (init) {Initialize model};
    \node [data, left of=init] (expert) {Read data};
    \node [block, below of=init] (search) {Define search space for $\tilde\beta$};
    \node [block, below of=search] (identify) {Generate value for $\tilde{\beta}$};
    \node [block, right of=identify] (solve) {Solve for electric field};
    \node [block, below of=solve] (evaluate) {Evaluate boundary conditions};
    \node [block, right of=evaluate] (newfreq) {Increment $\omega$};
    \node [left of=evaluate] (spacer) {};
    \node [block, left of=spacer, node distance=3cm] (update) {Refine $\tilde \beta$ search space};
    \node [decision, below of=evaluate] (BCtest) {Do BCs match?};
    \node [decision, left of=BCtest] (betterQ) {Is match best yet?};
    \node [inout, below of=BCtest, node distance=3cm] (write) {Write to file};
    \node [decision, right of=write] (doneQ) {Is\\ $\omega>\omega_{\text f}$?};
    \node [block, below of=doneQ] (end) {End};
    \path [line] (init) -- (search);
    \path [line] (search) -- (identify);
    \path [line] (identify) -- (solve);
    \path [line] (solve) -- (evaluate);
    \path [line] (evaluate) -- (BCtest);
    \path [line] (BCtest) -- node [near start]{no} (betterQ);
    \path [line] (betterQ) -| node [very near start]{yes} (update);
    \path [line] (update) |- (identify);
    \path [line] (betterQ) -- node [very near start]{no} (identify);
    \path [line] (BCtest) -- node [near start]{yes}(write);
    \path [line] (write) -- (doneQ);
    \path[line] (doneQ) -- node [very near start]{no} (newfreq);
    \path[line] (newfreq) |- (search);
    \path[line] (doneQ) -- node [near start] {yes} (end);
    \path [line,dashed] (expert) -- (init);
\end{tikzpicture}
\endgroup
\caption[A flow chart for the numerical solution of the doped waveguide.]{A flow chart depicting the program for numerically solving Maxwell's equations for a guide embedded with three-level $\Lambda$ atoms in the core. The dashed line indicates a conditional step, $\tilde\beta$ is the complex propagation constant, $\omega$ is the angular frequency and $\omega_{\text f}$ is the final angular frequency for which a solution is desired.}
\label{fig:numflow}
\end{figure}
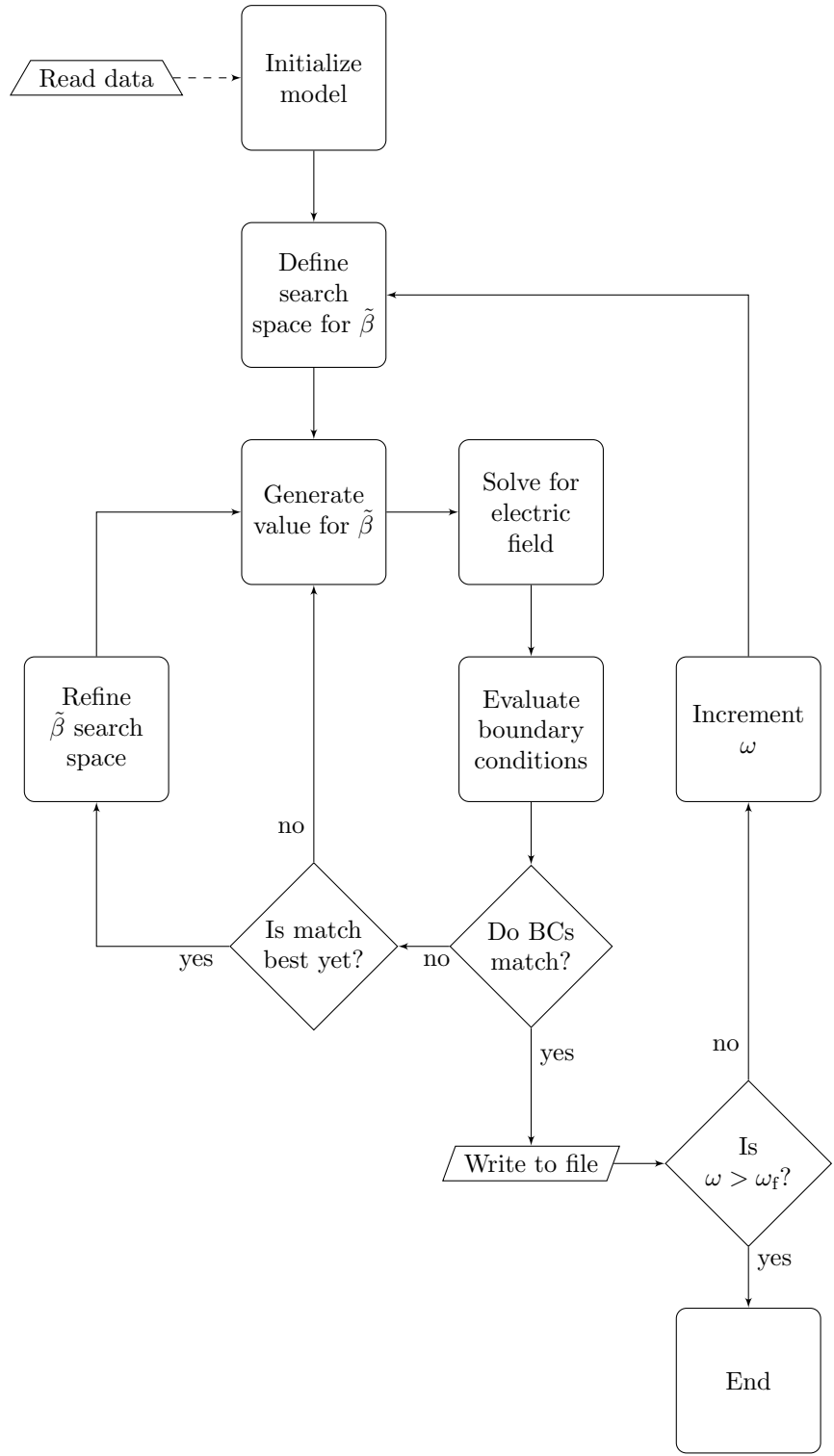

The waveguide core is embedded with three-level $\Lambda$ atoms in order to alter the index of refraction to become dependent on the intensity of the pump field, thereby enabling EIT. The pump field must travel in a waveguide mode, and a surface mode is chosen to increase the confinement. Surface modes, however, have an intensity that is dependent on the radial distance from the centre of the core. The resulting refractive index, then, is dependent on the radial distance from the centre as well. Thus, Maxwell's equations cannot be solved analytically, and the modes of the waveguide must be found numerically. A Mathematica\textsuperscript{\textregistered} program was written to solve Maxwell's equations for the electric field and propagation constant over a range of frequencies for the signal. The flow chart in Fig.~\ref{fig:numflow} shows the general flow of the program, the major steps of which are detailed in this section.

Waveguides can support a number of guided modes at a given frequency, each of which corresponds to a value of $\tilde\beta$ that satisfies the boundary conditions. That is, there may be multiple values of $\tilde\beta$ that are solutions to the dispersion relation with a single set of parameter values. As a result, care is needed to ensure that the value of $\tilde\beta$ determined at each frequency step corresponds to the appropriate mode.

One way to approach this problem is to simply find all the values of $\tilde\beta$ that correspond to an allowed mode at each frequency step. This method, however, is overly broad, as we are primarily interested in a single mode. Additionally, finding all of the $\tilde\beta$ that are solutions to the dispersion relation could potentially be time consuming, as there may be a large number. This is especially true at frequencies such that the wavelength of the electromagnetic field is much smaller than the raduis of the waveguide. 

The method employed here allows the values of $\tilde\beta$ for a single mode to be calculated for a number of adjacent frequency steps within an interval, without needing to calculate $\tilde\beta$ for any additional modes. This method takes advantage of a frequency where a known analytical solution or an approximately known value of $\tilde\beta$ exists. This frequency is chosen as the starting frequency. The $\tilde\beta$ corresponding to the desired mode is determined at each subsequent frequency using a guess based on the previous two $\tilde\beta$ values. The following is a breakdown of the program details.

The program begins with an initialization step, which defines all of the parameters and variables to be used, along with the name of the file that the data will be written to. During this step the program attempts to open the data file that is to be used and reads any existing data into the program. If no file exists one is created with the specified name. For the purposes of describing the flow of the program the data file will be assumed not to exist and must be created. If a file already exists, the program simply continues on from where the calculation was interrupted. 

A starting frequency $\omega$ is chosen such that $\tilde\beta$ is known, at least approximately, at this frequency. A complex number $\tilde\beta_{\omega}$ that is close to the known solution is chosen as an initial guess. The known value of $\tilde\beta$ can be used here for $\tilde\beta_{\omega}$, rather than choosing a nearby value.

A square search space is defined on the complex plane centred at $\tilde\beta_{\omega}$. The size of the search space is proportional to the value of the fitness function Eq.~(\ref{fitness}), which is manually given a value at the start, 1 for instance. The fitness function, and the search space, will be updated during the calculation. A random complex value is chosen from within the search space and Eq.~(\ref{TMvectwave}) is solved. The boundary conditions are then tested, and a value for the fitness $f$, Eq.~(\ref{fitness}), is determined. 

If the current value of $f$ is larger than the previous value, the process of selecting a trial $\tilde\beta_{\omega}$  is repeated. If the new $f$ is smaller, it replaces the old one. This serves to shrink the search space, which is also now shifted to be centred on the value for $\tilde\beta_{\omega}$ that corresponds to the smaller value of $f$. A new trial value for $\tilde\beta_{\omega}$ is then chosen from the refined search space, and Eq.~(\ref{TMvectwave}) is solved and the boundary conditions tested.

This process repeats, with the search space shrinking and shifting whenever a lower value of $f$ is calculated, until a value for $\tilde\beta_{\omega}$ is found that meets a preset tolerance level. The tolerance level used for the calculations done for this thesis is $f<10^{-9}$. For this value of $f$, the corresponding $\tilde\beta_{\omega}$ is considered to be a solution and the value of $\tilde\beta_{\omega}$ and the frequency $\omega$ are both recorded in the data file. The frequency is then incremented to the next adjacent frequency so the calculation can continue. 

The fact that the program starts near a known solution means the search for a solution to the dispersion relation can be restricted to a small area on the complex plane. This increases the probability of a random guess falling within the tolerance, $f<10^{-9}$, for a solution. By shrinking the search space as smaller values of $f$ are obtained, the number of trial values for $\tilde\beta$ that do not result in a solution is reduced. Thus, considerably reducing the time required to find a solution. As the area of the search space is reduced, it is possible for the search area to become too small and no longer encompass the solution. To ensure that the search space will include the solution, the search space is shifted to be centred on the current `best guess'.

Once the solution is obtained at the frequency $\omega$, the frequency is incremented by $\delta$ to the next adjacent frequency step $\omega+\delta$. The preceding process is performed again, with the search space initially centred on $\tilde\beta_{\omega}$. Trial values are chosen for $\tilde\beta_{\omega+\delta}$, and Eq.~(\ref{TMvectwave}) is solved and the boundary conditions tested for each trial $\tilde\beta_{\omega+\delta}$ until a solution is obtained.

When a value of $\tilde\beta$ is known for two adjacent frequencies, a linear extrapolation is done to improve the initial guess for $\tilde\beta$ at the subsequent frequency value. The extrapolation is done by simply taking $\tilde\beta_{\omega+2\delta}=2\tilde\beta_{\omega+\delta}-\tilde\beta_{\omega}$. The search space is then centred around $\tilde\beta_{\omega+2\delta}$ and the trial solution process as outlined above proceeds until a solution is found. The process is then repeated at each frequency step until the final value of $\tilde\beta$ has been calculated.

At every frequency step, the value of $\tilde\beta$ is written to a file, along with the corresponding frequency, when a solution is found. This data can then be used to calculate the group velocity and generate plots of the dispersion and attenuation of the mode near the EIT resonance.

There is a chance that the program can walk-off in the wrong direction if the landscape is such that there are decreasing, but not to within tolerance, values of $f$ in a direction leading away from the solution sought. If this is the case, eventually the program will reach a local minimum in $f$ and will become stuck. To prevent having to restart the program in this case, a counter tracks the number of sequential incorrect guesses. If the number of guesses in a row not resulting in a smaller value of $f$ exceeds a preset value, 100 for instance, then the program reverts back to the most recent solution, resets the value of $f$, and begins searching for the solution at the current frequency again.

If the unlikely case occurs that the program walks-off and finds a value of $f$ that is within the tolerance, the corresponding $\tilde\beta$ is a solution for a different mode. This type of error will likely cause the program to generate more than just a single error, as it will tend to solve for $\tilde\beta$ of the new mode thereafter. Due to the fact that this error rarely occurs, however, it is not necessary to write code to prevent it. A simple visual inspection of the dispersion and attenuation plots is sufficient to determine if this error has occurred. 

If it were necessary to prevent this error, one possible method would be to monitor the slope between successive solutions. For a frequency interval that is small enough, the slope between the latest two solutions should not differ considerably from the previous two. For solutions where the slope changes rapidly, the most recent value of $\tilde\beta$ can be recalculated, perhaps with a smaller search space to prevent another ``walk-off''. However, given the extremely rapid change in the dispersion and attenuation near the EIT resonance, the frequency interval required to reliably differentiate between the desired solution and one that corresponds to a different mode would be small. A small frequency interval would increase the number of solutions that are required and increase the calculation time. Additionally, calculating and comparing the derivatives for each solution would increase the number of calculations required, further slowing down the program.

I tested the convergence of the numerical routine by changing the step size and perturbing the initial conditions for three different values of the detuning, $\Delta=0\,\text{Hz},\:8\,\text{kHz},\:130\,\text{kHz},$ and $1\,\text{MHz}$. The computed $E_{z}$ from these calculations were then compared with $E_{z}$ calculated originally. If the results of these calculations agree with the original $E_{z}$ at the relevant detunings, then we can be confident that the numerical routine is not introducing significant numerical error into the calculations.

The default behaviour of the {\tt NDSolve[]} routine is to choose the step size adaptively to minimize errors. I compared $E_{z}$ computed using this default behaviour, along with three other calculations for $E_{z}$ done using the fixed step sizes $a/500,\:a/200,$ and $a/100$, with $a$ the radius of the waveguide core. These calculations were done at each of the four detunings listed above. The results from all four calculations agreed at each of the detunings.

The initial conditions used for this calculation are $E_{z}=1$ and $E'_{z}=0$, both at $r=0$. The value of $E_{z}$ at $r=0$ can be chosen freely, due to the presence of an integration constant in the solution at $r>0$, so it is not helpful to perturb this initial condition. However, $E'_{z}=0$ is required at $r=0$ due to the symmetry of the mode. We can perturb this initial condition from its correct value to test the numerical convergence.

I compared the calculated $E_{z}$ for $E'_{z}=0,\:0.01,\:0.05,$ and $0.1$ at each of the four detunings listed above. As with the step-size test, all four calculations agreed at each of the four detunings. Therefore, we can be confident that the numerical calculations performed converge.

To verify the validity of the solutions obtained with the numerical program, an approximation is made to the wave equation to make it solvable analytically. The approximate solutions, obtained analytically, can then be compared to the solutions calculated numerically. If the two solutions agree, the solutions generated by the numerical routine are correct.

\begin{figure}[t,b] 
       \centering
	\includegraphics[width=0.6\textwidth]{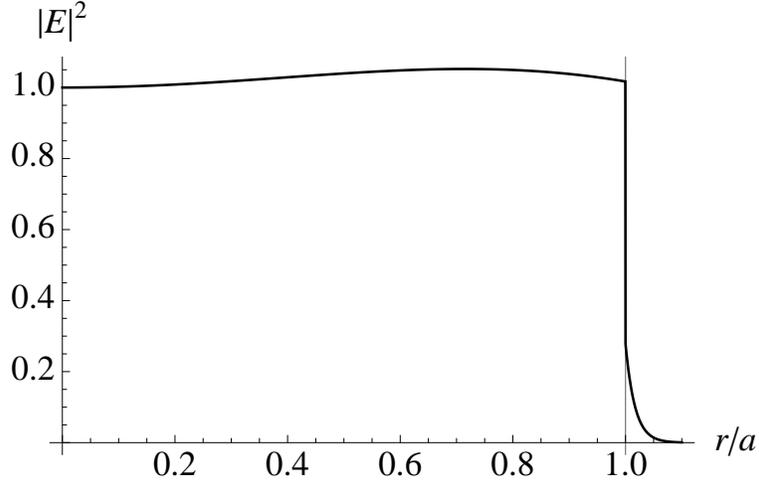}
	\caption[Plot of the $\left|\bm E\right|^{2}$ profile for the analytical approximation]{A plot of the profile of $\left|\bm E\right|^{2}$ (in arbirtrary units) for the mode that was used for the analytical approximation. The thin vertical line indicates the core-cladding boundary.}
     \label{fig:flatmode}   
\end{figure}

\begin{figure}[t,b] 
       \centering
       \subfloat{\label{fig:epsapproxre}\includegraphics[width=0.48\textwidth]{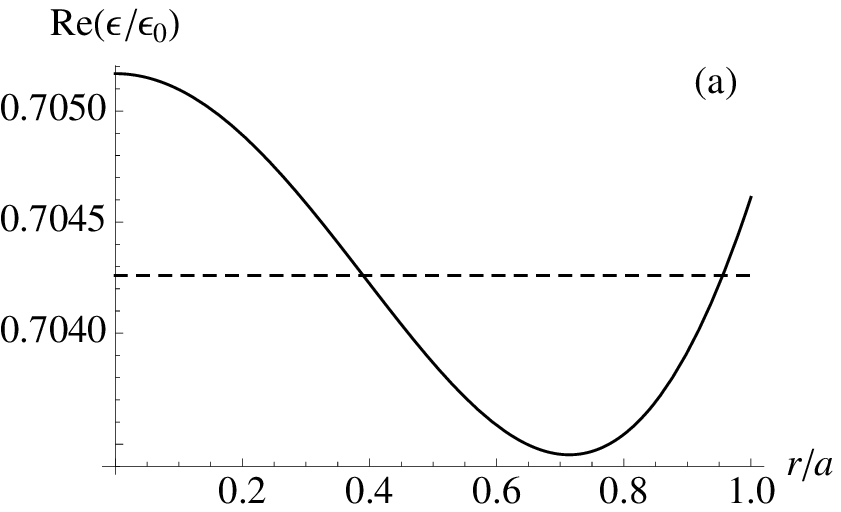}}\hfill
       \subfloat{\label{fig:epsapproxim}\includegraphics[width=0.48\textwidth]{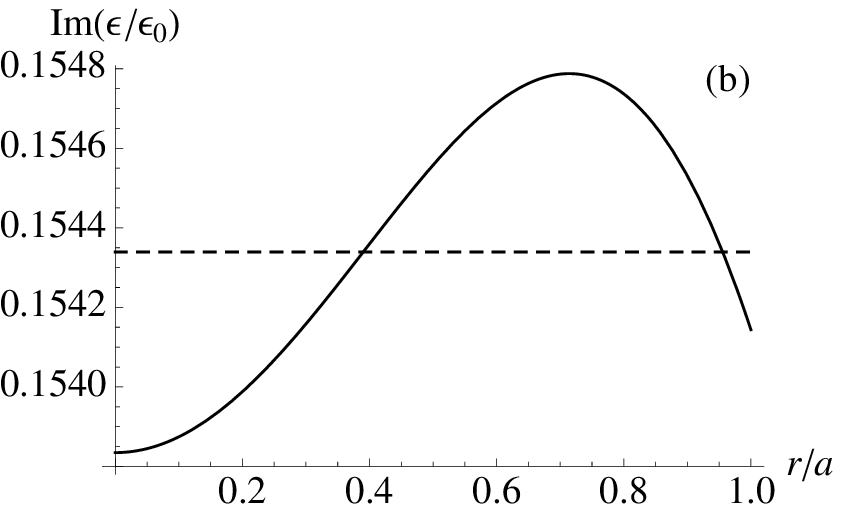}}
\caption[Plots of the permittivity approximation.]{Plots of the real parts \protect\subref{fig:epsapproxre} and imaginary parts \protect\subref{fig:epsapproxim} of the exact permittivity (solid lines) and the approximated permittivity (dashed lines) calculated for $\Delta=3\times10^{-9}\omega_{\text e}$.}
\label{fig:epsapprox}   
\end{figure}

Due to the steep gradient in the permittivity of the core for the pump frequency $\omega_{\text p}$, it is not possible to find an analytical approximation to the solution for more than a few special values of the signal frequency. For this reason, an analytical approximation is done at a new pump frequency $\omega'_{\text p}$ where the intensity of the pump mode has the flattest cross section (Fig.~\ref{fig:flatmode} shows the cross section of $\left|\bm E\right|^{2}$ for the mode used). 
Using a mode with a relatively flat intensity profile leads to the exact permittivity having a relatively flat gradient (Fig.~\ref{fig:epsapprox}), as the permittivity depends on the field intensity via the susceptibility of the EIT medium, which is given by Eq.~(\ref{EITchi}).

At this new pump frequency the permittivity of the dielectric core can be approximated by a permittivity that is $r$ invariant, but frequency dependent. The approximate permittivity is given by a median value of the exact permittivity,
\begin{equation}
\epsilon_{\text{approx}}(\omega)=\epsilon(\omega,b),
\end{equation}
 with $b$ chosen so that $\epsilon_{\text{approx}}(\omega)$ falls approximately half way between the maximum and minimum values of $\epsilon(\omega,r)$ for $0\leq r\leq a$. Figure~\ref{fig:epsapprox} shows the exact and approximated permittivity for the detuning $\Delta=3\times10^{-9}\omega_{\text e}$ as an example.

In order for Eq.~(\ref{TMvectwave}) to have an analytical solution, the criterion
\begin{equation}
\frac{1}{r}\gg\frac{\tilde{\beta}^{2}}{\kappa_{\text{eff}}^{2}(\omega,r)\epsilon_{\text{eff}}(\omega,r)}\frac{\text{d}}{\text{d}r}\epsilon_{\text{eff}}(\omega,r),\label{approxcond}
\end{equation}
must be satisfied for the exact permittivity. The analyitical approximation is done by first assuming that Eq.~(\ref{approxcond}) holds. Equation~(\ref{TMvectwave}) can then be written as
\begin{equation}
\frac{\text{d}^{2}}{\text{d}r^{2}}E_{z}+\frac{1}{r}\frac{\text{d}}{\text{d}r}E_{z}+\kappa_{\text{approx}}^{2}(\omega,r)E_{z}=0,\label{TMvectwaveapprox}
\end{equation}
with
\begin{equation}
\kappa_{\text{approx}}:=\sqrt{\omega^{2}\mu_{0}\epsilon_{\text{approx}}(\omega)-\tilde{\beta}^{2}}.
\end{equation}

\begin{figure}[t,b] 
       \centering
       \subfloat{\label{fig:approxdisp1}\includegraphics[width=0.48\textwidth]{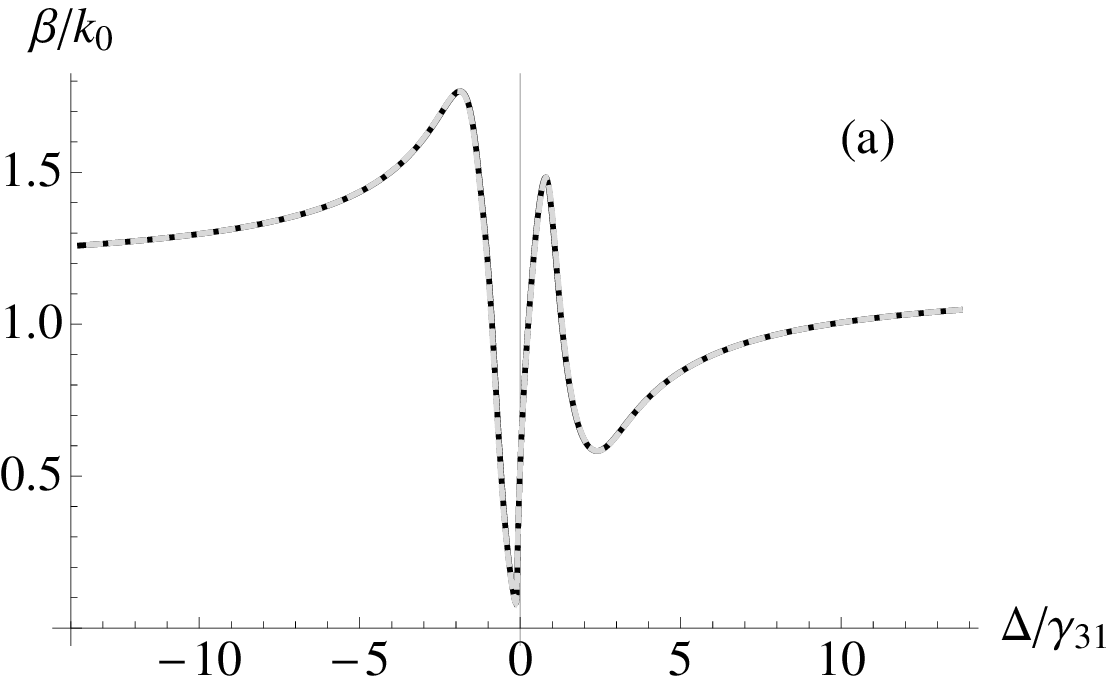}}\hfill
       \subfloat{\label{fig:approxloss1}\includegraphics[width=0.48\textwidth]{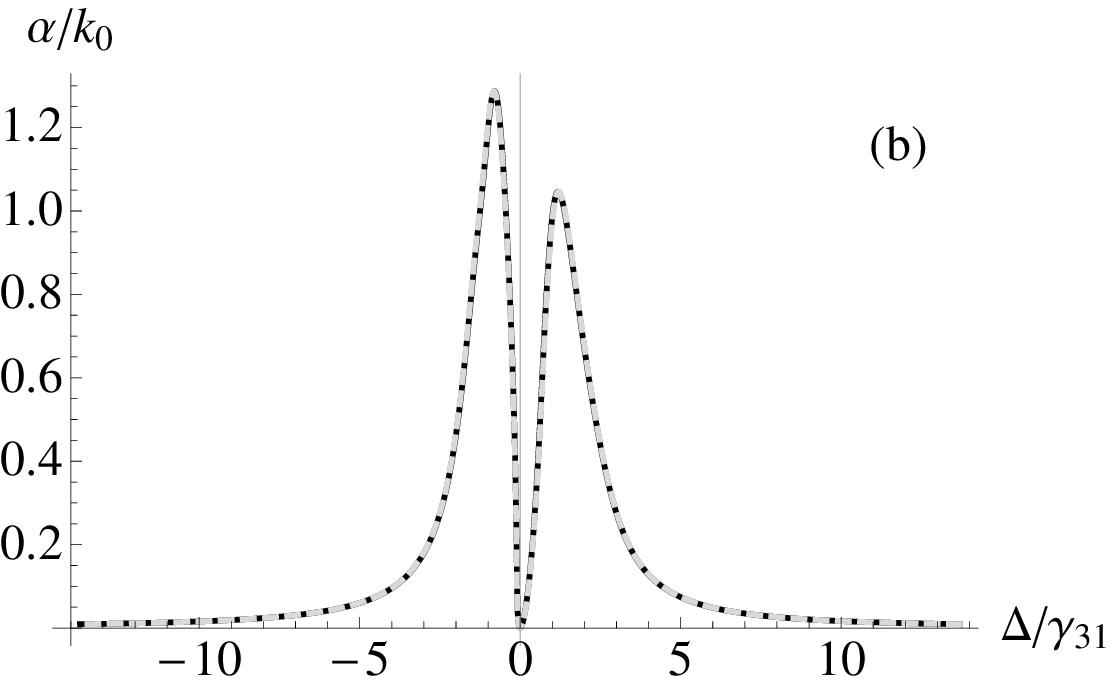}}\\
       \subfloat{\label{fig:approxdisp2}\includegraphics[width=0.48\textwidth]{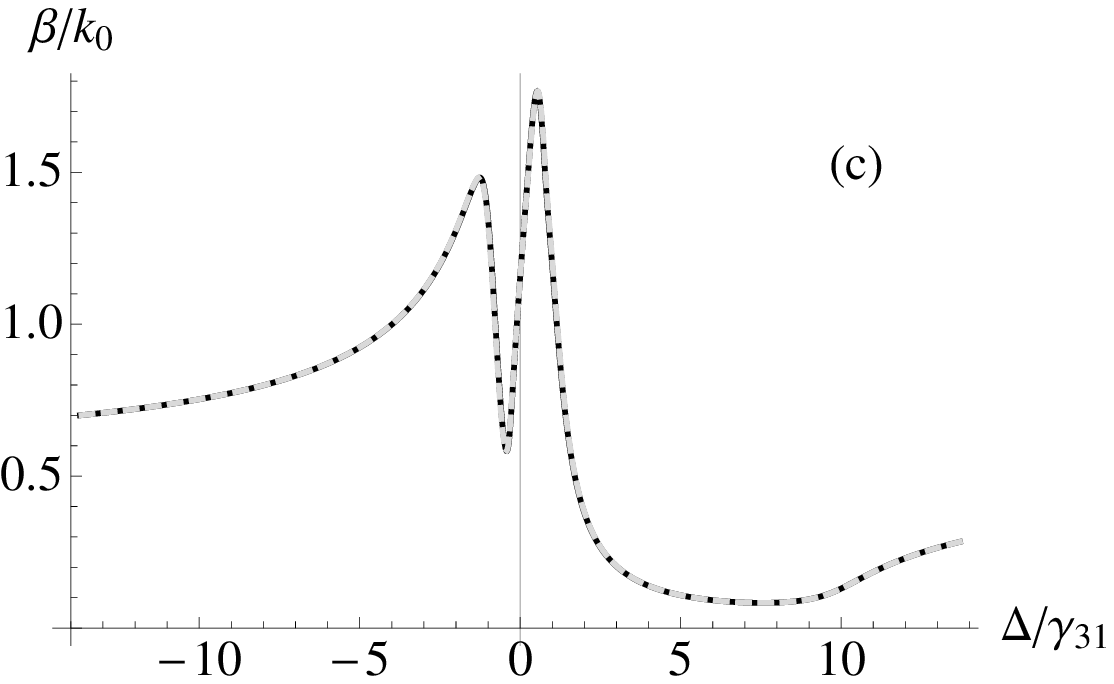}}\hfill
       \subfloat{\label{fig:approxloss2}\includegraphics[width=0.48\textwidth]{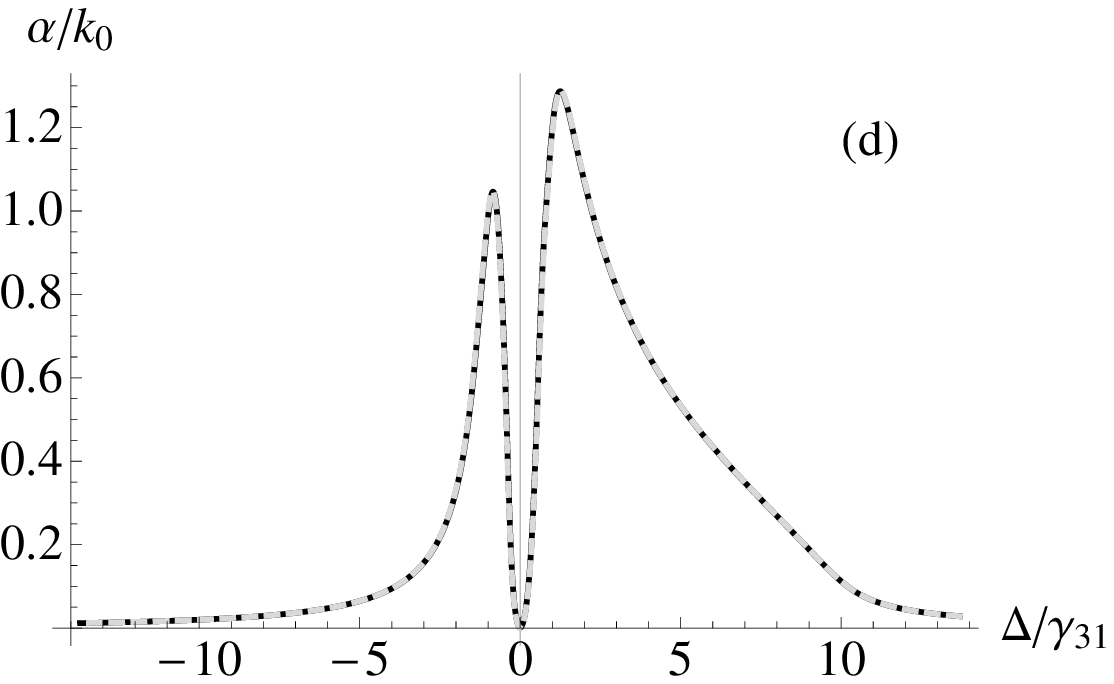}}
\caption[Plots of approximated modes.]{Plots of the dispersion \protect\subref{fig:approxdisp1} and \protect\subref{fig:approxdisp2}, and attenuation \protect\subref{fig:approxloss1} and \protect\subref{fig:approxloss2} for two modes with pump frequency $\omega'_{\text d}=0.19\omega_{\text e}$, showing the analytical approximation. The exact solution is the solid black line and the approximation is the dashed grey line. The vertical grey line indicates $\Delta=0$.}
\label{fig:approxmodes}   
\end{figure}

The criterion of Eq.~(\ref{approxcond}) is tested by solving Eq.~(\ref{TMvectwaveapprox}) for $\tilde\beta$ at the pump frequency $\omega'_{\text p}=0.19\omega_{\text e}$, and then using it to verify the validity of Eq.~(\ref{approxcond}). This is the best method for testing the validity of the approximation as $\tilde\beta$ cannot be assumed to be insignificant in Eq.~(\ref{approxcond}), and the values of $\tilde\beta$ can not be known before making the approximation. The numerical and approximated solutions are shown in Fig.~\ref{fig:approxmodes} for two different modes. 
From the figure it can be seen that the dispersion and attenuation curves for both modes match extremely well, suggesting that the numerical solutions are correct.

\section{Discussion}\label{ch:slowlight-discussion}

Attenuation of propagating fields in materials is characterized by the imaginary part of the refractive index $n=c\sqrt{\epsilon\mu}$. In most materials the relative permeability, defined as $\mu/\mu_{0}$, is real and equal to one; thus the attenuation is simply due to the imaginary part of the permittivity. In metamaterials, however, the permeability is also complex and frequency dependent, which changes the electromagnetic interaction. The refractive index of a metamaterial is written as a function of the complex permittivity and permeability in Eq.~(\ref{expandedn}).

In the frequency region where the low-loss mode exists, our parameters give $\left|\epsilon''\right|\ll\left|\epsilon'\right|$ and $\left|\mu''\right|\ll\left|\mu'\right|$, along with $\epsilon'<0$ and $\mu'>0$. The result is that the imaginary part of $n_{\text{MM}}$ is due primarily to the term $\epsilon'\mu'$, which is negative. This is analogous to a metal, in which the real part of $\epsilon$ is negative and the relative permeability is equal to one. In the case of the metamaterial, the relative permeability is less than one at these frequencies, which effectively reduces the magnitude of the imaginary part of $n_{\text{MM}}$.
 
The attenuation of propagating modes in a waveguide, however, is more complicated than propagation through bulk material. Simply reducing $\Re{\mu}$ is insufficient to reduce the attenuation of a metamaterial dielectric waveguide. The dissipation term $\Gamma_{\text{m}}$ plays a significant role as well~\cite{Lavoie:2012}. When $\Gamma_{\text{m}}=\Gamma_{\text{e}}$ the attenuation of a metamaterial-dielectric guide is comparable to a metal-dielectric guide, whereas reducing $\Gamma_{\text{m}}$ reduces the attenuation of a metamaterial-dielectric guide. For layered structures, such as the fishnet design, previous work~\cite{Penciu:2010} shows that $\Gamma_{\text{m}}\leq\Gamma_{\text{e}}$, and so a value of $\Gamma_{\text{m}}=\Gamma_{\text{e}}/3$ should be feasible with current metamaterial technology~\cite{Zhou:2008}.

The strong effect of $\Gamma_{\text{m}}$ on the attenuation is surprising, as it has virtually no effect on $\Im{n_{\text{MM}}}$ or on the effective guide width at these frequencies. Reducing $\Gamma_{\text{m}}$ by factors of 3 and 100 does, however, reduce $\Im{\mu_{\text{MM}}}$ by approximately the same amount (see table~\ref{gammaeffect}), indicating that attenuation in a waveguide depends on more than simply the imaginary part of the refractive index and the energy distribution. This is particularly surprising for the TM modes as $\mu_{\text{MM}}$ never appears in the dispersion relation separately from $\epsilon_{\text{MM}}$, but always as a product of the two, which is equal to $n_{\text{MM}}^{2}$.

\begin{table}[h,t,b]
\caption[Various quantities calculated for the metamaterial]{Values for the permittivity, permeability and refractive index of the metamaterial at $\omega=0.401\omega_{\text e}$ for different values of $\Gamma_{\text m}$.}
\centering
\begin{tabular}{cd{9}d{9}d{9}}
\hline\hline  & \multicolumn{1}{c}{$\Gamma_{\text m}=\Gamma_{\text e}$}&\multicolumn{1}{c}{$\Gamma_{\text m}=\Gamma_{\text e}/3$}
		&\multicolumn{1}{c}{$\Gamma_{\text m}=\Gamma_{\text e}/100$}\\
\hline
$\Re{\epsilon_{\text{MM}}/\epsilon_{0}}$ & -4.9487 & -4.9487 & -4.9487\\

$\Im{\epsilon_{\text{MM}}/\epsilon_{0}}$ & 2.891\times10^{-2} & 2.891\times10^{-2} & 2.891\times10^{-2}\\

$\Re{\mu_{\text{MM}}/\mu_{0}}$ & 0.3439 & 0.3439 & 0.3439\\

$\Im{\mu_{\text{MM}}/\mu_{0}}$ & 4.184\times10^{-3} & 1.395\times10^{-3} & 4.185\times10^{-5}\\

$\Re{n_{\text{MM}}}$ & -4.126\times10^{-3} & 1.165\times10^{-3} & 3.731\times10^{-3}\\

$\Im{n_{\text{MM}}}$ & 1.3046 & 1.3045 & 1.3045\\
\hline\hline
\end{tabular}
\label{gammaeffect}
\end{table}

In our calculation, we have treated the claddings as if they are infinitely thick. In practice, the TM surface modes for the pump and signal fields only penetrate into the metamaterial to some finite depth, so a finite thickness of the cladding regions is sufficient. The skin depth for both the signal and pump fields is about $0.1\lambda_{\text s}$, with $\lambda_{\text s}$ the wavelength of the signal at the EIT resonance. As typical fishnet metamaterial has thickness of $0.1\lambda_{\text s}$~\cite{Boltasseva:2008}, a few layers of fishnet metamaterial (with dielectric spacers in between) is sufficient for the cladding.

We find that the attenuation dip near $\omega_{\text s}$ (Fig.~\ref{fig:lowlosseit}) is not as pronounced in the presence of transverse field confinement in the cylindrical waveguide. The flat metamaterial-dielectric interface has near-zero attenuation near this frequency, whereas the attenuation dip in the cylindrical guide does not approach zero. The cause for the non-zero attenuation is that the transverse confinement limits the amount of electromagnetic energy that is ejected from the lossy metamaterial.

To illustrate why transverse confinement lessens the attenuation dip near $\omega_{\text s}$, consider a cylindrical guide in the limit of an infinite radius. In this limit, the interface that the surface wave propagates along becomes flat, and the dispersion relation for the cylindrical waveguide reduces to that for a flat interface waveguide. In the case of the flat interface, the energy can reside almost completely on the dielectric side, leading to near-zero loss~\cite{Moiseev:2010}.

Alternatively, for a cylindrical guide in the zero core radius limit, the core almost vanishes; hence, the energy propagates almost entirely in the lossy metamaterial, which leads to significant loss. For a waveguide with a finite radius, the fraction of the total energy that can be confined to the dielectric lies between 0 and 1; thus there is a non-negligible amount of attenuation due to the field interacting with the metamaterial cladding. 

We have considered a fishnet metamaterial for our scheme, but this scheme is not restricted to using fishnet metamaterial. Other types of metamaterial are applicable, as long as they have permittivity and permeability functions resembling equations~(\ref{epsilonMM}) and~(\ref{muMM}), with $\text{Re}(\epsilon_{\text{MM}})<0$, $0<\text{Re}(\mu_{\text{MM}})<1$, and $\Gamma_{\text m}<\Gamma_{\text e}$. Using other metamaterials is of interest, because when layered fishnet metamaterials are bent to form a cylindrical cladding, the metamaterial may not interact with the fields isotropically. There now exist various promising fabrication techniques to building bulk metamaterials instead of layered ones~\cite{Boltasseva:2008,Vignolini:2012}, which are isotropic. 

The system described herein is assumed to operate at a cryogenic temperature as this provides the advantage of mitigating decoherence effects that reduce EIT efficacy~\cite{Ham:1997}. An auxiliary advantage of the cryogenic temperatures is increased charge mobility in metals, thereby decreasing attenuation due to a portion of the electromagnetic fields propagating through the metamaterial cladding~\cite{Singh:2010}.

The major drawback of cryogenic temperatures is that this system needs to be contained within a cryostat, which will complicate its integration and use with other systems that need to operate at higher temperatures. We are, however, only suggesting the use of cryogenic temperatures in the interest of achieving a proof-of-principle design. Other effects of cryogenic temperatures to consider are a modified refractive index in the dielectric core and altered core dimensions. These changes, however, can be measured and compensated for based on the operational requirements.

Though our main result concerns controllable slow light, the use of EIT in this system means it is also capable of stopping the signal pulse completely by turning off the control field entirely~\cite{Turukhin:2001}. An important application of stopped light is optical quantum memories~\cite{Lvovsky:2009}, to which EIT is well suited due to its capacity for direct control. However, the narrow bandwidth restriction inherent in EIT-based schemes places a lower-bound restriction on the duration of the signal pulse. This has implications in terms of data rates for quantum communication and quantum computation schemes that would use this memory.

The imposed lower bound on pulse duration, which correlates directly with a minimum spatial extent, sets an upper limit on the pulse repetition rate. As the maximum allowable overlap for pulses to remain distinguishable restricts the minimum distance between pulses, the number of pulses that can be processed within a given time window is limited. The effects of the narrow bandwidth on the information storage capability could be avoided, however, by storing information in ways that are not affected by reduced bandwidth, such as amplitude and phase~\cite{Liu:2001,Appel:2008}.

The operational bandwidth of this device can easily be broadened by increasing the pump intensity, but comes at the expense of the group velocity reduction that can be achieved. This is because the spectral width of the EIT transparency window is related to the intensity of the pump field~\cite{Lambropoulos:2007}. Alternatively, the bandwidth can be broadened by using another technique to achieve slow light, such as Raman amplification~\cite{Sharping:2005} or photon echo techniques~\cite{Saglamyurek:2010}. However, such schemes come at the expense of increased group velocities and controllability. The Raman amplification technique does have the benefit of room temperature operation.

The analysis of the scheme presented in this thesis is done for controllable slow light and storage of light using three-level $\Lambda$ atoms. This is, however, only one design possibility for such a device. By using a dopant with a different level structure, for instance, the possible applications of this scheme can be expanded. 

Atoms with a five-level scheme are capable of double-EIT, that is EIT with two transparency windows at different frequencies~\cite{ZBWang:2006}. With double-EIT, two pulses would be able to propagate through the EIT medium without being absorbed, and both would be slowed due to the change in dispersion accompanying the transparency. In addition, the simultaneous interaction of both pulses with the EIT medium elicits a third-order nonlinear response that results in cross-phase modulation between the pulses. 

Both the reduced group velocity of the pulses, due to EIT, and strong confinement, provided by surface plasmon-polaritons, contribute to increasing the cross-phase modulation resulting from the nonlinear interaction. By increasing the local field intensity, and thereby increasing the strength of the nonlinear response of the material, the confinement provided by the surface plasmon-polaritons increases the rate at which the pulse phase is modulated. With the decreased group velocity allowing a longer interaction time within the same propagation distance, these two effects can elicit giant cross-phase modulation leading to large phase shifts, even for weak (few photon) fields~\cite{ZBWang:2006,Moiseev:2010}.

A complete analysis of the five-level atom scheme would need to be done in the setting of a waveguide to ensure that it performs as expected. If the results for the three-level system are any indication, however, the five-level system in a waveguide should perform similarly to the flat-interface scheme, only with slightly increased attenuation. The resulting giant cross-phase modulation that is possible with double-EIT would make the five-level atom scheme useful for an all-optical switch, quantum non-demolition measurements, or simply as a two-pulse quantum memory.



\section{Summary}\label{ch:slowlight-summary}



We have considered two cylindrical dielectric-core waveguides of equal size and dimensions, with one having a metal cladding and the other a fishnet metamaterial cladding; it is also possible to consider other types of metamaterial cladding. The permittivities of both claddings are given by the Drude model, but the metal cladding has constant permeability and the metamaterial cladding is given by the modified-Drude model. The signal field propagating through the waveguides is controllably slowed via electromagnetically induced transparency (EIT)  due to the three-level $\Lambda$ atoms homogeneously doped throughout the dielectric core of each waveguide. We have shown that the low-loss TM surface mode exists in the metamaterial-clad waveguide provided that the permeability of the metamaterial cladding has a real part less than unity, i.\e.\ $\Re{\mu}<1$, and the magnetic damping rate is less than the electric damping rate, i.\e.\ $\Gamma_{\text{m}}<\Gamma_{\text{e}}$. Previous work has shown that this condition is always true for certain optical metamaterials~\cite{Penciu:2010}. We predict that  such a cylindrical metamaterial-clad waveguide is capable of delivering the same slowing of light but with reduced loss compared to the cylindrical metal-clad waveguide.

%% file: Practicalconsid.tex
\chapter{Practical Considerations}\label{ch:pract-consid}

\section{Introduction}

Our work is a theoretical investigation of a dielectric waveguide doped with three-level $\Lambda$ atoms and having a metamaterial cladding. This system is proposed as a device for all-optical control of weak signals. We have shown that a metamaterial can provide increased performance, by means of reduced attenuation, under certain conditions. This project was undertaken with the anticipation that it could possibly be used as a platform on which to design physical devices. Our work, however, can only give limited insight into the actual behaviour of a realized device.

I therefore consider some construction methods and challenges that would accompany the fabrication of a metamaterial-clad waveguide device in Sec~\ref{sec:WGtesting}. Such a device would need to be tested, so I also consider how such test might be performed in an experimental setting. Additionally, in Sec.~\ref{sec:modechoice} I discuss the implication of the choice of modes for the signal and control field.

\section{Metamaterial-Clad Waveguide: Construction and Testing}\label{sec:WGtesting}

The applicability of the scheme presented in this thesis for controllable slowing and stopping of light in a metamaterial-clad waveguide is predicated on the ability to fabricate such waveguides. At the present, techniques for fabricating metamaterials are more mature for planar metamaterials than for true three-dimensional metamaterials~\cite{Boltasseva:2008}. There are methods for creating multi-layer planar metamaterials, but they still retain the anisotropy of the planar metamaterials they are built with. The analysis of metamaterial-clad waveguides done within this thesis relied on having an analytical model of the permittivity and permeability of the metamaterial. The permittivity and permeability functions corresponding to the optical fishnet metamaterial were chosen for two major reasons. 

First, theoretical studies have been done to determine the frequency dependence of the permittivity and permeability of optical fishnet metamaterials~\cite{Penciu:2010}, with these functions having been validated with experimental data~\cite{Xiao:2009}. Three-dimensional optical metamaterials are difficult to fabricate, however, so there have been few experimental investigations of these materials~\cite{Boltasseva:2008}. Additionally, because an effective and easily fabricated design has not been found, no theoretical analysis to determine the permittivity and permeability of three-dimensional metamaterials has been done. 

Second, the mechanism to achieve a negative permeability in three-dimensional optical metamaterials is similar to that of planar metamaterials. The only significant difference is that the sub-wavelength structures must form a three-dimensional lattice within a bulk material. Though the physical structures may not be exactly the same, the frequency dependent permeability of three-dimensional metamaterials must still be generated by the incident fields creating oscillating current loops, thereby generating magnetic fields. Thus, it is expected that the forms of the permeability for the two- and three-dimensional metamaterials will be similar. Using the permeability for two-dimensional metamaterials gives results that will be similar to what is expected with three-dimensional metamaterials.

Planar metamaterials are designed for fields that pass through the metamaterial perpendicular to the surface, not fields travelling along the surface. The response of a planar metamaterial to fields travelling along its surface has not been studied; though planar metamaterials could, in theory, be used as a flat interface to excite surface plasmon-polaritons for this purpose. The permittivity and, in particular, the permeability of a planar metamaterial for a surface plasmon-polariton are likely to differ from those resulting from fields propagating through the metamaterial. For this reason, the use of planar metamaterials, even if they can be formed into a cylindrical cladding, may not be suitable for the system envisioned herein. Unfortunately, fabrication techniques of three-dimensional metamaterials have not progressed enough to allow construction of waveguides. Despite the current fabrication limitations, however, it is still instructive to discuss metamaterial-clad waveguide construction.

The analysis of the system proposed in this thesis assumes the thickness of the cladding is effectively infinite, thereby eliminating the effects of the exterior surface of the cladding layer. As the metamaterial will act as the cladding, it does not need to be cylindrical, but simply needs to be large enough that the propagating fields are sufficiently small at the metamaterial's exterior surface. Thus, any three-dimensional metamaterial that is large enough will suffice for a cladding. There are a number of proposed methods to fabricate such three-dimensional metamaterials~\cite{Boltasseva:2008,Vignolini:2012}.

The problem, then, becomes how to create a cylindrical waveguide within the metamaterial. One simple solution is just to drill a hole through the metamaterial and fill it with the appropriate dielectric core material. Some issues arise with this method, however.

The core diameter needs to be on the order of a few wavelengths in order to provide adequate confinement of the fields, which may be challenging. Assuming that it is possible to drill a hole that small, a mechanical drill would leave the inner surface with both regular and irregular imperfections that would affect the propagation of the fields. It may be possible to avoid the mechanically created imperfections by burning a hole in the metamaterial using a high-power laser. Other problems could arise with this method, however, such as the effect that the residual heat would have on the optical properties of the remaining metamaterial. 

Once a hole has been formed in the metamaterial, it needs to be filled with a dielectric. Although metamaterial claddings are capable of guiding waves along a core consisting simply of air~\cite{Kim:2007}, a solid core is needed to act as a host for the three-level $\Lambda$ atoms. The simplest way to get a dielectric into the hole in the metamaterial would be to feed in a fibre core. This method would leave air gaps around the core, however, which would alter the propagation of the modes. By cooling the fibre and/or heating the metamaterial, the fibre could be made to fit more easily and would provide a better fit once the temperatures returned to normal. The imperfections on the metamaterial surface, however, would still remain and allow air to become trapped. 

Another option for metamaterial-clad waveguide construction is to start with the core and build a metamaterial around it. By starting with the core, the cladding layer could be made to fit around it, thereby eliminating most imperfections in the cladding and any trapped air. This method could prove challenging, however, as most three-dimensional metamaterials would begin with a solid material, or at least solid layers that are then stacked up to form the metamaterial~\cite{Boltasseva:2008}.

Self assembly techniques could provide a means to allow a metamaterial cladding to be built around a dielectric core~\cite{Boltasseva:2008,Vignolini:2012}. This type of metamaterial fabrication technique uses multiple compounds that form a solid structure through chemical reactions. Allowing the metamaterial to assemble around the core in this way would reduce cladding imperfections.  

As with the metamaterial cladding, the material and fabrication of the dielectric core must be considered. In light of future applications for the scheme presented herein, two different types of materials are considered for the dielectric core, a dielectric crystal, such as Y$_{2}$SiO$_{5}$, and a glass, such as silica. Both of these material types are viable candidates for a dielectric host, and have advantages and disadvantages for such an application.

Bulk dielectric crystals doped with rare-earth(RE) ions have been employed successfully to achieve EIT~\cite{Turukhin:2001,Longdell:2005,Baldit:2010}, and the long coherence times offered by RE-doped dielectric crystal waveguides are useful for photon-echo quantum memories~\cite{Saglamyurek:2010}. Dielectric crystal waveguides are typically fabricated on the surface of a bulk crystal by physically etching channels into the surface, or by using a second dopant to change the refractive index along a path near the crystal surface~\cite{Sinclair:2010,Thiel:2012}. These two fabrication methods, however, do not allow a metamaterial cladding to be placed around the entire waveguide; a restriction that undermines the usefulness of the metamaterial cladding. If either of these two waveguide constructions are to be employed, further analysis will be needed to determine the ability of these guides to perform the desired functions. To follow the scheme outlined here using a dielectric crystal host, a fabrication method that can form a crystal into a thin fibre would be needed.

Forming a dielectric crystal into a waveguide alters the system from that of a bulk crystal, and could affect the performance. Thiel et al. performed a comparison of dielectric crystals in both bulk and waveguide configurations using photon-echo and spectral hole burning techniques~\cite{Thiel:2012}. They found that there are some differences between the optical properties of bulk crystals and waveguides. However, if the waveguide fabrication is done carefully, these differences can be minimal, and the desirable properties of bulk crystal can be maintained in a waveguide configuration. 

The presence of atoms with nonzero nuclear or electron spin and slight irregularities in the crystal structure (some of which are caused by the presence of the dopant ions themselves) of dielectric crystals can contribute to the decoherence of EIT atoms~\cite{Staudt:2006,Thiel:2012}. The nonzero spin of atoms can occasionally change orientations, called a spin-flip, altering the field near the atom causing nearby EIT atoms to decohere. Additionally, phonons can easily travel through crystals, due to their regular structure, and can cause crystal deformations that alter the local electric field, thereby causing decoherence. Cooling the system to cryogenic temperatures, however, can significantly reduce these decoherence effects~\cite{Thiel:2012}. 

Glasses are easily drawn into and spliced with existing fibres, and are relatively cheap and easily fabricated. Additionally, because glasses are naturally amorphous, drawing them into fibres does not significantly change their optical properties. The amorphous nature of glass, however, is also a disadvantage for EIT applications. The lack of structure in RE-doped glass leads to ion clustering that causes homogeneous broadening due to ion-ion interactions.

Ion clustering can be reduced, but not eliminated, by co-doping with aluminum~\cite{Staudt:2006}. Additionally, coupling with so-called low-frequency tunnelling modes \cite{Staudt:2006,Macfarlane:2007}, which are not present in dielectric crystals, can further increase the linewidth of rare-earth ions. As with dielectric crystals, cryogenic temperatures help to reduce some of the decoherence effects present in RE-doped glass \cite{Staudt:2006}.

Despite the decreased coherence times observed in RE-doped glass, these systems are still useful for a variety of applications, including quantum memories \cite{Staudt:2006} and slow light \cite{Melle:2012}. These two applications of RE-doped glass are proposed using spectral hole burning, but they can still be used as a basis for discussing EIT-based schemes, as the main requirement for EIT, a long-lived intermediate state, is also a requirement for spectral hole burning.

The physical realization of the metamaterial-clad waveguide device, having a metamaterial cladding formed around a dielectric core that is doped with three-level $\Lambda$ atoms, will not be completely free of physical imperfections in both the core and cladding. Such imperfections can contribute to optical decoherence and may change the behaviour of the modes from the theoretical predictions. Even the existence of the core-clading boundary itself will affect the EIT response of the system, as the boundary conditions are responsible for guided modes within a waveguide. The presence of the cladding partially confines the electromagnetic fields to the core, thereby altering the field so that it does not have a uniform intensity profile.

The dispersion and attenuation experienced by a signal field in an EIT medium depends on the intensity of the pump field. As the pump field is travelling as a guided mode it does not have a spatially uniform intensity. The nonuniform pump intensity causes each atom of the EIT medium to react differently and generate different dispersion and attenuation effects for the signal field.

The resulting dispersion and attenuation of the signal field in the waveguide differ from those generated by a pump with a uniform intensity. Compare Figs.~\ref{fig:genEITplots} and \ref{fig:dopedbeta}, which show the frequency dependence of the dispersion and attenuation for a uniform pump and the waveguide system respectively. The asymmetry present in the plots for the waveguide system, which is not present for a free-space system with a uniform pump field, is an indicator that the two systems are not equivalent.

The work that has been done to characterize the metamaterial-clad waveguide and determine its ability to support pulses slowed via EIT needs to be verified through experimental investigations. Verifying the results will determine if metamaterials will retain their optical properties after being formed into a waveguide. As mentioned above, there are a number of possible issues that can arise and affect the performance. Unexpected issues may even arise, which can be identified and characterized using experimental data. These experiments will also reveal to what extent imperfections at the core-clading boundary, ion clustering within the core, and a nonuniform pump field affect the EIT response and the propagation characteristics of the waveguide system. 

A detailed experimental setup for testing the waveguide system will not be given, but the relevant measurements that need to be done will be discussed. The main results needed for a comparison of the actual device with the theoretical predictions are the complex propagation constant (i.e.\ the dispersion and attenuation) as a function of frequency, and the pulse delay as a function of pump intensity. In order to fully understand how the construction process affects the optical properties of the waveguide and the EIT response, two tests need to be done. First, a test of the metamaterial-clad waveguide without three-level atoms in the core to determine the effects of the fabrication process on the waveguide modes. Second, a test of the metamaterial-clad waveguide with three-level atoms embedded in the core to determine the EIT response. 

The restriction placed on the propagation constant by the Kramers-Kronig relations helps to simplify the experimental characterization of waveguides. An experimental test of the metamaterial-clad waveguide system (both with and without three-level atoms doped in the core) only needs to measure one of the dispersion or attenuation. By using the data gathered from one set of measurements, the corresponding values for the unmeasured quantity can be reconstructed using Eqs.~(\ref{eq:kramers-kronig}). 

For example, an experimental setup could be used to measure the dispersion of the metamaterial-clad waveguide by using beam-splitters. An optical signal is split, sending one part through the waveguide system, and the other through free space or a standard optical fibre. By comparing the phase of the two beams, the phase shift acquired by the signal that passes through the waveguide system can be determined.

The dispersion of an electromagnetic wave propagating through a medium, given by the real part of the propagation constant, determines the phase of the wave. If the phase difference is known, then the dispersion of the metamaterial-clad waveguide system can be inferred. Once the dispersion is known, the attenuation follows from the Kramers-Kronig relations. Alternatively, the attenuation could be measured, and the dispersion calculated using the Kramers-Kronig relations.

It is likely that there will be some amount of reflection of the beam as it enters and exits the metamaterial-clad waveguide. If this is the case, it would affect the measurements of the experimental test, particularly for attenuation measurements. If the reflected portion of the beam could be detected, the information about how much energy is reflected can be used to correct the measurements.

\section{Mode Choice}\label{sec:modechoice}

The analysis of the metamaterial-clad waveguide system doped with three-level atoms was done for TM modes due to the azimuthal invariance they possess, which allows the calculations to be simplified. Additionally, surface TM modes are supported for both metal- and metamaterial-clad waveguides. A metamaterial-clad guide can support a surface TE mode, albeit over a much smaller frequency range than for a TM surface mode, but a metal-clad guide cannot support TE surface modes. To allow a direct comparison between metal-clad and metamaterial-clad guides, the TM mode was chosen. 

A similar analysis to that done for TM modes can be done for TE and HE modes, though there will be some differences, especially for the HE modes. The TE modes will have different dispersion and attenuation curves, and a low-loss TE surface mode may or may not be supported by the guide. The effect of EIT on the TE modes, however, will be similar to that of the TM modes as the two mode types have similar intensity profiles. 

As with the TE modes, the dispersion and attenuation curves of the HE modes will be different from those of the TM modes. However, unlike the TE modes, the effect that EIT generated with an HE pump mode would have on an HE signal mode can not be inferred from the behaviour of the TM modes. The TE and TM modes suffer from an intensity that varies as a function of $r$, whereas the intensity of the HE modes depends on both $r$ and $\phi$. 

\begin{figure}[t,b] 
      \centering
	\subfloat{\label{fig:HE1sAmp}\includegraphics[width=0.5\textwidth]{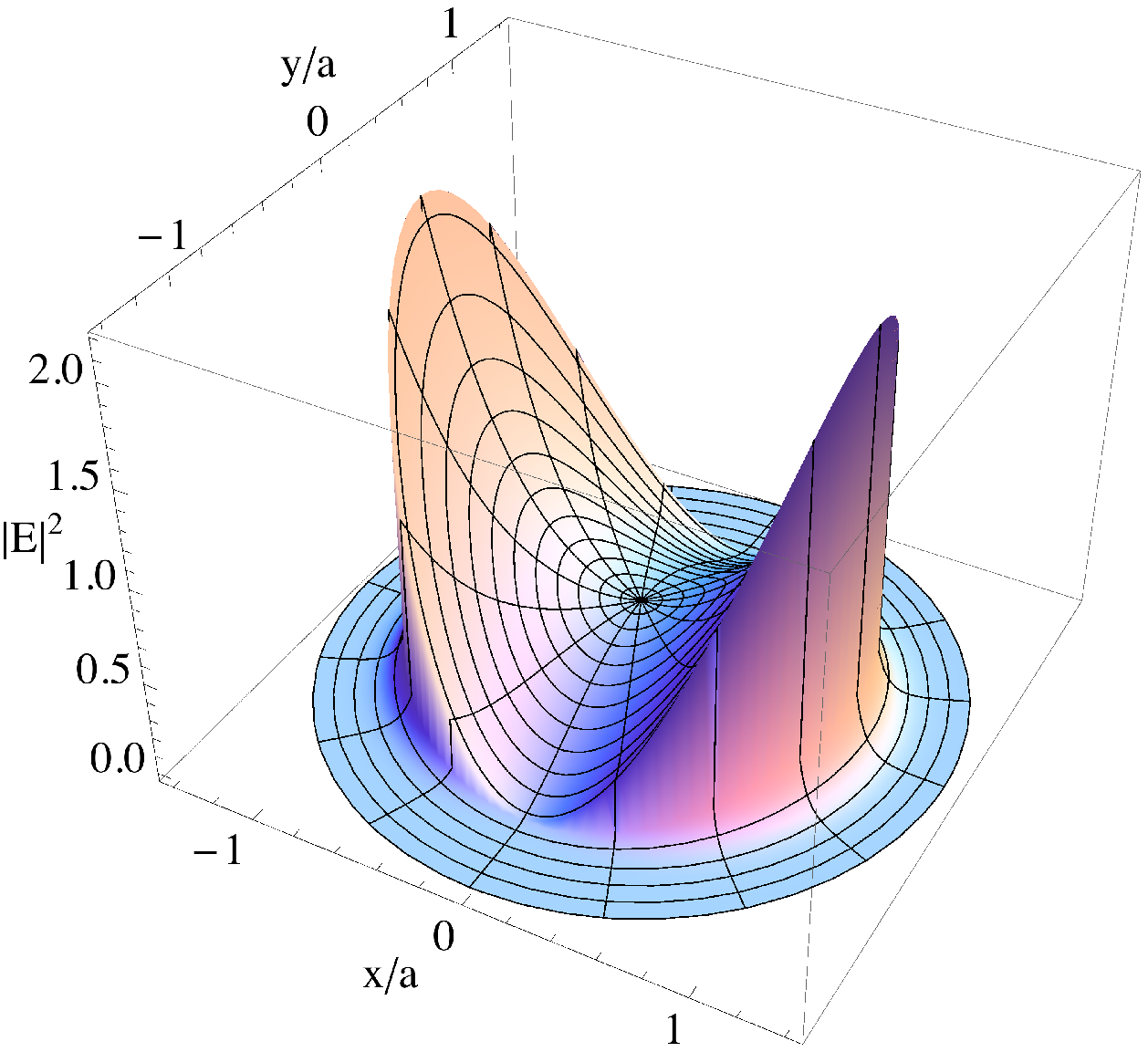}}
	\subfloat{\label{fig:HE2sAmp}\includegraphics[width=0.5\textwidth]{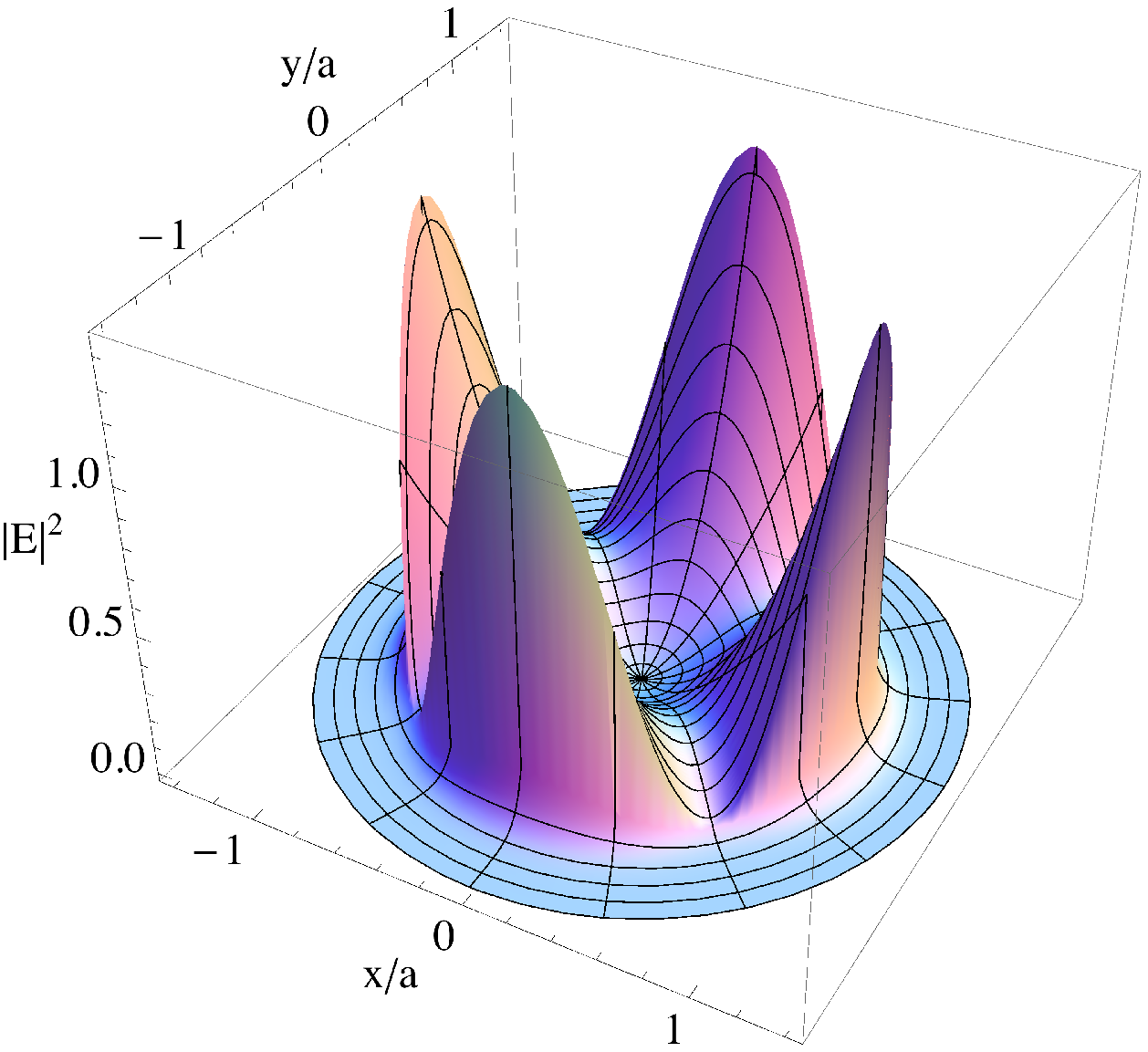}}	
	\caption[Plots of $|\bm E|^{2}$ for the HE$_{1\text s}$ and HE$_{2\text s}$ modes.]{Plots of $|\bm E|^{2}$ as a function of $r$ and $\phi$ for the HE$_{1\text s}$ mode \protect\subref{fig:HE1sAmp} and HE$_{2\text s}$ mode \protect\subref{fig:HE2sAmp}. The effect of the $\phi$ dependence on the fields for the HE modes causes $|\bm E|^{2}$ to vary significantly for the surface modes. Alignment between two co-propagating HE pulses would need to be maintained throughout the interaction region of the device to achieve the desired result. In the plots $x$ and $y$ satisfy $r=\sqrt{x^{2}+y^{2}}$.}
     \label{fig:HEsPlot}   
\end{figure}

The HE modes present a different set of challenges as they are not azimuthally invariant. The surface HE modes have a number of ``lobes'', depending on the order of the mode, consisting of high intensity regions separated by intensity minimums (see Fig.~\ref{fig:HEsPlot}). 
The variation of the intensity with $\phi$ will cause the EIT response to vary with $\phi$ and may cause some unanticipated behaviours. Slowing a single HE signal pulse using a TM pump should reduce these issues.

When two fields are interacting via the nonlinear Kerr effect, the intensity of each field affects the phase of the other. For two co-propagating HE pulses, the variation in the intensity means the relative orientation of each pulse will affect the resulting phase shift. There are two extreme cases that need to be considered; the case of both fields oriented in the same manner (high intensity regions are aligned, as are the low intensity regions), and the case of the fields oriented in a complementary manner (the high-intensity region of one field will coincide with the low-intensity region of the other). All other relative orientations fall somewhere between these two cases.

For the case where the high intensity regions of both pulses are aligned, the high-intensity regions will interact strongly with each other, inducing a large phase shift. The low-intensity regions, on the other hand, will only interact weakly, leading to a small phase shift. The case where the high- and low-intensity regions are aligned will result in a more uniform phase shift, though it will still not be completely uniform.

The self-phase modulation of the pulses in the high intensity region will partially compensate for the smaller cross-phase modulation due to the low-intensity regions. This is especially true if the two pulses have similar intensities. It is unclear, without further investigation, whether the different phase shifts will result in a net average phase shift, or will cause pulse distortion.

For the case of two co-propagating pulses where one is HE and one is TM, the orientation of the HE mode will not affect the interaction. However, the degree of interaction between the pulses will be reduced. This is due to the fact that the intensity of the TM pulse is evenly distributed about the guide axis, whereas the HE mode has regions of high and low intensity. In this case, the TM mode will effect uniform cross-phase modulation on the HE mode, but the opposite will not be true. The cross-phase modulation on the TM mode, due to the HE mode, will suffer from the same problems as the case for two HE modes.

Surface modes were chosen for their ability to confine electromagnetic energy to a small volume, thereby increasing the local energy density. Increased local energy density, analogous to increased local intensity, strengthens the nonlinear interaction between electromagnetic fields and matter. The usefulness of surface modes (plasmon-polaritons), however, is currently restricted by large losses. This is partly due to the fact that a large fraction of the plasmon-polariton energy propagates on the metal side of the surface, where it causes electrons to scatter and dissipate energy. 

Using ordinary modes may also be a possibility, as some ordinary modes exist that can provide good confinement. Although the confinement provided by these modes is much less than that of surface modes, ordinary modes have the advantage that they offer significantly reduced attenuation when compared to surface modes. The attenuation is further reduced when these modes propagate in waveguides with a dielectric cladding. As a metamaterial cladding does not offer any benefit to these modes, they were outside the scope of this thesis. However, considering the potential applications and the current metamaterial fabrication technology, a short discussion is warranted.

\begin{figure}[t,b] 
      \centering
	\subfloat{\label{fig:TM3confinement}\includegraphics[width=0.5\textwidth]{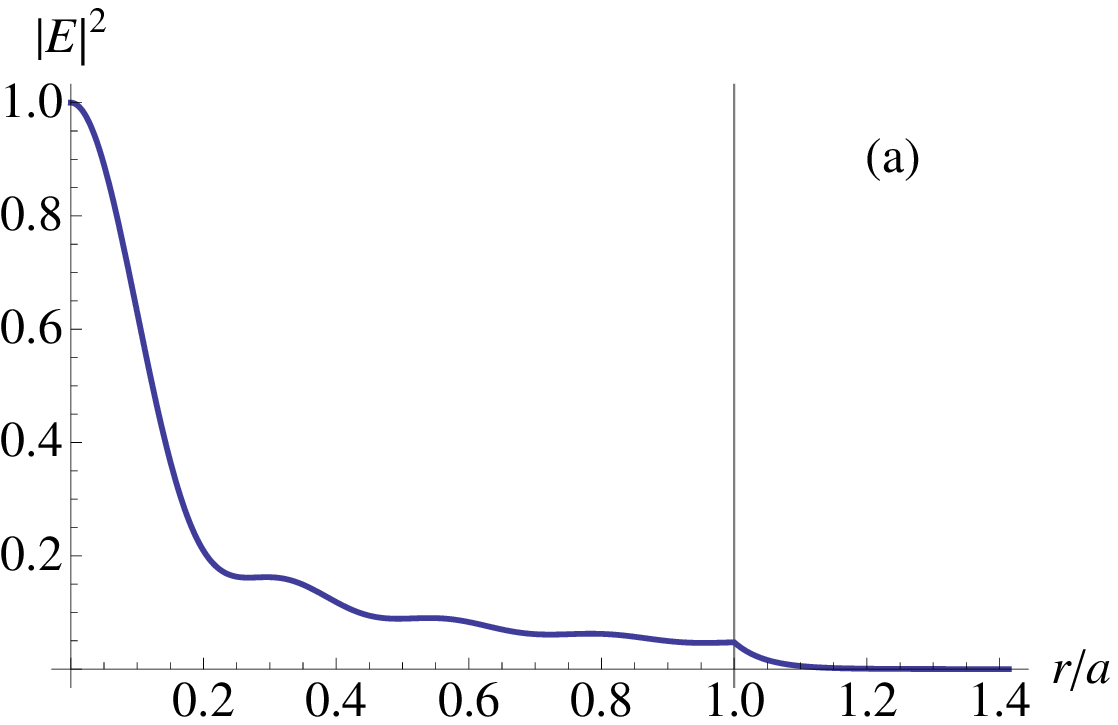}}
	\subfloat{\label{fig:TM4confinement}\includegraphics[width=0.5\textwidth]{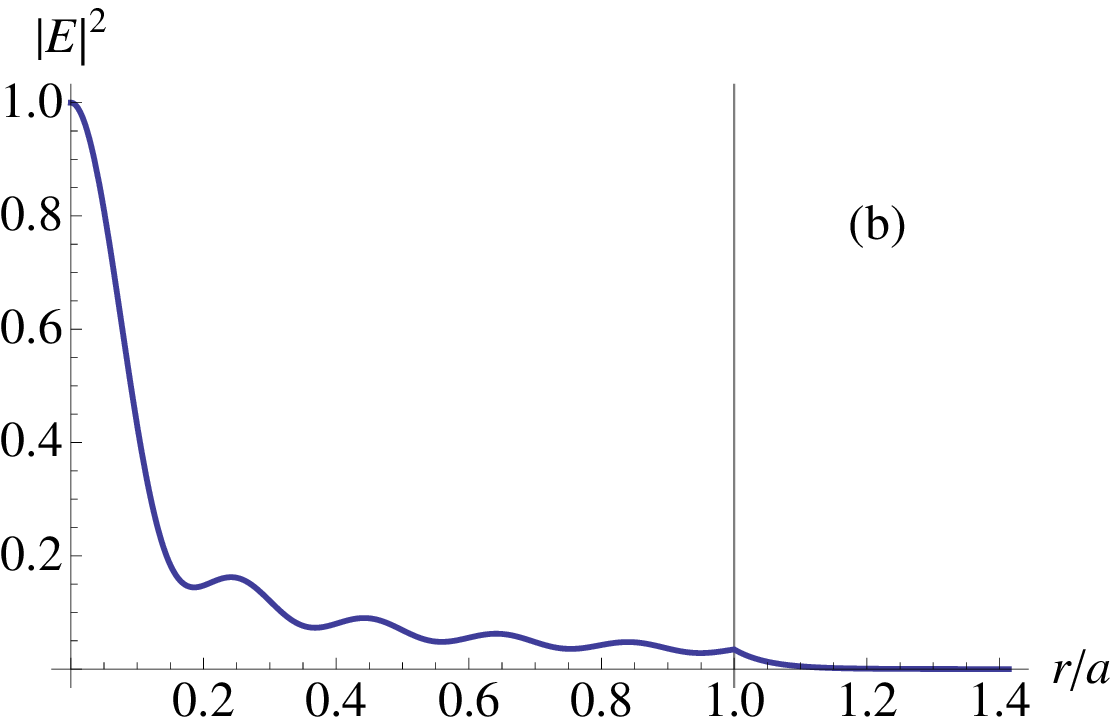}}	
	\caption[Plots of $|\bm E|^{2}$ for the TM$_{3}$ and TM$_{4}$ modes.]{Plots of $|\bm E|^{2}$ as a function of $r$ for the TM$_{3}$ mode \protect\subref{fig:TM3confinement} and TM$_{4}$ mode \protect\subref{fig:HE2sAmp}. The plots show how the decay of the amplitude with $r$ acts to concentrate the mode in the centre of the waveguide, thereby increasing the local intensity there. The thin vertical line represents the core-clading boundary.}
     \label{fig:TMconfinement}   
\end{figure} 

In a slab waveguide, the electromagnetic energy in an ordinary mode is distributed relatively evenly across the core of the waveguide. This is not the case with a cylindrical waveguide, where the fields in the core are described by the Bessel function $J_{\text m}(\kappa r)$. In a cylindrical waveguide the intensity generally decreases with $r$, though there are local maxima and minima. This means that for some modes, a large portion of the energy is located near the centre of the core. This is especially true for higher order modes, such as TM$_{3}$ and TM$_{4}$, as can be seen in Fig.~\ref{fig:TMconfinement}, which shows $|\bm E|^{2}$ as a function of $r$.

Some issues are likely to arise with the use of these modes. For instance, when a guide supports higher order modes, like TM$_{3}$ for a field of a given frequency, it will also support all of the lower order modes for that frequency. There is a possibility that some of the energy in the higher order modes can transfer into lower order modes and no longer contribute to the nonlinear response.

Additionally, the wavelength of the fields may be small compared to the guide width at frequencies above cut-off for these modes. This would have an impact on the ability of these modes to confine the fields. For the modes pictured in Fig.~\ref{fig:TMconfinement}, the wavelength of the fields are less than half the radius of the guides. If any of these other modes are to be considered as an alternative to TM surface modes, a full analysis, like the one presented here, will need to be done, and a realized prototype will have to be tested.

\section{Summary}

In order for any device to be used in real-world applications, it must undergo extensive testing to ensure it consistently performs as expected. Thus, the next logical step for this project is to build a metamaterial-clad waveguide and perform an experimental characterization. Testing a prototype design in a laboratory setting will serve to verify the theoretical analysis of metamaterial-clad waveguides, as well as provide insight into any unforeseen behaviours. 

Such behaviours could be directly related to the electromagnetic response of a specific metamaterial and how it affects the modes. The presence of imperfections that result from the fabrication process could also impact the behaviour of a realized device. Fabrication imperfections are unavoidable and their nature varies with the fabrication techniques used. For this reason, an experimental characterization is needed to determine the exact behaviour of a metamaterial-clad waveguide.

An experimental characterization would need to measure the propagation and attenuation constants of the metamaterial-clad waveguide. To do this, the Kramers-Kronig relations can be employed so that only one of two measurements needs to be done. The quantity not experimentally determined can be calculated from the measured quantity by using the appropriate Kramers-Kronig relation. Additionally, measuring the propagation delay would allow the group velocity of the pulse to be inferred.

The modes to be used must be considered as well. Within this thesis we concentrated on the TM$_{\text s}$ mode, as it provides good confinement of the electromagnetic fields and is azimuthally invariant. The TE$_{\text s}$ and HE$_{m\text{s}}$ modes are also capable of providing good field confinement. However, the intensities of the HE$_{m\text{s}}$ modes vary with $\phi$, the effects of which need to be explored separately. The intensity of the TE$_{\text s}$ mode does not depend on $\phi$, but this mode is not supported for as wide a range of frequencies as the TM$_{\text s}$ mode, and it is not clear if it will display the same low-loss behaviour as the TM$_{\text s}$ mode.

Higher-order TM ordinary modes of a cylindrical guide tend to concentrate the field energy at the centre of the core. The level of confinement provided by these modes is less than that of the surface modes, but they also have significantly less attenuation. Additionally, these modes do not require a metamaterial cladding, which would simplify device fabrication. As high-order modes are required, however, some energy can leak into lower-order modes, which propagate at different speeds, rendering the leaked energy unusable.

%% file: Summary.tex
\chapter{Summary}\label{ch:summary}

This project was motivated by the possibility of achieving giant cross-phase modulation with near-zero losses for single photon signals. Previous work using metamaterials as a means of guiding signals along a flat interface for this purpose showed promise~\cite{Kamli:2008,Moiseev:2010}. The intent of this work was to develop a theoretical framework, on which metamaterial-clad waveguides can be studied. This framework was then used to characterize a metamaterial-clad guide, and show that these guides do, in fact, display novel optical properties. 

The outcome of this project was anticipated to be a scheme for a device that could impart phase shifts of $\pi$ radians to single photons on demand. The ability to process single photons would make this scheme useful as a deterministic quantum controlled-not gate. It would also have uses for applications that do not require the ability to handle single photon signals. For example, quantum memories, and all-optical control devices, such as delay lines and switches.

During the course of the research for this thesis, however, it became apparent that this project would not produce the anticipated outcome. Since the project began, other research has found that imparting large phase shifts on single photons is significantly hindered by absorption~\cite{Gea-Banacloche:2010}. In fact, the probability of the photon being absorbed increases with the desired phase shift, making useful phase shifts on single photons unachievable. 

The absorption effect is greatest for a single photon as absorption results in a high probability of the photon being re-emitted into a different mode, rendering it lost. In the case of multiple photon pulses, however, it is more likely that an absorbed photon will be re-emitted into a populated mode through stimulated emission. For this reason, a pulse with a larger number of photons will have a higher probability of being preserved, with the probability increasing with the number of photons.

The limitation on single photon signals does not render this project without merit, however. Applications for all-optical control of weak fields could benefit from highly confined signals with near-zero loss. For instance, all-optical switching could replace conventional switching methods and significantly speed up data transfer in optical networks. 

The optical properties of waveguides depend strongly on the properties of the cladding layer, as it is the interface between the core and cladding that causes the guiding effect. By engineering this interface, for example by changing the cladding material, it is possible to alter the properties of the waveguide. It is this kind of boundary engineering that is used here to exploit the properties of metamaterials in an effort to overcome the large attenuation associated with strong confinement of propagating fields.

The work that sets the stage for this project showed that surface plasmon-polaritons travelling along a flat dielectric-metamaterial interface had drastically reduced losses for some frequencies. Additionally, it was shown that a large nonlinear interaction can be generated between two co-propagating weak pulses leading to a large phase shift by using EIT to simultaneously slow the low-loss SPPs. This foundational work, however, had the limitation that transverse confinement was not considered in the calculations, but assumed to be present through focusing~\cite{Kamli:2008}.

Transverse confinement was included in our analysis through the choice of waveguide structures. Although the slab geometry does not provide confinement in both transverse directions, it was studied because it contains elements of both the flat interface case and the cylindrical waveguide case. The cylindrical waveguide geometry was chosen as it inherently provides confinement in both transverse directions. Additionally, the dispersion equation for the cylindrical guide can be obtained analytically, and it is well studied due to the widespread use of optical fibres. 

For this thesis, we have assumed that the metamaterial comprising the waveguide cladding is isotropic. For the metamaterial permeability, we used the form for a fishnet metamaterial, despite the fact that this assumption is incompatible with two-dimensional metamaterials, as outlined in Sec.~\ref{sec:WGtesting}. This is a valid approximation, however, as the mechanism for achieving a frequency dependent permeability is similar in both two- and three-dimensional metamaterials. It was necessary to use the permeability of a two-dimensional metamaterial, as designs for three-dimensional metamaterials have not been fabricated and tested, largely due to fabrication limitations. However, new techniques, such as self-assembly~\cite{Vignolini:2012}, may prove useful for fabricating three-dimensional metamaterials.

Metamaterial-clad waveguides do not have as drastic a reduction in attenuation as flat interfaces, due to the mechanism responsible for the loss reduction. A flat interface supports a surface plasmon-polariton with near-zero loss when electromagnetic energy is expelled from the metamaterial and propagates entirely within the dielectric~\cite{Kamli:2008,Moiseev:2010}. Unlike flat interfaces, the structure of a slab or cylindrical waveguide does not allow for the electromagnetic fields to diverge arbitrarily far into the dielectric. The electromagnetic field expelled from the metamaterial can only spread out as far as the opposite interface, where it interacts with the metamaterial again. The consequence is that all of the energy cannot be expelled from the cladding, and the fields will continue to interact with the metamaterial, resulting in propagation losses.

We found that a metamaterial-clad waveguide supports a surface plasmon-polariton mode that, at certain frequencies and under a specific set of conditions, exhibits reduced propagation loss when compared with a metal-clad waveguide. The conditions are as follows: First, the metal-clad and metamaterial-clad waveguides have identical permittivity, which is described by Eq.~(\ref{epsilonMM}). Second, whereas a metal cladding naturally has a relative permeability that is constant and equal to 1, the metamaterial permeability is described by Eq.~(\ref{muMM}), which we call the modified-Drude model, and has a real part less than unity. Finally, the damping constant $\Gamma_{\text m}$ of the metamaterial permeability, which describes the energy dissipation due to magnetic interactions, must be less than its electric counterpart $\Gamma_{\text e}$ of the permittivity.

Having satisfied all of the above requirements, a metamaterial-clad slab or cylindrical waveguide will support a surface plasmon-polariton that exhibits less attenuation than an equivalent metal-clad guide. The level of attenuation reduction depends on the value of $\Gamma_{\text m}$. When $\Gamma_{\text m}=\Gamma_{\text e}/3$ the attenuation is reduced by 20\%, which should be achievable with current metamaterial fabrication techniques~\cite{Zhou:2008}. The attenuation can be reduced by as much as 40\% when $\Gamma_{\text m}=\Gamma_{\text e}/100$. Reducing the magnetic damping term far enough to achieve more than a 20\% attenuation reduction, however, is predicated on the ability to fabricate low-loss metamaterials. 

It may be possible to achieve a reduction in attenuation of more than 40\% if an active metamaterial is used for the cladding layer, as this would change the sign of $\Gamma_{\text m}$, $\Gamma_{\text e}$ or both. It is not clear, however, if the guided modes would be significantly altered, particularly if both $\Gamma_{\text m}$ and $\Gamma_{\text e}$ are made negative. 

In addition to a surface plasmon-polariton mode with reduced losses, we found that a metamaterial-clad waveguide can support a number of modes, which we call hybrid modes. These new modes display characteristics of both surface modes and ordinary guided modes. The energy distribution of hybrid modes suggests that a portion of the mode energy is transported within the electromagnetic fields propagating along the waveguide, and a portion is transported via electron oscillations at the core-cladding boundary. Hybrid modes have not been predicted for waveguides with other cladding materials, and are the result of the metamaterial cladding having a complex permeability. 

Hybrid modes are typically accompanied by long `tails' (slow evanescent decay into the cladding region) that travel on the metamaterial side of the interface. As the metamaterial cladding dissipates energy, via interaction with the fields, the long tails of hybrid modes lead to larger losses than predicted for the other mode types. The transition from an ordinary or surface mode to a hybrid mode generally happens over a narrow frequency range, with the increased attenuation appearing just as quickly. This sharp increase in attenuation as a function of frequency could enable hybrid modes to act as frequency filters. As these modes are accompanied by large losses and reduced field confinement, they are not good candidates for all-optical control. Thus, we did not study these modes in depth, but focussed on the surface mode instead.

The strong confinement in both transverse directions provided by the surface mode of a cylindrical guide, together with the reduced losses possible with a metamaterial cladding can be exploited for the purpose of all-optical control. An optical signal can be slowed using EIT via three-level $\Lambda$ atoms uniformly embedded in the dielectric core of a metamaterial-clad waveguide. Employing the low-loss surface mode results in highly confined, controllable slow, or stopped, light with reduced losses, when compared with light slowed in a metal-clad waveguide. The reduced losses do not come at the expense of the group velocity reduction or the confinement of the fields, which remain comparable to those of a metal-clad waveguide. 

As with the un-doped guide, however, the operation of such a device for slowing light is accompanied by restrictions. This project is driven by potential applications so the real-world operation of such a system must be considered. While it is true that some rare-earth ions, such as Pr$^{3+}$ and Er$^{3+}$, have an electronic level structure that is favourable for performing EIT, the presence of a host material can interfere. Cooling the system to cryogenic temperatures may help to mitigate some of these problems, but the need for a device to be cooled presents its own set of challenges. 

In order to cool a system down to cryogenic temperatures, it needs to be contained within a cryostat. This would present challenges for coupling the system with another that operates at higher temperatures. Also, the optical properties of the materials would be changed as a result of the reduced temperature, though this could be accounted for with material choice and design considerations. In addition to these restrictions, imperfections in the materials resulting from the fabrication process can lead to increased attenuation and reduced EIT efficiency. Determining the nature of the imperfections will require experimental testing of a working device.

The results obtained from this work raise some questions that are still unanswered. The exact role played by the permeability in mode propagation along a waveguide is still unknown. We showed that introducing an additional means of energy dissipation to the cladding (i.e.\ $\text{Im}(\mu)$) can lead to modes that have reduced losses. This problem raises a fundamental question about the connection between permeability and permittivity and their role in energy dissipation for propagating electromagnetic fields.


By engineering waveguide boundaries, through the use of a metamaterial cladding, we have shown that, under certain conditions, a metamaterial-clad waveguide can provide an incremental improvement over existing technology. Due to the restrictions on this result, however, it is my opinion that as an effective means of achieving low-loss all-optical control, this particular scheme does not provide a sufficient improvement to warrant further work. If active metamaterials become easily manufactured and readily available, however, then it is possible that a scheme, such as the one presented here, may provide a significant improvement.

%% file: appendix1.tex

\chapter{Computer Code for Numerical Calculations}\label{ap:1}

This chapter contains the Mathematica\textsuperscript{\textregistered} code I wrote to solve Eq.(\ref{TMvectwave}). This code was only intended to be used by myself to calculate $\tilde\beta$ for the doped metamaterial-clad waveguide and therefore is not optimized for readability. I have included comments describing certain portions of the code to aid the reader. This program calculates the real and imaginary parts of $\tilde\beta$ for a number of consecutive frequencies, and outputs them to separate files, along with the corresponding frequency. Plots of both these values, as functions of frequency, are also updated every 10 frequency steps in order to allow visual inspection of the solutions during the calculation.

\vspace{10 mm}
\noindent The code is as follows:

\vspace{10 mm}
\noindent\({\text{Module}[\{\epsilon ,\text{d$\epsilon $dr},\kappa ,\gamma ,\text{k0},\Delta ,\text{a0},\text{$\epsilon $nimm},\text{$\mu $nimm},\lambda,\text{$\Delta $R},\omega ,\text{$\Omega $csqr},\beta \},}\\
{\text{(* Defining various constants and quantities *)}}\\
{\text{$\omega $e}=\text{1.378$\times10^{16}$};}\\
{\text{$\gamma $e}=\text{2.73$\times10^{13}$};}\\
{\text{$\omega $0}=\text{$\omega $e}/5;}\\
{\text{Fm}=.5;}\\
{\text{$\gamma $m}=\text{$\gamma $e}/3;}\\
{\text{$\omega $31}=.409\text{$\omega $e};}\\
{\text{$\Delta $R}\text{:=}\Delta ;}\\
{\omega \text{:=}\Delta +\text{$\omega $31};}\\
{\text{$\gamma $21}=10^{-2};}\\
{\text{$\gamma $31}=10^5;}\\
{\hbar =\text{1.054571726$\times10^{-34}$};}\\
{\text{$\epsilon $0}=\text{8.854187817620$\times10^{-12}$};}\\
{\text{$\mu $0}=4\pi\times10^{-7};}\\
{c=1\left/\sqrt{\text{$\epsilon $0} \text{$\mu $0}}\right.;}\\
{\text{k0}\text{:=}\omega /c;}\\
{\lambda \text{:=}2\pi  c/\omega ;}\\
{\text{$\lambda $e}=2\pi  c/\text{$\omega $e};}\\
{R=2\text{$\lambda $e};}\\
{\Omega =8\times10^4;}\\
{n=\text{$\texttt{"}$filename$\texttt{"}$};}\\
{\delta =N\left[10^{-100}\right];}\\
{\text{$\beta $c}=1.156527+0.001297 i;}\\
{\text{$\omega $c}=\text{5.61823$\times10^{15}$};}\\
{\text{tol}=10^{-9};}\\
{\text{nsteps}=5000;}\\
{\text{basestep}=\text{1$\times10^{-4}$}\text{$\omega $e};}\\
{\delta \omega =\text{basestep};}\\
{\text{test}=.01;}\\
{\text{test2}=2+10^{-1}+i 10^{-1};}\\
{\text{pt1}=\text{Null};}\\
{\text{pt2}=\text{Null};}\\
{\text{sa}=.1;}\\
{p=40;}\\
{\text{a0}\text{:=}\displaystyle\frac{3 \pi  c^2}{1.3 \omega ^2}\text{1.26$\times10^{21}$};}\\
{\text{$\epsilon $nimm}\text{:=}1-\displaystyle\frac{\text{$\omega $e}^2}{\omega (\omega +i \text{$\gamma $e})} ;}\\
{\text{$\mu $nimm}\text{:=}1-\displaystyle\frac{\text{Fm}(\omega )^2}{\omega (\omega +i \text{$\gamma $m})-\text{$\omega $0}^2} ;}\\
{\text{$\kappa $lin}=\frac{\text{$\omega $c}}{c}\left(1.3-\text{$\beta $c}^2\right)^{1/2};}\\
{\text{$\gamma $lin}=\frac{\text{$\omega $c}}{c}\left(\text{$\beta $c}^2-1.3\right)^{1/2};}\\
{\text{$\Omega $csqr}=\displaystyle\Omega ^2\text{BesselJ}[0,\text{$\kappa $lin}\,r] \text{BesselJ}[0,\text{Conjugate}[\text{$\kappa $lin}] r]}\\
{\displaystyle\quad+\frac{\Omega ^2}{\text{$\kappa $lin}}\text{$\beta $c} \frac{\text{$\omega $c}}{c}\text{BesselJ}[1,\text{$\kappa $lin}\,r] \frac{1}{\text{Conjugate}[\text{$\kappa$lin}]}}\\
{\displaystyle\quad\times\text{Conjugate}[\text{$\beta $c}] \frac{\text{$\omega $c}}{c}\text{BesselJ}[1,\text{Conjugate}[\text{$\kappa $lin}] r];}\\
{\text{dispnumTM}={\text{Check}[}}\\
{\quad\text{Import}[\text{$\texttt{"}$filepath/filename-prefix1$\texttt{"}$}<> n<>\text{$\texttt{"}$.txt$\texttt{"}$},\text{$\texttt{"}$Table$\texttt{"}$]}}\\
,-1];\quad\text{(* Checking for existing file, $-1$ \text{is} \text{returned} \text{if} \text{file} \text{cannot} \text{be} \text{found} *)}\\
{\text{absnumTM}=\text{Check}[}\\
{\quad\text{Import}[\text{$\texttt{"}$filepath/filename-prefix2$\texttt{"}$}<> n<>\text{$\texttt{"}$.txt$\texttt{"}$},\text{$\texttt{"}$Table$\texttt{"}$]}}\\
,-1]; \\
{\text{If}[\text{dispnumTM}\text{===}-1,}\\
\text{(* \text{If} \text{file} \text{is} \text{not} \text{found}, initiate \text{calculation} \text{at} \text{detuning} $\Delta$  \text{using} \text{initial} guess  beta1 \text{*)}}\\
\text{(* \text{If} \text{the} \text{data} \text{files} \text{exist}, the \text{program} \text{skips} \text{to} \text{the} $\dagger$ \text{*)}}\\
{\quad\text{beta1}=0.6+2i;}\\
{\quad\Delta =\text{2.8$\times10^{5}$};}\\
{\quad\beta =\text{beta1};}\\
{\quad\epsilon =1.3+\displaystyle\frac{2\,\text{a0}}{\sqrt{1.3}\,\text{k0}}\frac{i \text{$\gamma $31}}{\text{$\gamma $31}-i \Delta +\text{$\Omega $csqr}(\text{$\gamma
$21}-i \Delta )^{-1}};}\\
{\quad\text{d$\epsilon $dr}=D[\epsilon ,r];}\\
{\quad\gamma =\displaystyle\frac{\omega }{c}\left(\beta ^2-\text{$\epsilon $nimm } \text{$\mu $nimm}\right)^{1/2};}\\
{\quad\kappa =\displaystyle\frac{\omega }{c}\left(\epsilon -\beta ^2\right)^{1/2};}\\
{\quad\text{ss}=\text{NDSolve}[\quad{\text{(* Numerically solving the wave equation *)}}}\\
{\qquad\displaystyle\left\{r A''[r]+\left(1-\frac{r \beta ^2 \text{k0}^2}{\epsilon  \kappa ^2}\text{d$\epsilon $dr}\right)A'[r]+r \kappa^2A[r]==0,A[\delta ]==1,A'[\delta ]==0\right\},}\\
{\qquad A,\{r,\delta ,R\},\text{MaxSteps}\to 100000];}\\
{\quad\text{const}=\text{Evaluate}[A[R]\text{/.}\text{ss}[[1]]]/\text{BesselK}[0,\gamma  R];\quad\text{(* Calculating boundary mismatch *)}}\\
{\quad\text{test}=\displaystyle\left(\text{Evaluate}\left[\frac{\epsilon }{\kappa ^2}A'[R]\text{/.}\text{ss}[[1]]\right]-\left(\frac{\text{const } \text{$\epsilon $nimm}}{\gamma }\text{  }\text{BesselK}[1,\gamma  R]\right)\right)}\\
{\qquad\displaystyle\left/\left(\frac{\text{const } \text{$\epsilon $nimm}}{\gamma }\text{  }\text{BesselK}[1,\gamma  R]\right)\right.\text{/.}r\to R; }\\
{\quad\text{test2}=\text{test};}\\
{\quad\text{If}[\text{Abs}[\text{test}]<\text{tol},\quad\text{(* \text{If} \text{mismatch} \text{is} \text{within} \text{tolerance}, keep \text{the} \text{solution} \text{*)}}}\\
{\quad\quad\text{beta}=\text{beta1};}\\
{\quad];}\\
{\quad\text{Monitor}[}\\
{\quad\quad\text{While}[\text{Abs}[\text{test}]>\text{tol},\quad\text{(* \text{If} \text{mismatch} \text{is} \text{not} \text{within} \text{tolerance}, \text{try} \text{again} \text{*)}}}\\
{\quad\quad\quad\text{beta}=\text{RandomComplex}[\quad\text{(* Choose value for beta *)}}\\
{\quad\quad\quad\quad\{\text{SetPrecision}[\text{beta1}-\text{sa } \text{Abs}[\text{test2}](1+i),p],}\\
{\quad\quad\quad\quad\text{SetPrecision}[\text{beta1}+\text{sa } \text{Abs}[\text{test2}](1+i),p]\},\text{WorkingPrecision}\to p];}\\
{\quad\quad\quad\text{If}[\text{Im}[\text{beta}]<0,\quad\text{(* \text{Ensuring} \text{that} \text{Im}(\text{beta}) $ > 0$ \text{*)}}}\\
{\quad\quad\quad\quad\text{beta}=\text{RandomComplex}[}\\
{\quad\quad\quad\quad\quad\{\text{SetPrecision}[\text{Re}[\text{beta1}]-\text{sa } \text{Abs}[\text{test2}]+0i,p],}\\
{\quad\quad\quad\quad\quad\text{SetPrecision}[\text{beta1}+\text{sa } \text{Abs}[\text{test2}](1+i),p]\},\text{WorkingPrecision}\to p];}\\
{\quad\quad\quad]; }\\
{\quad\quad\quad\beta =\text{beta};}\\
{\quad\quad\quad\epsilon =1.3+\displaystyle\frac{2\text{a0}}{\sqrt{1.3}\text{k0}}\frac{i \text{$\gamma $31}}{\text{$\gamma $31}-i \Delta +\text{$\Omega $csqr}(\text{$\gamma$21}-i \Delta )^{-1}};}\\
{\quad\quad\quad\text{d$\epsilon $dr}=D[\epsilon ,r];}\\
{\quad\quad\quad\gamma =\displaystyle\frac{\omega }{c}\left(\beta ^2-\text{$\epsilon $nimm } \text{$\mu $nimm}\right)^{1/2};}\\
{\quad\quad\quad\kappa =\displaystyle\frac{\omega }{c}\left(\epsilon -\beta ^2\right)^{1/2};}\\
{\quad\quad\quad\text{ss}=\text{NDSolve}[\quad\text{(* Numerically solving the wave equation *)}}\\
{\quad\quad\quad\quad\displaystyle\left\{r A''[r]+\left(1-\frac{r \beta ^2 \text{k0}^2}{\epsilon  \kappa ^2}\text{d$\epsilon $dr}\right)A'[r]+r \kappa ^2A[r]==0,A[\delta ]==1,A'[\delta ]==0\right\},}\\
{\quad\quad\quad\quad A,\{r,\delta ,R\},\text{MaxSteps}\to 100000];}\\
{\quad\quad\quad\text{const}=\text{Evaluate}[A[R]\text{/.}\text{ss}[[1]]]/\text{BesselK}[0,\gamma  R];\quad\text{(*  Calculating boundary mismatch *)}}\\
{}\\
{\quad\quad\quad\text{test}=\displaystyle\left(\text{Evaluate}\left[\frac{\epsilon}{\kappa ^2}A'[R]\text{/.}\text{ss}[[1]]\right]-\left(\frac{\text{const } \text{$\epsilon $nimm}}{\gamma }\text{BesselK}[1,\gamma  R]\right)\right)}\\
{\quad\quad\quad\quad\displaystyle\left/\left(\frac{\text{const } \text{$\epsilon $nimm}}{\gamma}\text{BesselK}[1,\gamma  R]\right)\right.\text{/.}r\to R;}\\
{\quad\quad\quad\text{If}[\text{Abs}[\text{test}]<\text{Abs}[\text{test2}], \text{(* \text{Determining} \text{if} \text{the} \text{new} \text{solution} \text{is} \text{better} *)} }\\
{\quad\quad\quad\quad\text{test2}=\text{test};}\\
{\quad\quad\quad\quad\text{beta1}=\text{beta};}\\
{\quad\quad\quad];}\\
{\quad\quad];}\\
{\quad\left.,\left\{\Delta \left/10.^6\right., \text{Abs}[\text{test2}],\text{beta1},\text{Abs}[\text{test}],\text{beta}\right\}\right];}\\
{\quad\text{betaold}=\text{beta};}\\
{\quad\text{betalast}=\text{beta};}\\
{\quad\text{dispnumTM}=\{\{\Delta /\text{$\omega $e},\text{Re}[\text{beta}]\}\};\quad\text{(* \text{Defining} \text{arrays} for \text{the} \text{calculated} results \text{*)}}}\\
{\quad\text{absnumTM}=\{\{\Delta /\text{$\omega $e},\text{Im}[\text{beta}]\}\};}\\
{\quad\text{PutAppend}[\quad\text{(* Creating data files and writing the data to them *)}}\\
{\quad\quad\text{OutputForm}[\text{ToString}[\text{FortranForm}[\Delta /\text{$\omega $e}]]<>\text{$\texttt{"}$   $\texttt{"}$}<>\text{ToString}[\text{FortranForm}[\text{Re}[\text{beta}]]]],}\\
{\quad\quad\text{$\texttt{"}$filepath/filename-prefix1$\texttt{"}$}<>n<>\text{$\texttt{"}$.txt$\texttt{"}$}]; }\\
{\quad\text{PutAppend}[}\\
{\quad\quad\text{OutputForm}[\text{ToString}[\text{FortranForm}[\Delta /\text{$\omega $e}]]<>\text{$\texttt{"}$   $\texttt{"}$}<>\text{ToString}[\text{FortranForm}[\text{Im}[\text{beta}]]]],}\\
{\quad\quad\text{$\texttt{"}$filepath/filename-prefix2$\texttt{"}$}<>n<>\text{$\texttt{"}$.txt$\texttt{"}$}];}\\
{,\quad\text{(* $\dagger$ \text{If} \text{the} \text{data} \text{files} already \text{exist}, this \text{is} \text{where} \text{the} \text{program} \text{starts} the calculations \text{*)}}}\\
{\quad\Delta =\text{dispnumTM}[[\text{Length}[\text{dispnumTM}],1]]\text{$\omega $e};}\\
{\quad\text{(* Set detuning value from the data array *)}}\\
{\quad\text{(* \text{If} \text{more} \text{than} \text{one} \text{value} \text{of} \text{beta} \text{has} \text{been} \text{calculated}, a linear \text{extrapolation} \text{is} \text{done} \text{to}}}\\
{\quad\text{ make a \text{better} initial \text{guess} \text{*)}}}\\
{\quad\text{If}[\text{Length}[\text{dispnumTM}]>1,\quad\text{(* Reading the last two values of beta *)}}\\
{\quad\quad\text{betalast}=\text{dispnumTM}[[\text{Length}[\text{dispnumTM}],2]]+i \text{ absnumTM}[[\text{Length}[\text{absnumTM}],2]]; }\\
{\quad\quad\text{betaold}=\text{dispnumTM}[[\text{Length}[\text{dispnumTM}]-1,2]]}\\
{\quad\quad\quad+i \text{ absnumTM}[[\text{Length}[\text{absnumTM}]-1,2]];}\\
{\quad,\quad\text{(* \text{Values} \text{to} \text{be} \text{set} \text{if} \text{only} \text{one} \text{value} \text{of} \text{beta} }\\
\text{has \text{been} \text{calculated} \text{*)}}}\\
{\quad\quad\text{betalast}=\text{dispnumTM}[[\text{Length}[\text{dispnumTM}],2]]+i \text{ absnumTM}[[\text{Length}[\text{absnumTM}],2]]; }\\
{\quad\quad\text{betaold}=\text{betalast};}\\
{\quad];}\\
{\quad\text{(* \text{Define} \text{plots} \text{to} \text{visually} \text{monitor} \text{the} \text{modes} during \text{calculation} \text{*)}}}\\
{\quad\text{plots}=\text{GraphicsArray}[}\\
{\quad\quad\{\text{ListPlot}[\text{dispnumTM},\text{Joined}\to \text{True},\text{PlotRange}\to \text{All},\text{ImageSize}\to 500],}\\
{\quad\quad\text{ListPlot}[\text{absnumTM},\text{Joined}\to \text{True},\text{PlotRange}\to \text{All},\text{ImageSize}\to 500]\}}\\
{\quad];}\\
{];}\\
{i=1;\quad\text{(* Set counter for plotting *)}}\\
{\text{Monitor}[}\\
{\quad\text{Monitor}[}\\
{\quad\quad\text{While}[\Delta \geq -\text{2.8}\times10^{5},\quad\text{(* Set the end of the frequency interval *)}}\\
{\quad\quad\text{\text{(*} \text{If} \text{needed}, \text{the} \text{search} \text{area} \text{and} \text{step} \text{size} \text{can} \text{be} \text{adjusted} \text{for} \text{different}}}\\
{\quad\quad\text{\text{frequency} \text{intervals} \text{*)}}}\\
{\quad\quad\quad\text{Which}[}\\
{\quad\quad\quad\quad-\text{5$\times10^{-7}$}\text{$\omega $e}\leq \Delta \leq \text{5$\times10^{-7}$}\text{$\omega $e}}\\
{\quad\quad\quad\quad\delta \omega =-\text{1$\times10^{-9}$}\text{$\omega $e};}\\
{\quad\quad\quad\quad\text{sa}=\text{1.5$\times10^{-1}$};}\\
{\quad\quad\quad\quad\text{sa1}=\text{sa};}\\
{\quad\quad\quad,}\\
{\quad\quad\quad\quad-\text{5$\times10^{-6}$}\text{$\omega $e}\leq \Delta \leq \text{5$\times10^{-6}$}\text{$\omega $e}}\\
{\quad\quad\quad,}\\
{\quad\quad\quad\quad\delta \omega =-\text{1$\times10^{-8}$}\text{$\omega $e};}\\
{\quad\quad\quad\quad\text{sa}=.1;}\\
{\quad\quad\quad\quad\text{sa1}=\text{sa};}\\
{\quad\quad\quad,}\\
{\quad\quad\quad\quad-\text{5$\times10^{-5}$}\text{$\omega $e}\leq \Delta \leq \text{5$\times10^{-5}$}\text{$\omega $e}}\\
{\quad\quad\quad,}\\
{\quad\quad\quad\quad\delta \omega =-\text{1$\times10^{-7}$}\text{$\omega $e};}\\
{\quad\quad\quad\quad\text{sa}=.1;}\\
{\quad\quad\quad\quad\text{sa1}=\text{sa};}\\
{\quad\quad\quad,}\\
{\quad\quad\quad\quad-\text{1$\times10^{-4}$}\text{$\omega $e}\leq \Delta \leq \text{1$\times10^{-4}$}\text{$\omega $e}}\\
{\quad\quad\quad,}\\
{\quad\quad\quad\quad\delta \omega =-\text{1$\times10^{-6}$}\text{$\omega $e};}\\
{\quad\quad\quad\quad\text{sa}=.000001;}\\
{\quad\quad\quad\quad\text{sa1}=\text{sa};}\\
{\quad\quad\quad,}\\
{\quad\quad\quad\quad-\text{1$\times10^{-3}$}\text{$\omega $e}\leq \Delta \leq \text{1$\times10^{-3}$}\text{$\omega $e}}\\
{\quad\quad\quad,}\\
{\quad\quad\quad\quad\delta \omega =-\text{1$\times10^{-5}$}\text{$\omega $e};}\\
{\quad\quad\quad\quad\text{sa}=.005;}\\
{\quad\quad\quad\quad\text{sa1}=\text{sa};}\\
{\quad\quad\quad,}\\
{\quad\quad\quad\quad\text{True}}\\
{\quad\quad\quad,}\\
{\quad\quad\quad\quad\delta \omega =\text{basestep};}\\
{\quad\quad\quad\quad\text{sa}=.1;}\\
{\quad\quad\quad\quad\text{sa1}=\text{sa};}\\
{\quad\quad\quad];}\\
{\quad\quad\Delta =\Delta +\delta \omega ;\quad\text{(* Incrementing the frequecy *)}}\\
{\quad\quad\text{test}=1.1;}\\
{\quad\quad\text{If}[\text{betalast}==\text{betaold},\quad\text{(* Setting values to prevent errors *)}}\\
{\quad\quad\quad\text{test2}=1.1;}\\
{\quad\quad\quad\text{sa}=.1;}\\
{\quad\quad\quad\text{sa1}=\text{sa};}\\
{\quad\quad];}\\
{\quad\quad j=1;\quad\text{(* Define counters to monitor calculation *)}}\\
{\quad\quad k=0;}\\
{\quad\quad\text{\text{(*} \text{Calculating} \text{various} \text{quantities} \text{at} \text{new} }\\
\text{\text{frequency} value \text{*)}}}\\
{\quad\quad\text{beta1}=\text{SetPrecision}[(2\text{betalast}-\text{betaold}),p];}\\
{\quad\quad\beta =\text{beta1};}\\
{\quad\quad\displaystyle\epsilon =1.3+\frac{2\text{a0}}{\sqrt{1.3}\text{k0}}\frac{i \text{$\gamma $31}}{\text{$\gamma $31}-i \Delta +\text{$\Omega $csqr}(\text{$\gamma
$21}-i \Delta )^{-1}};}\\
{\quad\quad\text{d$\epsilon $dr}=D[\epsilon ,r];}\\
{\quad\quad\displaystyle\gamma =\frac{\omega }{c}\left(\beta ^2-\text{$\epsilon $nimm} \text{$\mu $nimm}\right)^{1/2};}\\
{\quad\quad\displaystyle\kappa =\frac{\omega }{c}\left(\epsilon -\beta ^2\right)^{1/2};}\\
{\quad\quad\text{ss}=\text{NDSolve}[\quad\text{(* Numerically solving the wave equation *)}}\\
{\quad\quad\quad\displaystyle\left\{r A''[r]+\left(1-\frac{r \beta ^2 \text{k0}^2}{\epsilon  \kappa ^2}\text{d$\epsilon $dr}\right)A'[r]+r \kappa ^2A[r]==0,A[\delta ]==1,A'[\delta ]==0\right\},}\\
{\quad\quad\quad A,\{r,\delta ,R\},\text{MaxSteps}\to 100000];}\\
{\quad\quad\text{const}=\text{Evaluate}[A[R]\text{/.}\text{ss}[[1]]]/\text{BesselK}[0,\gamma  R];\quad\text{(* Calculating the boundary mismatch *)}}\\
{\quad\quad\text{test}=\displaystyle\left(\text{Evaluate}\left[\frac{\epsilon  }{\kappa ^2}A'[R]\text{/.}\text{ss}[[1]]\right]-\left(\frac{\text{const } \text{$\epsilon $nimm}}{\gamma }\text{BesselK}[1,\gamma  R]\right)\right)}\\
{\quad\quad\quad\displaystyle\left/\left(\frac{\text{const } \text{$\epsilon $nimm}}{\gamma }\text{  }\text{BesselK}[1,\gamma  R]\right)\right.\text{/.}r\to R;}\\
{\quad\quad\text{test2}=\text{test};}\\
{\quad\quad\text{testold}=\text{test};}\\
{\quad\quad\text{If}[\text{Abs}[\text{test}]<\text{tol},}\\
{\quad\quad\quad\text{beta}=\text{beta1};}\\
{\quad\quad];}\\
{\quad\quad\text{While}[\text{Abs}[\text{test}]>\text{tol},}\\
{\quad\quad\quad\text{search}=\text{sa } \text{Abs}[\text{test2}];\quad\text{\text{(*} \text{Define} \text{search} \text{area} \text{in} \text{terms} \text{of} \text{boundary} \text{mismatch} \text{*)}}}\\
{\quad\quad\quad\text{If}[j>500,\quad\text{\text{(*} \text{If} \text{no} \text{solution} \text{found} \text{after} 500 \text{iterations} \text{adjust} \text{search} \text{size} \text{*)}}}\\
{\quad\quad\quad\quad\text{If}\left[\text{search}>10^{-20},\right.\quad\text{\text{(*} \text{If} \text{search} \text{space} \text{is} \text{large,} \text{its} \text{size} \text{is} \text{reduced} \text{*)} }}\\
{\quad\quad\quad\quad\quad\text{sa}=.2\text{sa};}\\
{\quad\quad\quad\quad,}\\
{\quad\quad\quad\quad\quad\text{sa}=16\text{sa1};\quad\text{\text{(*} \text{Otherwise} \text{size} \text{of} \text{search} \text{space} \text{is} }\\
\text{\text{increased} \text{*)}}}\\
{\quad\quad\quad\quad\quad k\text{++};}\\
{\quad\quad\quad\quad\quad\text{If}[k>2,\text{\text{(*} \text{If} \text{this} \text{is} \text{done} \text{too} \text{many} \text{times,} the \text{values} are \text{reset,}}}\\
{\quad\quad\quad\quad\quad\text{ \text{and} the calculation is started \text{again} \text{*)}}}\\
{\quad\quad\quad\quad\quad\quad\text{beta1}=\text{SetPrecision}[(2\text{betalast}-\text{betaold}),p];}\\
{\quad\quad\quad\quad\quad\quad\text{test2}=\text{testold};}\\
{\quad\quad\quad\quad\quad\quad k=0;}\\
{\quad\quad\quad\quad\quad];}\\
{\quad\quad\quad\quad];}\\
{\quad\quad\quad\quad j=0;}\\
{\quad\quad\quad];}\\
{\quad\quad\quad\text{beta}=\text{RandomComplex}[\quad\text{(* Choosing a random value for beta *)}}\\
{\quad\quad\quad\quad\{\text{SetPrecision}[\text{beta1}-\text{search}(1+i),p],}\\
{\quad\quad\quad\quad\text{SetPrecision}[\text{beta1}+\text{search}(1+i),p]\},\text{WorkingPrecision}\to p];}\\
{\quad\quad\quad\text{If}[\text{Im}[\text{beta}]<0,\quad\text{\text{(*} \text{ensuring} \text{that} \text{Im}$($\text{beta}$) > 0$ \text{*)}}}\\
{\quad\quad\quad\quad\text{beta}=\text{RandomComplex}[}\\
{\quad\quad\quad\quad\quad\{\text{SetPrecision}[\text{Re}[\text{beta1}]-\text{search}+0i,p],}\\
{\quad\quad\quad\quad\quad\text{SetPrecision}[\text{beta1}+\text{search}(1+i),p]\},\text{WorkingPrecision}\to p];}\\
{\quad\quad\quad];}\\
{\quad\quad\quad\text{(* Calculating quantities for new test beta *)}}\\
{\quad\quad\quad\beta =\text{beta};}\\
{\quad\quad\quad\displaystyle\epsilon =1.3+\frac{2\text{a0}}{\sqrt{1.3}\text{k0}}\frac{i \text{$\gamma $31}}{\text{$\gamma $31}-i \Delta +\text{$\Omega $csqr}(\text{$\gamma
$21}-i \Delta )^{-1}};}\\
{\quad\quad\quad\text{d$\epsilon $dr}=D[\epsilon ,r];}\\
{\quad\quad\quad\displaystyle\gamma =\frac{\omega }{c}\left(\beta ^2-\text{$\epsilon $nimm } \text{$\mu $nimm}\right)^{1/2};}\\
{\quad\quad\quad\displaystyle\kappa =\frac{\omega }{c}\left(\epsilon -\beta ^2\right)^{1/2};}\\
{\quad\quad\quad\text{ss}=\text{NDSolve}[\quad\text{(* Numerically solving the wave equation *)}}\\
{\quad\quad\quad\quad\displaystyle\left\{r A''[r]+\left(1-\frac{r \beta ^2 \text{k0}^2}{\epsilon  \kappa ^2}\text{d$\epsilon $dr}\right)A'[r]+r \kappa ^2A[r]==0,A[\delta ]==1,A'[\delta ]==0\right\},}\\
{\quad\quad\quad\quad A,\{r,\delta ,R\},\text{MaxSteps}\to 100000];}\\
{\quad\quad\quad\text{const}=\text{Evaluate}[A[R]\text{/.}\text{ss}[[1]]]/\text{BesselK}[0,\gamma  R];\quad\text{(* Calculating the boundary mismatch *)}}\\
{\quad\quad\quad\displaystyle\text{test}=\left(\text{Evaluate}\left[\frac{\text{  }\epsilon  }{\kappa ^2}A'[R]\text{/.}\text{ss}[[1]]\right]-\left(\frac{\text{const } \text{$\epsilon $nimm}}{\gamma }\text{  }\text{BesselK}[1,\gamma  R]\right)\right)}\\
{\quad\quad\quad\quad\displaystyle\left/\left(\frac{\text{const } \text{$\epsilon $nimm}}{\gamma }\text{  }\text{BesselK}[1,\gamma  R]\right)\right.\text{/.}r\to R;}\\
{\quad\quad\quad\text{If}[\text{Abs}[\text{test}]<\text{Abs}[\text{test2}],\quad\text{\text{(*} \text{If} \text{boundary} \text{mismatch} \text{within} \text{tolerance,}}}\\
{\quad\quad\quad\text{ reset \text{some} \text{values} \text{*)}}}\\
{\quad\quad\quad\quad\text{test2}=\text{test};}\\
{\quad\quad\quad\quad\text{beta1}=\text{beta};}\\
{\quad\quad\quad\quad\text{If}[\text{sa}>\text{sa1},}\\
{\quad\quad\quad\quad\quad\text{sa}=\text{sa1};}\\
{\quad\quad\quad\quad];}\\
{\quad\quad\quad\quad j=0;}\\
{\quad\quad\quad];}\\
{\quad\quad\quad j\text{++};}\\
{\quad\quad];}\\
{\quad\quad\text{betaold}=\text{betalast};}\\
{\quad\quad\text{betalast}=\text{beta};}\\
{\quad\quad\text{(* Write beta to data arrays and data files *)}}\\
{\quad\quad\text{dispnumTM}=\text{Append}[\text{dispnumTM},\{\Delta /\text{$\omega $e},\text{Re}[\text{beta}]\}];}\\
{\quad\quad\text{absnumTM}=\text{Append}[\text{absnumTM},\{\Delta /\text{$\omega $e},\text{Im}[\text{beta}]\}];}\\
{\quad\quad\text{PutAppend}[}\\
{\quad\quad\quad\text{OutputForm}[\text{ToString}[\text{FortranForm}[\Delta /\text{$\omega $e}]]<>\text{$\texttt{"}$   $\texttt{"}$}<>\text{ToString}[\text{FortranForm}[\text{Re}[\text{beta}]]]],}\\
{\quad\quad\quad\text{$\texttt{"}$filepath/filename-prefix1$\texttt{"}$}<>n<>\text{$\texttt{"}$.txt$\texttt{"}$}];}\\
{\quad\quad\text{PutAppend}[}\\
{\quad\quad\quad\text{OutputForm}[\text{ToString}[\text{FortranForm}[\Delta /\text{$\omega $e}]]<>\text{$\texttt{"}$   $\texttt{"}$}<>\text{ToString}[\text{FortranForm}[\text{Im}[\text{beta}]]]],}\\
{\quad\quad\quad\text{$\texttt{"}$filepath/filename-prefix2$\texttt{"}$}<>n<>\text{$\texttt{"}$.txt$\texttt{"}$}];}\\
{\quad\quad\text{If}[i==1\lor \text{IntegerQ}[i/10],\quad\text{\text{(*} Plot beta every ten solutions found \text{*)}}}\\
{\quad\quad\text{plots}=\text{GraphicsArray}[}\\
{\quad\quad\quad\quad\{\text{ListPlot}[\text{dispnumTM},\text{Joined}\to \text{True},\text{PlotRange}\to \text{All},\text{ImageSize}\to 500],}\\
{\quad\quad\quad\quad\text{ListPlot}[\text{absnumTM},\text{Joined}\to \text{True},\text{PlotRange}\to \text{All},\text{ImageSize}\to 500]\}}\\
{\quad\quad\quad];}\\
{\quad\quad];}\\
{\quad\quad i\text{++};\quad\text{(* Track solutions for plotting *)}}\\
{\quad];}\\
{\quad,\text{plots}]\quad\text{(* Show plots *)}}\\
{\text{(* Quantities that are continuously monitored *)}}\\
{\left.\left.,\left\{\Delta \left/10.^6\right.,\text{Abs}[\text{test2}],\text{beta1},j,i,k,\text{sa}\right\}\right]\right];}\)